\documentclass[review]{elsarticle}
\usepackage{lineno,hyperref}
\usepackage{color}
\usepackage{amsmath}
\usepackage{subfigure}
\usepackage{float}
\usepackage{bm}
\usepackage{CJK}
\usepackage{hyperref}
\usepackage{array, threeparttable}
\usepackage[left=2.5cm,right=2.5cm,top=3.5cm,bottom=3.5cm]{geometry}

\modulolinenumbers[5]

\journal{Journal of Computers and Mathematics with Applications}

\begin{document}
	
	
	\begin{frontmatter}
		
		\title{A gas-surface interaction algorithm for discrete velocity methods in predicting rarefied and multi-scale flows: For Maxwell boundary model}
		
		\author[address1]{Jianfeng Chen}
		\ead{chenjf@mail.nwpu.edu.cn}
		
		\author[address1,address2,address3]{Sha Liu\corref{mycorrespondingauthor}}
		\cortext[mycorrespondingauthor]{Corresponding author}
		\ead{shaliu@nwpu.edu.cn}
		
		\author[address1,address4]{Yong Wang}
		\ead{wongyung@mail.nwpu.edu.cn}
		
		\author[address1,address2,address3]{Congshan Zhuo}
		\ead{zhuocs@nwpu.edu.cn}
		
		\author[address5]{Yanguang Yang}
		\ead{yangyanguang@cardc.cn}
		
		\author[address1,address2,address3]{Chengwen Zhong}
		\ead{zhongcw@nwpu.edu.cn}
		
		\address[address1]{School of Aeronautics, Northwestern Polytechnical University, Xi'an, Shaanxi 710072, China}
		\address[address2]{Institute of Extreme Mechanics, Northwestern Polytechnical University, Xi'an, Shaanxi 710072, China}
		\address[address3]{National Key Laboratory of Aircraft Configuration Design, Northwestern Polytechnical University, Xi'an, Shaanxi 710072, China}
		\address[address4]{State Key Laboratory of High Temperature Gas Dynamics, Institute of Mechanics, Chinese Academy of Sciences, Beijing 100190, China}
		\address[address5]{China Aerodynamics Research and Development Center, Mianyang, Sichuan 621000, China}
		
		\begin{abstract}
			The discrete velocity method (DVM) for rarefied flows and the unified methods (based on the DVM framework) for flows in all regimes, from continuum one to free molecular one, have worked well as precise flow solvers over the past decades and have been successfully extended to other important physical fields. Both DVM and unified methods endeavor to model the gas-gas interaction physically. However, for the gas-surface interaction (GSI) at the wall boundary, they have only use the full accommodation boundary up to now, which can be viewed as a rough Maxwell boundary with a fixed accommodation coefficient (AC) at unity, deviating from the real value. For example, the AC for metal materials typically falls in the range of 0.8 to 0.9. To overcome this bottleneck and extend the DVM and unified methods to more physical boundary conditions, an algorithm for Maxwell boundary with an adjustable AC is established into the DVM framework. The Maxwell boundary model splits the distribution of the bounce-back molecules into specular ones and Maxwellian (normal) ones. Since the bounce-back molecules after the spectral reflection does not math with the discrete velocity space (DVS), both macroscopic conservation (from numerical quadrature) and microscopic consistency in the DVS are hard to achieve in the DVM framework. In this work, this problem is addressed by employing a combination of interpolation methods for mismatch points in DVS and an efficient numerical error correction method for micro-macro consistency. On the other hand, the current Maxwell boundary for DVM takes the generality into consideration, accommodating both the recently developed efficient unstructured velocity space and the traditional Cartesian velocity space. Moreover, the proposed algorithm allows for calculations of both monatomic gases and diatomic gases with internal degrees in DVS. Finally, by being integrated with the unified gas-kinetic scheme within the DVM framework, the performance of the present GSI algorithm is validated through a series of benchmark numerical tests across a wide range of Knudsen numbers.
		
		\end{abstract}
		
		\begin{keyword}
			Discrete velocity method; Unified method; Multi-scale flows; Gas-surface interaction; Accommodation coefficient
		\end{keyword}
		
	\end{frontmatter}
	
	
	\section{Introduction}
	A profound comprehension of rarefied and multi-scale hypersonic flows is essential for the aerodynamic design of spacecraft, ultra-low orbital vehicles, and rarefied plumes related to rocket exhausts and spacecraft propulsion systems.
	Typically, the degree of gas rarefaction is quantified by the Knudsen number ($Kn$), defined as $Kn = \lambda / L$, where $\lambda$ is the molecular mean free path and $L$ is the characteristic length of the flow. According to $Kn$, the flow can be divided into continuum ($Kn<0.001$), slip ($0.001<Kn<0.1$), transitional ($0.1<Kn<10$) and free molecular flows ($Kn>10$).
	Multi-scale flows are prevalent in engineering applications, particularly in the study of fluid dynamics for near-space vehicles and micro-/nano-electro-mechanical systems (MEMS/NEMS). In such scenarios, a variety of flow regimes—ranging from continuum flow to slip flow and even free molecular flow—can coexist within the same computational domain.
	As essential aspects of fluid dynamics, gas-gas interactions (GGI) and gas-surface interactions (GSI) are handled by flow solvers and boundary conditions, respectively.
	
	In terms of flow solvers, traditional computational fluid dynamics methods based on the Navier-Stokes (N-S) equations are well-suited for continuum flows in aerospace and at the macro scale. Conversely, the model molecular approach, such as the direct simulation Monte Carlo (DSMC) method \cite{bird1994molecular}, is effective for simulating rarefied gas flows.	
	However, each of these methods faces challenges when simulating multi-scale flows \cite{bird1994molecular,chen2012unified,zhu2016discrete}.
	Motivated by the success of the N-S method in continuum flow and the DSMC method in rarefied flow, researchers have developed a hybrid framework that integrates both methods to simulate multi-scale flows \cite{schwartzentruber2006hybrid,schwartzentruber2008hybrid,burt2009hybrid}.
	In this hybrid framework, the computational domain is divided into continuum and rarefied regions. Each sub-domain is then simulated using its corresponding solver. Typically, empirical or semi-empirical criteria are employed to partition the computational domain \cite{sun2004hybrid, schwartzentruber2007modular}. However, accurately defining the domain boundaries and validating the methods at the interface between these regions remains a significant challenge \cite{chen2012unified}.	
	In recent years, a class of multi-scale asymptotic preserving \cite{jin1999efficient} schemes based on kinetic theory have been proposed, making it possible to solve multi-scale flow problems using a unified numerical method. With a local analytical integral solution to the kinetic model equation for the connected particle transport and collision in the numerical flow, the unified gas-kinetic scheme (UGKS) developed by Xu et al. \cite{xu2010unified,xu2014direct} is a multi-scale technique.
	Therefore, the cell size and time step are not constrained by the molecular mean free path and collision time. 
	The discrete unified gas-kinetic scheme (DUGKS) proposed by Guo et al. is another multi-scale approach based on the same physical process as UGKS \cite{guo2013discrete,guo2015discrete}. Instead of relying on an analytical integral solution, the DUGKS employs a characteristic difference solution to solve the kinetic model equation in both space and time. This method couples molecular transport and collision effects within a numerical time step, enabling the reconstruction of multi-scale numerical fluxes at cell interfaces.
	After a decade of development, many numerical techniques have been developed and implemented in the UGKS and DUGKS to increase the computational efficiency and reduce memory cost, such as unstructured mesh computation \cite{zhu2016discrete,sun2017multidimensional}, moving grids \cite{wang2019arbitrary,chen2012unified}, velocity space adaptation \cite{chen2012unified}, memory reduction \cite{chen2017unified}, wave-particle adaptation \cite{zhu2019unified}, implicit schemes \cite{zhu2016implicit,yuan2020conservative,rui_zhang_conservative_2024}, parallelization algorithm \cite{zhang2022unified_parallelization}, and further simplification and modification \cite{chen2016simplification,liu2012modified,zhong2020simplified,zhong2021simplified}. With these improvement, the UGKS and DUGKS have been successfully applied to a variety of flow problems in different flow regimes, such as micro flows \cite{zhu2017numerical}, compressible flows \cite{guo2015discrete,zhang2023unified}, jet flow\cite{chen2020compressible,guang_zhao_application_2022,guang_zhao_numerical_2023}, multi-phase flows \cite{yang2019phase}, gas-solid flows \cite{tao2018combined}, and gas mixture systems \cite{zhang2018discrete_3}. Besides multi-scale flow problems, the UGKS and DUGKS were also extended to multi-scale transport problems such as radiative transfer \cite{sun2015asymptotic_gray}, phonon heat transfer \cite{guo2016discrete}, plasma physics \cite{pan2018unified}, neutron transport \cite{shuang2019parallel}, granular flow \cite{liu2019unified}, and turbulent flow \cite{wang2016comparison,zhang2020large}.	
	On the other hand, because of the wide spreading of the particle velocity distribution in high-speed flows and the narrow-kernel distribution function in cases with a large $Kn$, the velocity mesh must cover a huge domain with high resolution. The storage and computational demand will be unbearable if a conventional structured velocity mesh is used \cite{titarev2017numerical}. To reduce the computational cost, an adaptive velocity mesh in phase space was proposed in the literature \cite{chen2012unified,baranger2012locally,arslanbekov2013kinetic}. However, compared to conventional velocity mesh approaches, adaptive velocity mesh approaches are more challenging and complex. Titarev et al. \cite{titarev2017numerical} and Yuan et al.\cite{yuan2020conservative} provided a relatively simple method for producing a non-uniform unstructured DVS, and Chen et al. \cite{chen2019conserved,chen2020compressible} adopted this method to conserved DUGKS. The benefit of unstructured DVS is that it enables flexible grid point placement, which decreases the number of grid points and prevents ray effects.
	
	In terms of the GSI model, which serves as a wall boundary, it has a significant influence on aerodynamic forces and heat transfer, with its impact generally intensifying as gas rarefaction increases. Therefore, choosing a suitable GSI model is essential for accurately predicting hypersonic rarefied flows \cite{santos2004dsmc}. The first and fundamental explanation of the GSI model, encompassing the incidence and reflection processes, was proposed by Maxwell \cite{bird1994molecular,kennard1938kinetic}. He introduced two classical reflection models: the specular reflection model and the diffuse reflection model. These models represent scenarios where a gas molecule collides with either a perfectly reflecting or a fully accommodating solid surface \cite{jin2020effects}.
	In the diffuse reflection model, gas molecules that are adsorbed near the wall are uniformly re-emitted into the half-space following a Maxwellian distribution upon desorption.
	In the specular reflection model, when a gas molecule collides directly with a wall without being absorbed, it rebounds much like a perfectly elastic ball. The velocity component normal to the wall can be reversed, but the component parallel to the wall remains unchanged. The reflection angle mirrors the incidence angle.
	The diffuse reflection model is commonly applied in natural scenarios. The duration for gas molecules to be adsorbed by the surface, especially in continuum flows, is typically much longer than in rarefied gas flows or flows over smooth surfaces.
	It's important to note that either diffuse or specular reflection models should not be exclusively employed in situations where the adsorption time is neither very long nor very short.
	Maxwell combined the diffuse and specular reflection models by introducing the accommodation coefficient (AC) $\sigma$ \cite{maxwell1879vii}. In this model, a fraction ($\sigma$) of incident molecules reflects diffusely, while the rest reflect specularly \cite{santos2004dsmc,jin2020effects}.
	However, the Maxwell model, with its single parameter, struggles to accurately describe both momentum and energy transport simultaneously. Determining the AC $\sigma$ is also challenging, as the fraction of molecules undergoing diffuse reflection depends on various physical variables of the gas and surface \cite{cao2009molecular}. Cercignani and Lampis \cite{cercignani1971kinetic} developed a phenomenological model (CL model) to characterize the GSI in accordance with the reciprocal law, where two parameters pertaining to the tangential momentum and normal energy transport were incorporated. Later, Lord \cite{lord1991some,lord1995some} expanded and updated the CL model (known as the CLL model in the DSMC), making it a popular tool for theoretical and computational studies of rarefied gas flows. The CLL model enhances the consistency of the reflected gas molecule velocity distributions, closely resembling the lobular distribution. However, it does not align with experimental findings using a molecular beam. Some models describe GSI by fitting empirical parameters to experimental results, among which the Nocilla model \cite{nocilla1962surface, hurlbut1968application} and the multi-flux model \cite{wadsworth2003gas} are the most typical.
		
	Due to its straightforward implementation, the GSI model in DSMC is widely applied \cite{hedahl1995comparisons,jin2020effects,wadsworth2003gas,yamanishi1999multistage}. However, because of its relative complexity, the GSI model in the discrete velocity method (DVM) is less frequently used \cite{zhangyd2018discrete}.
	Since DVM employs a discrete velocity space (DVS) that remains fixed in time and space, the specular reflection model can only be applied when the DVS is symmetric about the wall boundary. Consequently, in prior research, specular reflection has been exclusively implemented on straight boundaries \cite{zhangyd2018discrete, lim2002application}. Moreover, establishing a DVS that maintains symmetry across all boundaries is challenging, particularly when the boundaries are not straight.
	Therefore, developing a GSI algorithm compatible with both the general DVM framework and various boundary geometries is essential.
	In this paper, we present, for the first time, an algorithm for the Maxwell GSI model specifically designed for the DVM framework. This advancement allows a range of DVM methods to adopt a more precise wall boundary condition, where the AC can be adjusted to match the properties of different solid materials.
	The algorithm ensures macroscopic conservation, such as preventing flow penetration through solid walls, and achieves microscopic consistency through interpolations of the discrete distribution function and the recently developed error correction method for DVS.
	Designed for arbitrary solid boundary geometries and unstructured DVS, this algorithm addresses the limitations of earlier research on the Maxwell boundary in DVM, which focused solely on straight boundary geometries and did not involve the pivotal mismatch of DVS points.
	Furthermore, this new boundary algorithm can be applied to both monatomic gases and diatomic gases with thermal non-equilibrium, characterized by different relaxation rates of translational and rotational energy.
	Finally, the algorithm has been implemented in the multi-scale UGKS method and rigorously tested in a series of typical multi-scale benchmark cases.	
		
	The remainder of the paper is organized as follows: Sec.~\ref{sec:numerical method} introduces the UGKS based on the Boltzmann-Shakhov and Boltzmann-Rykov model equations. The GSI algorithm based on unstructured DVS is detailed in Sec.~\ref{sec:GSI boundary}. In Sec.~\ref{sec:test cases}, we present the numerical simulation and comparison of supersonic flows over various geometries, including the sharp flat plate, circular cylinder, blunt wedge, blunt circular cylinder, and truncated flat plate. Finally, Sec.~\ref{sec:conclusion} provides the concluding remarks of this paper.
	
	\section{Numerical method}\label{sec:numerical method}
	The suggested GSI algorithm is applied within numerical methods. To offer a concise overview, let's introduce the basic kinetic model, the treatment of distribution functions, and the core algorithm utilized in these numerical methods.
	
	\subsection{Gas-kinetic models}
	The UGKS adopts the gas-kinetic relaxation model equation in the following form:
	\begin{equation}
		\frac{\partial f}{\partial t}+\bm{\xi }\cdot \nabla f=\Omega,
	\end{equation}
	where $f=f\left( \bm{x},\bm{\xi },\bm{\eta },e,t \right)$ is the velocity distribution function for particles moving in D-dimensional physical space with a velocity of $\bm{\xi }=\left( \xi _{1}^{{}},...,\xi _{D}^{{}} \right)$ at position $\bm{x}=\left( x_{1}^{{}},...,x_{D}^{{}} \right)$ and time $t$. Here, $\bm{\eta }=\left( \xi _{D+1}^{{}},...,\xi _{3}^{{}} \right)$ is the dummy velocity (with the degree of freedom $L=3-D$) consisting of the remaining components of the translational velocity of particles in three-dimensional space; $e$ is a vector of $K$ elements representing the internal degree of freedom of molecules; $\Omega$ is the collision operator. Many kinetic models, including the Bhatnagar-Gross-Krook (BGK) collision model \cite{bhatnagar1954model}, the Shakhov model \cite{shakhov1968generalization}, the ellipsoidal statistical model (ES model) \cite{holway1966new}, and the Rykov model \cite{rykov1975model}, have been proposed and used in the research of rarefied flows to simplify the complex collision term of the full Boltzmann equation. In this paper, the Shakhov and Rykov models are employed to describe the flow of monoatomic and diatomic gases, respectively. The control equation of the Boltzmann-BGK type model is:
	\begin{equation}
		\frac{\partial f}{\partial t}+\bm{\xi }\cdot \nabla f=\frac{{{g}^{*}}-f}{\tau },
	\end{equation}
	where ${g}^{*}$ is the equilibrium distribution function and $\tau$ is the relaxation time for the translational degree of freedom determined by the dynamic viscosity $\mu$ and translational pressure
	$p_t$ with $\tau=\mu/p_t$, and $p_t$ is determined by the translational temperature $T_t$. 
	In classical statistical mechanics, the equipartition theorem is a general formula that relates the temperature of a system with its average energies:
	\begin{equation}\label{equipartition_theorem}
		T=\frac{3T_t+KT_r}{3 + K},
	\end{equation}
	where $T_r$ is the rotation temperature and $K$ is 0 or 2 for monoatomic or diatomic molecules. The variable hard sphere (VHS) model is adopted in this study to determine viscosity:
	\begin{equation}
		\mu ={{\mu }_{0}}{{\left( \frac{{{T}_{t}}}{{{T}_{0}}} \right)}^{\omega }},
	\end{equation}
	where $\omega$ is the viscosity index and ${\mu }_{0}$ is the reference viscosity at the reference temperature $T_0$ . According to the inter-molecular interaction model, $\omega$ is set to 0.5, 0.81, and 0.74 for the hard sphere model, ideal argon, and nitrogen, respectively. The viscosity of the freestream flow $\mu$ is correlated with the gas mean free path $\lambda $ in the following way:
	\begin{equation}
		\lambda =\frac{2\mu \left( 5-2\omega  \right)\left( 7-2\omega  \right)}{15\rho {{\left( 2\pi RT \right)}^{1/2}}},
	\end{equation}
	where $\rho$ and $R$ are the density and the specific gas constant, respectively. It can be inferred from the definition of $Kn$ that:
	\begin{equation}
		Kn=\frac{\lambda }{{{L}}}\text{=}\frac{2\mu \left( 5-2\omega  \right)\left( 7-2\omega  \right)}{15\rho {{L}}{{\left( 2\pi RT \right)}^{1/2}}}=\sqrt{\frac{2\gamma }{\pi }}\frac{\left( 5-2\omega  \right)\left( 7-2\omega  \right)}{15}\frac{Ma}{\operatorname{Re}},
	\end{equation}
	where ${L}$ is the characteristic length, and $Ma$ and $Re$ are the Mach number and Reynolds number, respectively. 
	
	\subsection{Reduced distribution function}
	The transport process of the distribution function depends only on the D-dimensional particle velocity $\bm{\xi }$ and is irrelevant to $\bm{\eta }$ and $e$ (for diatomic gases). The following reduced distribution functions \cite{yang1995rarefied} are introduced to save  computational memory and cost: 
	\begin{equation}
		\begin{aligned}
			& G\left( t,\bm{x},\bm{\xi } \right)=m\int{fded\bm{\eta }}, \\ 
			& H\left( t,\bm{x},\bm{\xi } \right)=m\int{{{\eta }^{2}}fded\bm{\eta }}, \\ 
			& R\left( t,\bm{x},\bm{\xi } \right)=\int{efded\bm{\eta }}. \\ 
		\end{aligned}	
	\end{equation}
	Note that $R$ is only for diatomic gases and $H$ will inevitably vanish in three dimensions. The macroscopic variables can be solved
	by the moments of the reduced distribution functions as follows:
	\begin{equation}
		\bm{W}=\left( \begin{matrix}
			\rho   \\
			\rho \bm{u}  \\
			\rho E  \\
			\rho {{E}_{r}}  \\
		\end{matrix} \right)=\int{\left( \begin{matrix}
				G  \\
				\bm{\xi }G  \\
				\frac{1}{2}\left( {{\xi }^{2}}G+H \right)\text{+}R  \\
				R  \\
			\end{matrix} \right)}d\bm{\xi },
	\end{equation}
	where $\rho E$ and $\rho {{E}_{r}}$ are the total energy density and the rotational energy density. The translational, rotational and total heat fluxes are expressed as:
	\begin{equation}
		\begin{aligned}
			& {{\bm{q}}_{t}}=\frac{1}{2}\int{\bm{c}\left( {{c}^{2}}G+H \right)}d\bm{\xi }, \\ 
			& {{\bm{q}}_{r}}=\int{\bm{c}R}d\bm{\xi }, \\ 
			& \bm{q}={{\bm{q}}_{t}}+{{\bm{q}}_{r}}, \\ 
		\end{aligned}	
	\end{equation}
	where $\bm{c}=\bm{\xi }-\bm{u}$ is the peculiar velocity. 
	
	The governing equation of the reduced distribution function is:
	\begin{equation}\label{eq_reduced}
		\begin{aligned}
			& \frac{\partial G}{\partial t}+\bm{\xi }\cdot \frac{\partial G}{\partial \bm{x}}=\frac{{{g}^{G}}-G}{\tau }, \\ 
			& \frac{\partial H}{\partial t}+\bm{\xi }\cdot \frac{\partial H}{\partial \bm{x}}=\frac{{{g}^{H}}-H}{\tau }, \\ 
			& \frac{\partial R}{\partial t}+\bm{\xi }\cdot \frac{\partial R}{\partial \bm{x}}=\frac{{{g}^{R}}-R}{\tau }. \\ 
		\end{aligned}
	\end{equation}
	For monatomic gases flow, the Shakhov equilibrium is:
	\begin{equation}
		\begin{aligned}
			& {{g}^{G}}=g_{{}}^{eq}+G_{\Pr }^{{}}, \\ 
			& {{g}^{H}}=H_{{}}^{eq}+H_{\Pr }^{{}}, \\ 
		\end{aligned}
	\end{equation}
	with
	\begin{equation}
		G_{\Pr }^{{}}=\left( 1-\Pr  \right)\frac{\bm{c}\cdot \bm{q}}{5pRT}\left( \frac{c_{{}}^{2}}{RT}-D-2 \right)g_{{}}^{eq},
	\end{equation}
	\begin{equation}
		H_{{}}^{eq}=\left( K+3-D \right)RTg_{{}}^{eq},
	\end{equation}
	\begin{equation}
		H_{\Pr }^{{}}=\left( 1-\Pr  \right)\frac{\bm{c}\cdot \bm{q}}{5pRT}\left[ \left( \frac{c_{{}}^{2}}{RT}-D \right)\left( K+3-D \right)-2K \right]RTg_{{}}^{eq},
	\end{equation}
	where the Prandtl (Pr) number equals 2/3. ${g}^{eq}$ represents the Maxwell equilibrium:
	\begin{equation}
		{{g}^{eq}\left( {{T}} \right)}=\frac{\rho }{{{\left( 2\pi RT \right)}^{D/2}}}\exp \left( -\frac{{\bm{c}^{2}}}{2RT} \right).
	\end{equation}
	For diatomic gases flow, the Rykov equilibrium is:
	\begin{equation}
		\begin{aligned}
			& {{g}^{G}}=\left( 1-\frac{1}{Zr} \right){{G}^{t}}+\frac{1}{Zr}{{G}^{r}}, \\ 
			& {{g}^{H}}=\left( 1-\frac{1}{Zr} \right){{H}^{t}}+\frac{1}{Zr}{{H}^{r}}, \\ 
			& {{g}^{R}}=\left( 1-\frac{1}{Zr} \right){{R}^{t}}+\frac{1}{Zr}{{R}^{r}}, \\ 
		\end{aligned}
	\end{equation}
	with
	\begin{equation}
		\begin{aligned}
			& {{G}^{t}}={{g}^{eq}}\left( {{T}_{t}} \right)\left[ 1+\frac{\bm{c}\cdot {{\bm{q}}_{t}}}{15{{p}_{t}}R{{T}_{t}}}\left( \frac{{{c}^{2}}}{R{{T}_{t}}}-D-2 \right) \right], \\ 
			& {{G}^{r}}={{g}^{eq}}\left( T \right)\left[ 1+{{\omega }_{0}}\frac{\bm{c}\cdot {{\bm{q}}_{t}}}{15pRT}\left( \frac{{{c}^{2}}}{RT}-D-2 \right) \right], \\ 
		\end{aligned}	
	\end{equation}
	\begin{equation}
		\begin{aligned}
			& {{H}^{t}}=R{{T}_{t}}{{g}^{eq}}\left( {{T}_{t}} \right) \left( 3-D \right)\left[\left( 1+\frac{\bm{c}\cdot {{\bm{q}}_{t}}}{15{{p}_{t}}R{{T}_{t}}}\left( \frac{{{c}^{2}}}{R{{T}_{t}}}-D \right) \right) \right], \\ 
			& {{H}^{r}}=RT{{g}^{eq}}\left( T \right) \left( 3-D \right)\left[\left( 1+{{\omega }_{0}}\frac{\bm{c}\cdot {{\bm{q}}_{t}}}{15pRT}\left( \frac{{{c}^{2}}}{RT}-D \right) \right) \right], \\ 
		\end{aligned}
	\end{equation}
	\begin{equation}
		\begin{aligned}
			& {{R}^{t}}=R{{T}_{r}}\left[ {{G}^{t}}+\left( 1-\delta  \right)\frac{\bm{c}\cdot {{\bm{q}}_{r}}}{{{p}_{t}}R{{T}_{t}}}{{g}^{eq}}\left( {{T}_{t}} \right) \right], \\ 
			& {{R}^{r}}=RT\left[ {{G}^{r}}+{{\omega }_{1}}\left( 1-\delta  \right)\frac{\bm{c}\cdot {{\bm{q}}_{r}}}{pRT}{{g}^{eq}}\left( T \right) \right], \\ 
		\end{aligned}	
	\end{equation}
	where the coefficients are set as \cite{xu2014direct,yuan2020conservative}: $\delta =1/1.55$, ${{\omega }_{0}}=0.2354$, and ${{\omega }_{1}}=0.3049$ for nitrogen in this study. $Zr$ is the rotational relaxation collision number accounting for the ratio of the slower inelastic translation-rotation energy relaxation relative to the elastic translational relaxation. All the above equilibrium distribution functions can be obtained by macroscopic variables.
	
	\subsection{UGKS with simplified multi-scale numerical flux}
	Generally, four independent characteristic variables are introduced in the non-dimensional reference system, namely, reference length $L_{ref}=L_{c}$, reference temperature $T_{ref}=T_{\infty}$, reference density $\rho_{ref}=\rho_{\infty}$ and reference speed $U_{ref}=\sqrt{2RT_{ref}}$, where $L_{c}$ is the characteristic length scale of the flow, $T_{\infty}$ and $\rho_{\infty}$ are temperature and density of the freestream, respectively.
	Thus, the following basic non-dimensional quantities can be obtained:
	\begin{equation}\label{non-dimensional}
		\hat{L}=\frac{L}{L_{ref}},   \hat{T}=\frac{T}{T_{ref}}, \hat{\rho}=\frac{\rho}{\rho_{ref}}, \hat{U}=\frac{U}{U_{ref}}.
	\end{equation}
	One can obtain a complete non-dimensional system by employing these basic quantities. Unless declared otherwise, all variables in the following that lack a ``hat" are non-dimensional quantities for simplicity's sake.	
	
\begin{figure}
		\centering
		\includegraphics[width=0.7\textwidth]{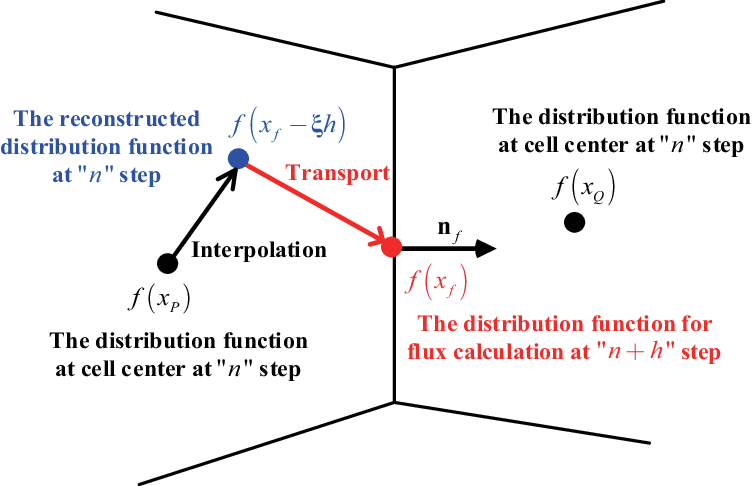}
		\caption{\label{fig01}The sketch of constructing the distribution function at cell interface (which is travel from the reconstructed distribution function along particle characteristic-line).}
	\end{figure}

	In this study, the macroscopic variables and distribution functions are updated sequentially by the UGKS with simplified multi-scale numerical flux \cite{zhang2023unified}. For the sake of simplicity, $G$, $H$, and $R$ can be replaced with the new symbol $\phi$ in the algorithm because they share the same updating process, as shown in Eq. (\ref{eq_reduced}). Then, the governing equation can be rewritten as:
	\begin{equation}\label{eq_phi}
		\frac{\partial \phi }{\partial t}+\bm{\xi }\cdot \nabla \phi =\Omega _{\phi }^{{}}\equiv \frac{{{g}^{\phi }}-\phi }{\tau }.
	\end{equation}
	Integrating Eq. (\ref{eq_phi}) over control volume $V_{j}$ from time $t_{n}$ to $t_{n+1}$ and then the discrete governing equation can be written as:
	\begin{equation}\label{eq_dphi}
		\phi _{j}^{n+1}\left( \bm{\xi } \right)-\phi _{j}^{n}\left( \bm{\xi } \right)+\frac{\Delta t}{\left| V_{j}^{{}} \right|}F_{j}^{n+1/2}\left( \bm{\xi } \right)=\frac{\Delta t}{2}\left[ \Omega _{j}^{n+1}\left( \bm{\xi } \right)+\Omega _{j}^{n}\left( \bm{\xi } \right) \right],
	\end{equation}
	where $\left| V_{j} \right|$ and ${\Delta t=t_{n+1}-t_{n}}$ denote the volume of $V_{j}$ and the time interval. $F_{j}^{n+1/2}$ is the microscopic flux across the cell interface:
	\begin{equation}
		F_{j}^{n+1/2}\left( \bm{\xi } \right)=\int_{\partial V_{j}^{{}}}{\left( \bm{\xi }\cdot \bm{n}_f \right)}\phi \left( {\bm{{x}}_{f}},\bm{\xi },t_{n+1/2}^{{}} \right)dS,
	\end{equation}
	where $\partial V_{j}^{{}}$ is the surface of the control volume $V_{j}$,  $\bm{x}_{f}$ and $\bm{n}_f$ are the interface's midpoint and the external normal unit vector of $dS$ (an infinitesimal element of $\partial V_{j}^{{}}$), respectively. 
	Take the moment of Eq. (\ref{eq_dphi}) we can derive the macroscopic governing equation for a control volume $V_{j}$\cite{liu2018conserved,chen2019conserved}:
	\begin{equation}\label{eq_W}
		\bm{W}_{j}^{n+1}=\bm{W}_{j}^{n}-\frac{\Delta t}{\left| {{V}_{j}} \right|}\bm{F}_{j}^{n+1/2}+S,
	\end{equation}
	where $\bm{F}_{j}^{n+1/2}$ is the microscopic flux across the cell interface:
	\begin{equation}\label{macroflux}
		\bm{F}_{j}^{n+1/2}=\int{\int_{\partial {{V}_{j}}}{\left( \bm{\xi }\cdot \bm{n}_f \right)}\left( \begin{matrix}
				G\left( {\bm{x}_{f}},\bm{\xi },t_{n+1/2}^{{}} \right)  \\
				\bm{\xi }G\left( {\bm{x}_{f}},\bm{\xi },t_{n+1/2}^{{}} \right)  \\
				\frac{1}{2}\left[ \xi _{{}}^{2}G\left( {\bm{x}_{f}},\bm{\xi },t_{n+1/2}^{{}} \right)+H\left( {\bm{x}_{f}},\bm{\xi },t_{n+1/2}^{{}} \right) \right]  \\
				R\left( {\bm{x}_{f}},\bm{\xi },t_{n+1/2}^{{}} \right)  \\
			\end{matrix} \right)dSd\bm{\xi }}.
	\end{equation}
	$S={{\left( 0,0,0,{{S}_{r}} \right)}^{T}}$ is the source term and only the rotation term is not zero:
	\begin{equation}
		{{S}_{r}}=\frac{\Delta t}{2}\int{\left[ \Omega _{r}^{n}+\Omega _{r}^{n+1} \right]}d\bm{\xi }=\frac{\Delta t}{2Z_r{{\tau }^{n}}}\left( {{\left[ \rho RT \right]}^{n}}-{{\left[ \rho R{{T}_{r}} \right]}^{n}} \right)+\frac{\Delta t}{2Z_r{{\tau }^{n+1}}}\left( {{\left[ \rho RT \right]}^{n+1}}-{{\left[ \rho R{{T}_{r}} \right]}^{n+1}} \right).
	\end{equation}
	Afterward, the implicit equation Eq. (\ref{eq_dphi}) can be transformed into the following explicit equation:
	\begin{equation}\label{eq_update_phi}
		\phi _{j}^{n+1}\left( \bm{\xi } \right)=\left( 1+\frac{\Delta t}{2\tau _{j}^{n+1}} \right)_{{}}^{-1}\left[ \phi _{j}^{n}\left( \bm{\xi } \right)-\frac{\Delta t}{\left| V_{j}^{{}} \right|}F_{j}^{n+1/2}\left( \bm{\xi } \right)+\frac{\Delta t}{2}\left( \frac{g_{j}^{\phi ,n+1}\left( \bm{\xi } \right)}{\tau _{j}^{n+1}}+\frac{g_{j}^{\phi ,n}\left( \bm{\xi } \right)-\phi _{j}^{n}\left( \bm{\xi } \right)}{\tau _{j}^{n}} \right) \right].
	\end{equation}
	The evolution equations for the microscopic distribution functions and the macroscopic conservative variables are shown in Eqs. (\ref{eq_W}) and (\ref{eq_update_phi}), respectively. The entire scheme may be established after the distribution functions $\phi \left( {\bm{x}_{f}},\bm{\xi },t_{n+1/2}^{{}} \right)$ at the cell interface are obtained.
	In this study, the simplified multi-scale flux is built by using the characteristic difference solution of the Boltzmann model equation \cite{zhang2023unified}. The Eq. (\ref{eq_phi}) is integrated along the characteristic line (in the direction of particle velocity) from $t_{n}$ to $t_{n+1/2}$ and the characteristic line ends at the midpoint of the cell interface, as shown in Fig. \ref{fig01}. This procedure is described by the following equation:
	\begin{equation}
		{{\phi }}\left( {{\bm{x}}_{f}},\bm{\xi },{{t}_{n}}+h \right)-{{\phi }}\left( {{\bm{x}}_{f}}-\bm{\xi }h,\bm{\xi },{{t}_{n}} \right)=h\frac{{{g}^{\phi }}\left( {{\bm{x}}_{f}},\bm{\xi },{{t}_{n}}+h \right)-{{\phi }}\left( {{\bm{x}}_{f}},\bm{\xi },{{t}_{n}}+h \right)}{{{\tau }^{n+1/2}}},
	\end{equation}
	where $h=\frac{\Delta t}{2}$ denotes a half-time step. Therefore, the interface distribution function $\phi \left( {\bm{x}_{f}},\bm{\xi },t_{n+1/2}^{{}} \right)$ can be expressed as:
	\begin{equation}\label{eq_hf}
		{{\phi }}\left( {{\bm{x}}_{f}},\bm{\xi },{{t}_{n}}+h \right)=\frac{{{\tau }^{n+1/2}}}{{{\tau }^{n+1/2}}+h}{{\phi }}\left( {{\bm{x}}_{f}}-\bm{\xi }h,\bm{\xi },{{t}_{n}} \right)+\frac{h}{{{\tau }^{n+1/2}}+h}{{g}^{\phi }}\left( {{\bm{x}}_{f}},\bm{\xi },{{t}_{n}}+h \right),
	\end{equation}
	where ${{\phi }}\left( {{\bm{x}}_{f}},\bm{\xi },{{t}_{n}}+h \right)$ can be integrated into the velocity space to produce the macroscopic quantity at the half-time step, on which ${{g}^{\phi }}\left( {{\bm{x}}_{f}},\bm{\xi },{{t}_{n}}+h \right)$ depends. Meanwhile, ${{\phi }}\left( {{\bm{x}}_{f}}-\bm{\xi }h,\bm{\xi },{{t}_{n}} \right)$ can be calculated from the following reconstruction through the Taylor expansion at the control volume:
	\begin{equation}
		\phi \left( \bm{x}_{f}^{{}}-\bm{\xi }h,\bm{\xi },t_{n}^{{}} \right)={{\phi }_{C}}\left( \bm{x}_{C}^{{}},\bm{\xi },t_{n}^{{}} \right)+\Psi \left( \bm{x}_{C}^{{}},\bm{\xi },t_{n}^{{}} \right)\nabla {{\phi }_{C}}\left( \bm{x}_{C}^{{}},\bm{\xi },t_{n}^{{}} \right)\cdot \left( \bm{x}_{f}^{{}}-\bm{\xi }h-\bm{x}_{C}^{{}} \right), \bm{x}_{f}^{{}}-\bm{\xi }h\in V_{C}^{{}},
	\end{equation}
	where ${{V}_{c}}$ represents the control volume which is centered at point C (Fig. \ref{fig01}). If $\bm{\xi }\cdot \bm{n}_{f}^{{}}\ge 0$, point C is P (the center of the left cell) in Fig. \ref{fig01}, otherwise, point C is Q (the center of the right cell). $\nabla{\mathop{\phi _{C}}}\,\left( \bm{x}_{C}^{{}},\bm{\xi },t_{n}^{{}} \right)$ is the gradient of distribution functions at point C, which is calculated by the least-square method in this study, and $\Psi \left( \bm{x}_{C}^{{}},\bm{\xi },t_{n}^{{}} \right)$ is the gradient limiter used to suppress the numerical oscillations. The Venkatakrishnan limiter \cite{1995Convergence} is chosen in this paper.	
	
	\section{Maxwell boundary condition for DVM framework}\label{sec:GSI boundary}
	For flows with solid walls, appropriate boundary conditions, i.e. GSI models, should be specified for the discrete distribution functions at the wall surface \cite{guo2021progress}.	The most widely used model is the Maxwell model, which is based on classical thermodynamics and assumes that molecules will either reflect diffusely with complete energy accommodation or reflect specularly with no change in energy \cite{hedahl1995comparisons}. 
	The fraction of molecules that will be scattered diffusely is specified by the AC $\sigma$, and the reflected distribution function from the wall to the flow field is described as:
	\begin{equation}
		\phi \left( {{\bm{x}}_{w}},{{\bm{\xi }}},{{t}^{n}}+h \right)=\sigma {{\phi }_{d}}\left( {{\bm{\xi }}};{{\rho }_{w}},{{\bm{u}}_{w}},{{T}_{w}} \right)+\left( 1-\sigma  \right){\phi}_{s} \left( {{\bm{x}}_{w}},{{\bm{\xi}}},{{t}^{n}}+h \right),\ {{\bm{\xi}}}\cdot {{\bm{n}}_{w}}<0,
	\end{equation}
	where ${\bm{x}}_{w}$, ${\bm{\rho}}_{w}$, ${\bm{u}}_{w}$, ${{T}}_{w}$, and ${\bm{n}}_{w}$ are respectively the center position, density, velocity, temperature, and unit normal vector of the wall, and the direction of ${\bm{n}}_{w}$ is pointing to wall.
	${{\phi }_{d}}\left( {{\bm{\xi }}};{{\rho }_{w}},{{\bm{u}}_{w}},{{T}_{w}} \right)$ and ${\phi}_{s} \left( {{\bm{x}}_{w}},{{\bm{\xi }}},{{t}^{n}}+h \right)$ are the distribution functions for diffuse and specular reflections, respectively.
	
	\subsection{Diffuse-scattering boundary condition}
	The diffuse reflection model, commonly referred to as the diffuse-scattering rule, serves as a general boundary condition, assuming that the distribution function for reflected molecules adheres to the Maxwellian distribution.
	The distribution function for the $i$-th reflected molecules at the half-time step ${t}^{n}+h$ is expressed as:
	\begin{equation}
		{\phi}_{d} \left( {{\bm{x}}_{w}},{{\bm{\xi }}_{i}},{{t}^{n}}+h \right)={{\phi }^{eq}}\left( {{\bm{\xi }}_{i}};{{\rho }_{w}},{{\bm{u}}_{w}},{{T}_{w}} \right),\ {{\bm{\xi}}_{i}}\cdot {{\bm{n}}_{w}}<0.
	\end{equation}
	The wall density $\rho_{w}$ is determined based on the no-penetration condition, ensuring that no molecules can pass through the wall:
	\begin{equation}
		\sum\limits_{\bm{\xi}_{i}\cdot \bm{n}_{w}<0}{{{w}_{i}}\left( {{\bm{\xi }}_{i}}\cdot \bm{n} \right)}{{\phi }^{eq}}\left( {{\bm{\xi }}_{i}};{{\rho }_{w}},{{\bm{u}}_{w}},{{T}_{w}} \right)+\sum\limits_{\bm{\xi}_{i}\cdot \bm{n}_{w}>0}{{{w}_{i}}\left( {{\bm{\xi }}_{i}}\cdot \bm{n} \right)}{\phi}_{d} \left( {{\bm{x}}_{w}},{{\bm{\xi }}_{i}},{{t}^{n}}+h \right)=0.
	\end{equation}
	By solving Eq. (\ref{eq_hf}) for the incidence distribution function, the wall density can be determined:
	\begin{equation}
		{{\rho }_{w}}=-\sum\limits_{\bm{\xi}_{i}\cdot \bm{n}_{w}>0}{{{w}_{i}}\left( \bm{\xi }\cdot \bm{n} \right)}{\phi}_{d} \left( {{\bm{x}}_{w}},{{\bm{\xi }}_{i}},{{t}^{n}}+h \right)/\sum\limits_{\bm{\xi}_{i}\cdot \bm{n}_{w}<0}{{{w}_{i}}\left( \bm{\xi }\cdot \bm{n} \right)}{{\phi }^{eq}}\left( \bm{\xi };1,{{\bm{u}}_{w}},{{T}_{w}} \right).
	\end{equation}

	\subsection{Specular reflection boundary condition}
\begin{figure}
		\centering
		\includegraphics[width=0.75\textwidth]{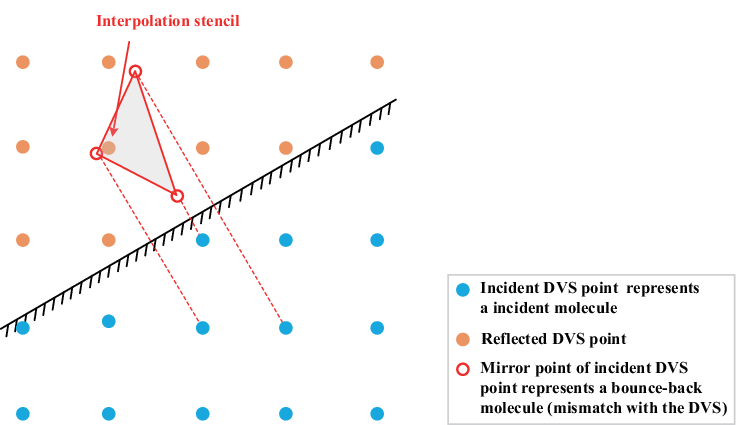}
		\caption{\label{fig02a}{Interpolation of the reflected distribution function (Solid point: mesh point; Hollow point: mirror point).}}
	\end{figure}

	According to the theory of specular reflection, a gas molecule reflects like an ideal sphere, preserving a constant tangential velocity component and reversing its normal velocity component when the reflection angle equals the incidence angle.
	When the discrete velocity mesh is asymmetric about the wall, it becomes challenging for the reflected gas molecules to land on a specific point within the DVM mesh.
	Therefore, interpolation methods are necessary to calculate the reflected distribution function.
	As depicted in Fig. \ref{fig02a}, solid points represent the DVS mesh points (distribution functions) used in the DVM. Without loss of generality, the blue solid points in the incident zone and the red hollow points in the reflection zone correspond to the incident discrete velocity points and their mirror points in the case of a oblique wall. The incident point $\bm{\xi }_{i}$ and the wall normal vector $\bm{n}_{w}$ can be employed to determine the position of the mirror point $\bm{\xi }_{i}^{'}$ in the velocity space:
	\begin{equation}
		\bm{\xi }_{i}^{'}={{\bm{\xi }}_{i}}-2{{\bm{n}}_{w}}\cdot \left( {{\xi}_{i}}\cdot {{\bm{n}}_{w}} \right).
	\end{equation}
	Equation (\ref{eq_hf}) can be used to calculate the distribution function of the incident points. Additionally, the distribution functions of the mirror points (${\phi}_{s} \left( {{\bm{x}}_{w}},{{\bm{\xi '}_{i}}},{{t}^{n}}+h \right)$) are the same as those of the corresponding incident points (${\phi}_{s} \left( {{\bm{x}}_{w}},{{\bm{\xi}_{i}}},{{t}^{n}}+h \right)$). The reflected distribution function of the orange solid points is then determined through interpolation. Representing the distribution functions of the unknown orange solid points and the $i$-th known surrounding red hollow points as ${\phi_o}$ and ${\phi_{ri}}$, respectively, yields the following equation:
	\begin{equation}\label{interplation}
		\phi_{ri} = \phi_o + \nabla {\phi }_o \cdot \Delta x_{ri}.
	\end{equation}
	In this equation, $\nabla {\phi }_o$ represents the gradient of the distribution function at the orange solid point, and $\Delta x_{ri}$ is the distance between the $i$-th surrounding red hollow point and the orange solid point. Determining ${\phi_o}$ involves selecting a sufficient number of nearby red hollow points to create a system of equations.
	The interpolation template for this paper includes at least three of the closest red hollow points.
	It is important to note that extrapolation is also not recommended when interpolation is not feasible. In such cases, it is permissible to assume that the reflected distribution function is identical to that of the nearest mirrored point.
	Within the UGKS framework, both the reflected distribution function and macroscopic flux must satisfy the specular reflection boundary condition.
	
	\subsubsection{Reflected macroscopic flux}
		Instead of integrating the reflected microscopic flux in the DVS, a more practical approach is to directly apply specular reflection to the macroscopic flux using the incident macroscopic flux.
	In the specular reflection model, the reflected normal momentum flux equals the incident normal momentum flux, while the mass, tangential momentum, and energy flux at the wall surface are all zero.	
	Let the macroscopic incident flux and reflected flux be denoted as $\bm{F}^{I}$ and $\bm{F}^{R}$, respectively. The reflected macroscopic flux can be expressed as:
	\begin{equation}\label{eq_FR}
		\begin{aligned}
			& F_{\rho }^{R}=-F_{\rho }^{I}, \\ 
			& F_{\rho E}^{R}=-F_{\rho E}^{I}, \\ 
			& F_{mn}^{R}=F_{mn}^{I}, \\ 
			& F_{mt}^{R}=-F_{mt}^{I}, \\ 
		\end{aligned}
	\end{equation}
	where $F_{\rho }$, $F_{\rho E}$, $F_{mn}$ and $F_{mt}$ represent the reflected flux of mass, energy, normal momentum and tangential momentum, respectively. Generally, employing vectors in a global coordinate system is more advantageous for programming. Therefore, the reflected momentum flux can be alternatively expressed as:
	\begin{equation}
		\begin{aligned}
			& F_{\rho U}^{R}=-\left[ \left( 1-2n_{x}^{2} \right)F_{\rho U}^{I}+\left( -2{{n}_{x}}{{n}_{y}} \right)F_{\rho V}^{I} \right], \\ 
			& F_{\rho V}^{R}=-\left[ \left( -2{{n}_{x}}{{n}_{y}} \right)F_{\rho U}^{I}+\left( 1-2n_{y}^{2} \right)F_{\rho V}^{I} \right], \\ 
		\end{aligned}
	\end{equation}
	where $F_{\rho U}$ and $F_{\rho V}$ represent the flux of momentum in x-directional and y-directiona, respectively.
		
	\subsubsection{Micro-macro consistency}
	It is evident that numerical errors are inherent in any interpolation process, and the accuracy of the interpolation methods dictates the extent of these errors. Given that the reflected distribution function is obtained through interpolation, implementing a correction technique is crucial to ensure both conservation and consistency between microscopic and macroscopic scales.
	The interpolated distribution function ${{\phi }_{o}}$ can be represented as the sum of the actual reflected distribution function ${{\phi }_{act}}$ and the erroneous distribution function ${{\phi }_{err}}$:
	\begin{equation}\label{miflux_error}
		{{\phi }_{o}}={{\phi }_{act}}+{{\phi }_{err}},\ {{\bm{\xi}}_{i}}\cdot {{\bm{n}}_{w}}<0.
	\end{equation}
	It is clear that the distribution function of boundary cell updated by Eq.  \ref{eq_update_phi} is influenced by the numerical error of the reflected distribution function. By integrating Eq. \ref{miflux_error} in DVS, one can establish the relationship of reflected macroscopic flux:
	\begin{equation}\label{maflux_error}
		{\bm{F}^{R}_{o}}={{\bm{F}^{R}}_{act}}+{{\bm{F}^{R}}_{err}},\ {{\bm{\xi}}_{i}}\cdot {{\bm{n}}_{w}}<0.
	\end{equation}
	Here, ${\bm{F}^{R}}_{act}$ should satisfy Eq. \ref{eq_FR}. Therefore, the erroneous reflected macroscopic flux ${\bm{F}^{R}}_{err}$ should be applied to correct the interpolated distribution function ${{\phi }_{o}}$.
	It is worth noting that this correction process may involve complex and expensive computations.
	
	It's important to note that the macroscopic variables of the control volume, updated by Eq. \ref{eq_W}, are accurate. This precision is ensured by the exact macroscopic flux obtained from Eq. \ref{eq_FR}, which strictly adheres to the specular reflection boundary. Consequently, the deviation of the equilibrium distribution function can be utilized to finely adjust the distribution function of the control volume, ensuring micro-macro consistency:
	\begin{equation}
		{{\phi}^{n+1}_{j}}={{\overline{\phi}}^{n+1}_{j}}+[g\left( {{\bm{W}}^{n+1}_{j}} \right)-\overline{g}\left( {{\overline{\bm{W}}}^{n+1}_{j}} \right)],
	\end{equation}
	where ${\overline{\phi}}^{n+1}_{j}$ is the distribution function of control volume updated by Eq. \ref{eq_update_phi}, and the macroscopic variables of control volume  ${\overline{\bm{W}}}^{n+1}_{j}$ (with numerical error) can be obtained by integrating ${\overline{\phi}}^{n+1}_{j}$ in DVS. ${\bm{W}}^{n+1}_{j}$ is the macroscopic variables of control volume determined by accurately updating the exact macroscopic flux.  It is worth noting that if numerical error correction was not used in the simulations of hypersonic flow past a cylinder, the program would crash very quickly.
	
	The key idea behind the proposed GSI algorithm is the combination of interpolation methods for mismatch points in DVS and an efficient numerical error correction method for micro-macro consistency. This approach serves as a cornerstone for implementing the Maxwell boundary condition, specifically designed for rarefied and multi-scale flows and other physical fields within the unstructured DVS.
		
	\section{Numerical results and discussions}\label{sec:test cases}
	In this section, several test cases are conducted to validate the extended application of the Maxwell GSI model in the DVM framework. The simulation results from the DS2V software \cite{bird1994molecular} and experimental data are used as reference benchmarks.
	
	\subsection{Verification of monatomic flows}
	\subsubsection{Supersonic flow over a sharp flat plate in the slip regime}
\begin{figure}
		\centering
		\subfigure[]{\label{fig03a}\includegraphics[width=0.45\textwidth]{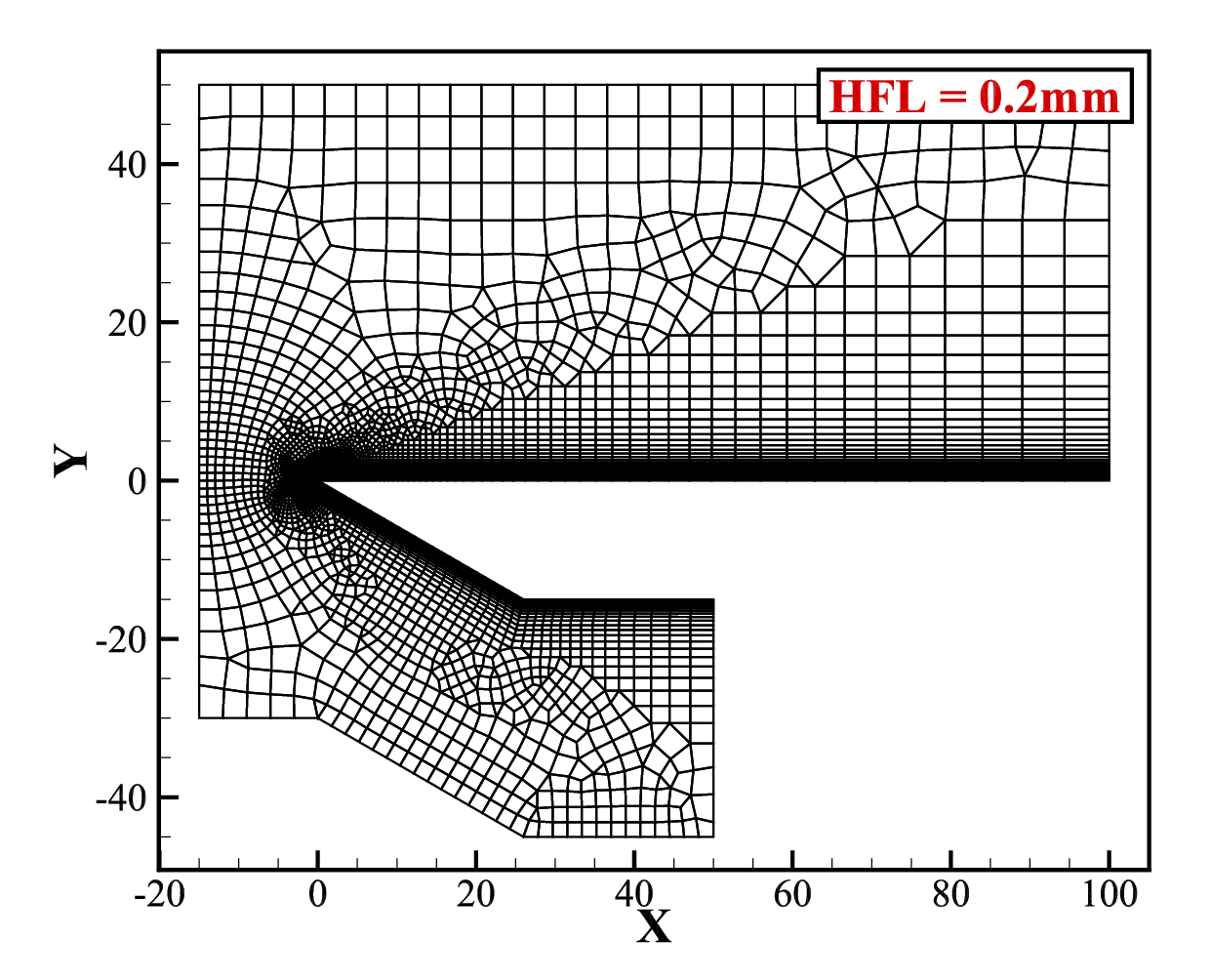}}
		\subfigure[]{\label{fig03b}\includegraphics[width=0.37\textwidth]{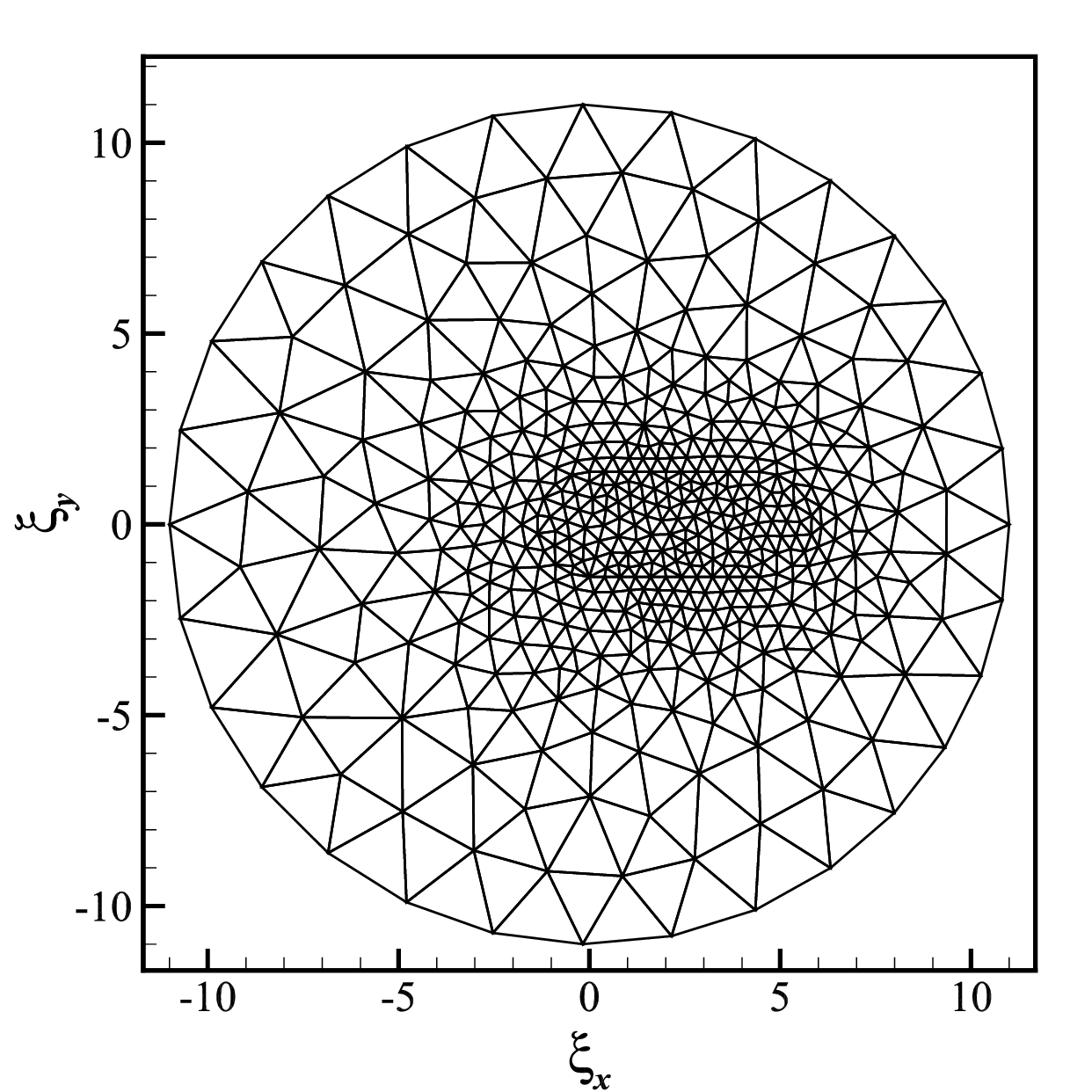}}
		\caption{\label{fig03}{The (a) unstructured physical mesh and (b) unstructured velocity mesh for the supersonic flow passing a flat plate ($Ma$ = 4.89, $Kn$ = 0.0078, $T_{\infty}$ = 116 K, $T_{w}$ = 290 K).}}
	\end{figure}

	To assess the accuracy of the GSI model in monatomic UGKS, the initial numerical simulation investigates a supersonic flow of argon gas passing over a sharp flat plate. Fig. \ref{fig03a} illustrates the geometric shape of the flat plate and the unstructured physical mesh. The flat plate has dimensions of 100 mm in length, 15 mm in thickness, and a sharp angle of 30 degrees. Following the wind tunnel test run34\cite{tsuboi2005experimental}, the freestream has a Mach number of 4.89 and a temperature of 116 K. The hard sphere molecular model with $\omega$ = 0.5 is employed, and the temperature of the plate surface is set to 290 K. The Knudsen number for the freestream, with the plate's length as the characteristic length, is 0.0078. The height of the first layer (HFL) of the mesh on the surface of the plate is 0.2 mm. The unstructured velocity mesh, depicted in Fig. \ref{fig03b}, comprises 896 cells.
	
\begin{figure}
		\centering
		\subfigure[]{\label{fig04a}\includegraphics[width=0.45\textwidth]{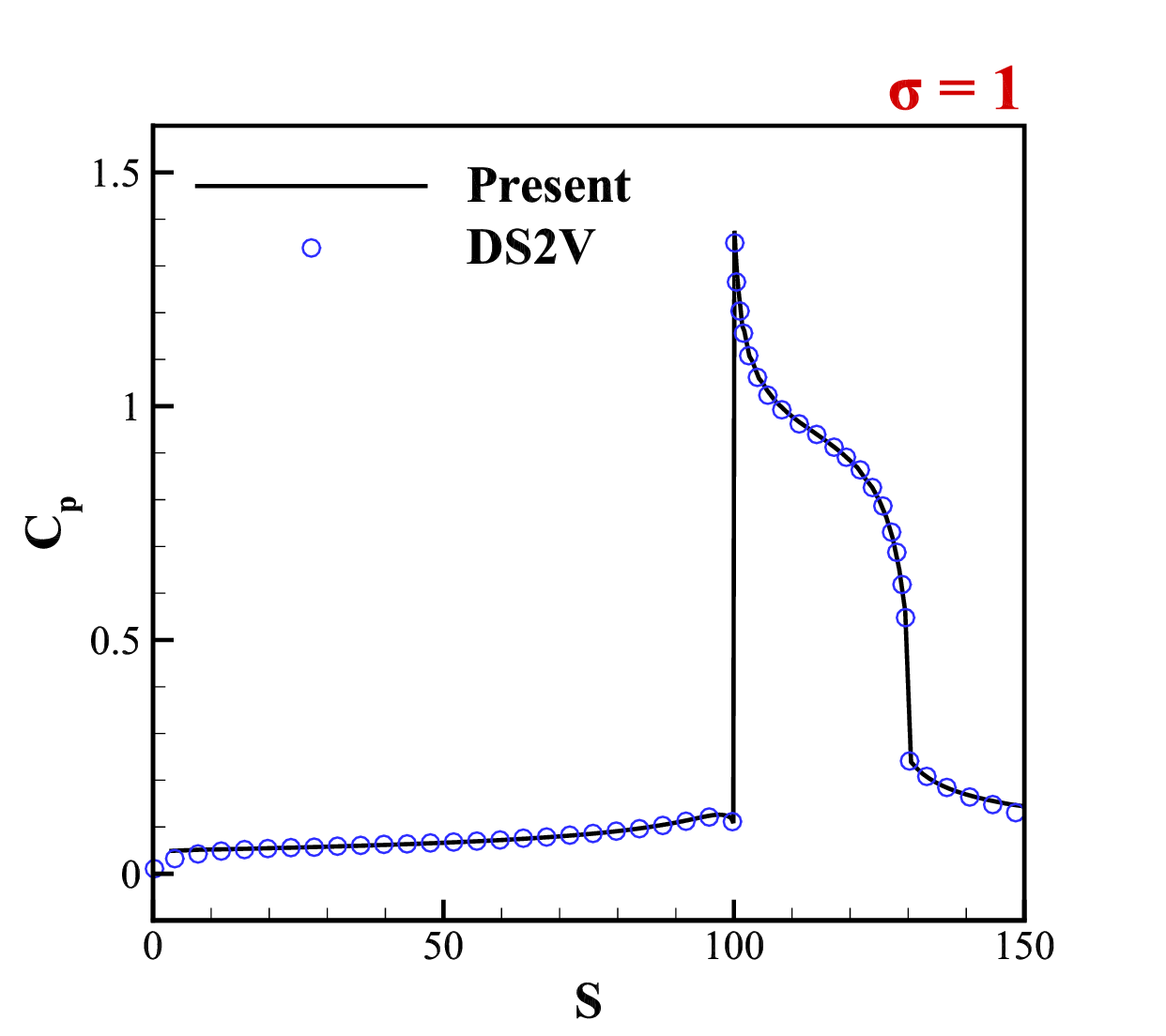}}
		\subfigure[]{\label{fig04b}\includegraphics[width=0.45\textwidth]{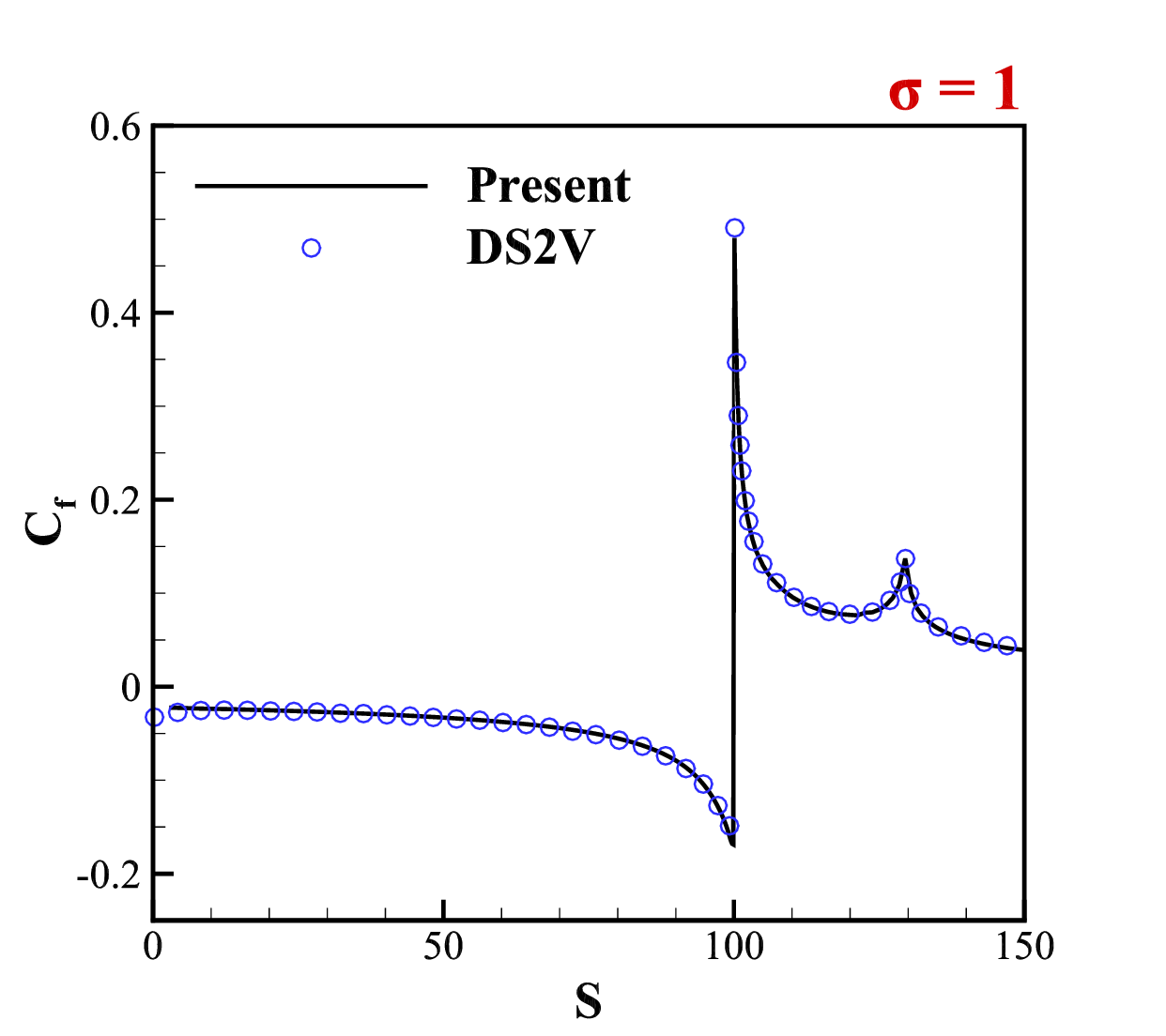}}
		\subfigure[]{\label{fig04c}\includegraphics[width=0.45\textwidth]{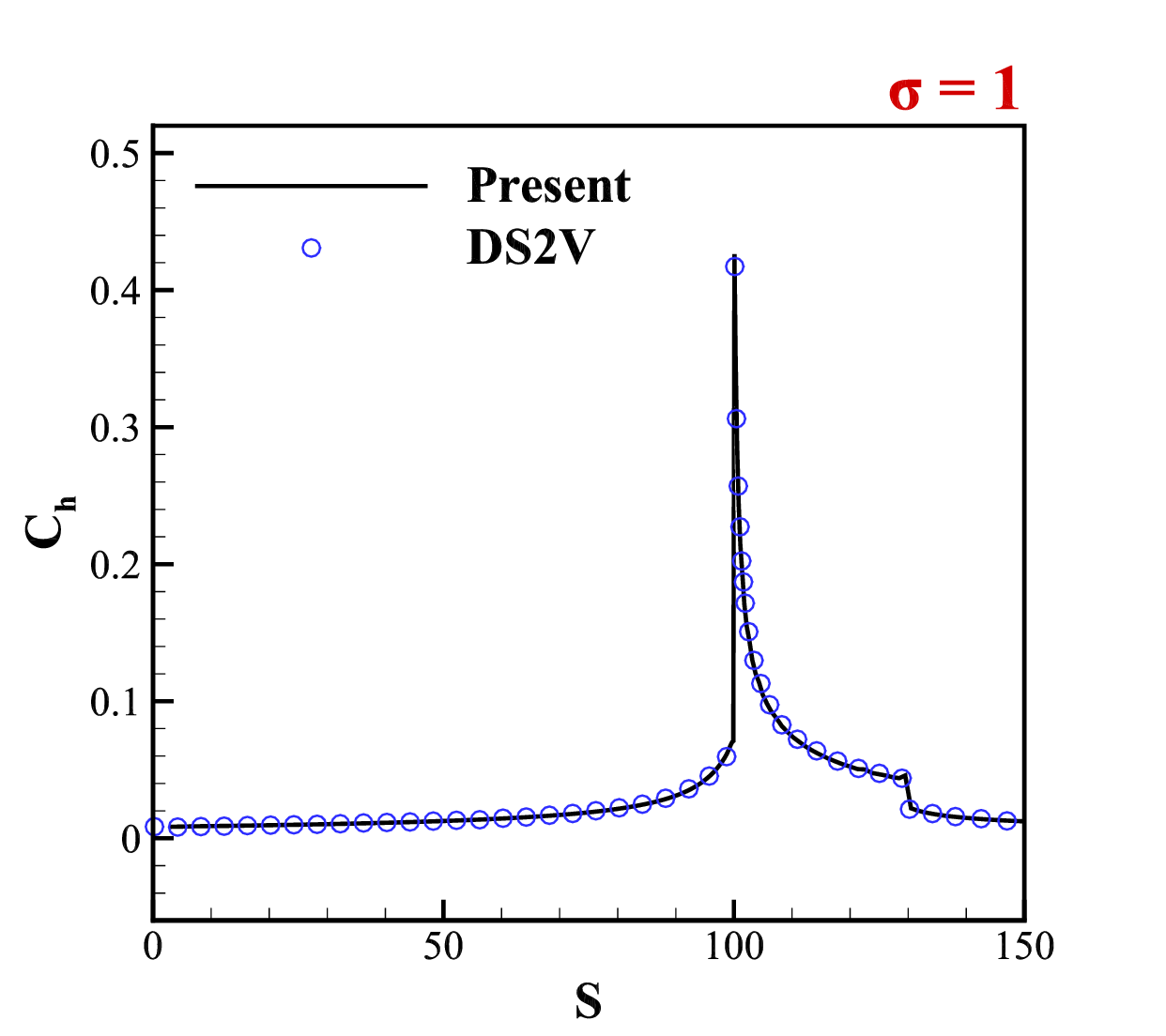}}
		\caption{\label{fig04}{Comparison of the (a) pressure coefficient, (b) skin friction coefficient, and (c) heat transfer coefficient on the surface of flat plate with $\sigma$ = 1 ($Ma$ = 4.89, $Kn$ = 0.0078, $T_{\infty}$ = 116 K, $T_{w}$ = 290 K).}}
	\end{figure}

\begin{figure}
		\centering
		\subfigure[]{\label{fig05a}\includegraphics[width=0.45\textwidth]{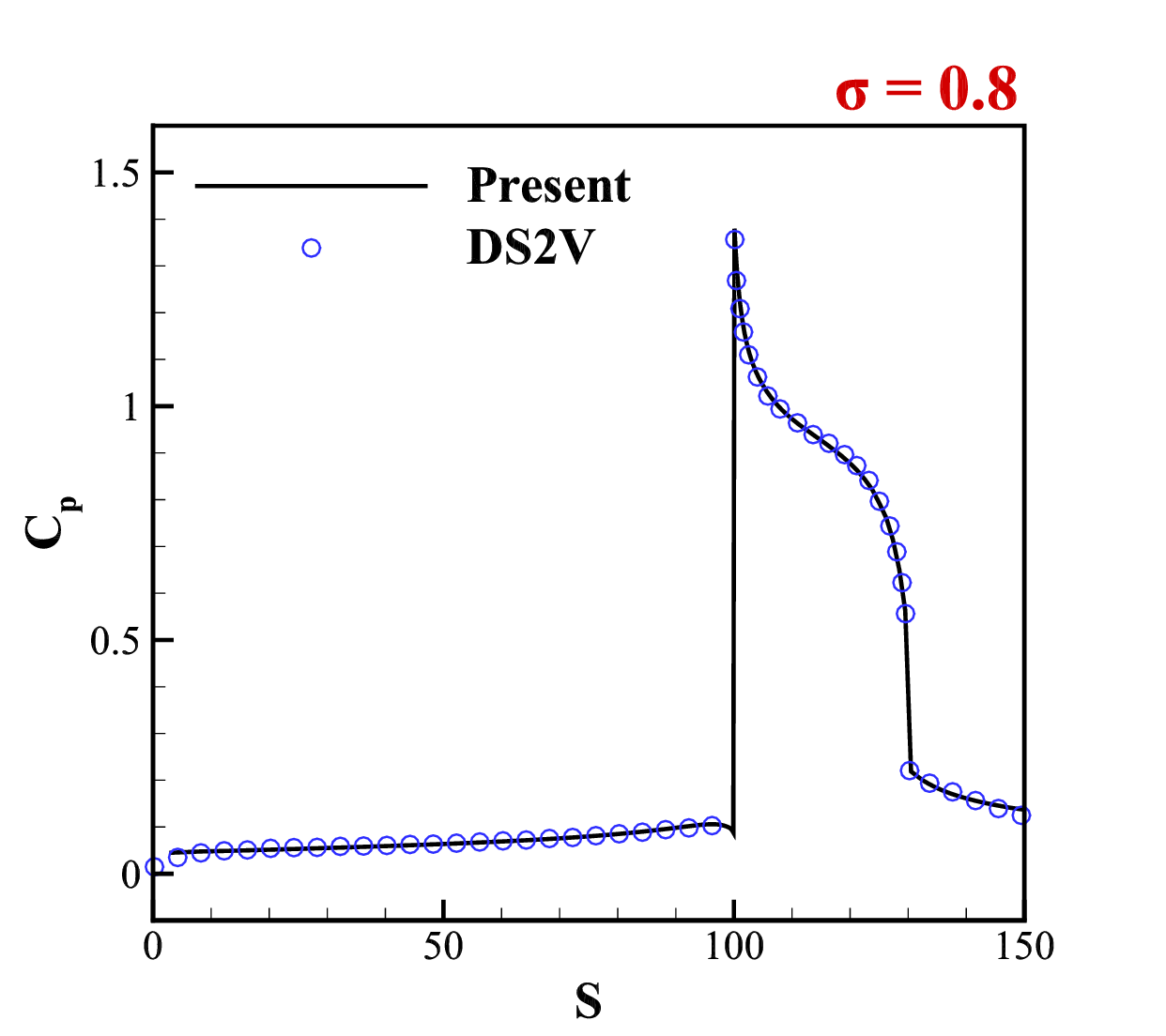}}
		\subfigure[]{\label{fig05b}\includegraphics[width=0.45\textwidth]{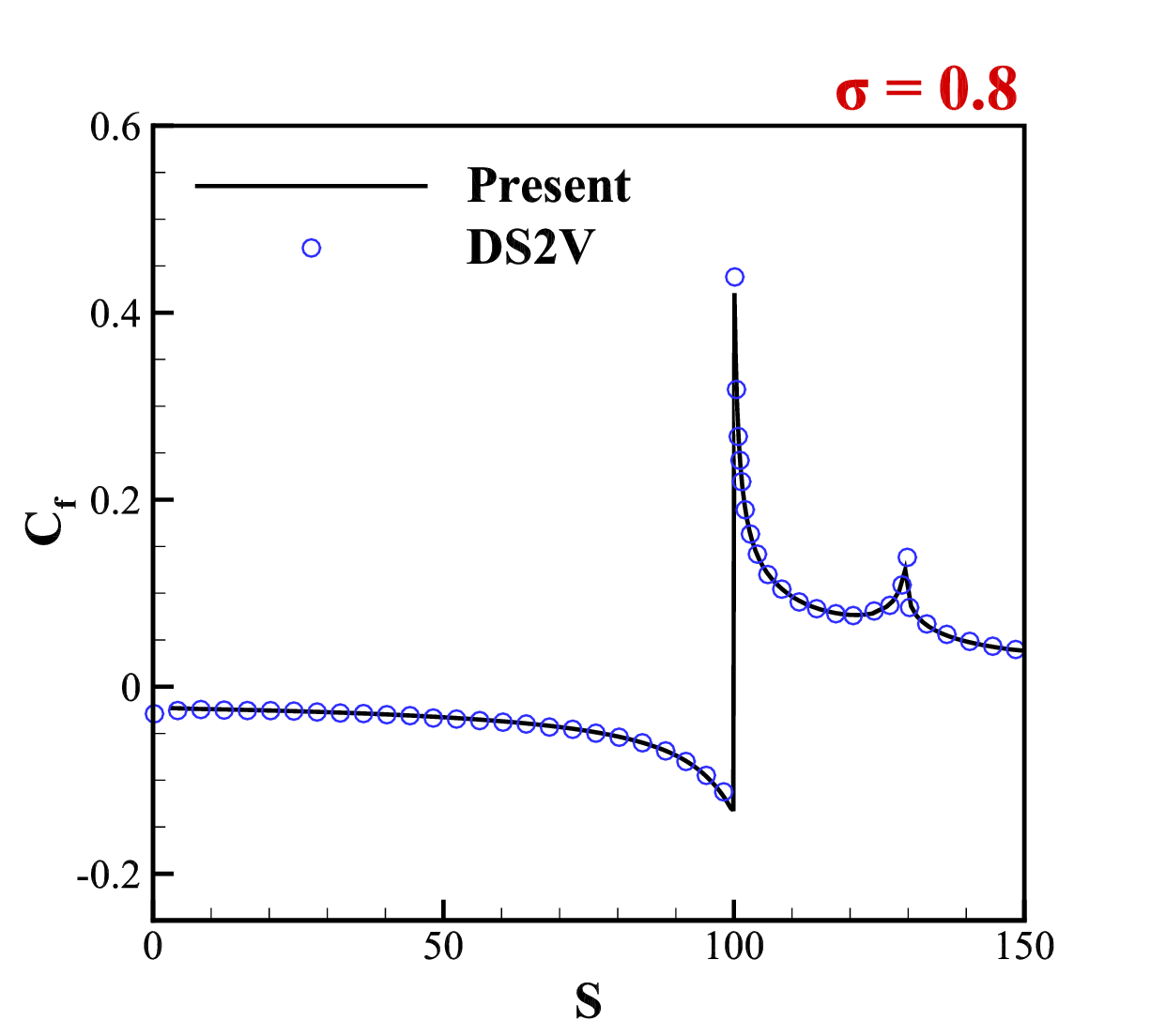}}
		\subfigure[]{\label{fig05c}\includegraphics[width=0.45\textwidth]{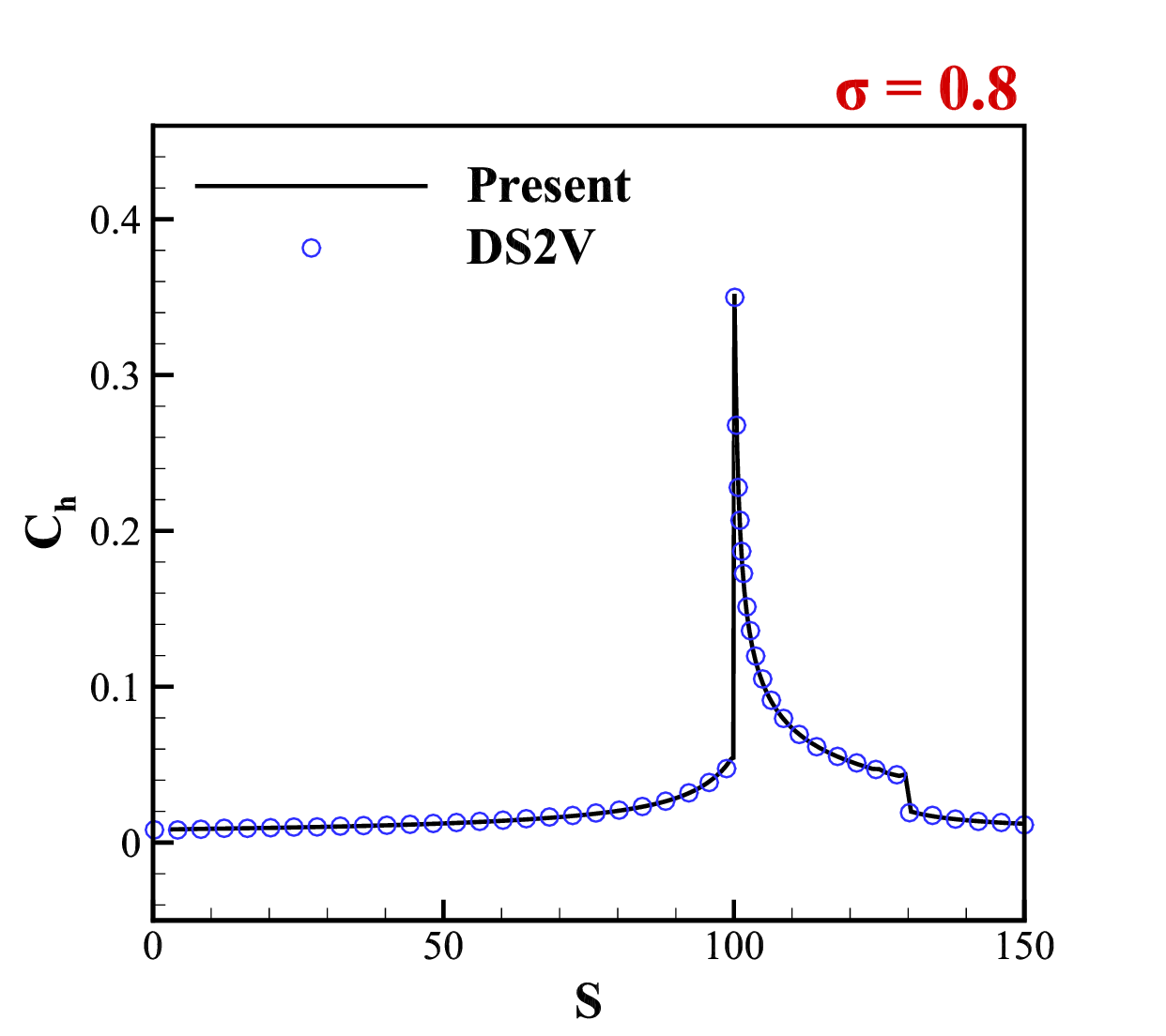}}
		\caption{\label{fig05}{Comparison of the (a) pressure coefficient, (b) skin friction coefficient, and (c) heat transfer coefficient on the surface of flat plate with $\sigma$ = 0.8 ($Ma$ = 4.89, $Kn$ = 0.0078, $T_{\infty}$ = 116 K, $T_{w}$ = 290 K).}}
	\end{figure}

\begin{figure}
		\centering
		\subfigure[]{\label{fig06a}\includegraphics[width=0.45\textwidth]{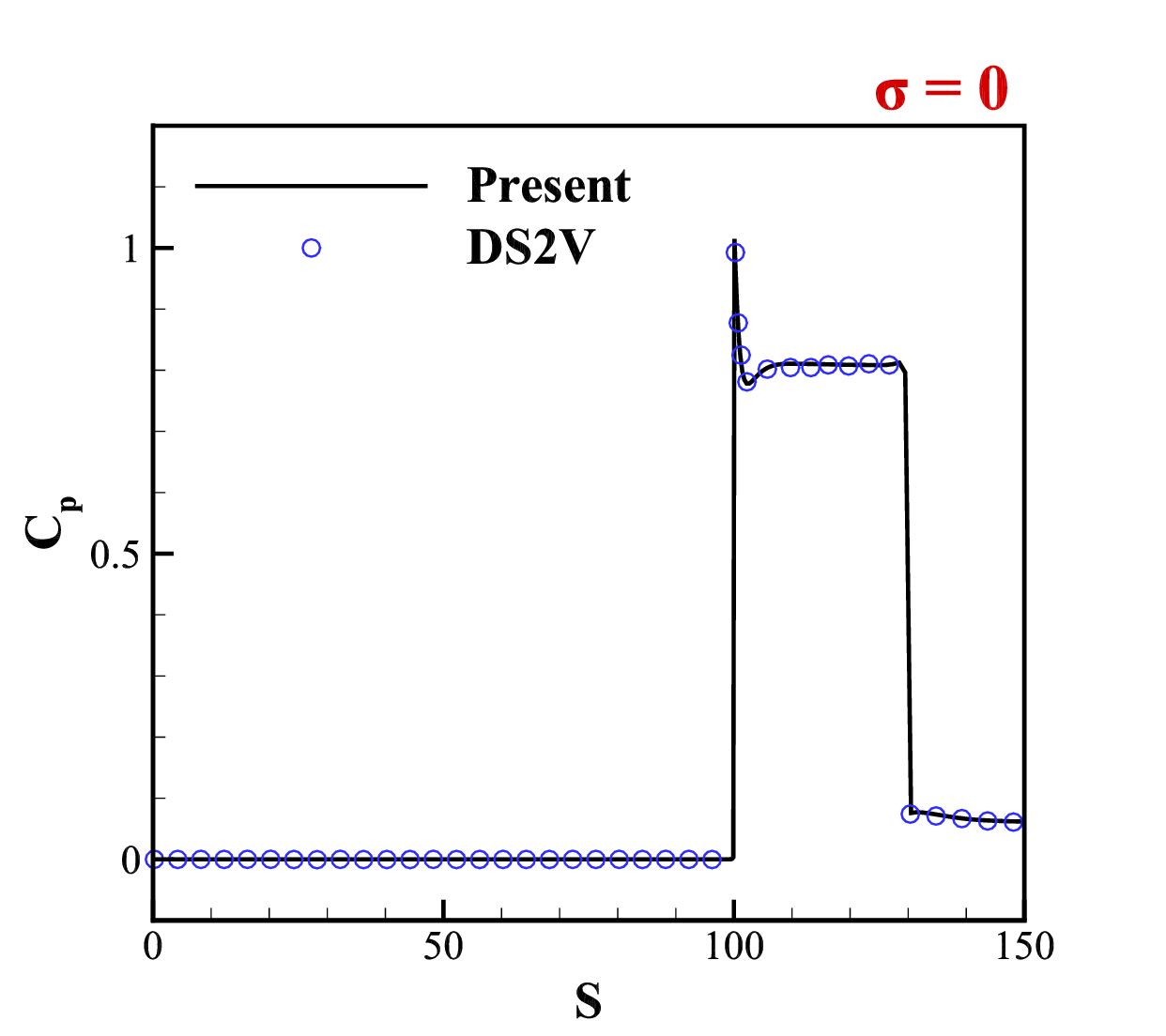}}
		\subfigure[]{\label{fig06b}\includegraphics[width=0.45\textwidth]{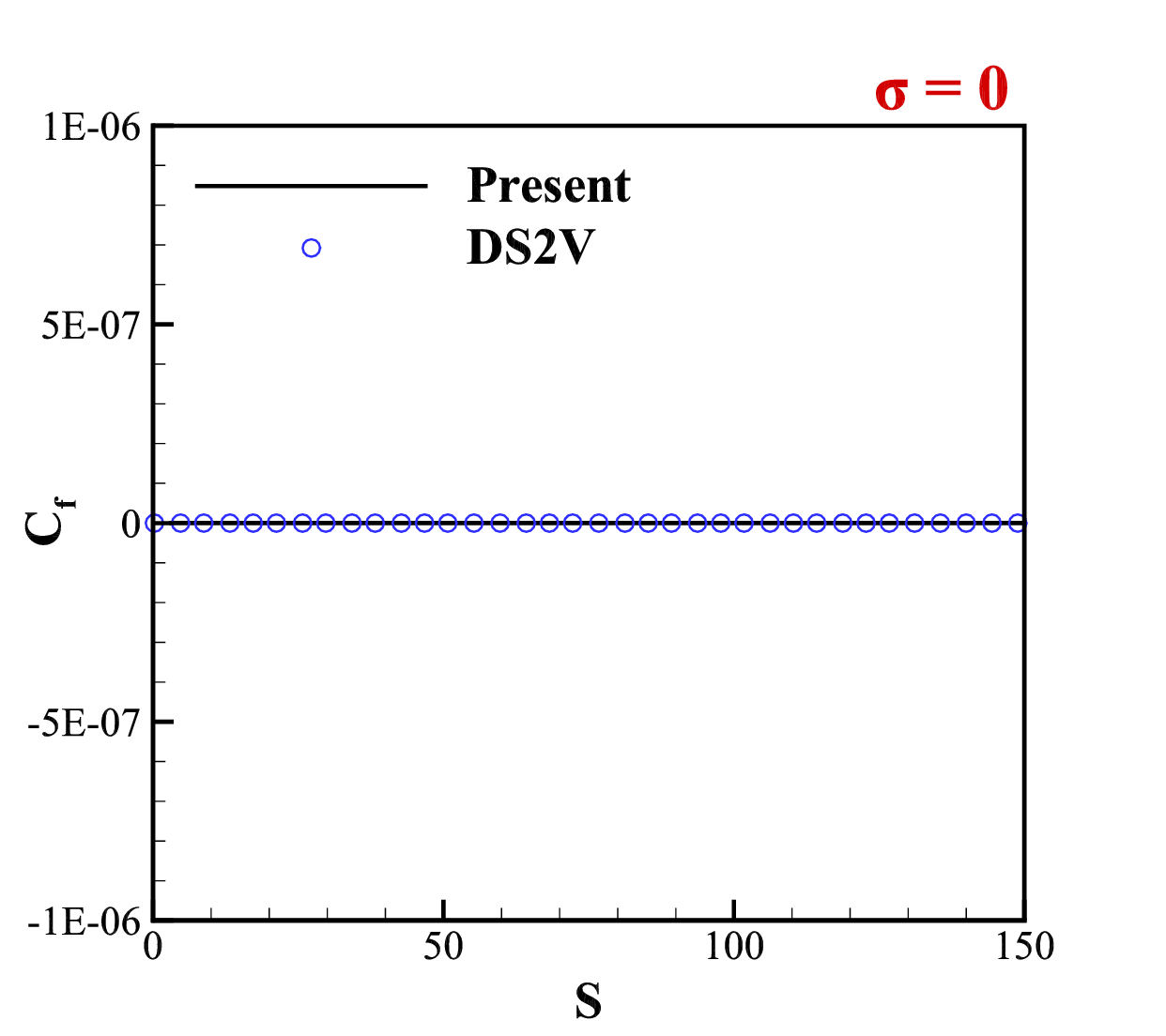}}
		\subfigure[]{\label{fig06c}\includegraphics[width=0.45\textwidth]{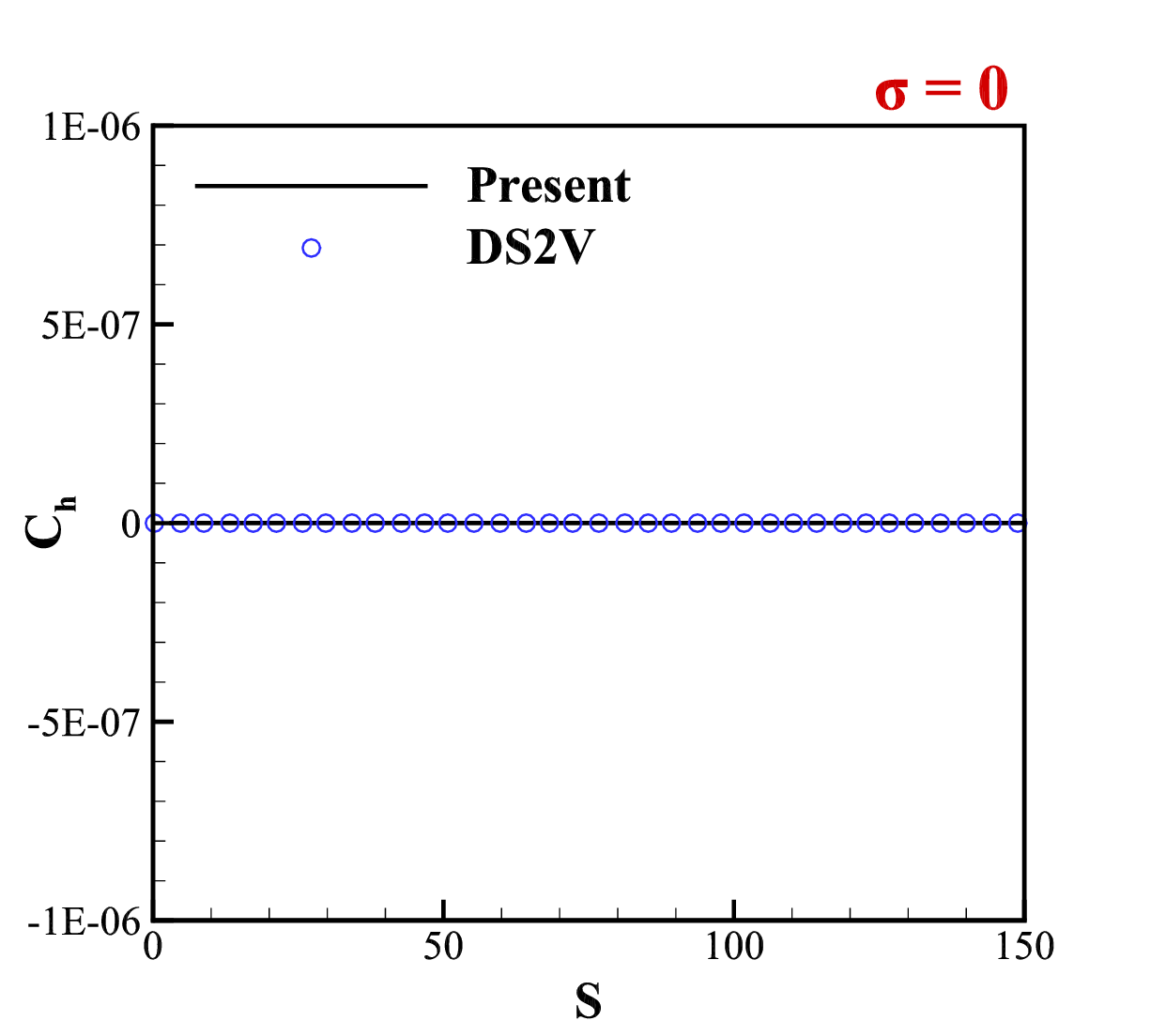}}
		\caption{\label{fig06}{Comparison of the (a) pressure coefficient, (b) skin friction coefficient, and (c) heat transfer coefficient on the surface of flat plate with $\sigma$ = 0 ($Ma$ = 4.89, $Kn$ = 0.0078, $T_{\infty}$ = 116 K, $T_{w}$ = 290 K).}}
	\end{figure}

\begin{figure}
		\centering
		\subfigure[]{\label{fig06add1_a}\includegraphics[width=0.45\textwidth]{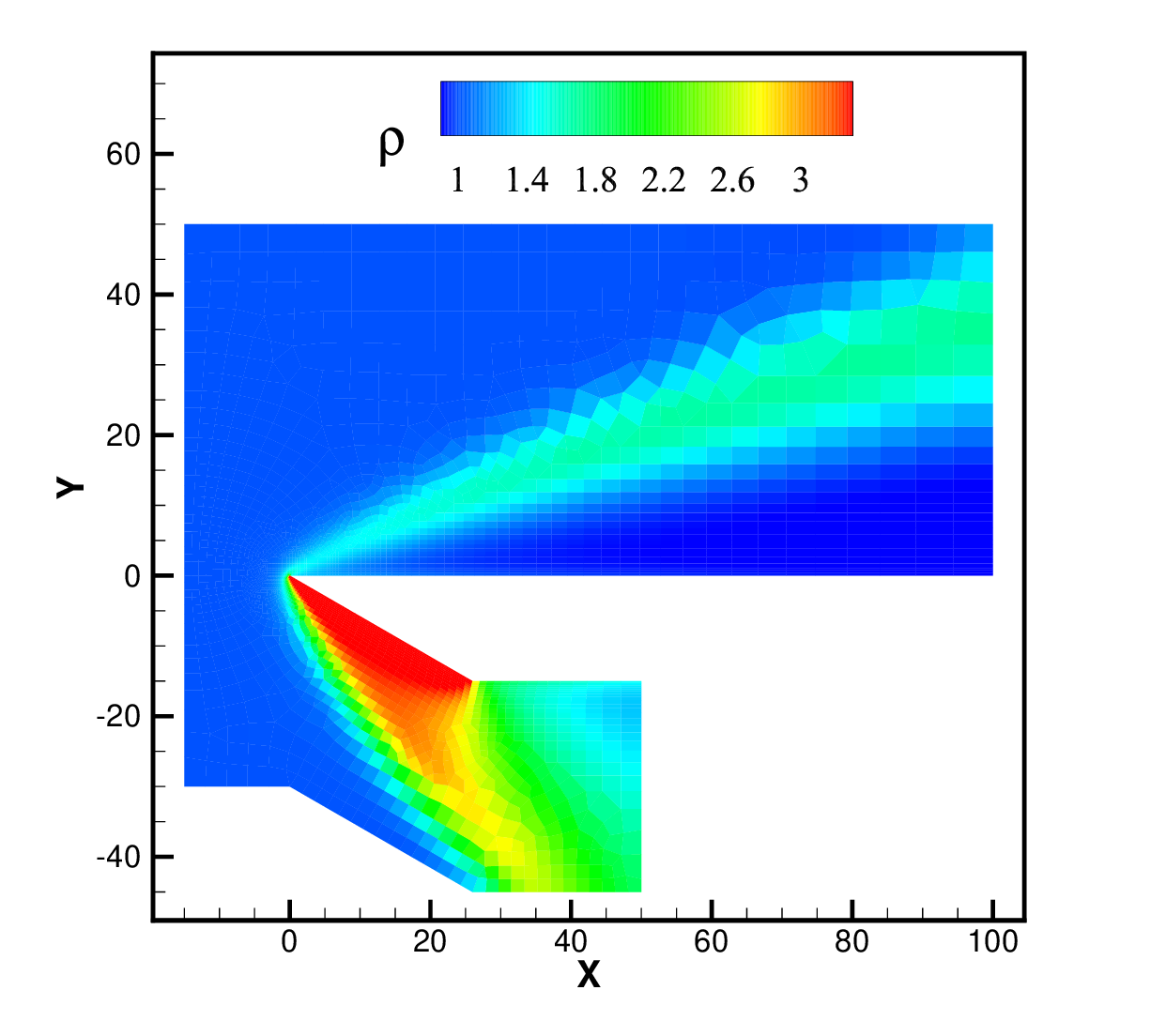}}
		\subfigure[]{\label{fig06add1_b}\includegraphics[width=0.45\textwidth]{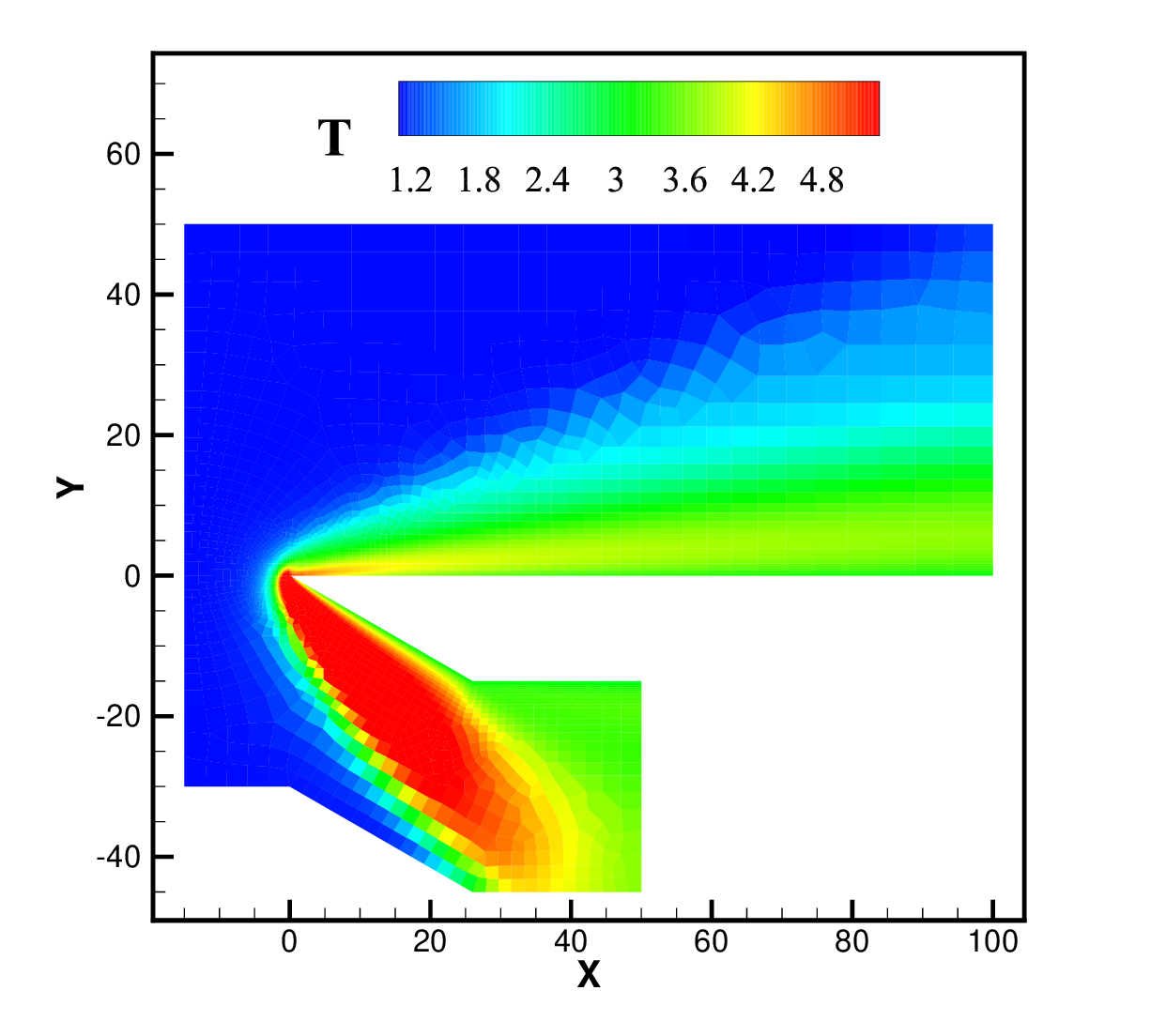}}
		\subfigure[]{\label{fig06add1_c}\includegraphics[width=0.45\textwidth]{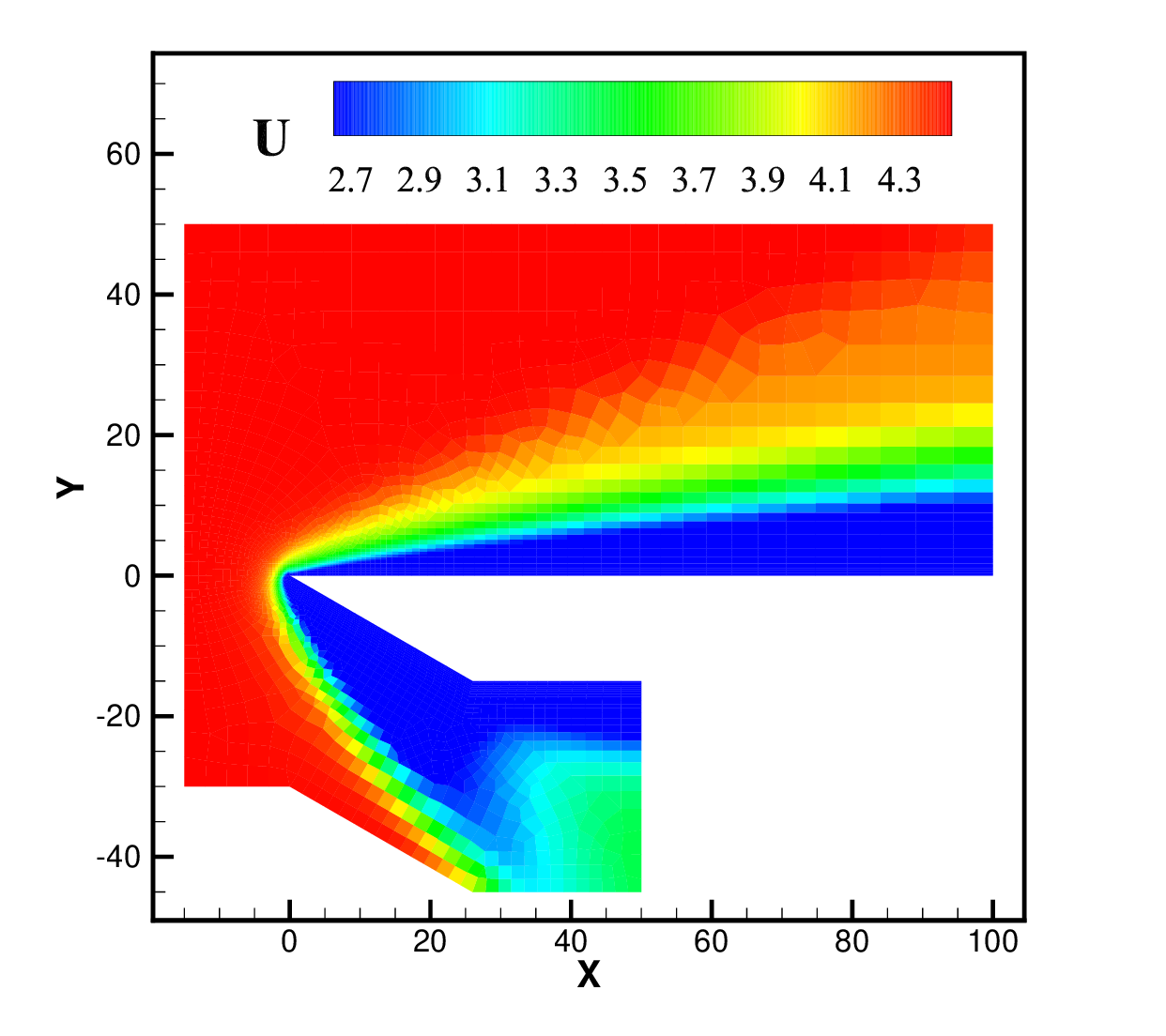}}
		\caption{\label{fig06add1}{The contour charts of the supersonic flow over a flat plate with diffuse wall ($Ma$ = 4.89, $Kn$ = 0.0078, $T_{\infty}$ = 116 K, $T_{w}$ = 290 K). (a) Density, (b) temperature, and (c) horizontal velocity.}}
	\end{figure}

\begin{figure}
		\centering
		\subfigure[]{\label{fig06add2_a}\includegraphics[width=0.45\textwidth]{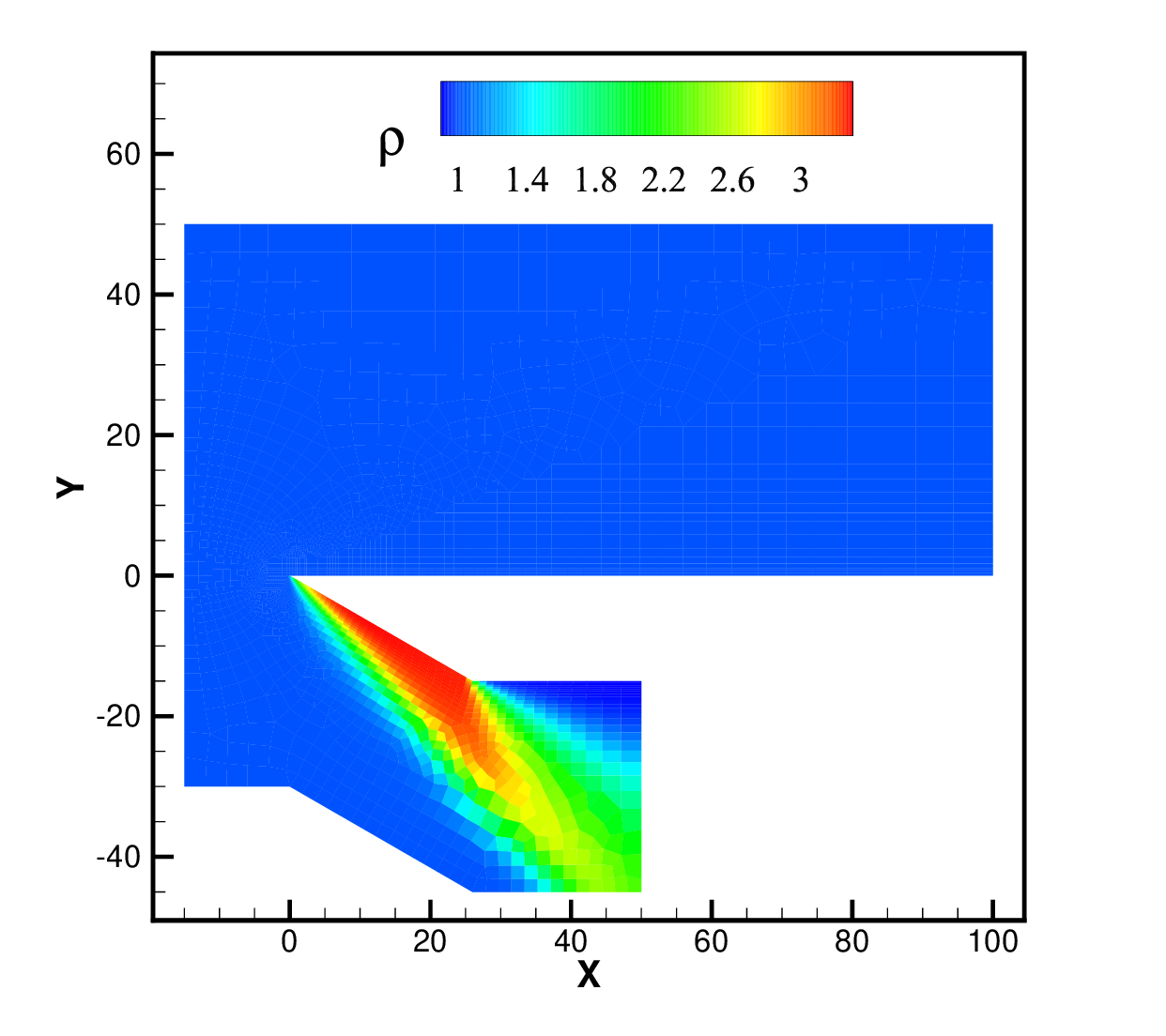}}
		\subfigure[]{\label{fig06add2_b}\includegraphics[width=0.45\textwidth]{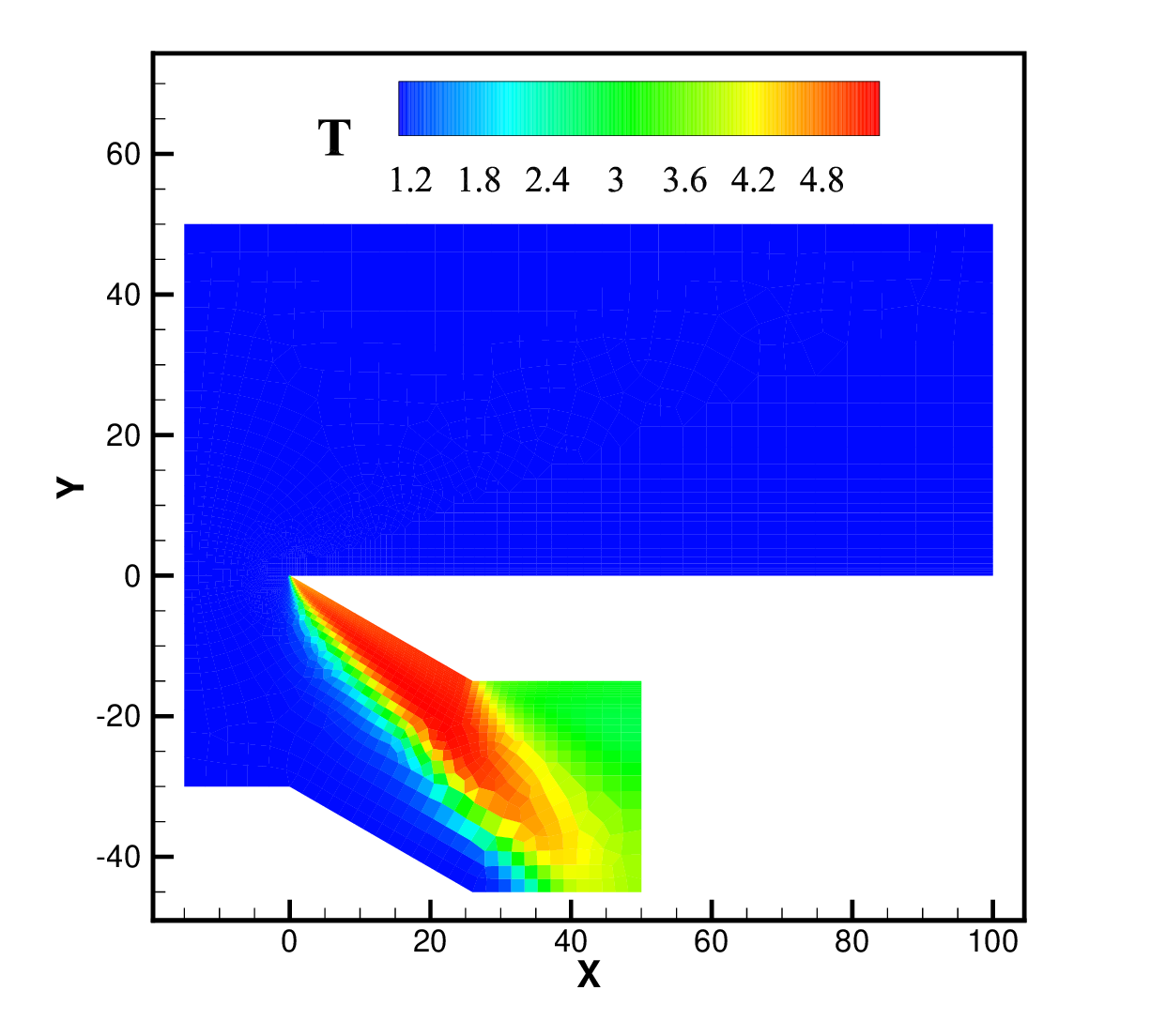}}
		\subfigure[]{\label{fig06add2_c}\includegraphics[width=0.45\textwidth]{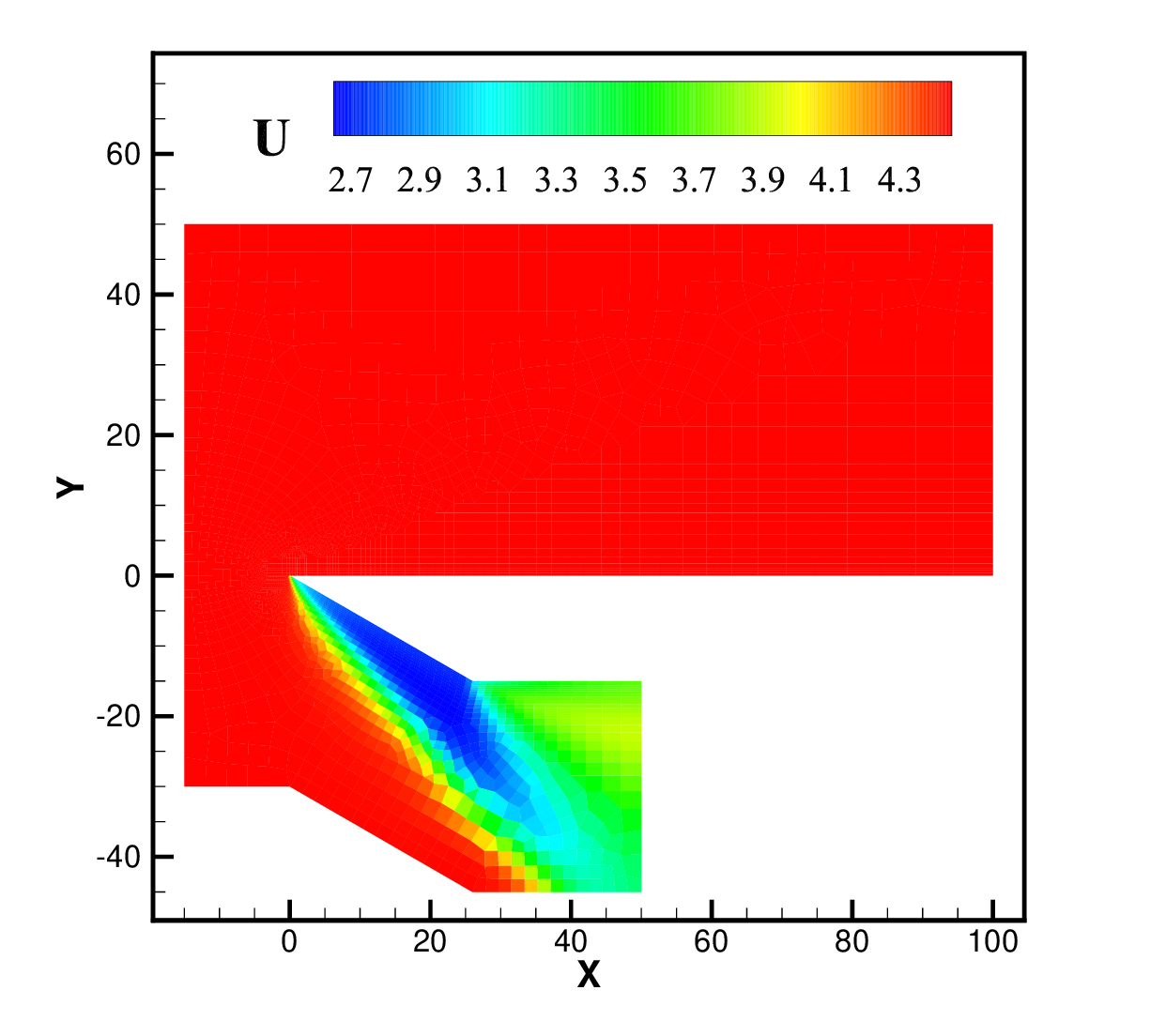}}
		\caption{\label{fig06add2}{The contour charts of the supersonic flow over a flat plate with specular wall ($Ma$ = 4.89, $Kn$ = 0.0078, $T_{\infty}$ = 116 K, $T_{w}$ = 290 K). (a) Density, (b) temperature, and (c) horizontal velocity.}}
	\end{figure}

\begin{figure}
		\centering
		\subfigure[]{\label{fig06add4_a}\includegraphics[width=0.45\textwidth]{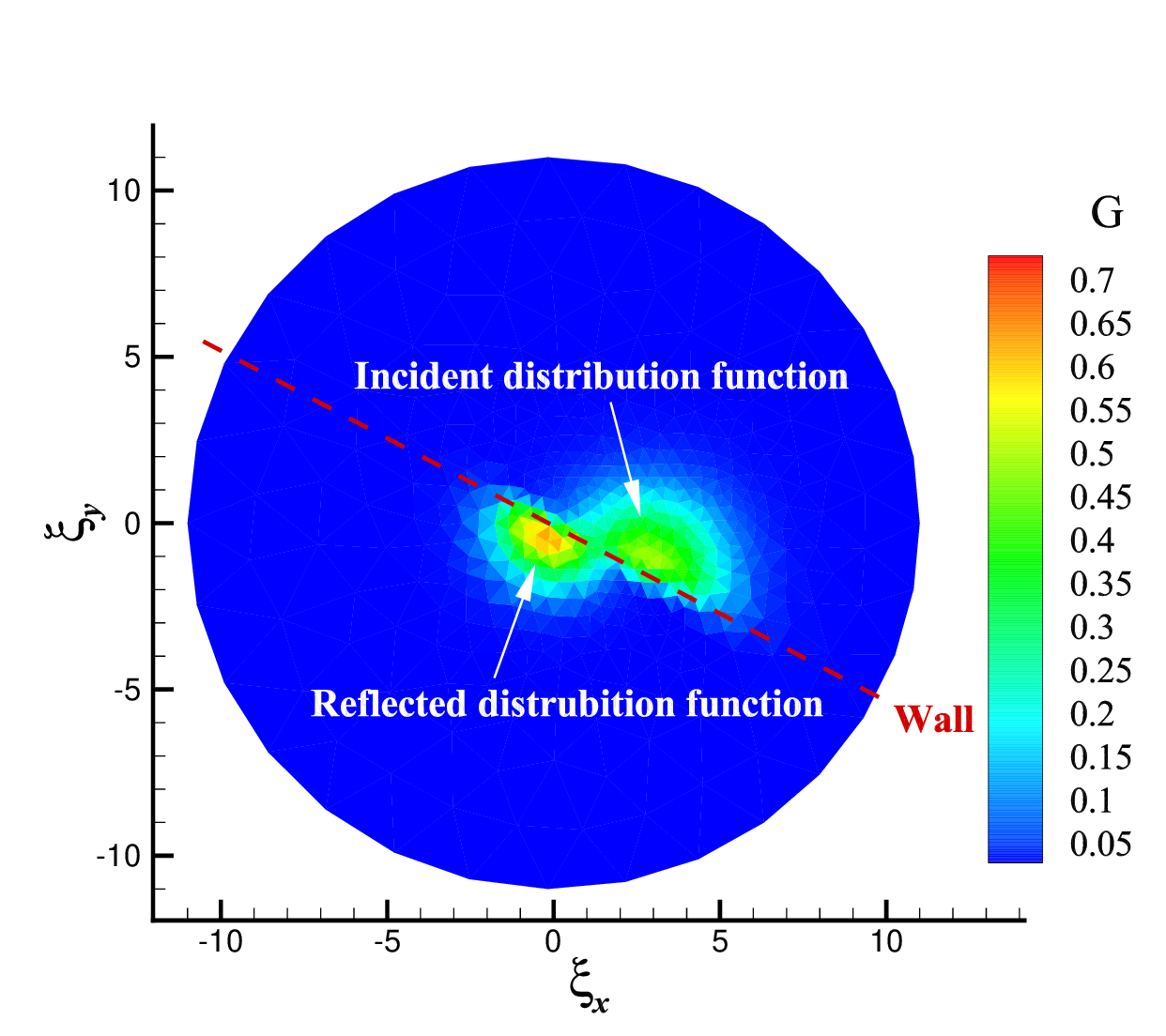}}
		\subfigure[]{\label{fig06add4_b}\includegraphics[width=0.45\textwidth]{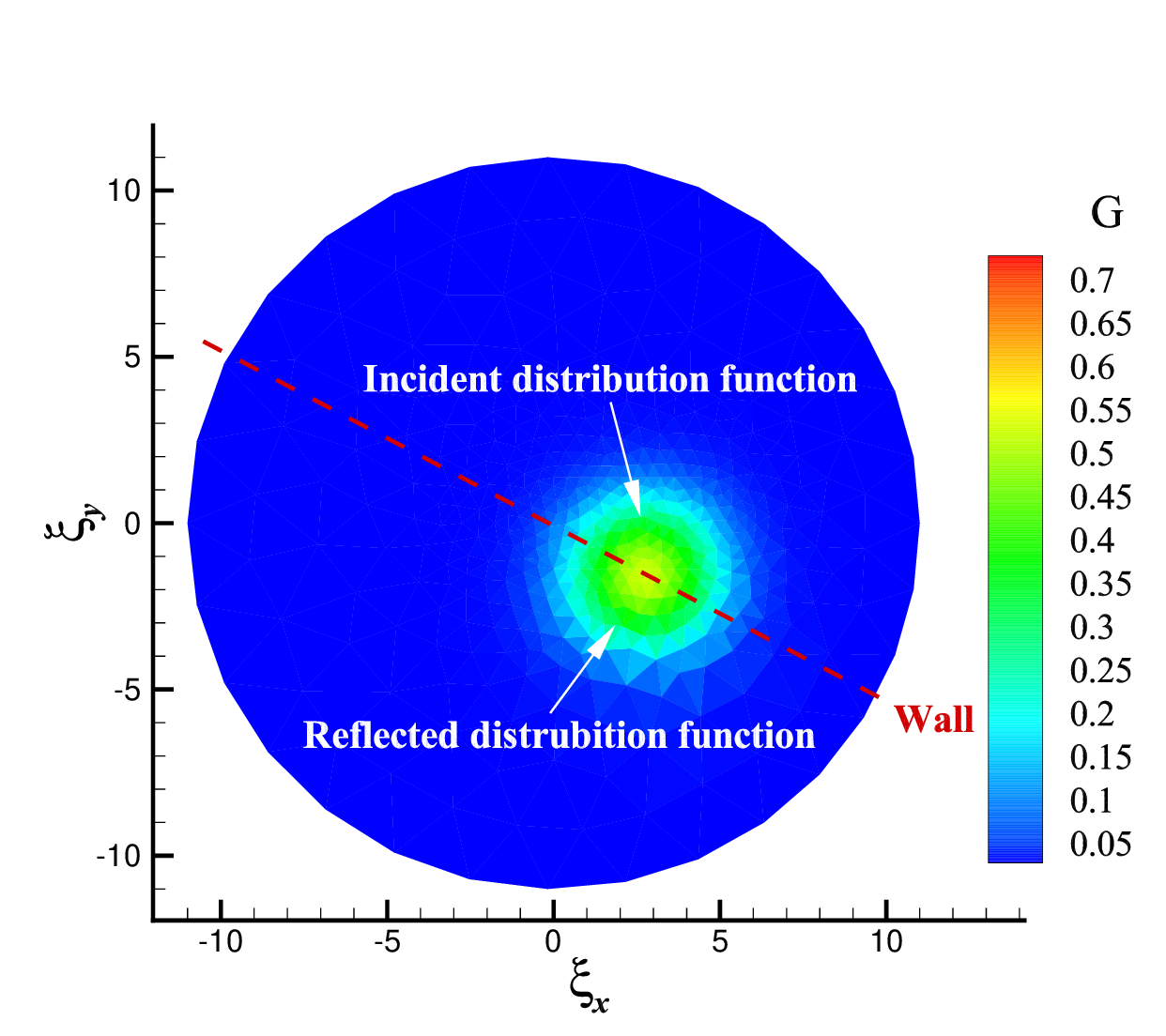}}
		\caption{\label{fig06add4}{The distribution function on the bevel of the flat ($Ma$ = 4.89, $Kn$ = 0.0078, $T_{\infty}$ = 116 K, $T_{w}$ = 290 K). (a) Diffuse wall, and (b) specular wall.}}
	\end{figure}

\begin{figure}
		\centering
		\subfigure[]{\label{fig07a}\includegraphics[width=0.45\textwidth]{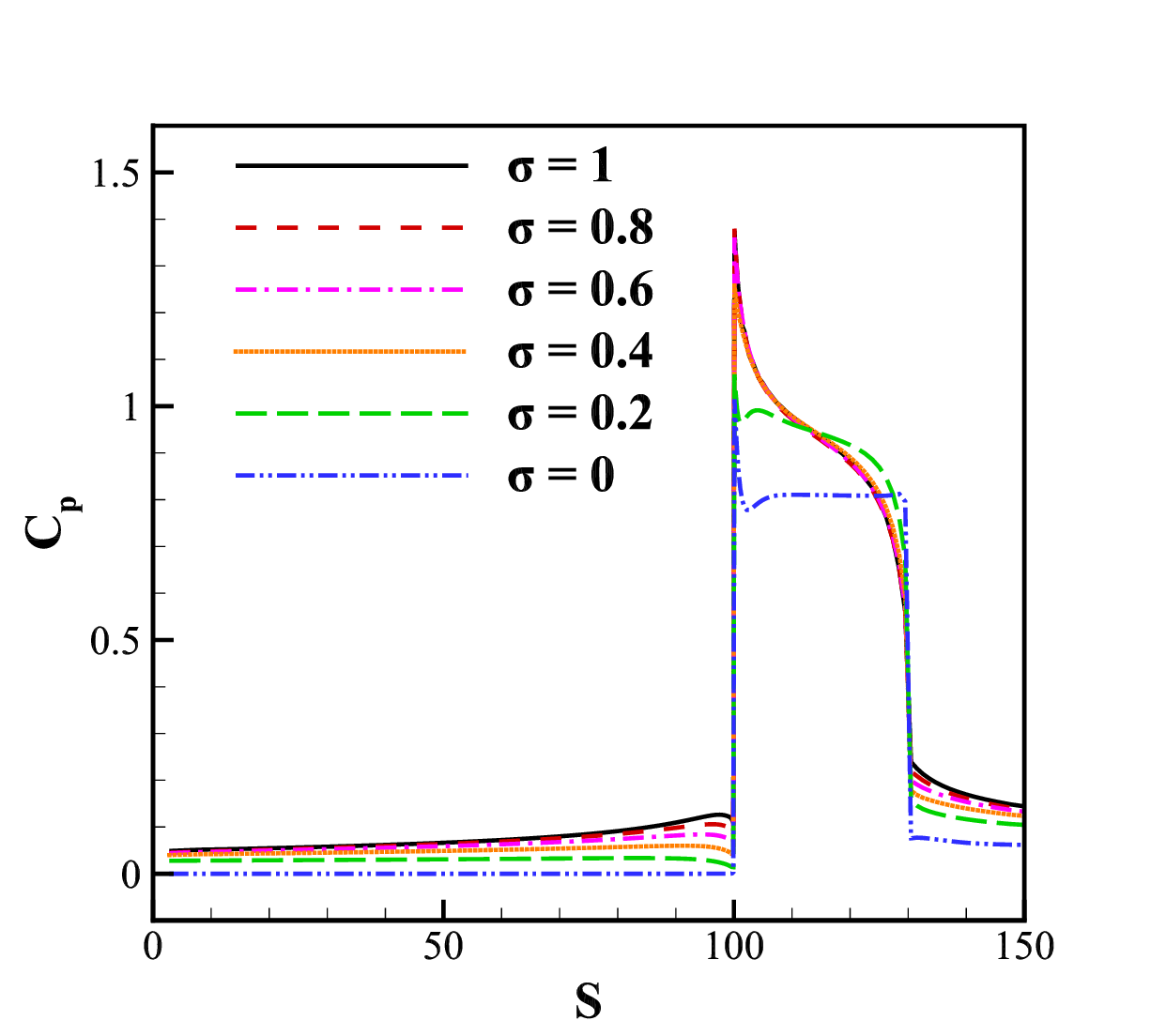}}
		\subfigure[]{\label{fig07b}\includegraphics[width=0.45\textwidth]{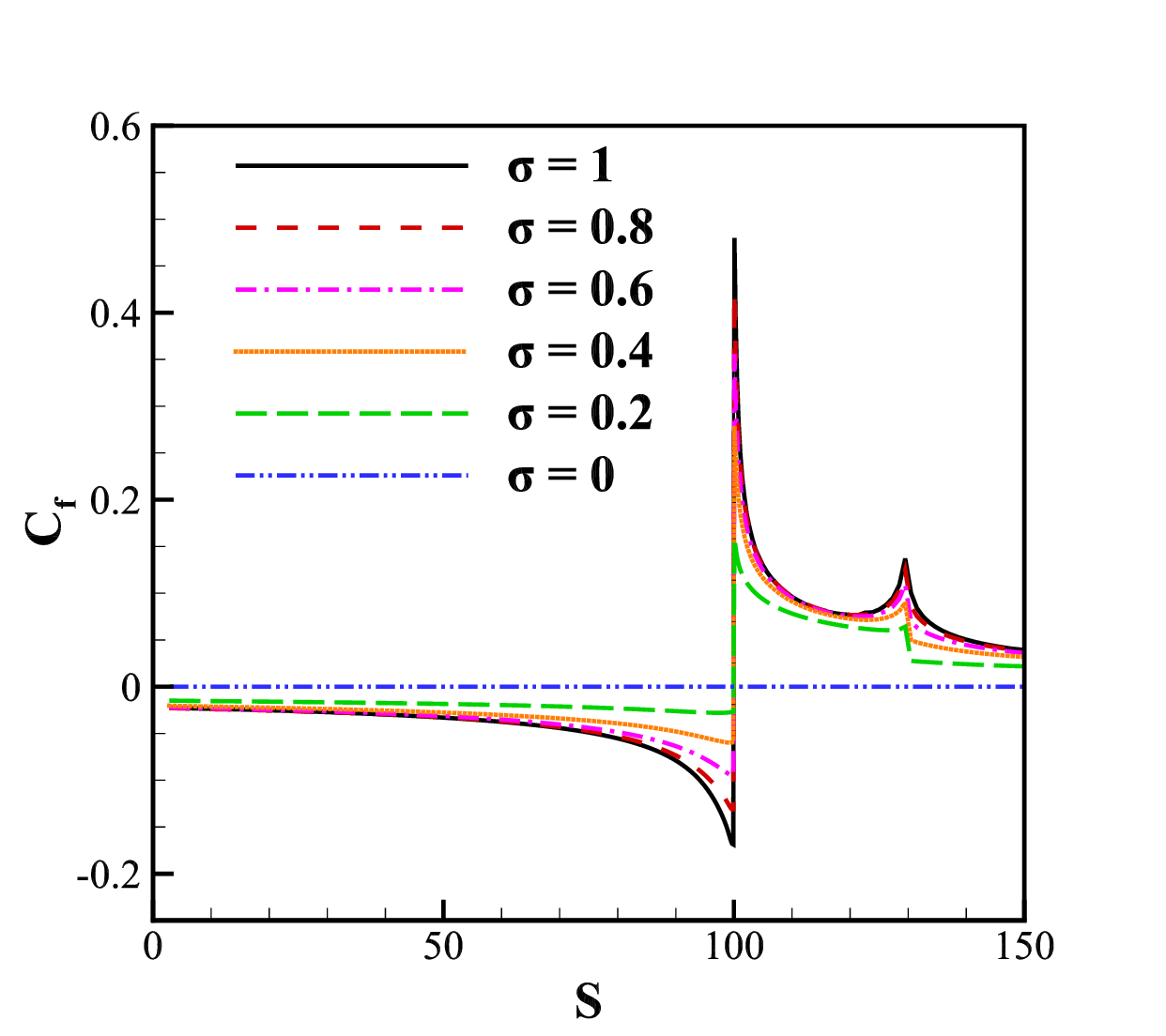}}
		\subfigure[]{\label{fig07c}\includegraphics[width=0.45\textwidth]{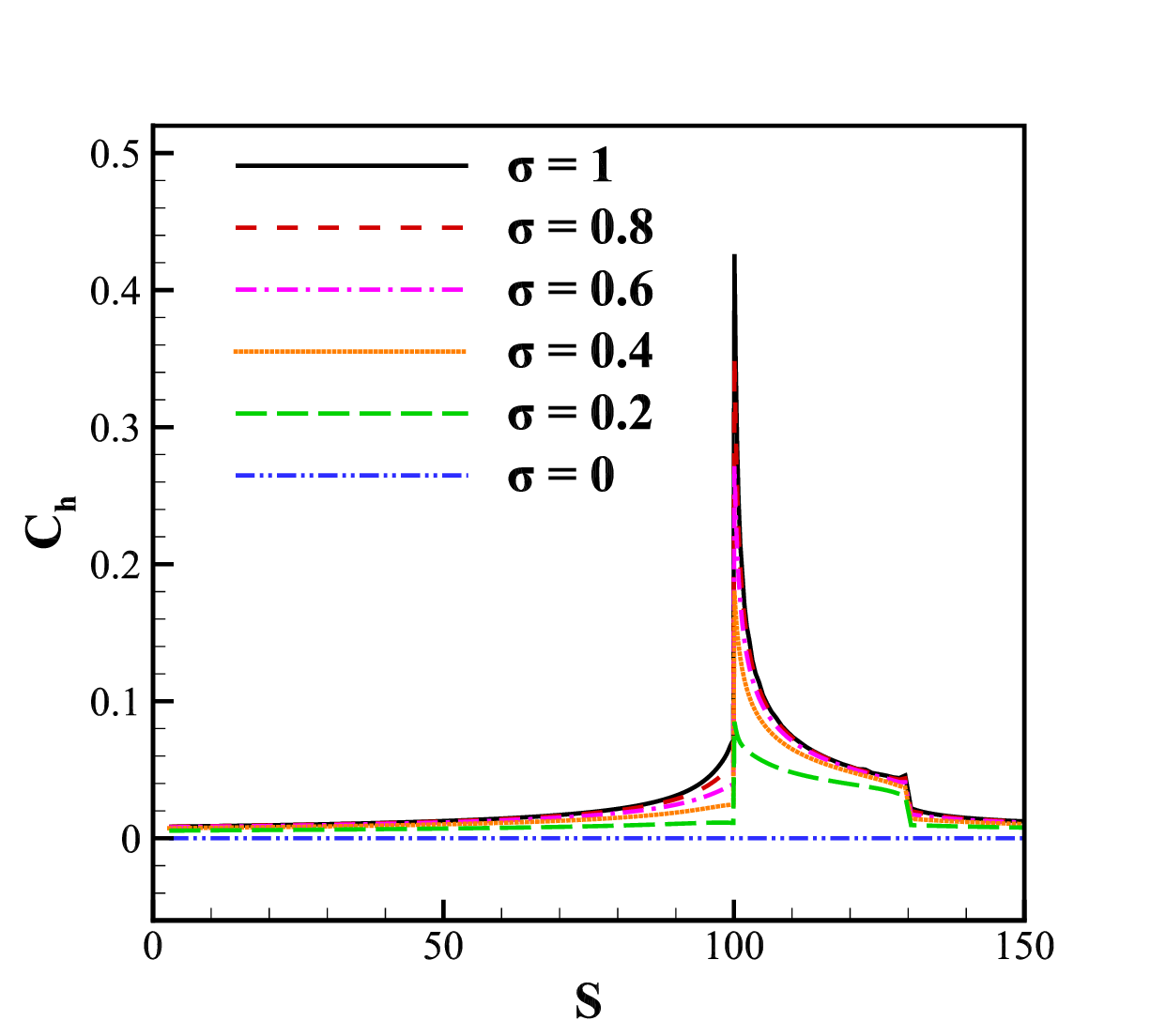}}
		\caption{\label{fig07}{Comparison of the (a) pressure coefficient, (b) skin friction coefficient, and (c) heat transfer coefficient on the surface of flat plate with different $\sigma$ ($Ma$ = 4.89, $Kn$ = 0.0078, $T_{\infty}$ = 116 K, $T_{w}$ = 290 K).}}
	\end{figure}

	When $\sigma$ equals 1, indicating a completely diffuse reflection, the distributions of pressure, skin friction, and heat transfer coefficients on the wall are presented in Fig. \ref{fig04}. The obtained results closely align with those from DS2V. In the comparison of variables on the solid wall for the flat plate simulation, the horizontal coordinate is denoted as S, with the trailing edge of the flat plate's upper surface serving as the starting point in this coordinate. The distributions of pressure, skin friction, and heat transfer coefficients on the wall at $\sigma$ = 0.8 are depicted in Fig. \ref{fig05}. Fig. \ref{fig06} illustrates the distributions of pressure, skin friction, and heat transfer coefficients on the wall at $\sigma$ = 0, i.e., complete specular reflection. The comparison of these results indicates that the Maxwell GSI model based on the unstructured discrete velocity mesh is accurate and well-aligned with DS2V calculations.	
	The density, temperature, and horizontal velocity contours for diffuse reflection and specular reflection from the simulation are depicted in Fig. \ref{fig06add1} and Fig. \ref{fig06add2}, respectively. It is evident that the specular reflection boundary, differing significantly from the diffuse reflection boundary, is comparable to the inviscid boundary and has no hysteresis impact on the tangential flow. Fig. \ref{fig06add4} illustrates the distribution function contours on the bevel of the flat plate. The distribution function for a diffuse reflection wall follows the Maxwellian distribution, while the reflection distribution function for a specular reflection wall is symmetric with the incidence distribution function about the boundary. To further investigate the impact of the GSI model during the change from complete diffuse reflection to complete specular reflection, Fig. \ref{fig07} presents the distribution of physical quantities on the flat plate's surface for $\sigma$ values of 1, 0.8, 0.6, 0.4, 0.2, and 0, respectively.

	\subsubsection{Mach 5 rarefied gas flow over a circular cylinder in the transitional regime}
\begin{figure}
		\centering
		\subfigure[]{\label{fig08a}\includegraphics[width=0.45\textwidth]{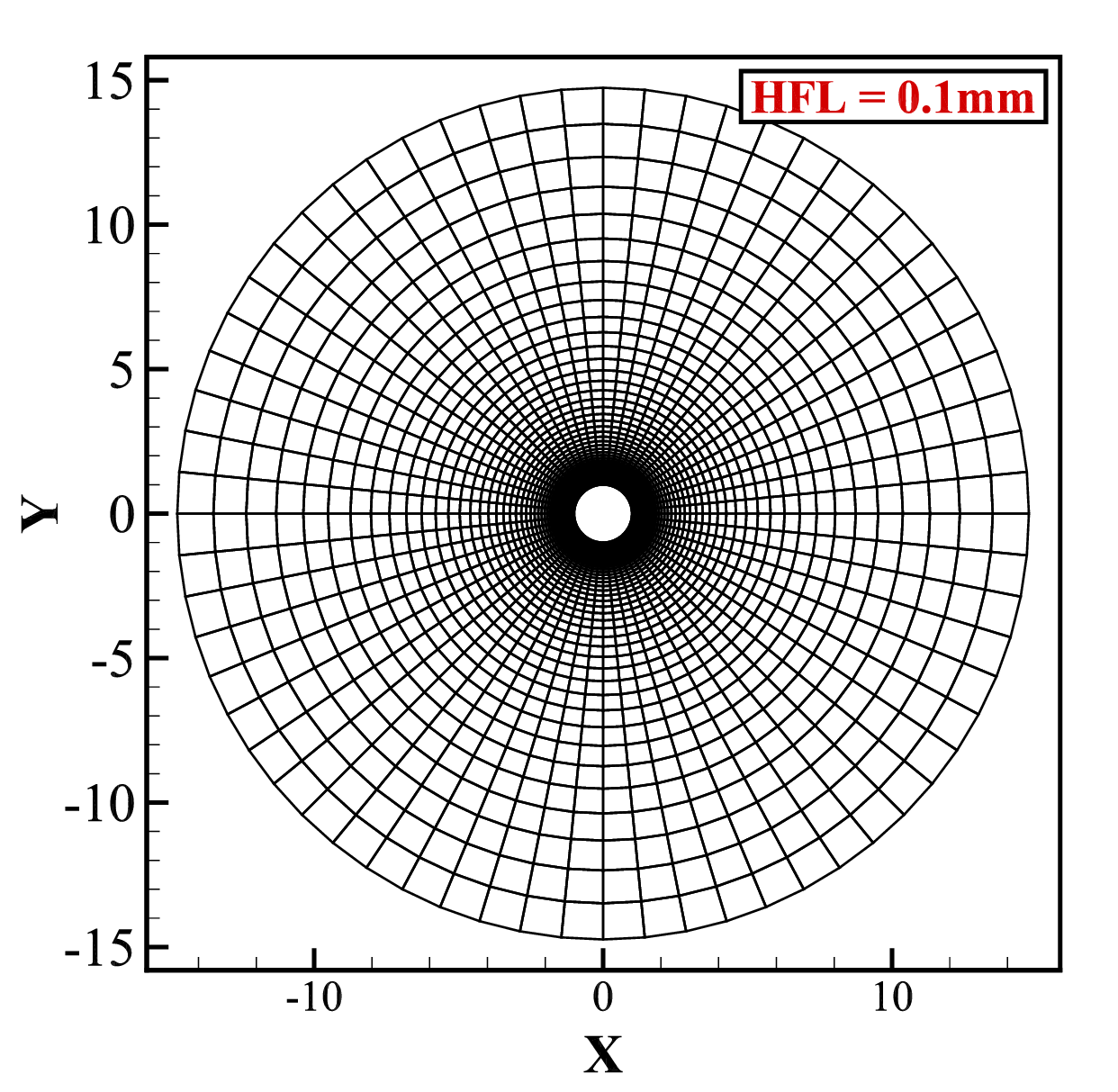}}
		\subfigure[]{\label{fig08b}\includegraphics[width=0.44\textwidth]{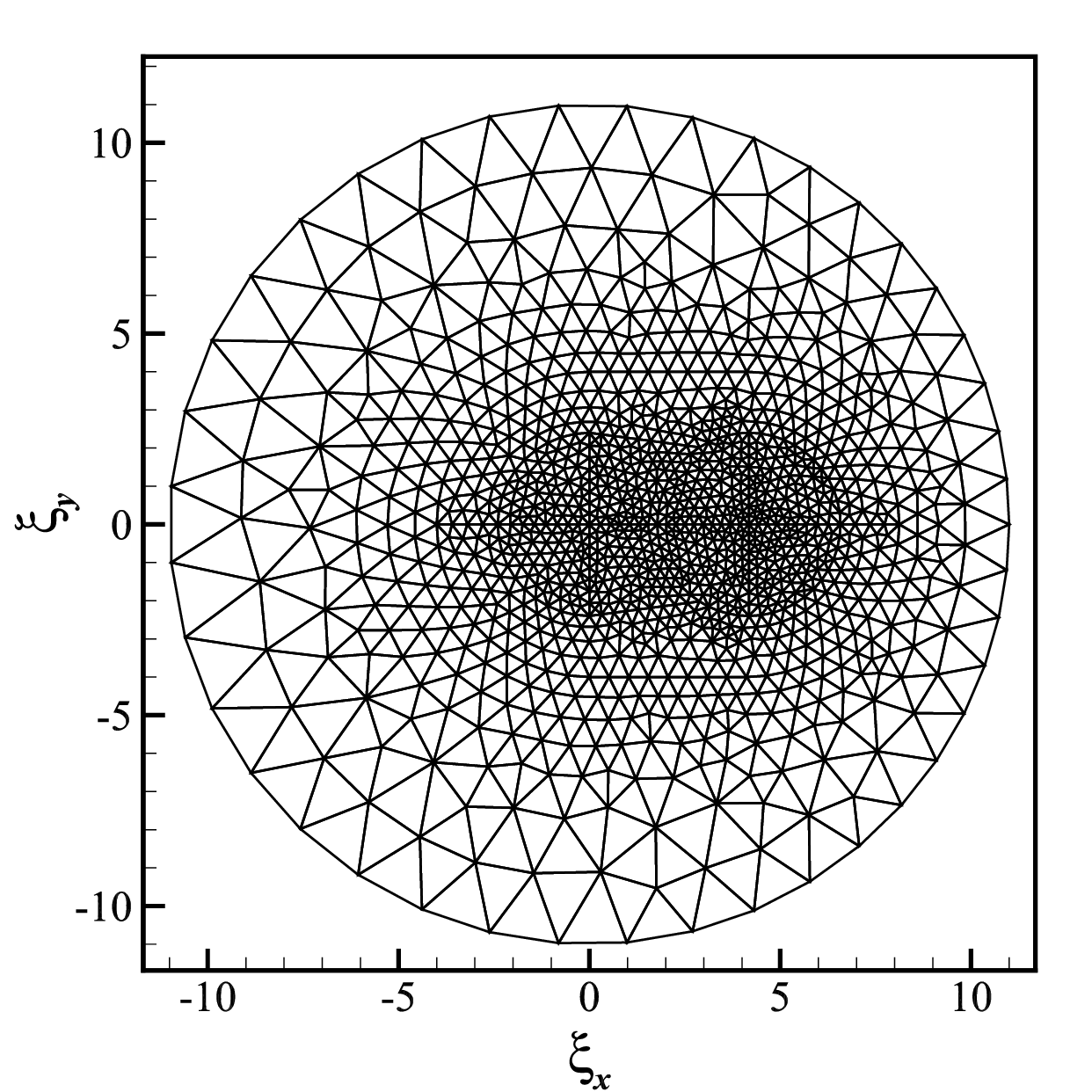}}
		\caption{\label{fig08}{The (a) unstructured physical mesh and (b) unstructured velocity mesh for the supersonic flow over a cylinder ($Ma$ = 5.0, $Kn$ = 1.0, $T_{\infty}$ = 273 K, $T_{w}$ = 273 K).}}
	\end{figure}

	The performance of the current GSI algorithm on a curved surface is demonstrated through simulating high-speed flow over a circular cylinder. The cylinder's radius is indicated as $R$ = 0.01 m. In relation to the cylinder radius, the Knudsen number of the freestream is specified as $Kn$ = 1.0. Argon is selected as the working gas, and the VHS molecular model with $\omega$ = 0.81 is employed. Operating at a Mach number of 5.0 and a temperature of 273 K, the freestream exhibits a velocity of 1539.3 m/s. The temperature of the cylinder surface is maintained at 273 K.
	The computational domain is discretized into 64x61 control volumes, forming a circular region with a center at (0,0) and a radius of 15$R$ (see Fig. \ref{fig08a}). The mesh quality along the wall must satisfy specific criteria to precisely represent the heat flux on the cylindrical surface. In this depiction, the HFL of the mesh on the cylinder's surface is set at 0.1 mm. The unstructured discrete velocity mesh employed in this example consists of 2391 cells, as illustrated in Fig. \ref{fig08b}.
	
\begin{figure}
		\centering
		\subfigure[]{\label{fig09a}\includegraphics[width=0.45\textwidth]{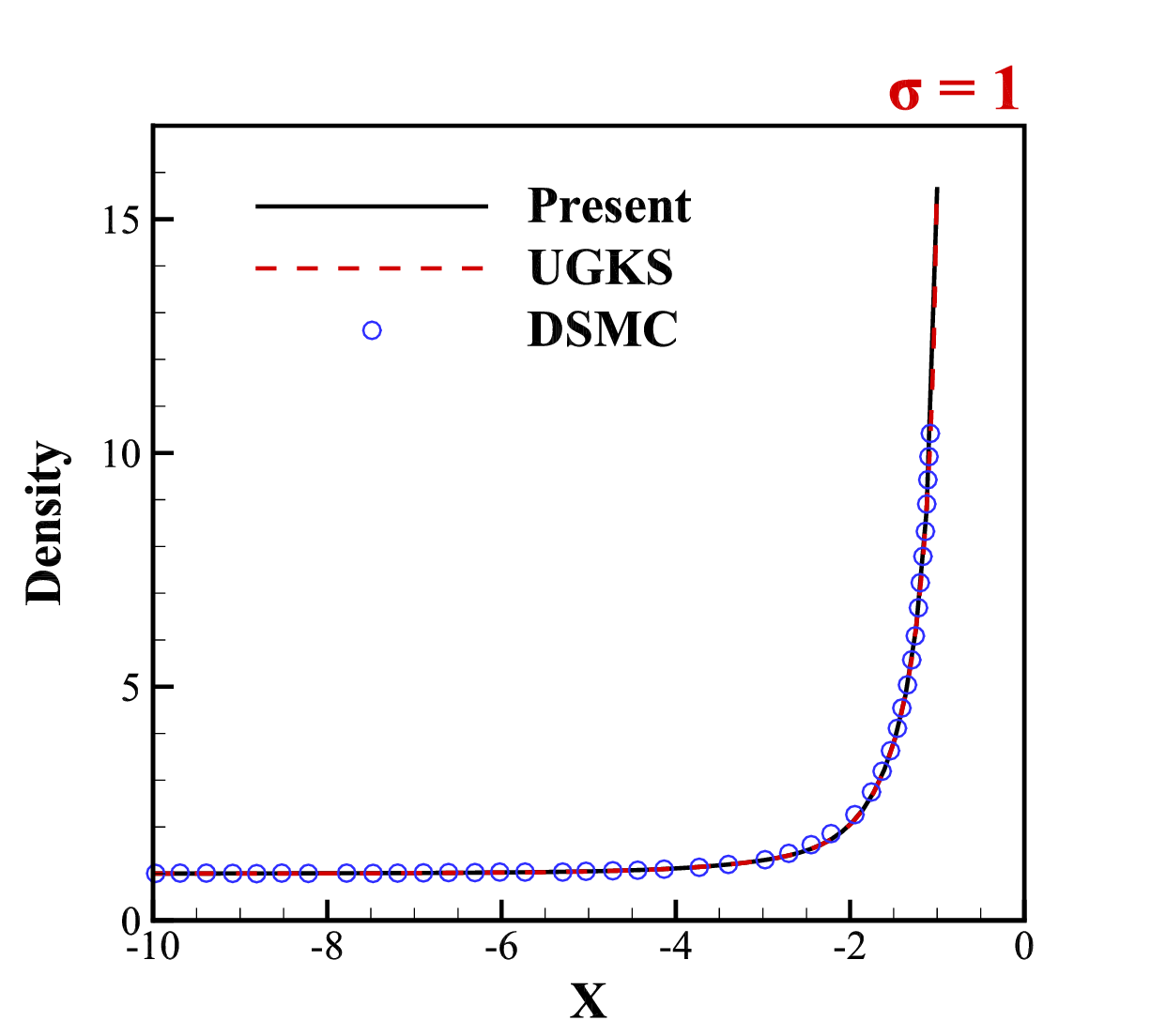}}
		\subfigure[]{\label{fig09b}\includegraphics[width=0.45\textwidth]{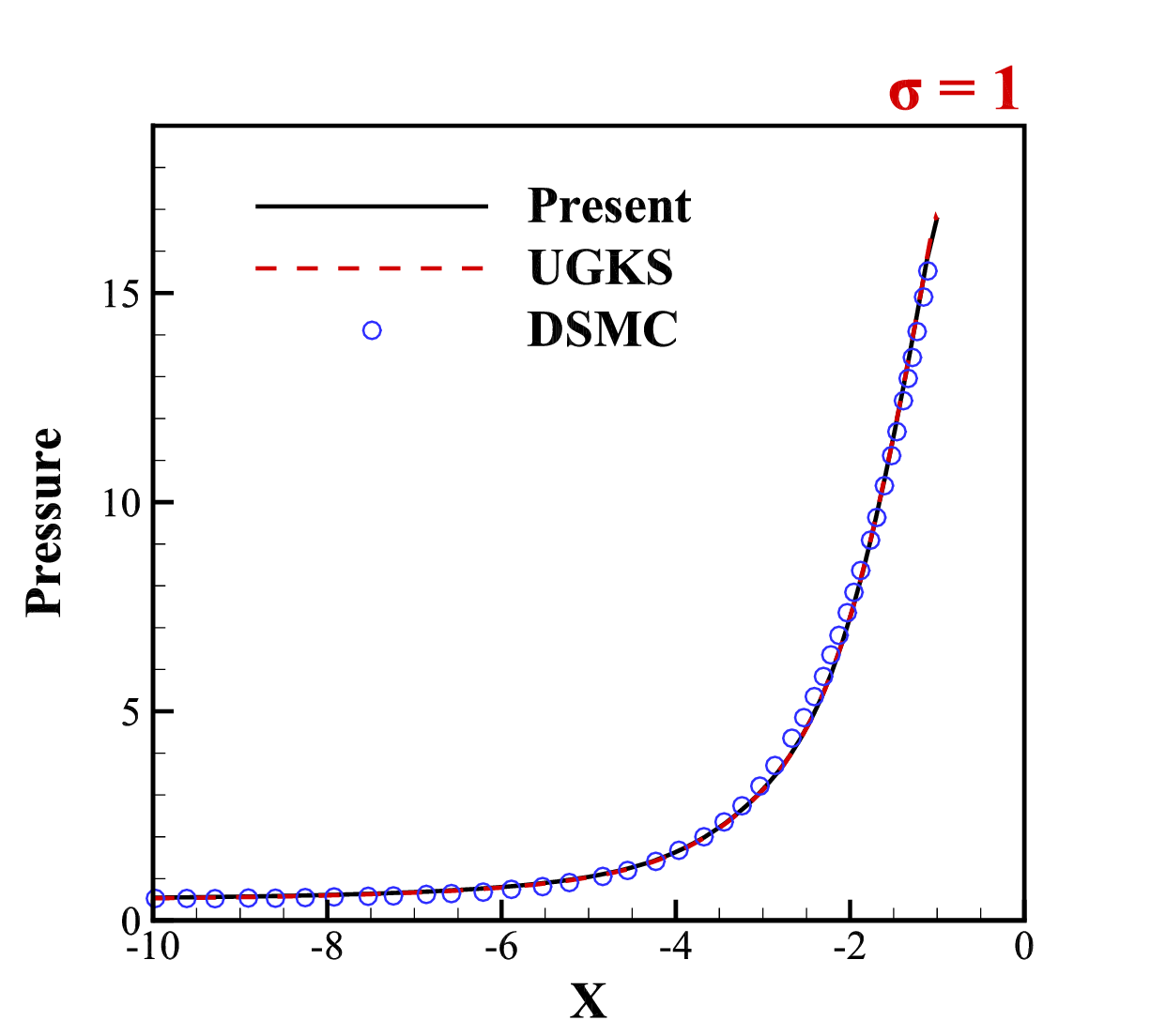}}
		\subfigure[]{\label{fig09c}\includegraphics[width=0.45\textwidth]{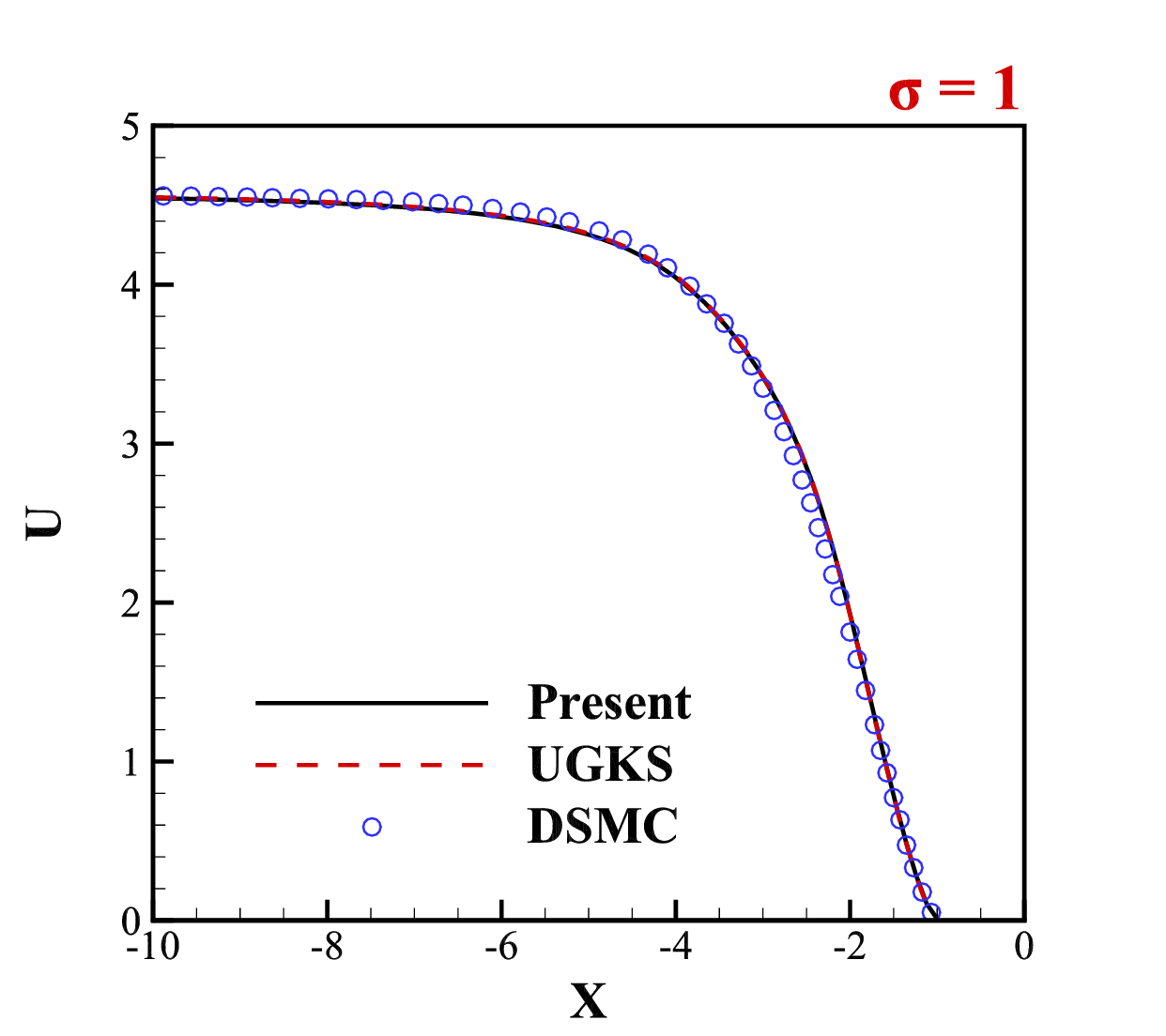}}
		\subfigure[]{\label{fig09d}\includegraphics[width=0.45\textwidth]{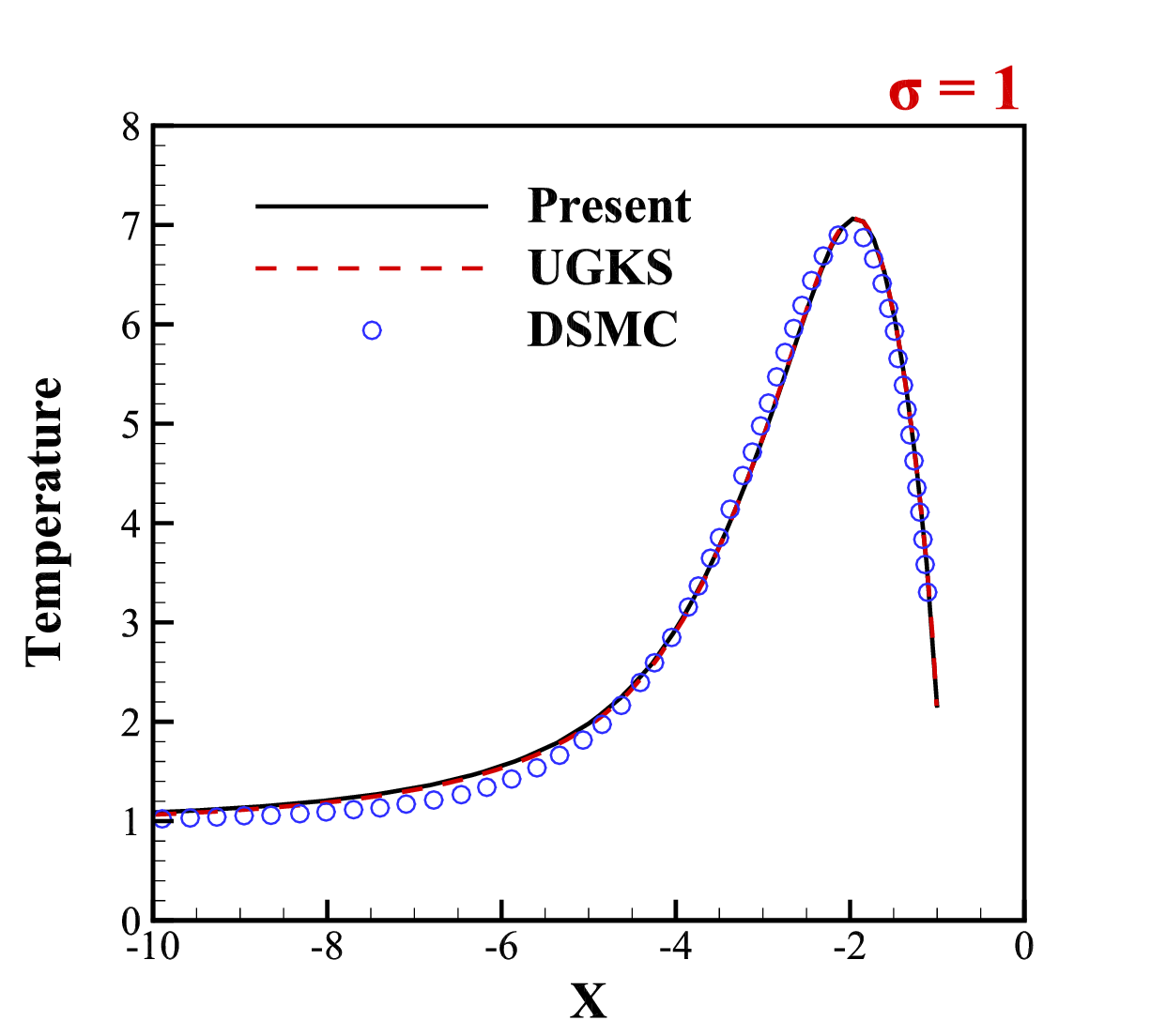}}
		\caption{\label{fig09}{Comparison of the (a) density, (b) pressure, (c) horizontal velocity, and (d) temperature along the stationary line in front of the cylinder with $\sigma$ = 1 ($Ma$ = 5.0, $Kn$ = 1.0, $T_{\infty}$ = 273 K, $T_{w}$ = 273 K).}}
	\end{figure}

\begin{figure}
		\centering
		\subfigure[]{\label{fig10a}\includegraphics[width=0.45\textwidth]{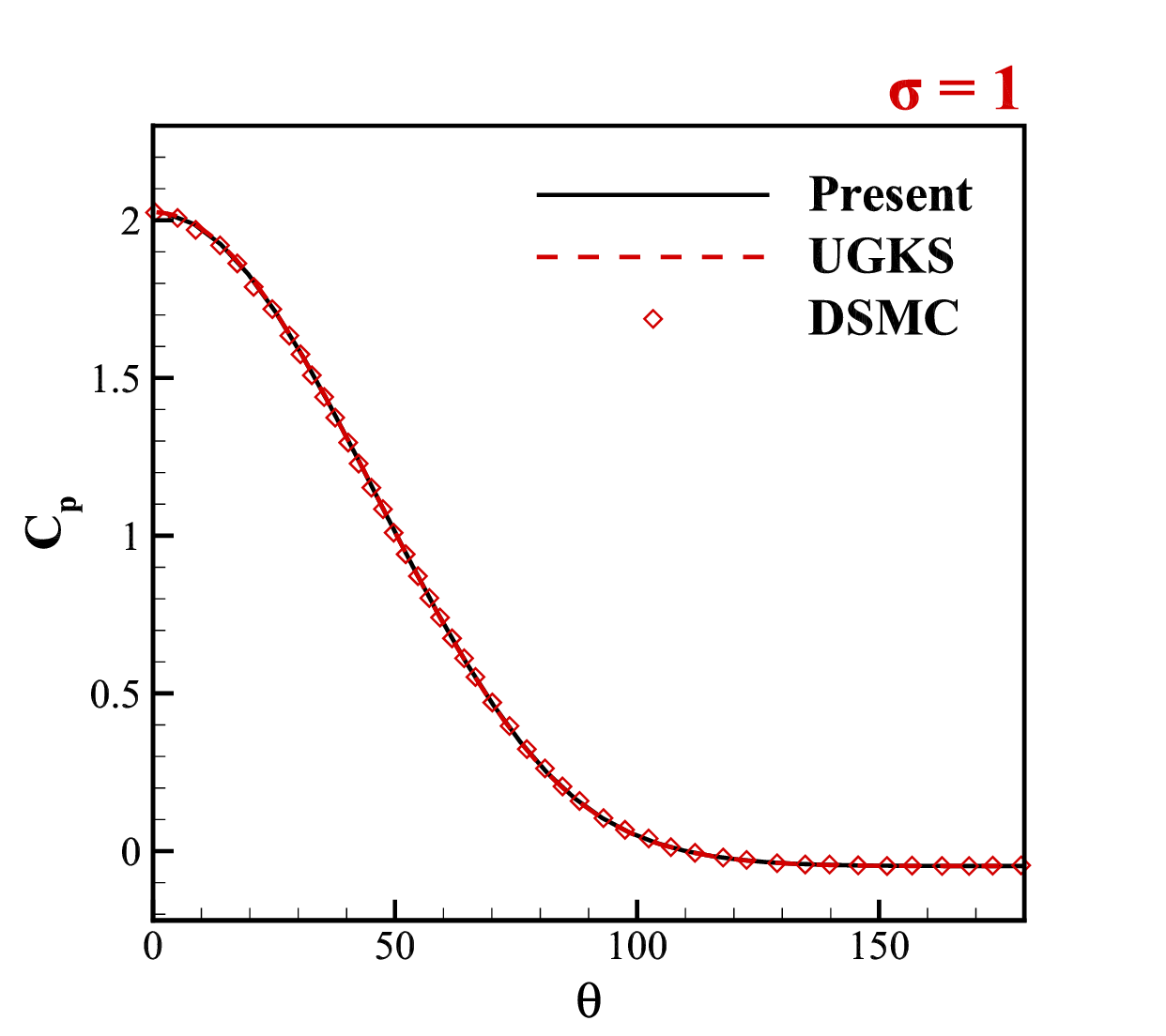}}
		\subfigure[]{\label{fig10b}\includegraphics[width=0.45\textwidth]{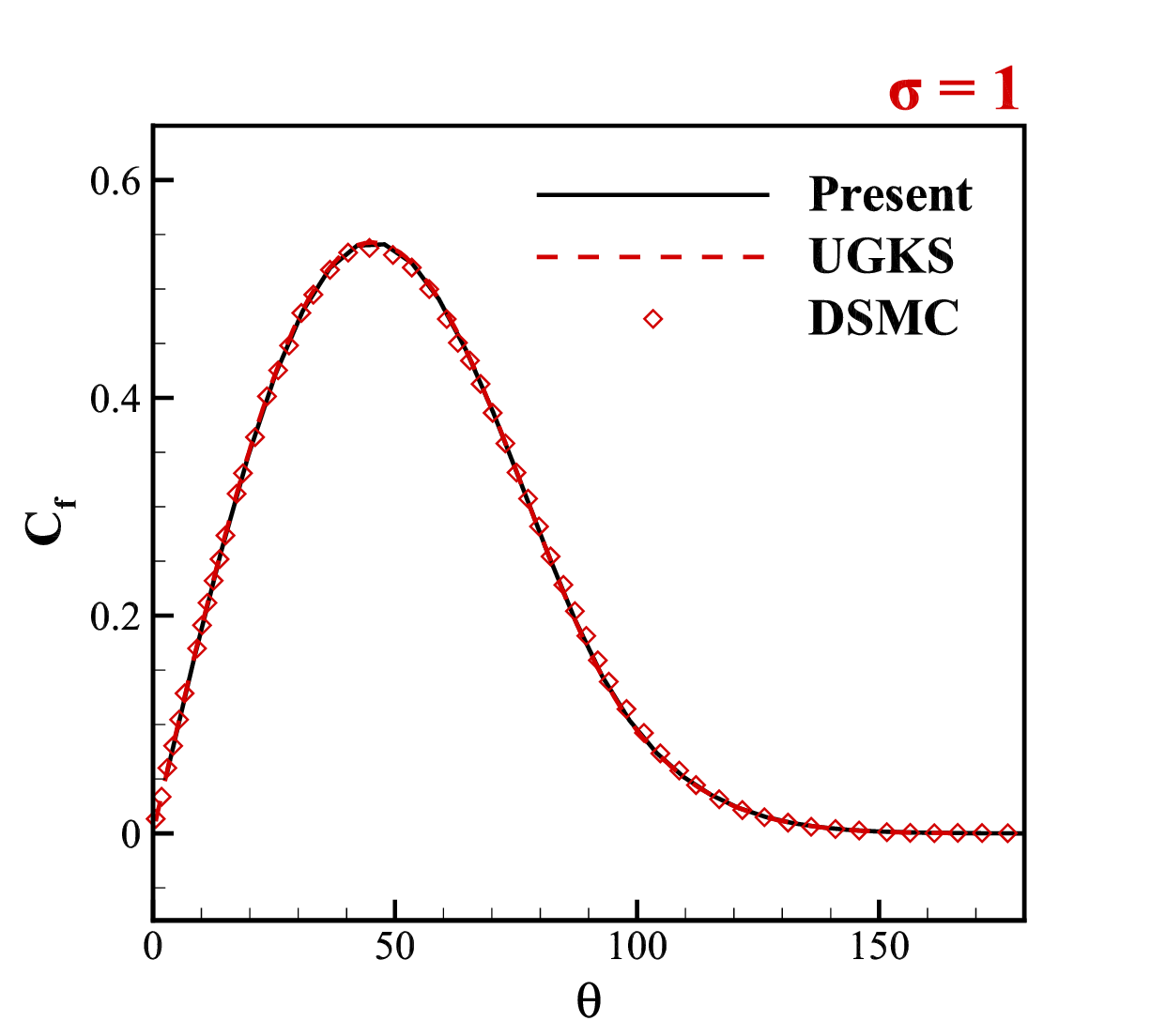}}
		\subfigure[]{\label{fig10c}\includegraphics[width=0.45\textwidth]{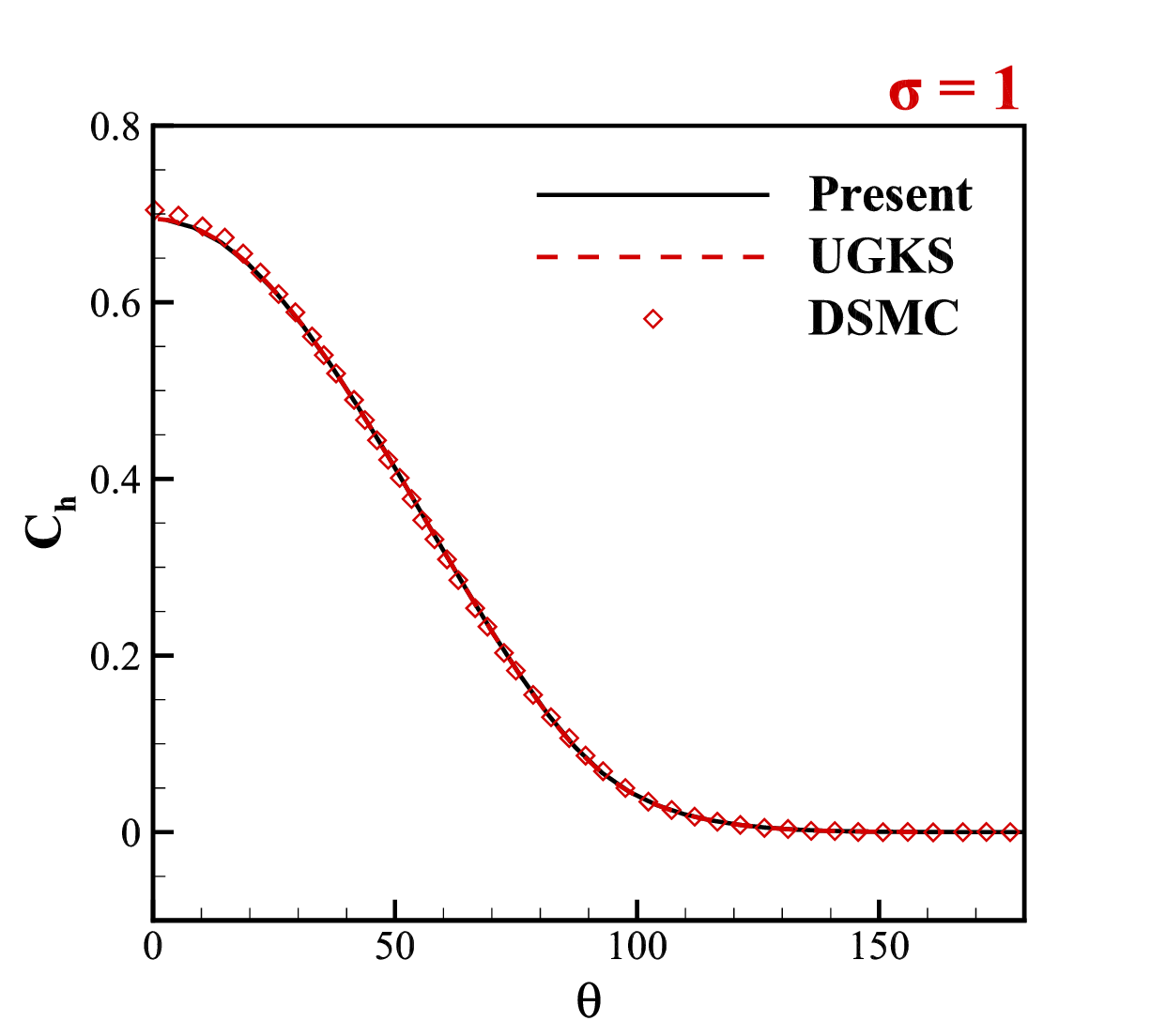}}
		\caption{\label{fig10}{Comparison of the (a) pressure coefficient, (b) skin friction coefficient, and (c) heat transfer coefficient on the surface of cylinder with $\sigma$ = 1 ($Ma$ = 5.0, $Kn$ = 1.0, $T_{\infty}$ = 273 K, $T_{w}$ = 273 K).}}
	\end{figure}

\begin{figure}
		\centering
		\subfigure[]{\label{fig11a}\includegraphics[width=0.45\textwidth]{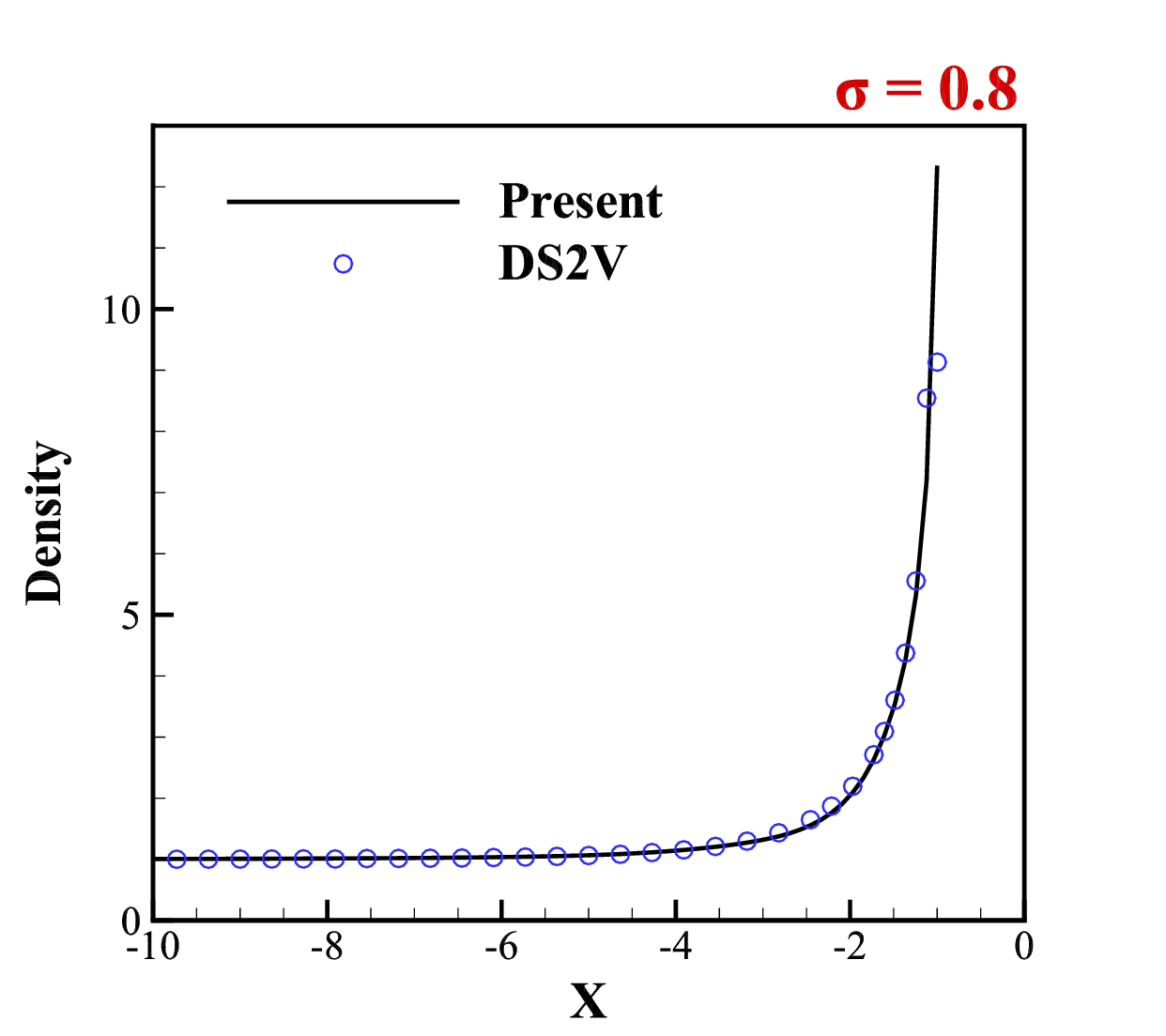}}
		\subfigure[]{\label{fig11b}\includegraphics[width=0.45\textwidth]{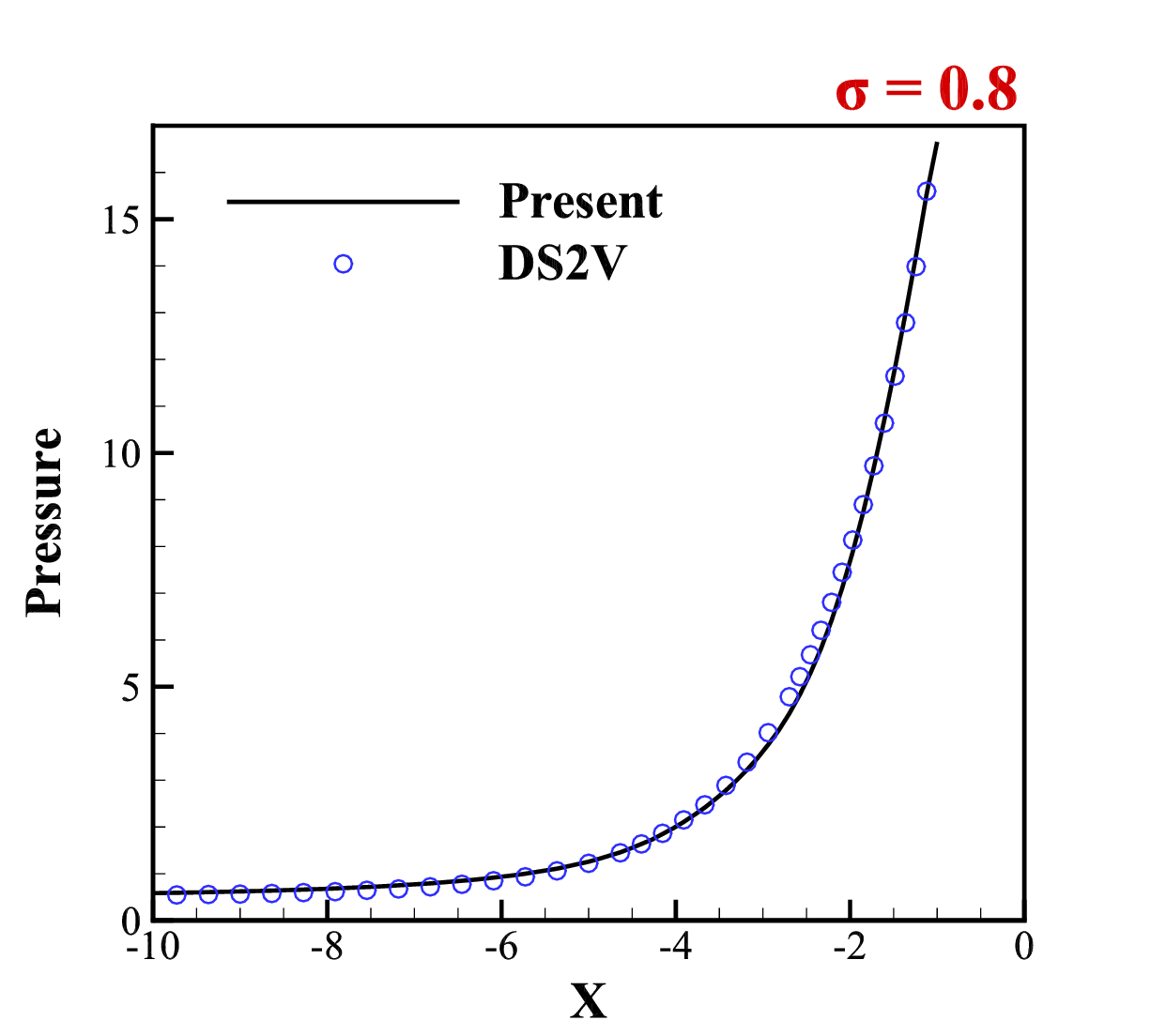}}
		\subfigure[]{\label{fig11c}\includegraphics[width=0.45\textwidth]{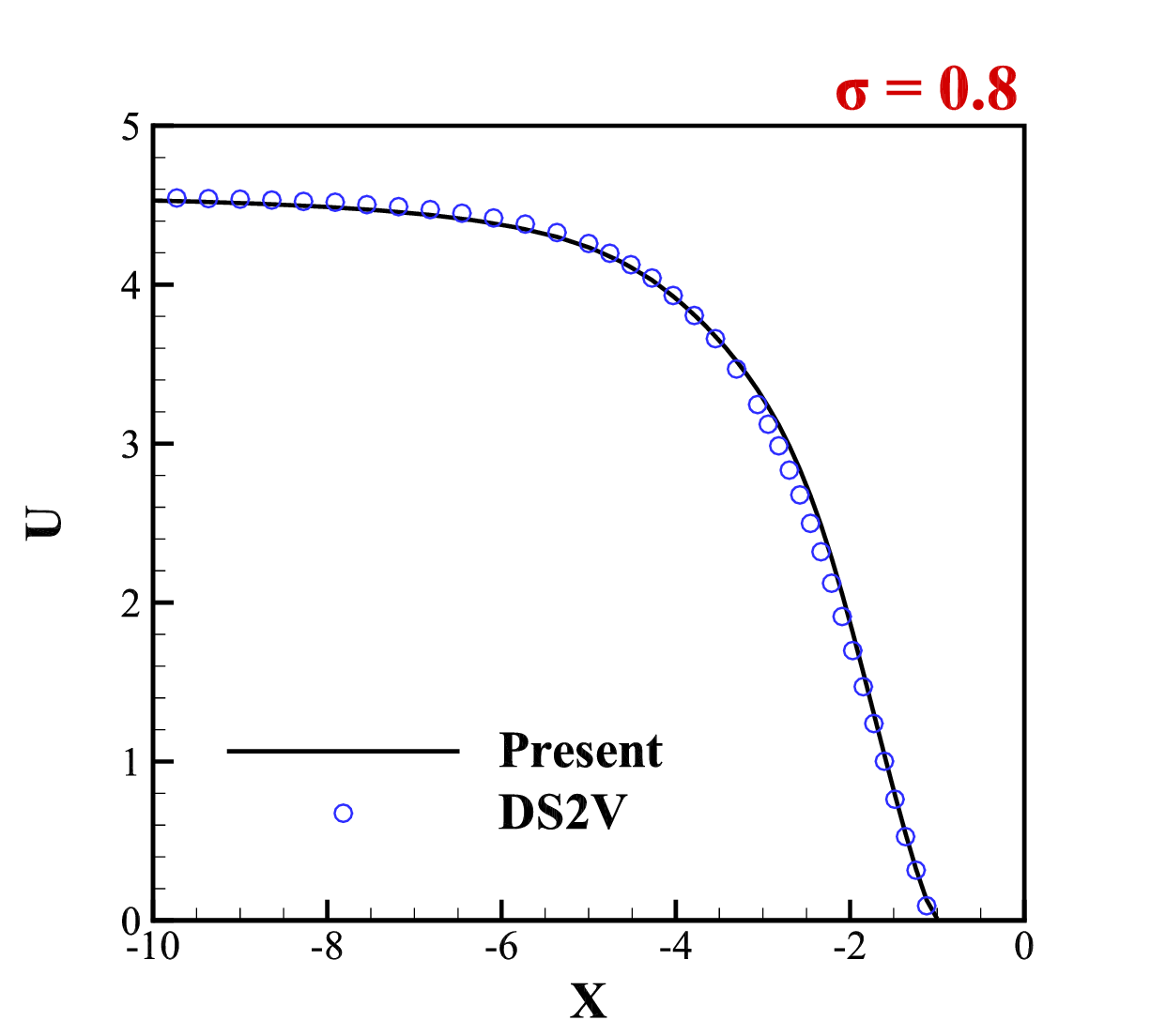}}
		\subfigure[]{\label{fig11d}\includegraphics[width=0.45\textwidth]{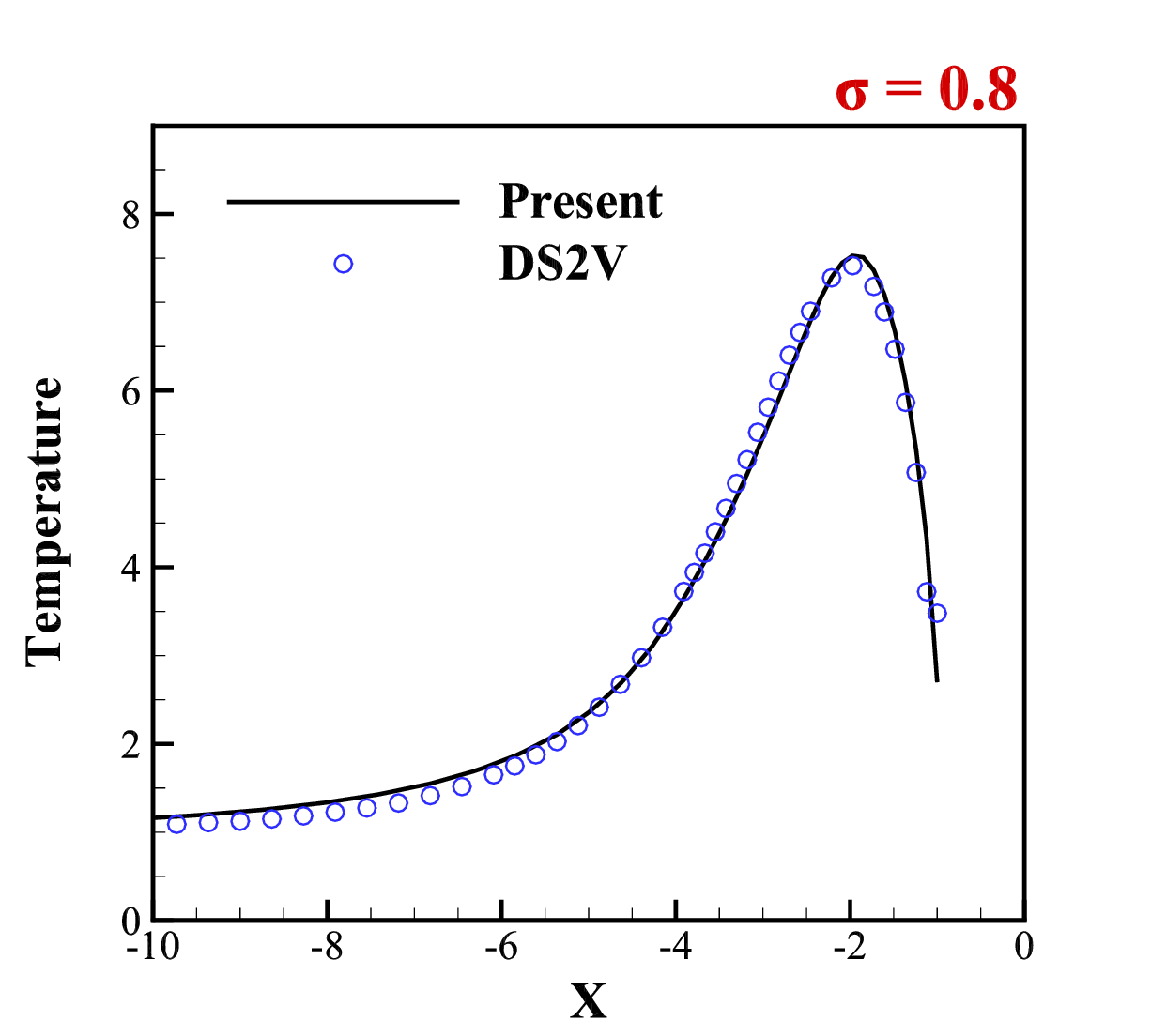}}
		\caption{\label{fig11}{Comparison of the (a) density, (b) pressure, (c) horizontal velocity, and (d) temperature along the stationary line in front of the cylinder with $\sigma$ = 0.8 ($Ma$ = 5.0, $Kn$ = 1.0, $T_{\infty}$ = 273 K, $T_{w}$ = 273 K).}}
	\end{figure}

\begin{figure}
		\centering
		\subfigure[]{\label{fig12a}\includegraphics[width=0.45\textwidth]{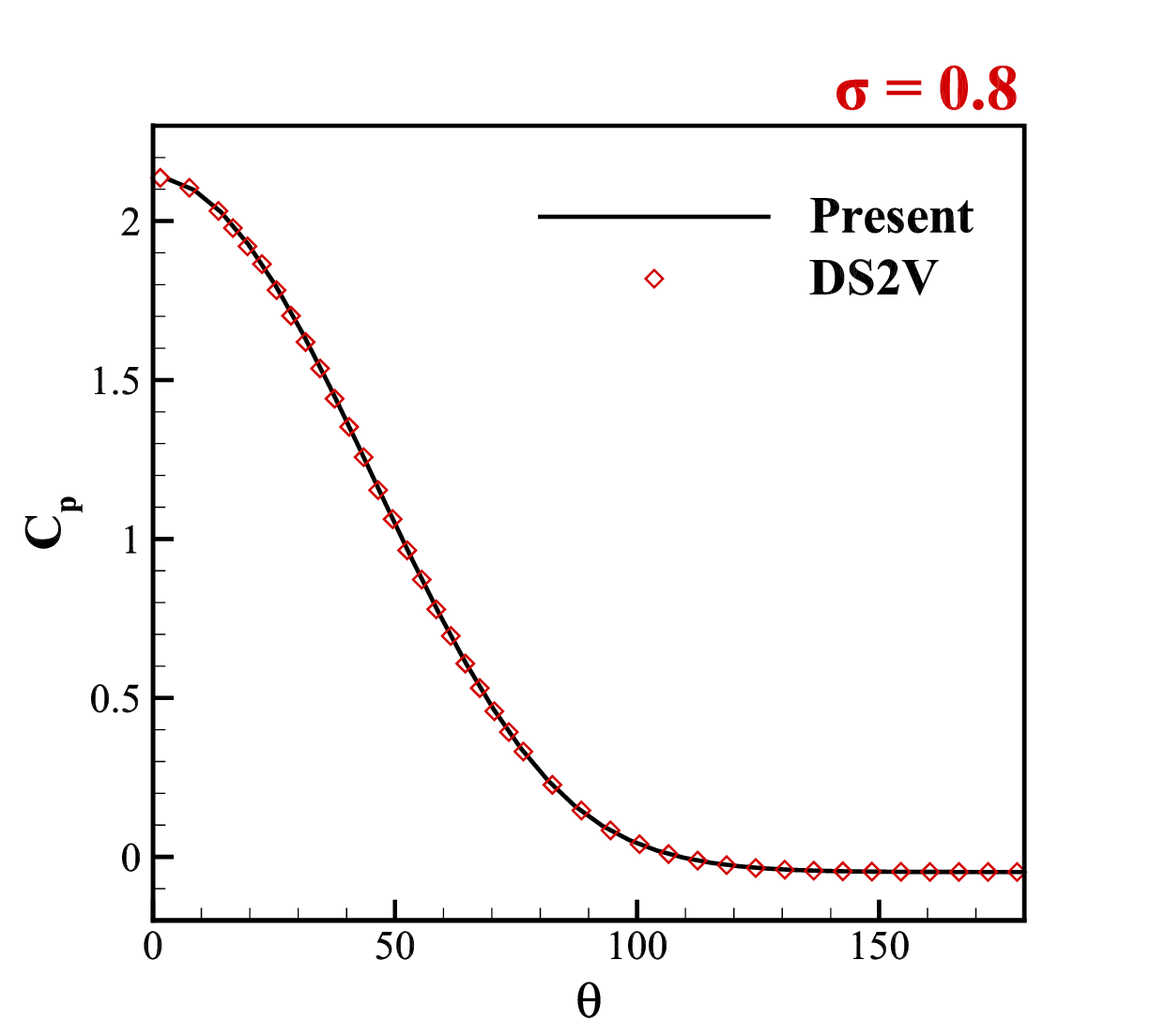}}
		\subfigure[]{\label{fig12b}\includegraphics[width=0.45\textwidth]{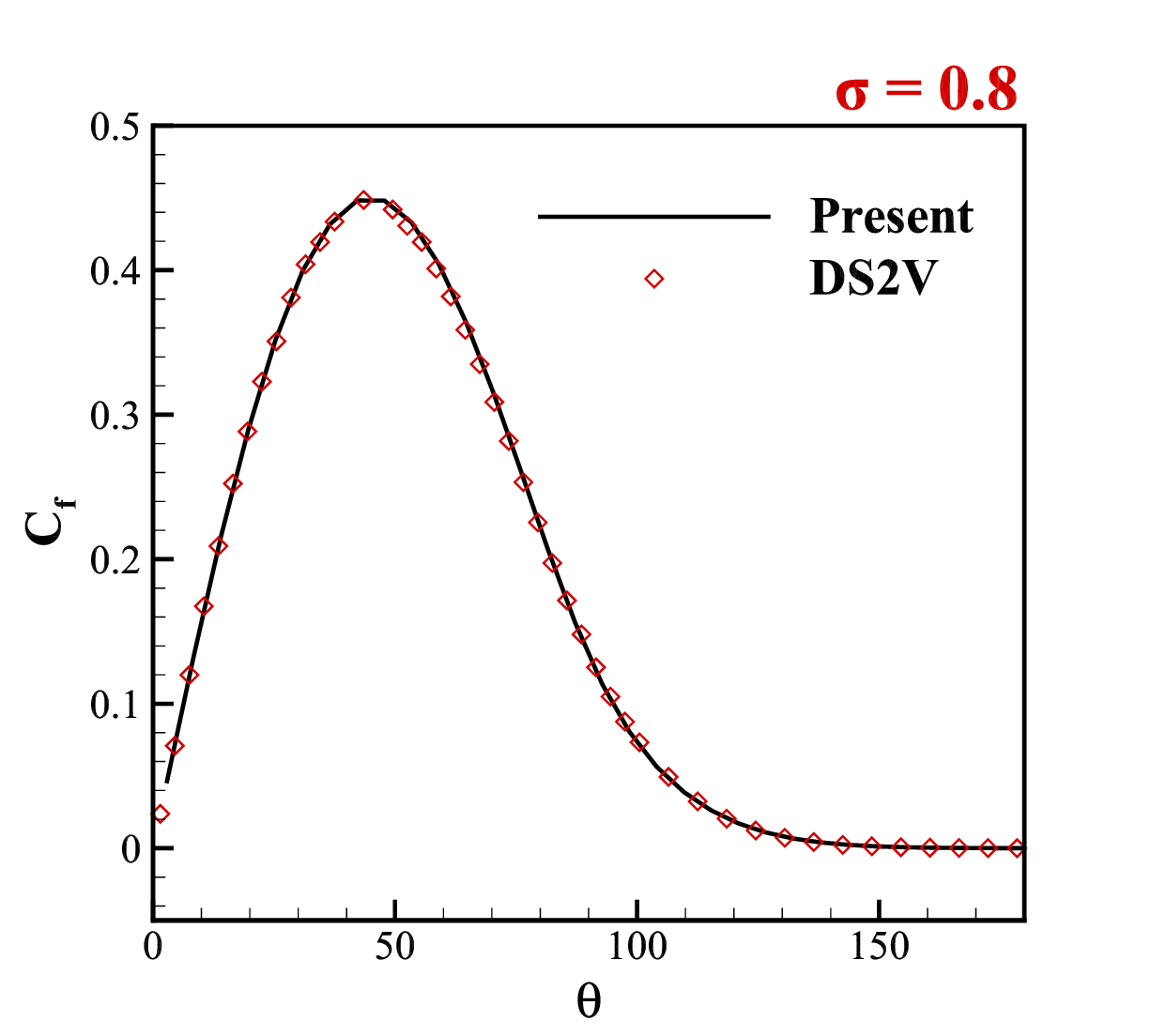}}
		\subfigure[]{\label{fig12c}\includegraphics[width=0.45\textwidth]{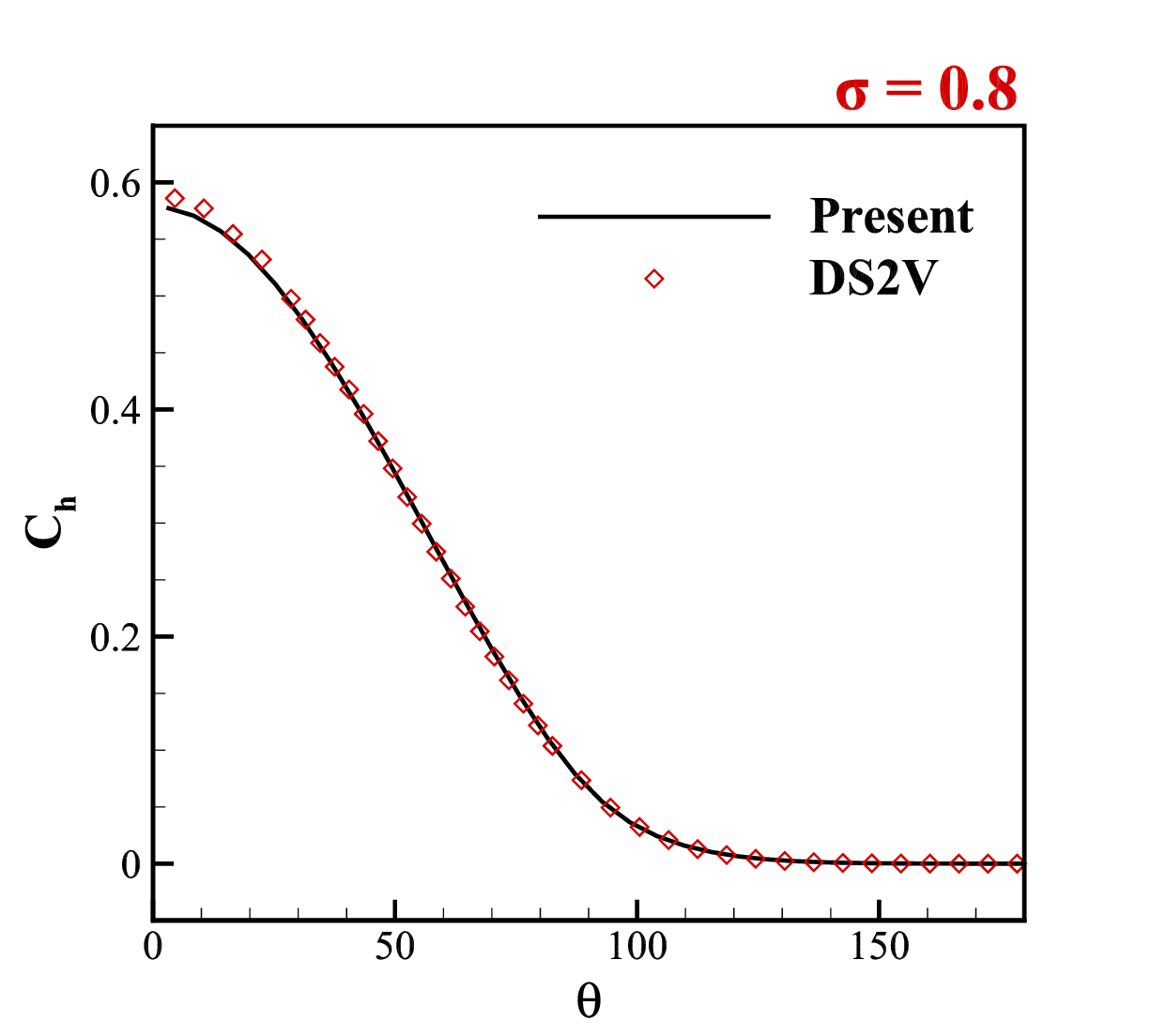}}
		\caption{\label{fig12}{Comparison of the (a) pressure coefficient, (b) skin friction coefficient, and (c) heat transfer coefficient on the surface of cylinder with $\sigma$ = 0.8 ($Ma$ = 5.0, $Kn$ = 1.0, $T_{\infty}$ = 273 K, $T_{w}$ = 273 K).}}
	\end{figure}

\begin{figure}
		\centering
		\subfigure[]{\label{fig13a}\includegraphics[width=0.45\textwidth]{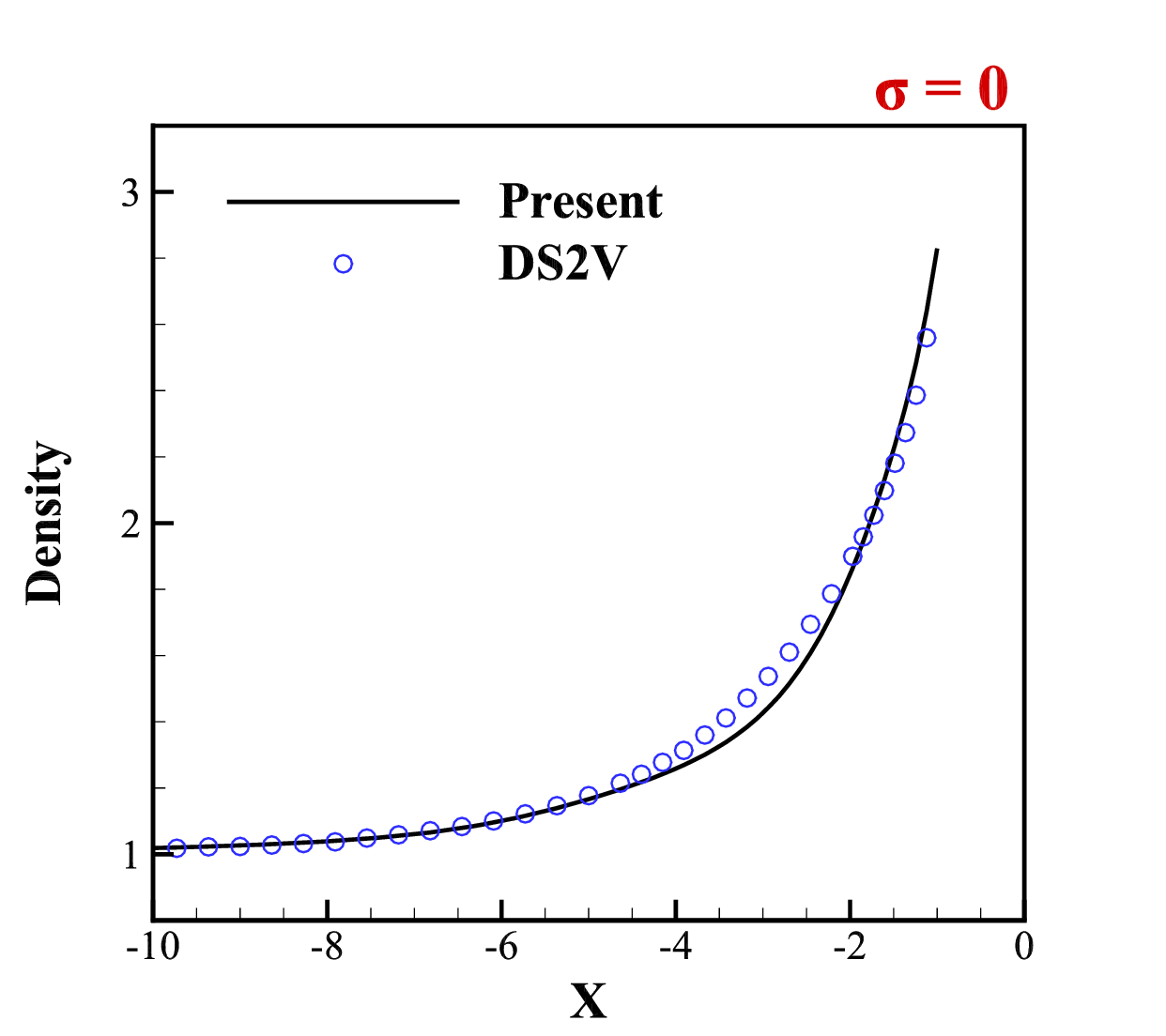}}
		\subfigure[]{\label{fig13b}\includegraphics[width=0.45\textwidth]{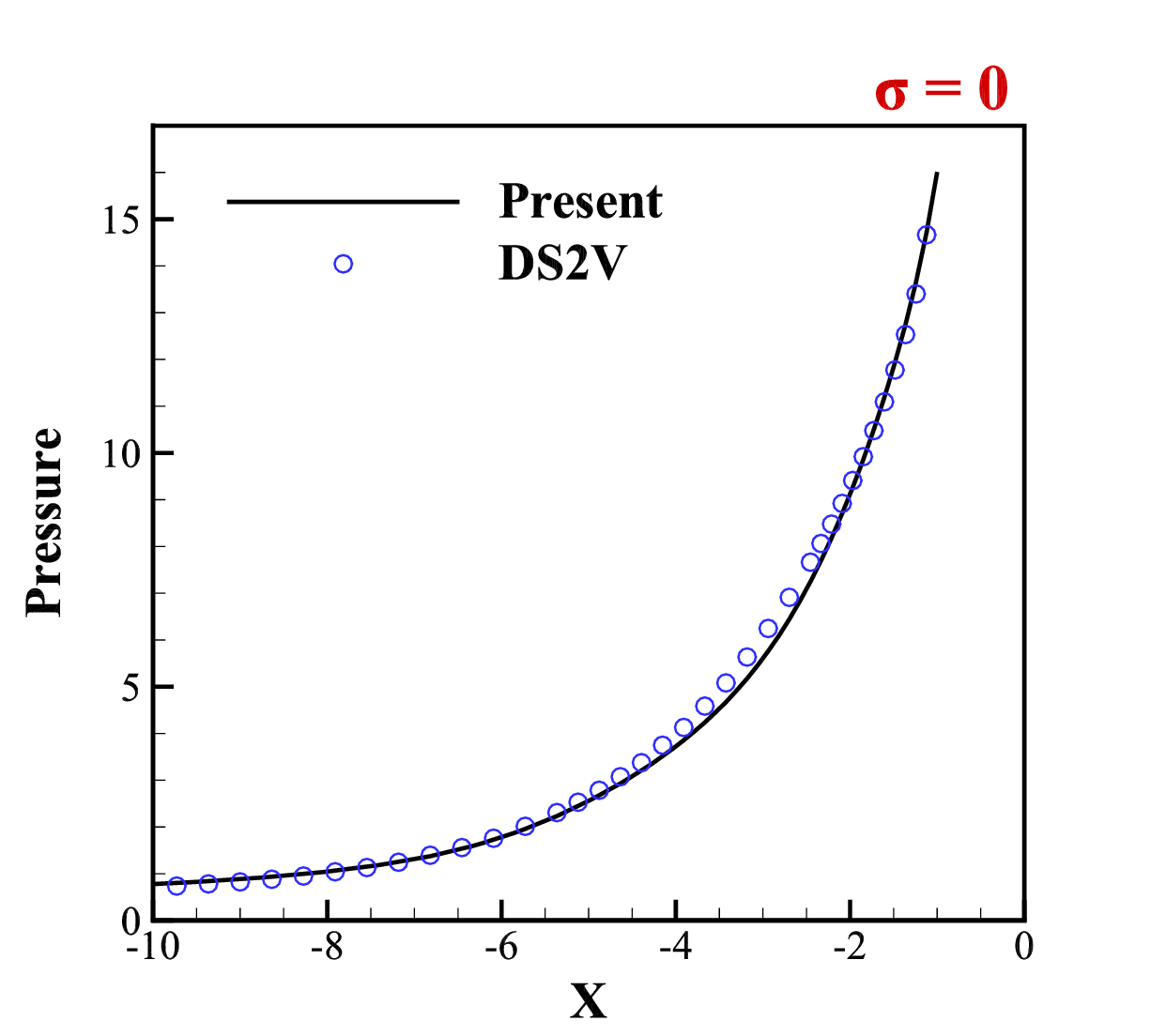}}
		\subfigure[]{\label{fig13c}\includegraphics[width=0.45\textwidth]{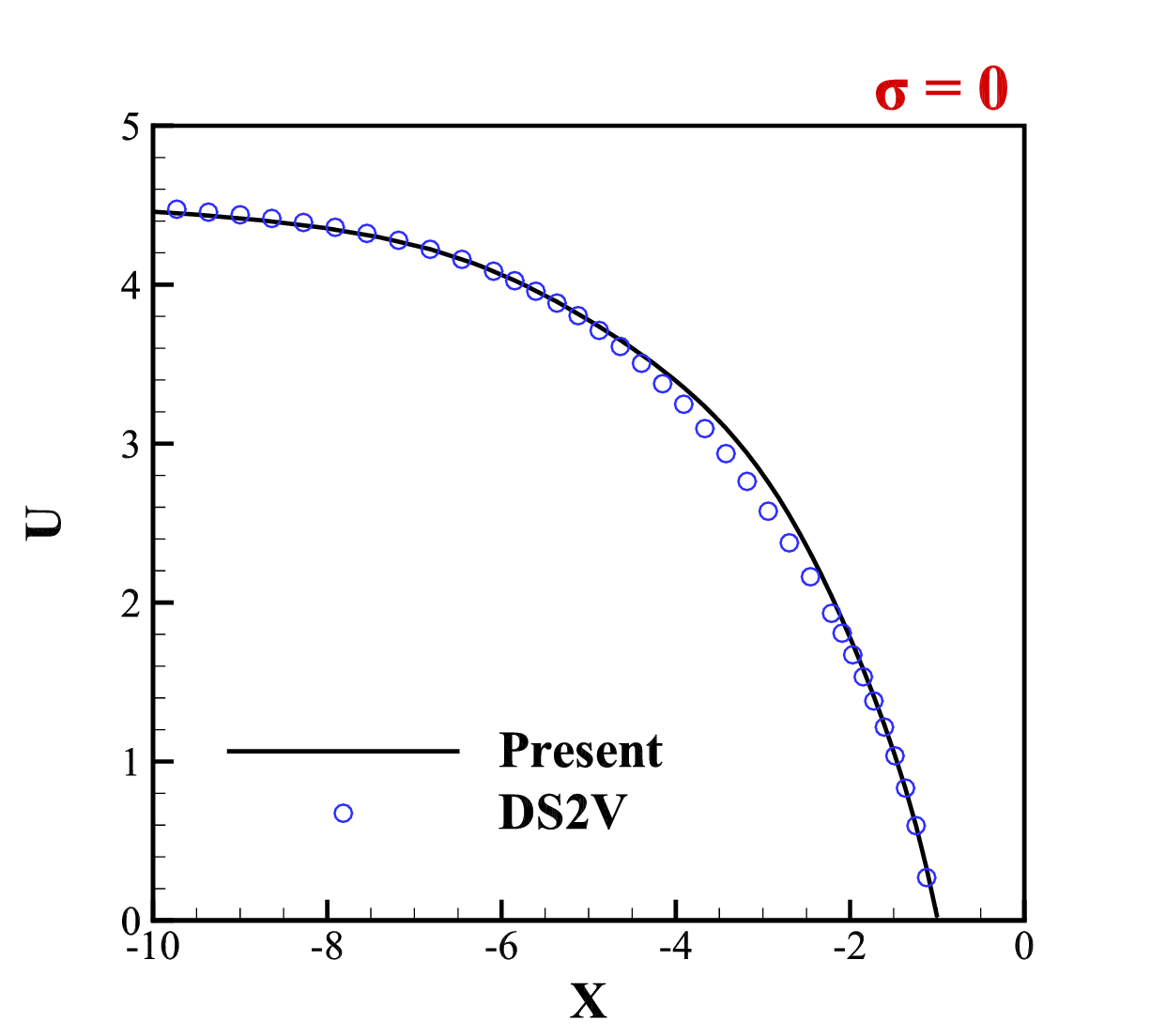}}
		\subfigure[]{\label{fig13d}\includegraphics[width=0.45\textwidth]{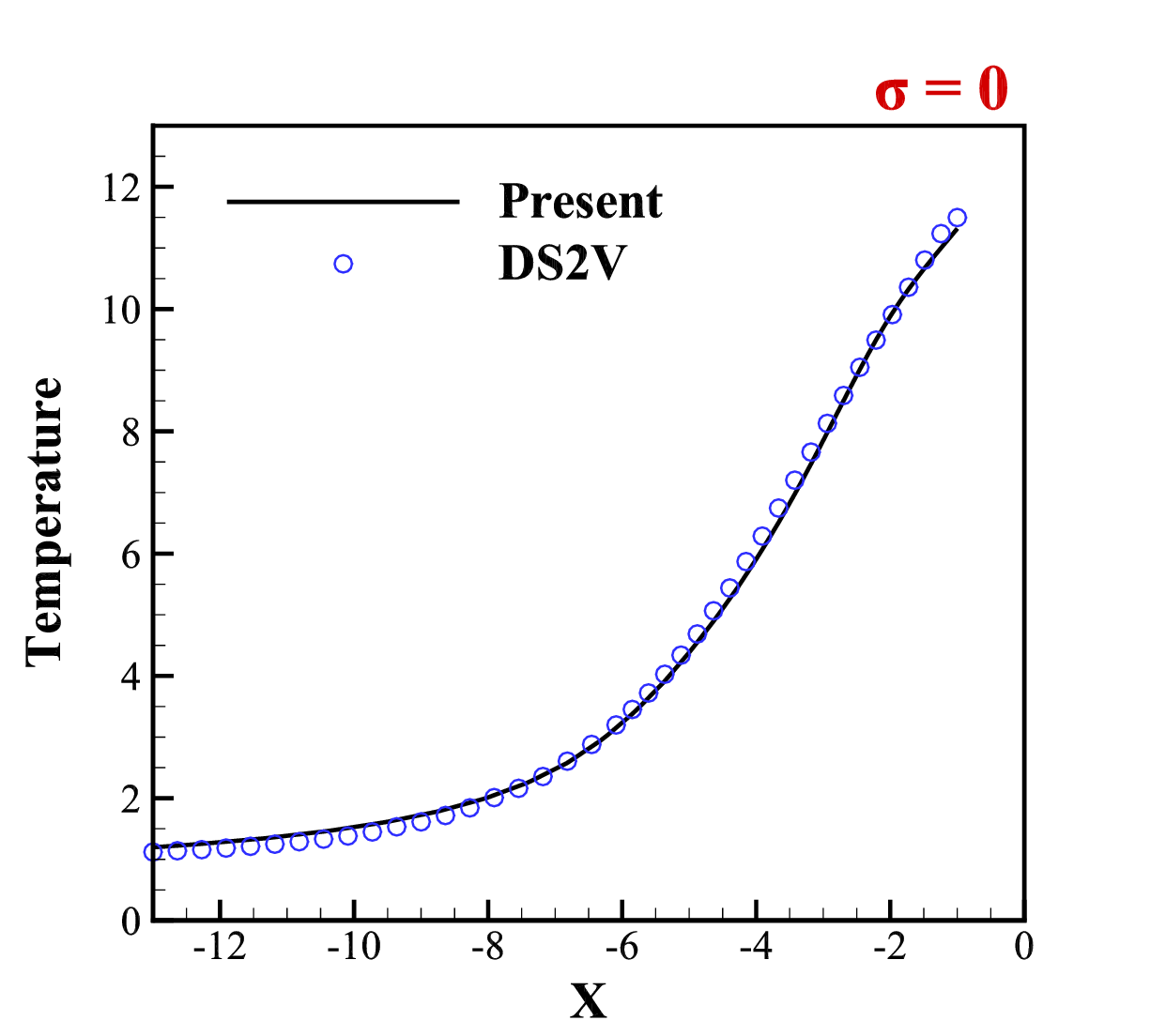}}
		\caption{\label{fig13}{Comparison of the (a) density, (b) pressure, (c) horizontal velocity, and (d) temperature along the stationary line in front of the cylinder with $\sigma$ = 0 ($Ma$ = 5.0, $Kn$ = 1.0, $T_{\infty}$ = 273 K, $T_{w}$ = 273 K).}}
	\end{figure}

\begin{figure}
		\centering
		\subfigure[]{\label{fig14a}\includegraphics[width=0.45\textwidth]{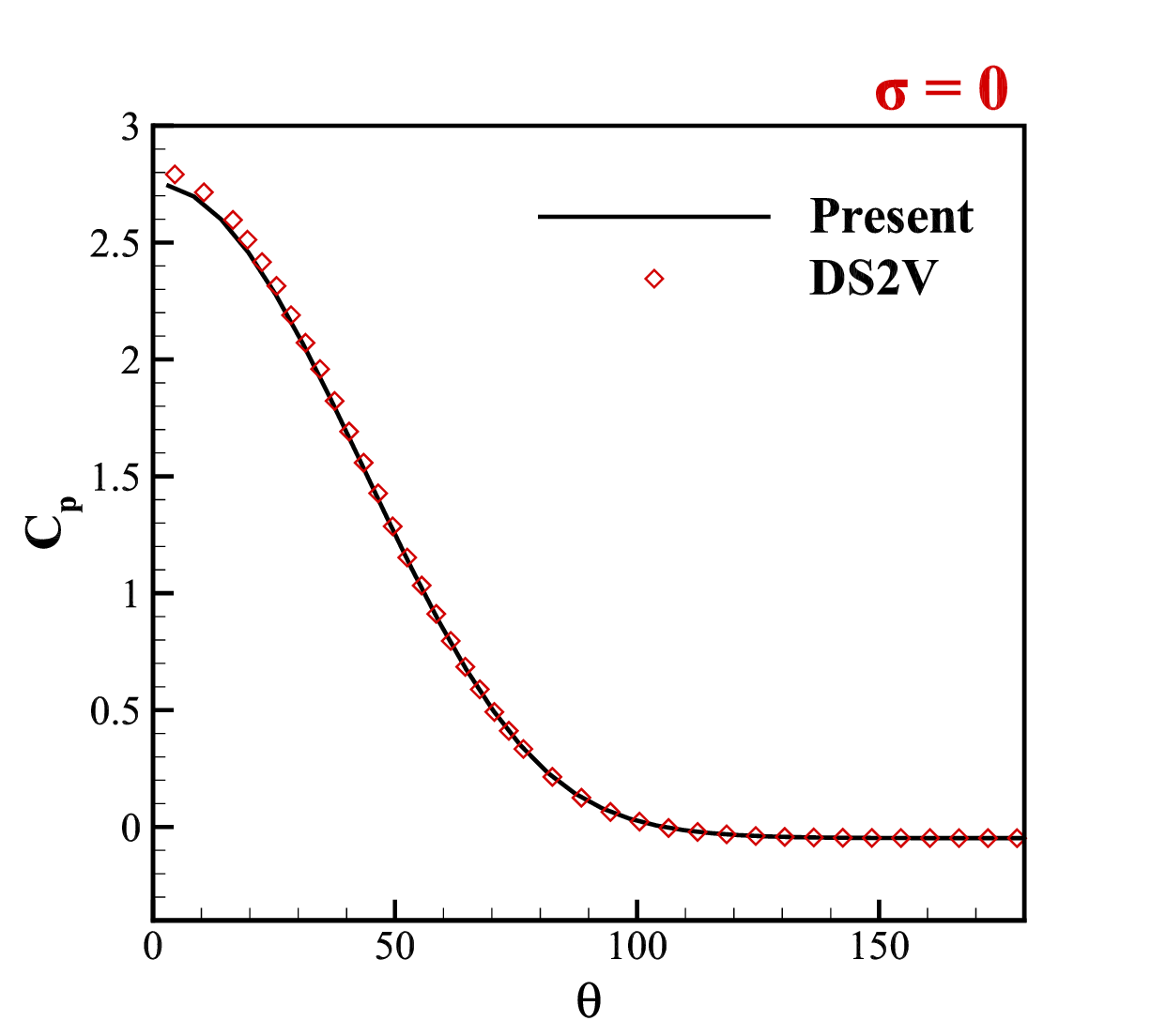}}
		\subfigure[]{\label{fig14b}\includegraphics[width=0.45\textwidth]{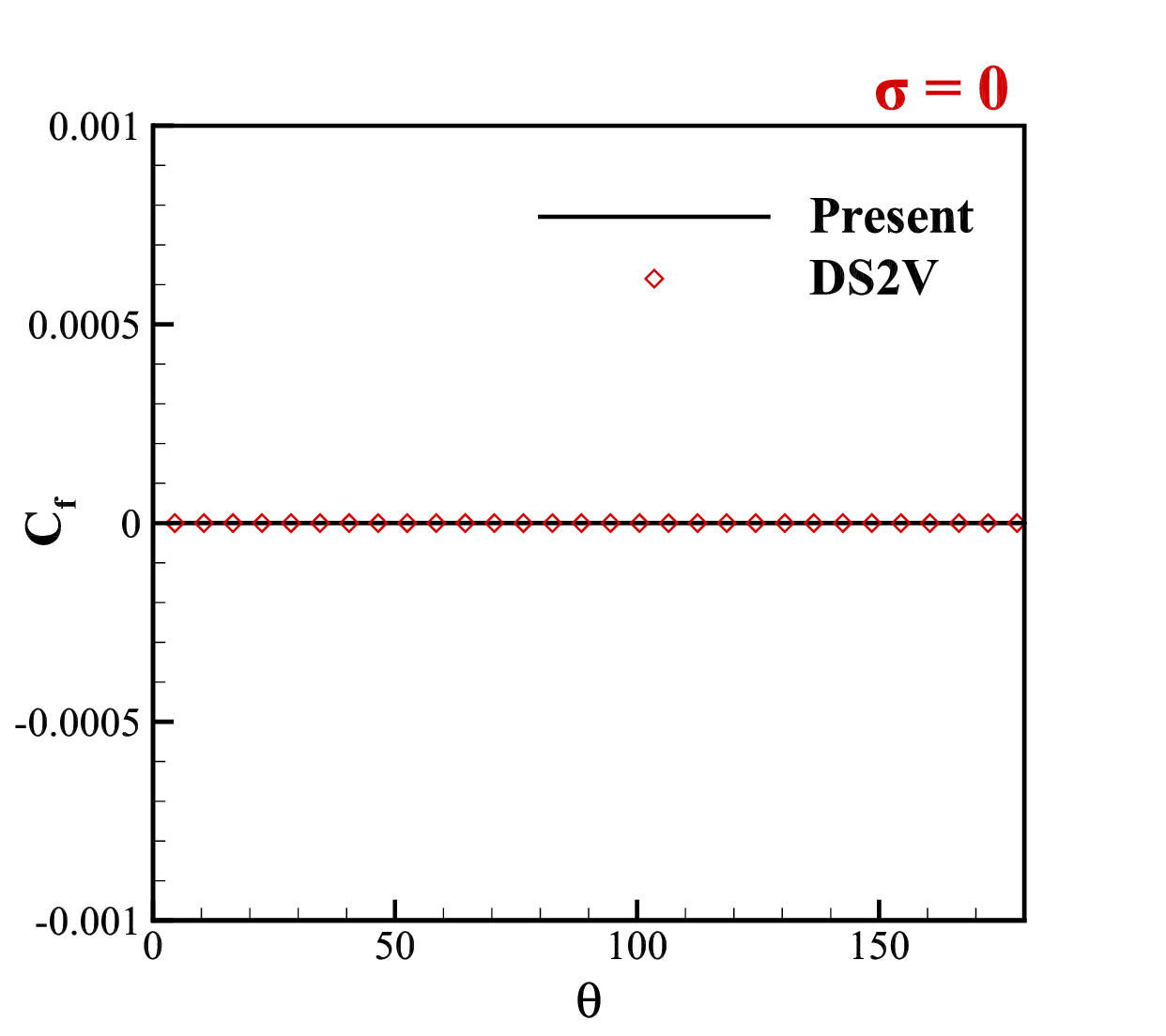}}
		\subfigure[]{\label{fig14c}\includegraphics[width=0.45\textwidth]{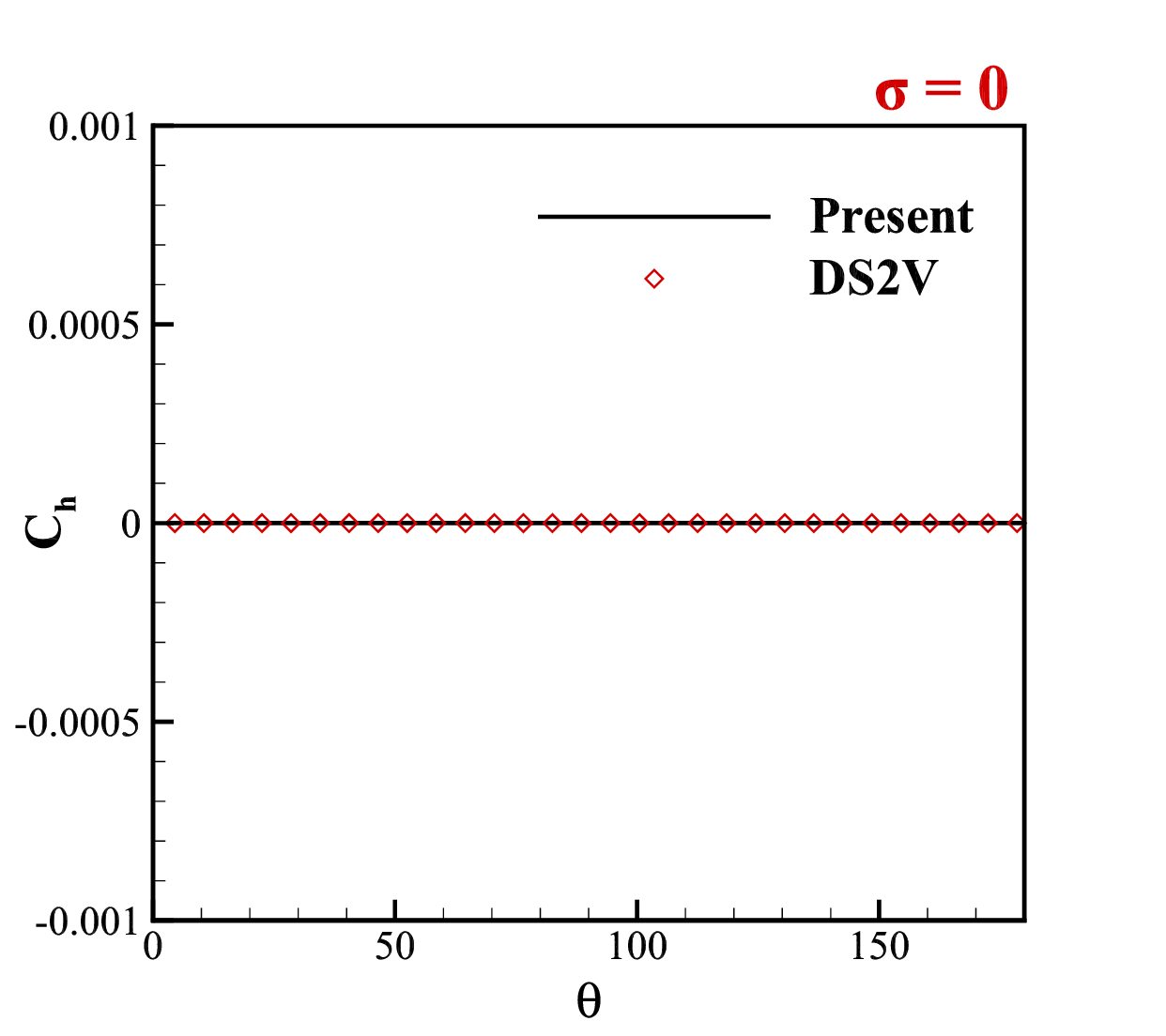}}
		\caption{\label{fig14}{Comparison of the (a) pressure coefficient, (b) skin friction coefficient, and (c) heat transfer coefficient on the surface of cylinder with $\sigma$ = 0 ($Ma$ = 5.0, $Kn$ = 1.0, $T_{\infty}$ = 273 K, $T_{w}$ = 273 K).}}
	\end{figure}

\begin{figure}
		\centering
		\subfigure[]{\label{fig15a}\includegraphics[width=0.45\textwidth]{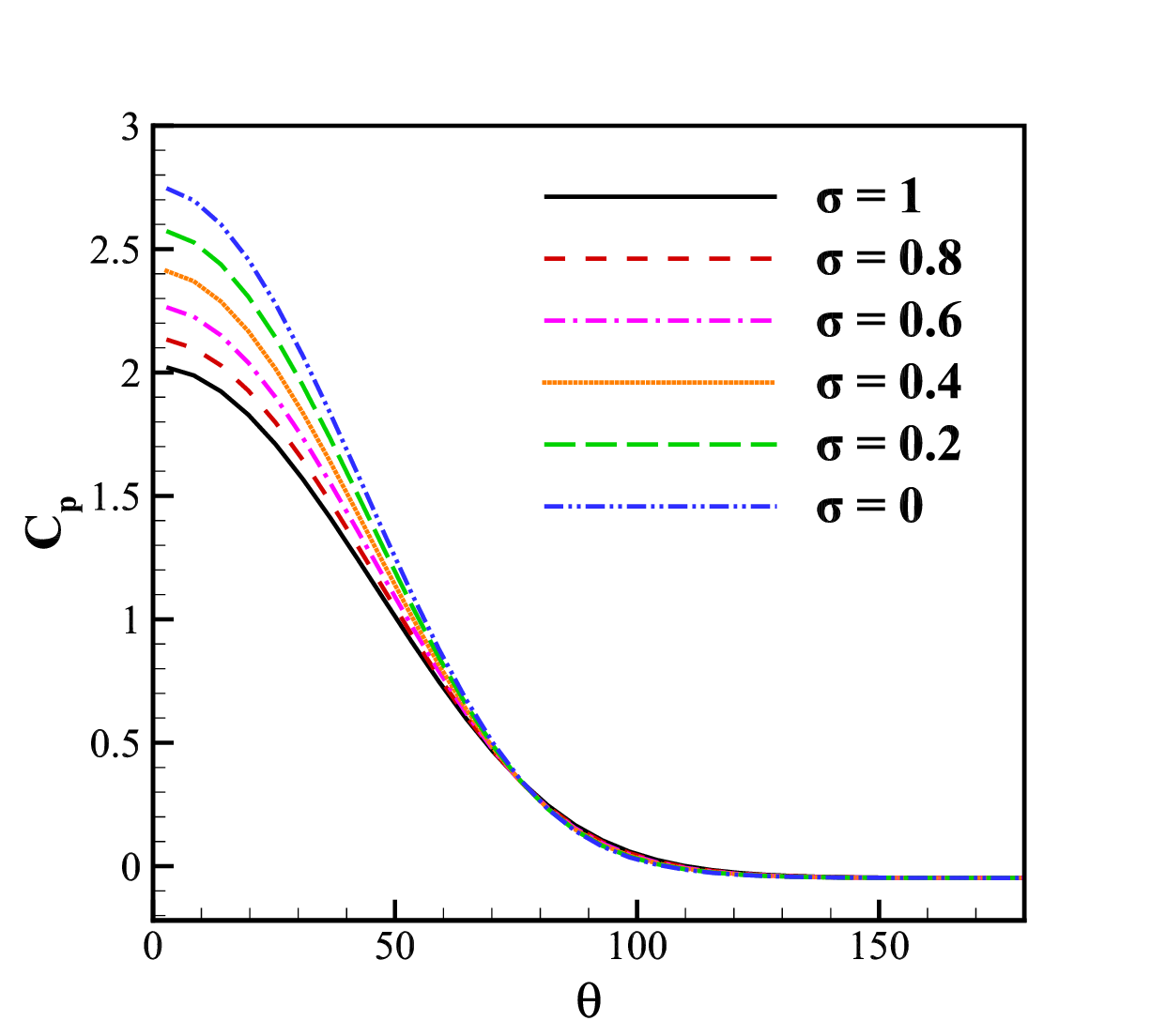}}
		\subfigure[]{\label{fig15b}\includegraphics[width=0.45\textwidth]{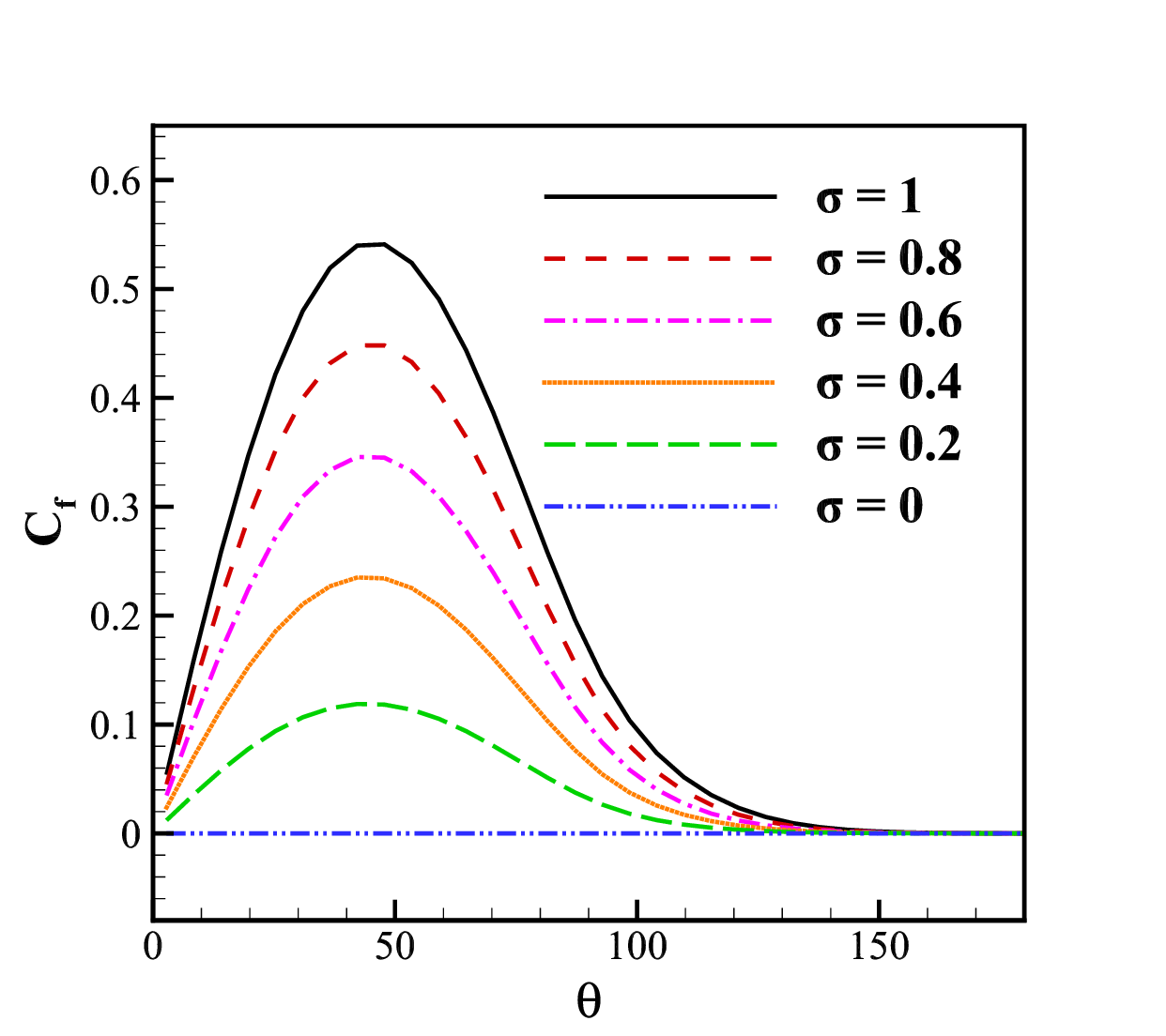}}
		\subfigure[]{\label{fig15c}\includegraphics[width=0.45\textwidth]{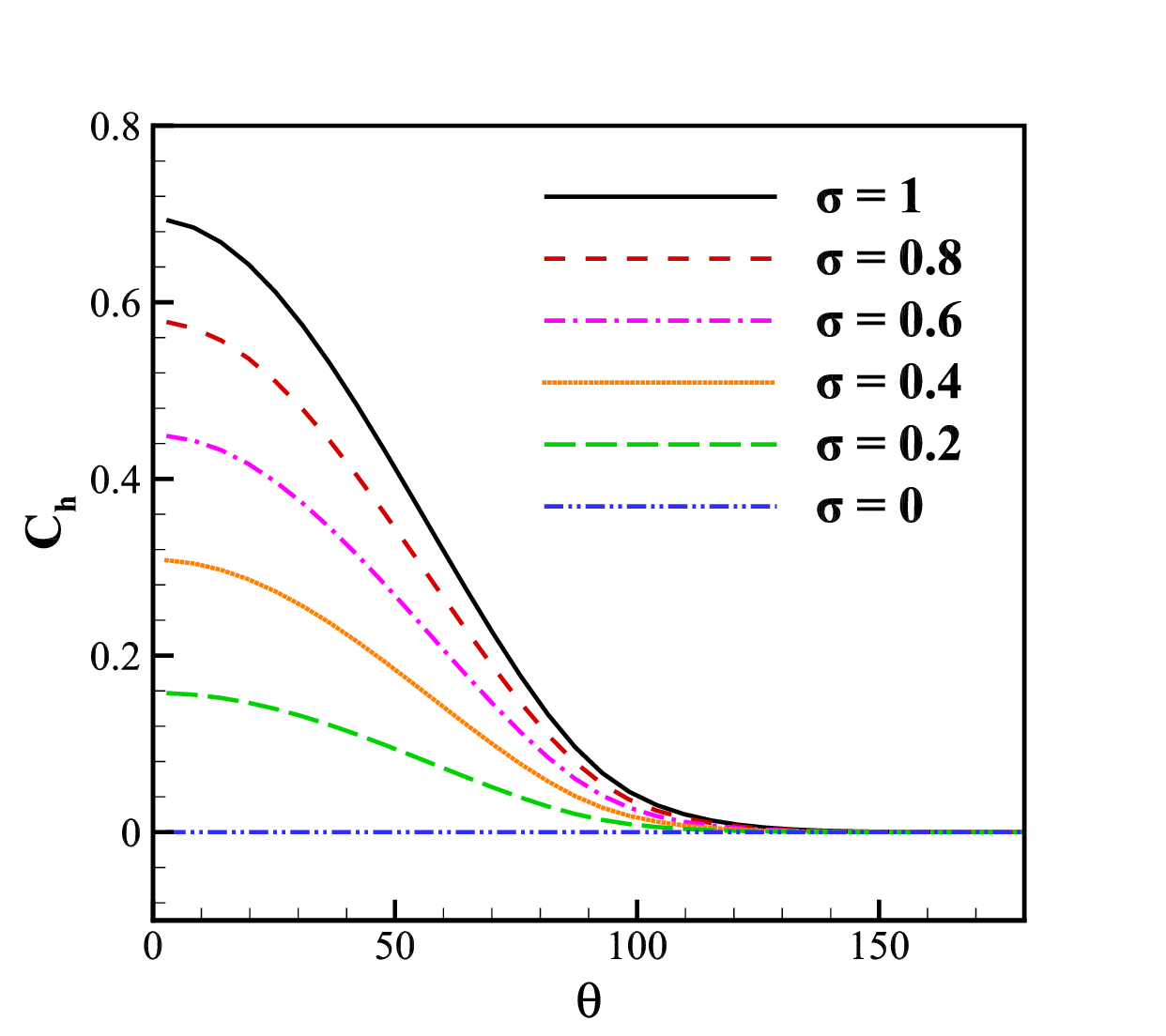}}
		\caption{\label{fig15}{Comparison of the (a) pressure coefficient, (b) skin friction coefficient, and (c) heat transfer coefficient on the surface of cylinder with different $\sigma$ ($Ma$ = 5.0, $Kn$ = 1.0, $T_{\infty}$ = 273 K, $T_{w}$ = 273 K).}}
	\end{figure}

	In Fig. \ref{fig09}, the distribution of density, pressure, temperature, and horizontal velocity on the stationary line in front of the cylinder is presented for the case where $\sigma$ is set to 1. Notably, the results obtained from the UGKS and DSMC methods exhibit a high degree of consistency with the current calculations. The application of the Shakhov model results in a temperature distribution on the stationary line that is slightly higher than the DSMC results before the bow shock, a phenomenon in line with observations described in the literature \cite{liu2014investigation} regarding shock wave structure. Fig. \ref{fig10} provides a detailed view of the distribution of pressure, skin friction, and heat transfer coefficients on the surface of the cylinder when $\sigma$ is 1. Impressively, all calculated results align perfectly with those obtained from UGKS and DSMC methods. In Fig. \ref{fig11}, the distribution of density, pressure, temperature, and horizontal velocity on the stationary line in front of the cylinder is illustrated for the scenario where $\sigma$ is set to 0.8. When $\sigma$ is 0.8, the distribution of pressure, skin friction, and heat transfer coefficients on the cylinder's surface is depicted in Fig. \ref{fig12}. Additionally, Fig. \ref{fig13} displays the distribution of density, pressure, temperature, and horizontal velocity on the stationary line in front of the cylinder when $\sigma$ is set to 0. Subsequently, Fig. \ref{fig14} presents the distribution of pressure, skin friction, and heat transfer coefficients on the cylinder's surface under the condition of $\sigma$ being 0. It is evident from the presented figures that the calculated results under $\sigma$ values of 1, 0.8, and 0 closely correspond to the results obtained through the DSMC method.	
	In Fig. \ref{fig15}, the distribution of pressure, skin friction, and heat transfer coefficients on the cylinder's surface is presented for various values of $\sigma$. It is observed that the results of hypersonic cylinder flow are influenced to varying degrees by the ACs in the GSI model.	When the GSI model is treated with total specular reflection, the wall does not alter the tangential motion of gas molecules, resembling inviscid flow. Specifically, at $\sigma$ = 0, the pressure at the front of the cylinder is higher compared to the case at $\sigma$ = 1. This discrepancy is attributed to the complete reflection of normal motion of gas molecules, resembling a scenario where the wall imparts a stronger force on the gas molecules than in the case of completely diffuse reflection.	Table \ref{tab:cylinder coefficients} provides a comparison of drag coefficients at various $\sigma$. The drag coefficients obtained in the current study closely resemble those from DS2V calculations. The drag force on the cylinder results from the combined effects of pressure and friction. Interestingly, the drag coefficient of the cylinder exhibits minimal variation under different $\sigma$ values, indicating an increase in wall pressure and a decrease in wall skin friction as $\sigma$ decreases. Future investigations will explore the impact of the GSI model on the cylinder's resistance under various Knudsen numbers.
	
	\begin{table}
		\centering 
		\caption{\label{tab:cylinder coefficients}The drag coefficients of cylinder.}
		\begin{tabular}{p{30pt}<{\centering} p{80pt}<{\centering} p{80pt}<{\centering} p{110pt}<{\centering}}
			\hline
			\hline
			$\sigma$ & Present & DS2V & Relative Error \\
			\hline
			1	 & 1.925	& 1.917	& 0.42$\%$   \\
			\hline
			0.8	 & 1.914	& 1.910	& 0.26$\%$   \\
			\hline
			0.6	 & 1.910	& 1.912	& -0.10$\%$  \\
			\hline
			0.4	 & 1.912	& 1.924	& -0.62$\%$  \\
			\hline
			0.2	 & 1.921	& 1.946	& -1.28$\%$  \\
			\hline
			0	 & 1.938	& 1.970	& -1.62$\%$  \\
			\hline
			\hline
		\end{tabular}
	\end{table}

	\subsubsection{Hypersonic flow over a blunt wedge in the transitional regime}
\begin{figure}
		\centering
		\includegraphics[width=0.7\textwidth]{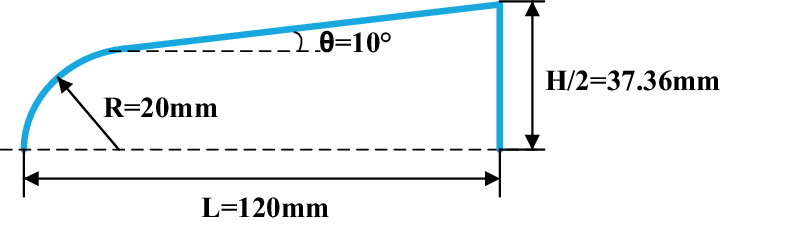}
		\caption{\label{fig16}Geometric of blunt wedge (half-model).}
	\end{figure}

\begin{figure}
		\centering
		\subfigure[]{\label{fig17a}\includegraphics[width=0.49\textwidth]{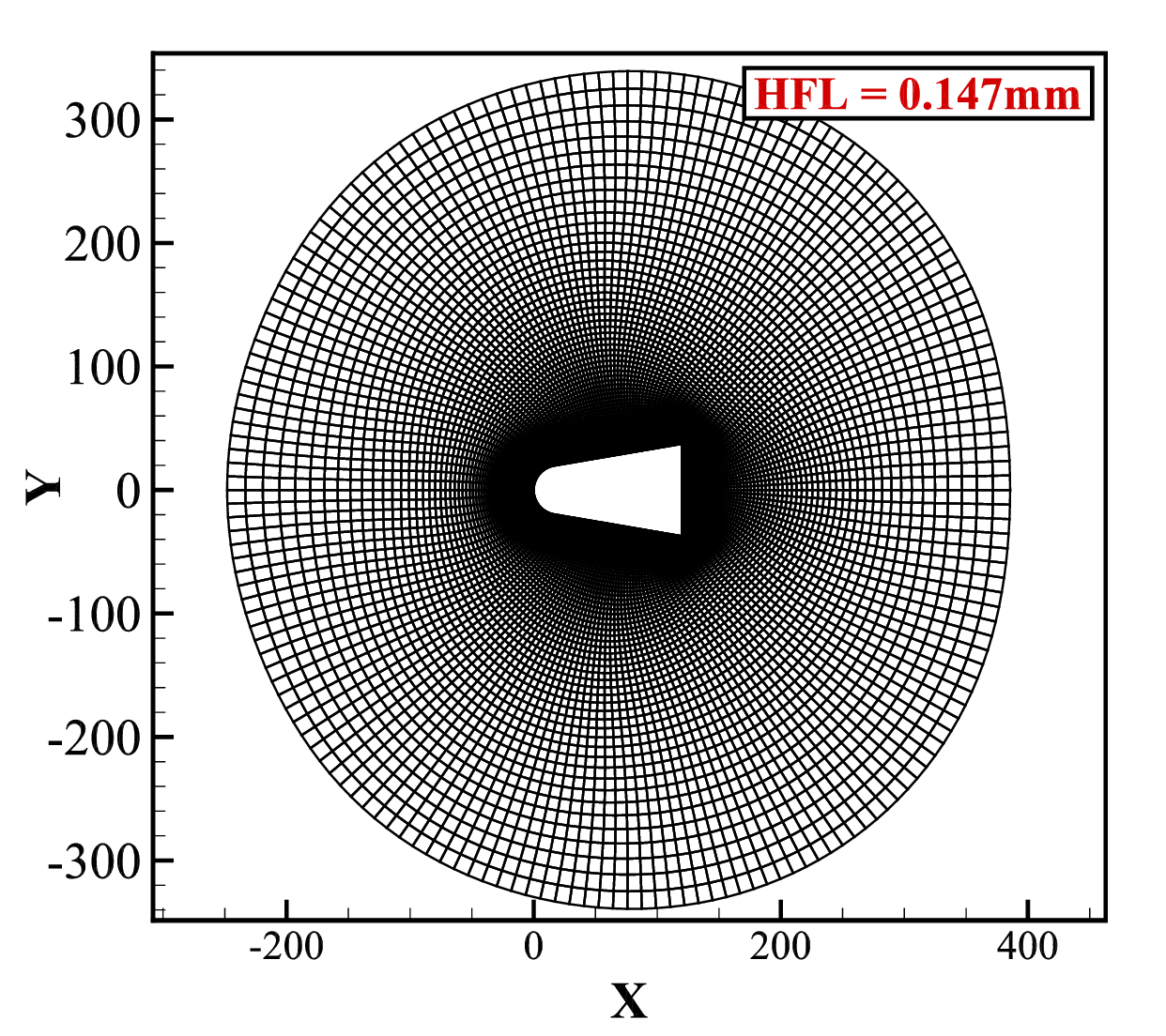}}
		\subfigure[]{\label{fig17b}\includegraphics[width=0.44\textwidth]{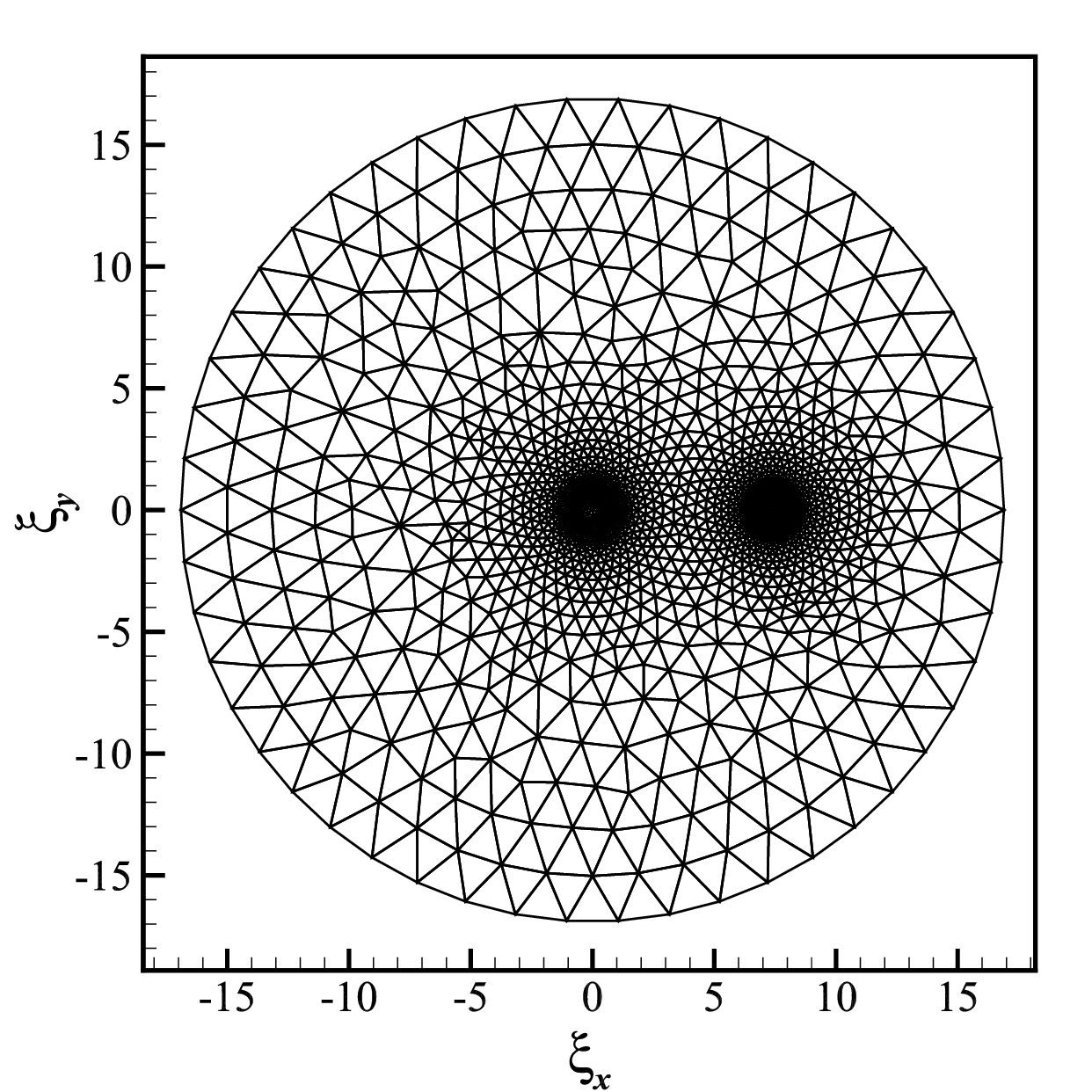}}
		\caption{\label{fig17}{The (a) unstructured physical mesh and (b) unstructured velocity mesh for the hypersonic flow passing a blunt wedge ($Ma$ = 8.1, $Kn$ = 0.338, $T_{\infty}$ = 189 K, $T_{w}$ = 273 K).}}
	\end{figure}

	To further showcase the performance of the current GSI algorithm in hypersonic non-equilibrium flows, we simulate dilute expansion flow over a blunt wedge, referencing a configuration from the literature \cite{jiang2019implicit}. 	
	Fig. \ref{fig16} depicts the geometry of the blunt wedge, characterized by a length of $L$ = 120 mm, a head radius of $R$ = 20 mm, a bottom height of $H$ = 74.72 mm, and a body slope of $\theta$ = 10 degrees. Argon is employed as the working gas, and the VHS molecular model with $\omega$ = 0.81 is utilized.	
	The freestream parameters include a Mach number of $Ma$ = 8.1, a temperature of $T_{\infty}$ = 189 K (equivalent to an altitude of 85 Km), and an angle of attack (AOA) of 0 degrees.	The surface temperature is maintained at 273 K, and the Knudsen numbers with $R$ and $H$ as characteristic lengths are 0.338 and 0.090, respectively. For this simulation, the unstructured physical mesh comprises 14790 cells, while the unstructured velocity mesh consists of 3056 cells, as illustrated in Figs. \ref{fig17a} and \ref{fig17b}, respectively. The HFL of the mesh on the surface of the blunt wedge is set at 0.147 mm.
	
\begin{figure}
		\centering
		\subfigure[]{\label{fig18a}\includegraphics[width=0.45\textwidth]{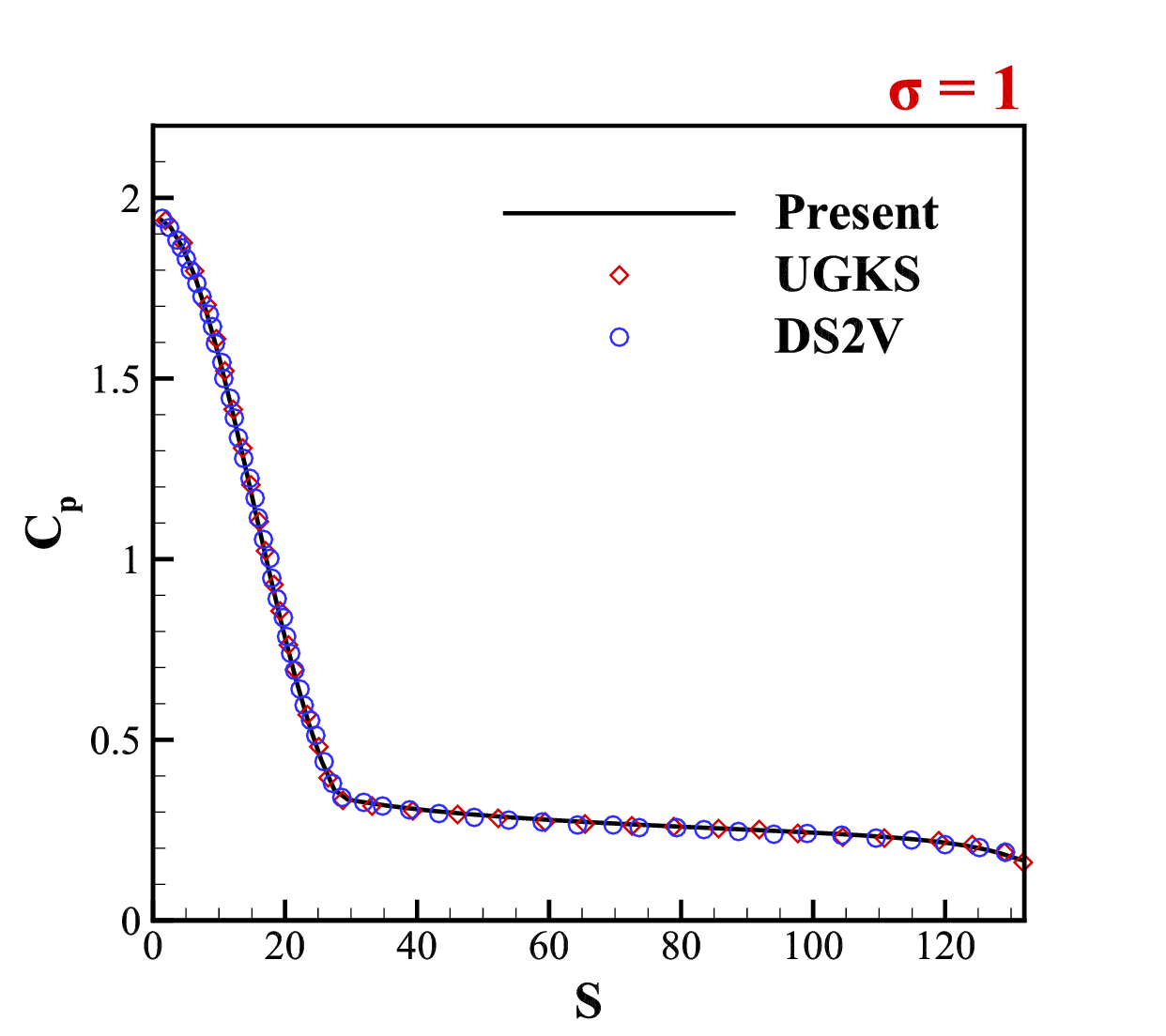}}
		\subfigure[]{\label{fig18b}\includegraphics[width=0.45\textwidth]{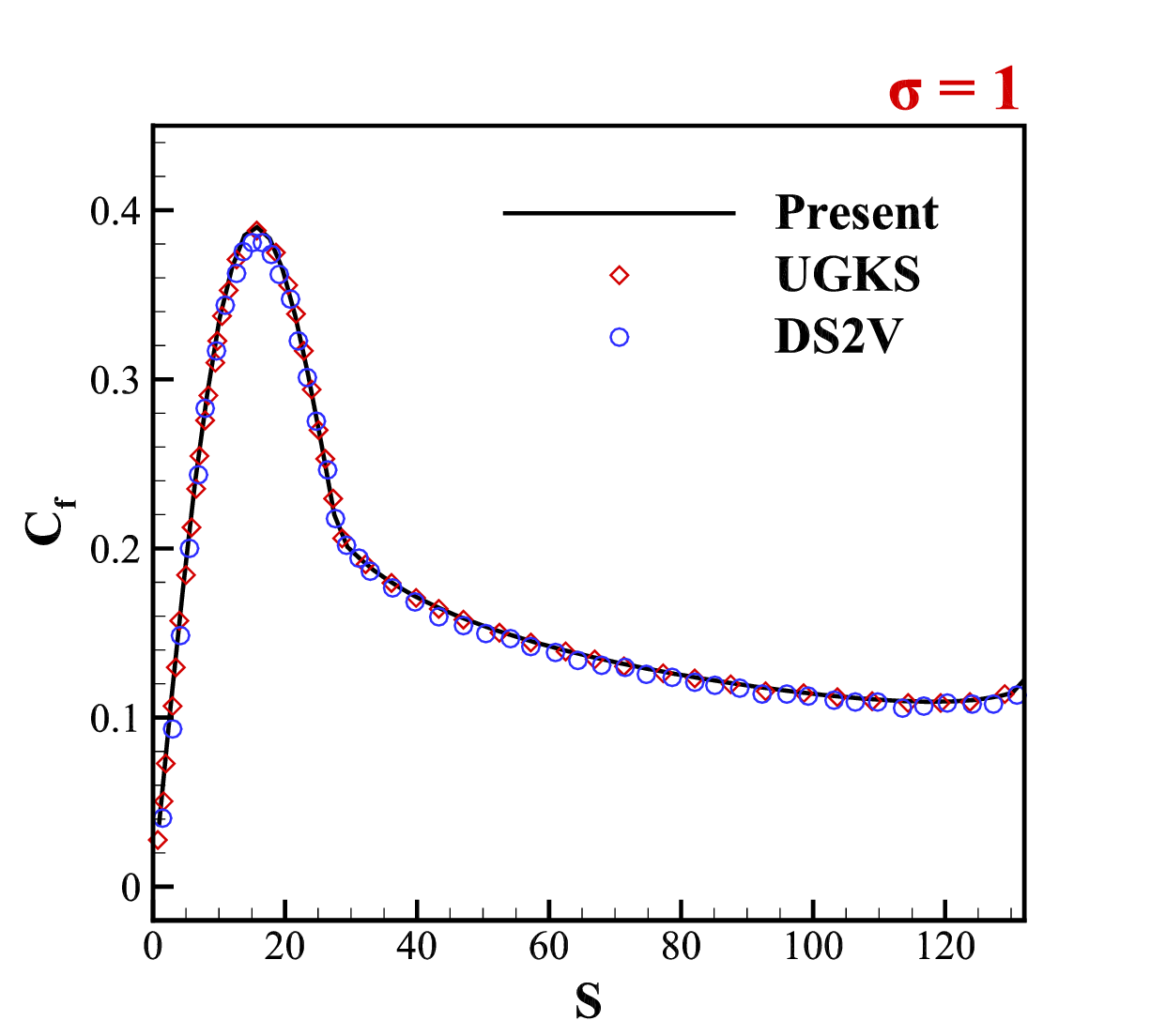}}
		\subfigure[] {\label{fig18c}\includegraphics[width=0.45\textwidth]{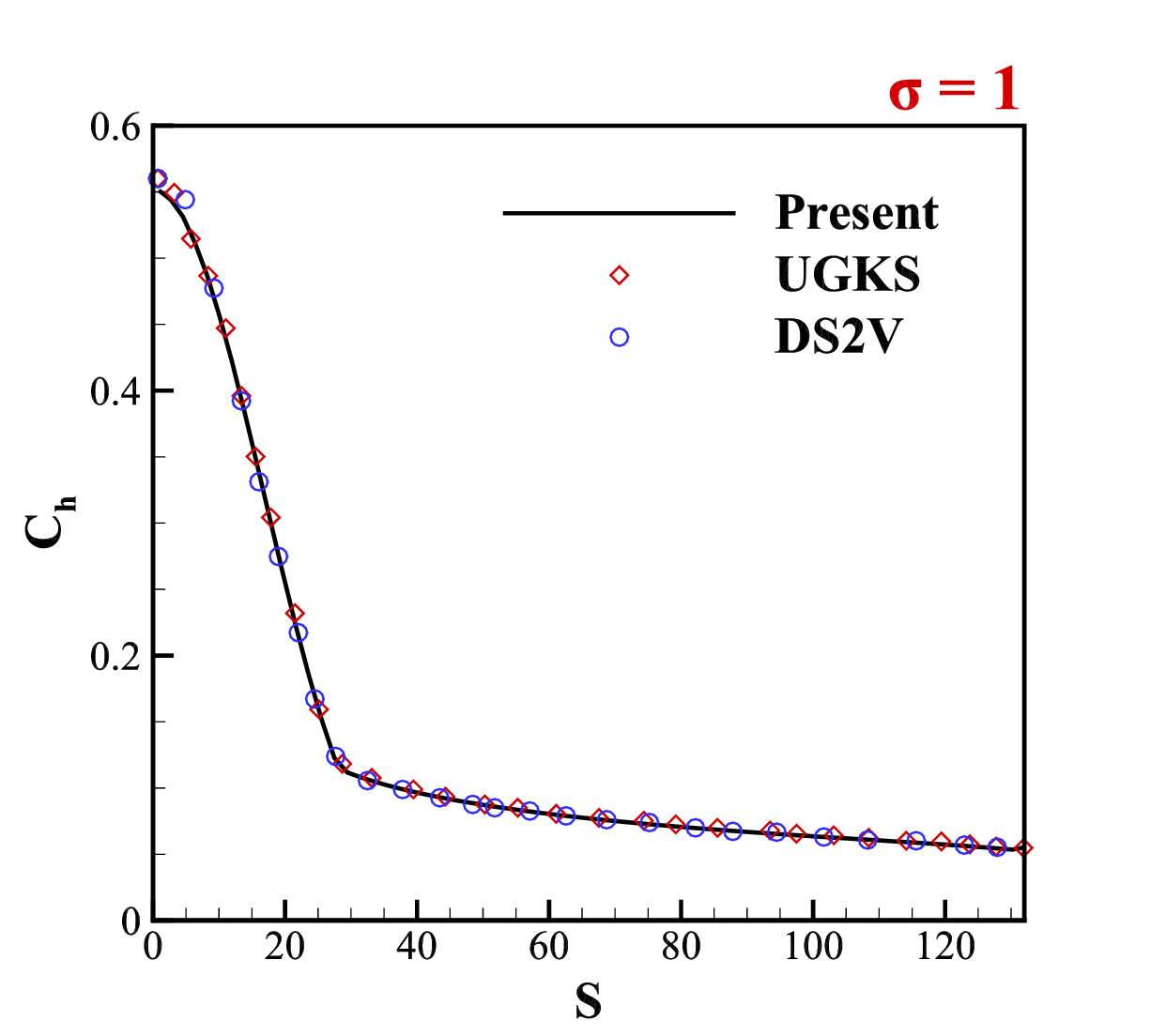}}
		\subfigure[]{\label{fig18d}\includegraphics[width=0.45\textwidth]{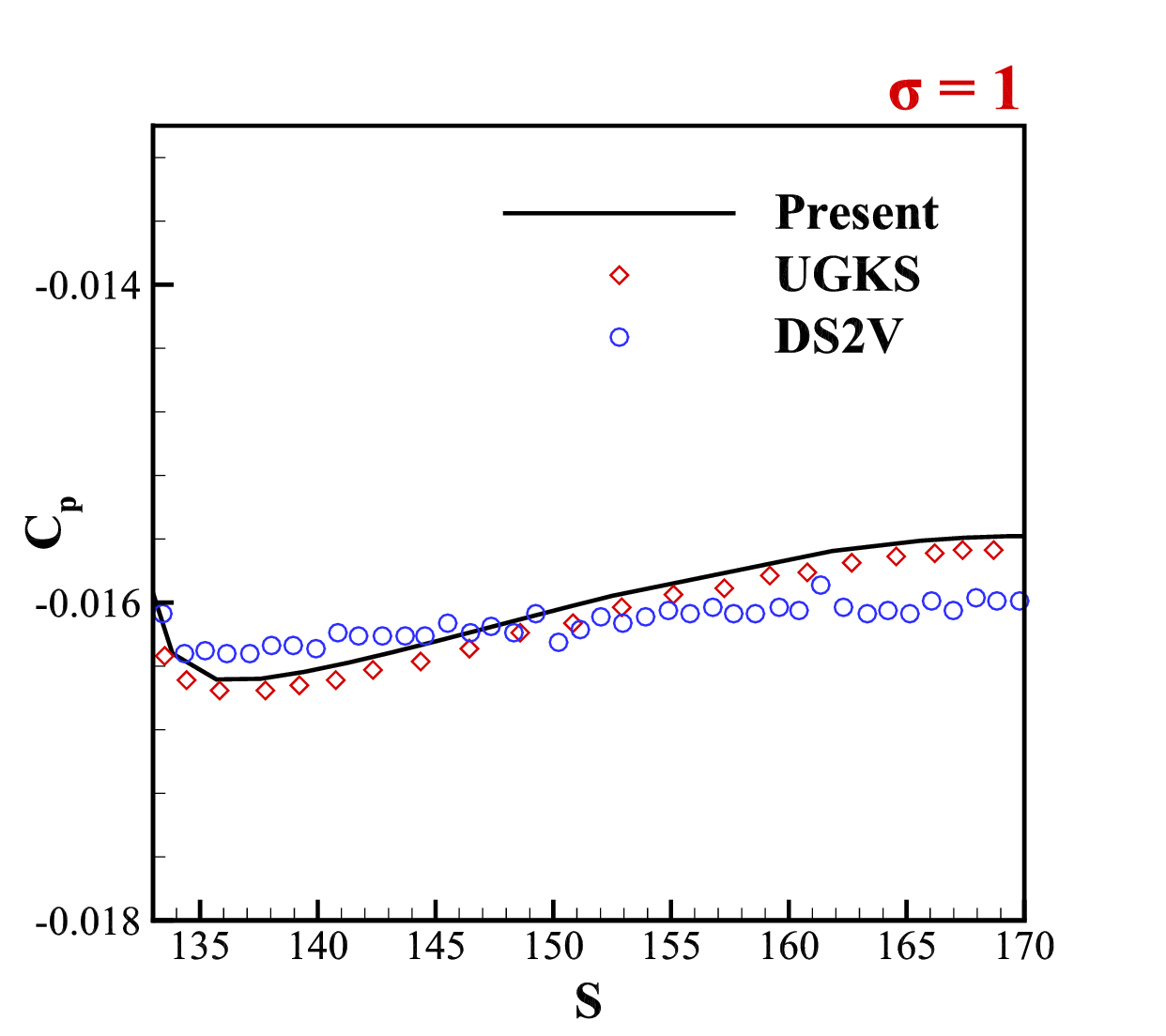}}
		\subfigure[]{\label{fig18e}\includegraphics[width=0.45\textwidth]{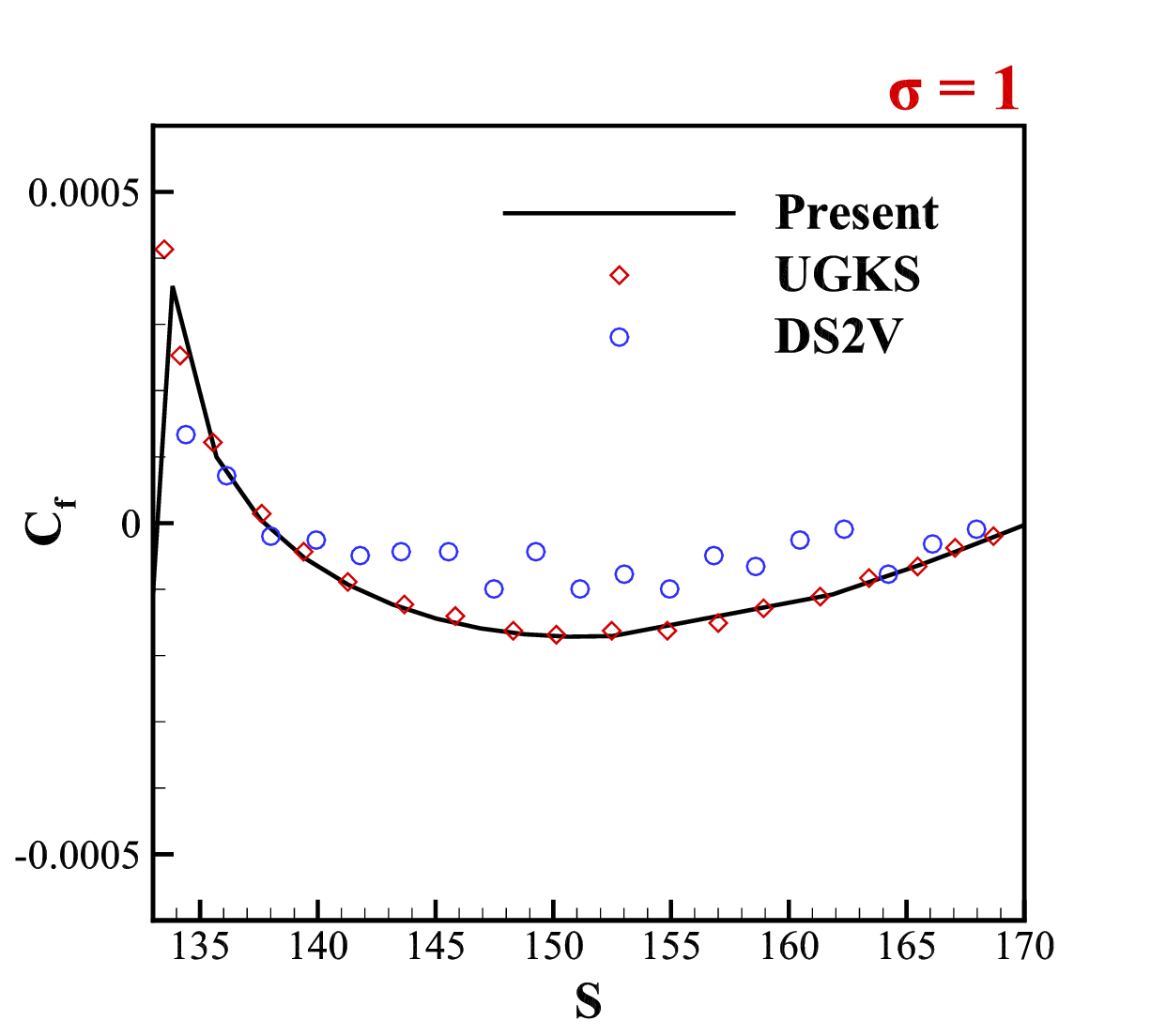}}
		\subfigure[]{\label{fig18f}\includegraphics[width=0.45\textwidth]{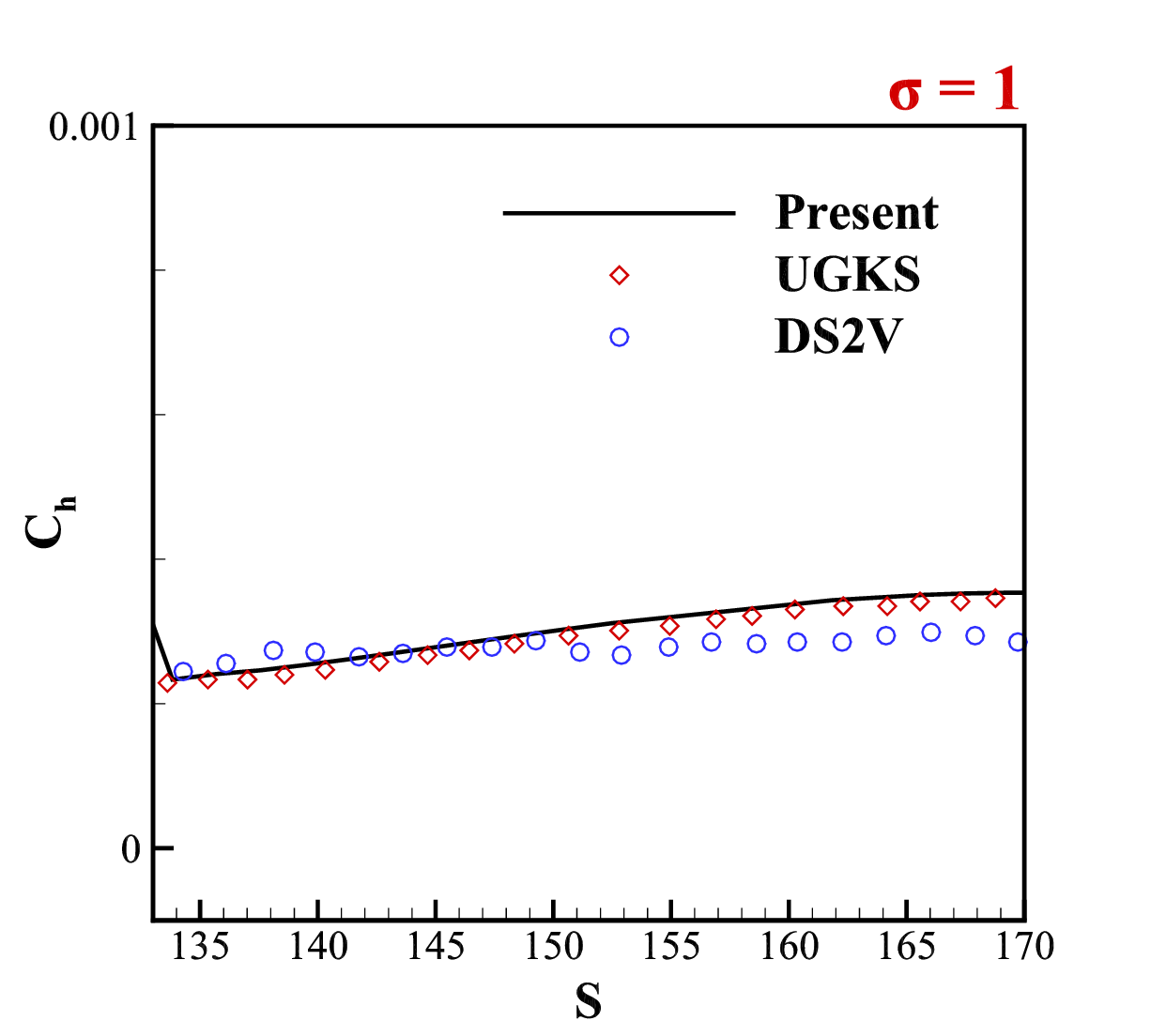}}
		\caption{\label{fig18}{Comparison of the (a) body pressure coefficient, (b) body skin friction coefficient, (c) body heat transfer, (d) bottom pressure coefficient, (e) bottom skin friction coefficient, and (f) bottom heat transfer coefficient on the surface of blunt wedge with $\sigma$ = 1 ($Ma$ = 8.1, $Kn$ = 0.338, $T_{\infty}$ = 189 K, $T_{w}$ = 273 K).}}
	\end{figure}

\begin{figure}
		\centering
		\subfigure[]{\label{fig19a}\includegraphics[width=0.45\textwidth]{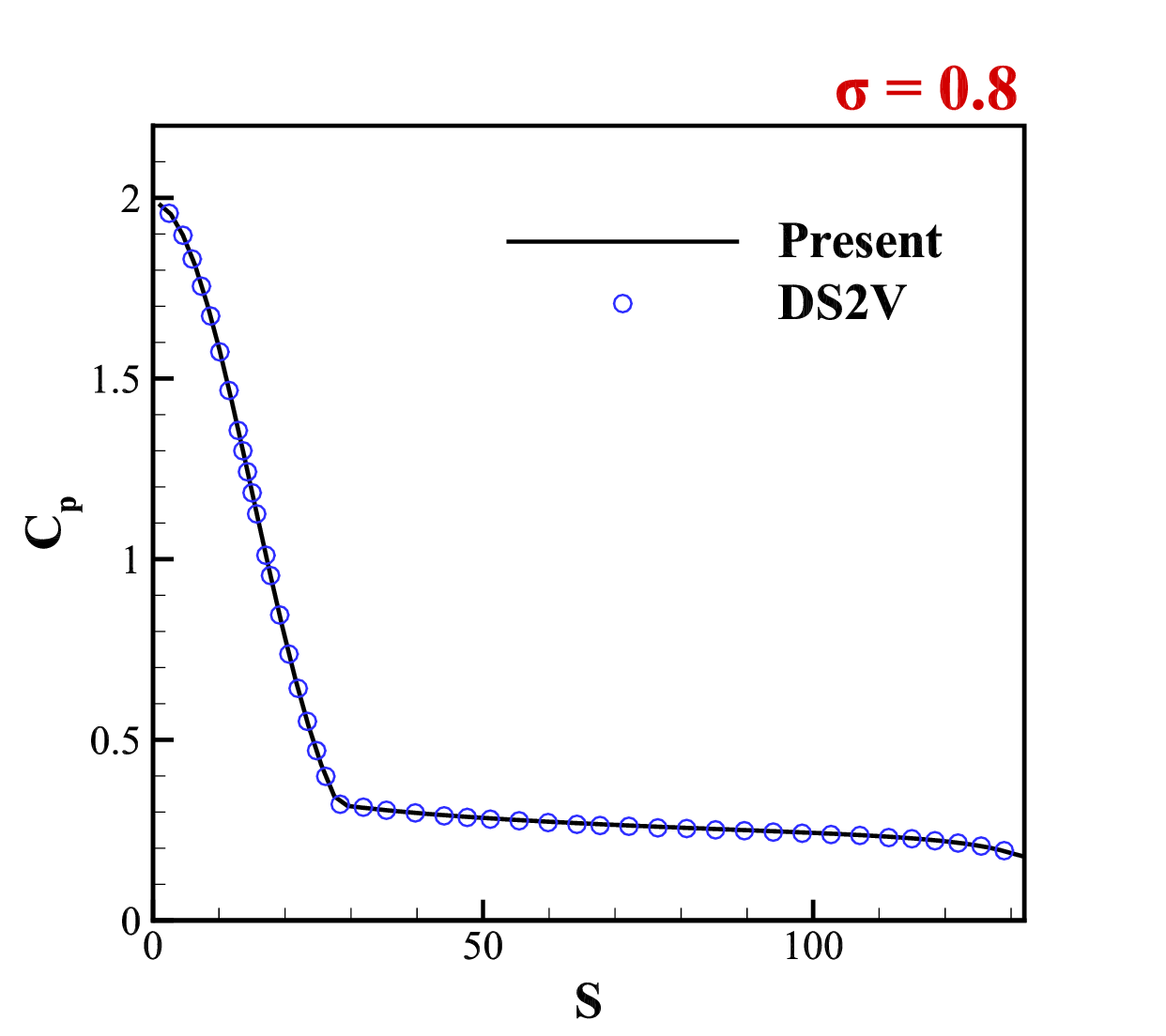}}
		\subfigure[]{\label{fig19b}\includegraphics[width=0.45\textwidth]{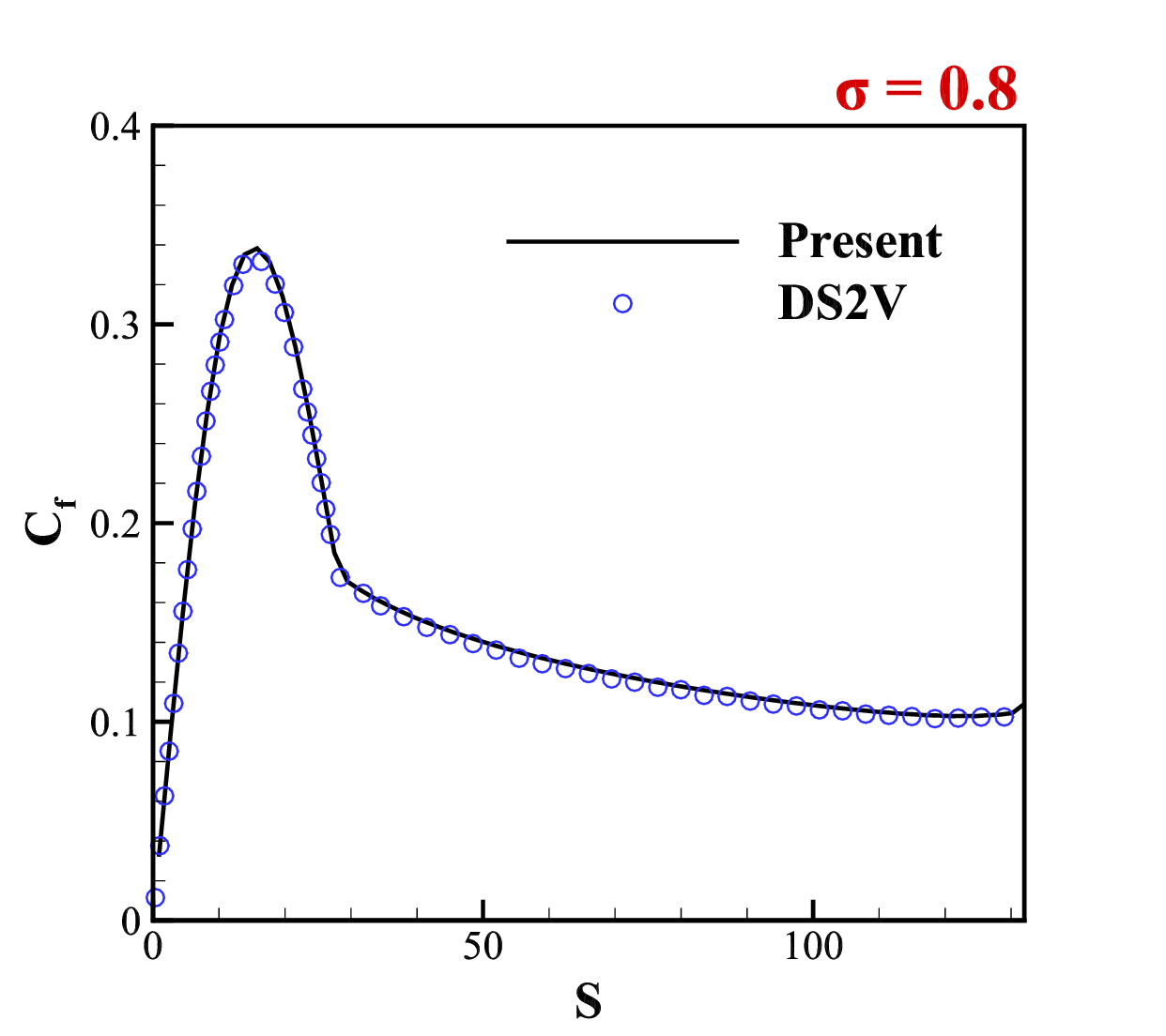}}
		\subfigure[]{\label{fig19c}\includegraphics[width=0.45\textwidth]{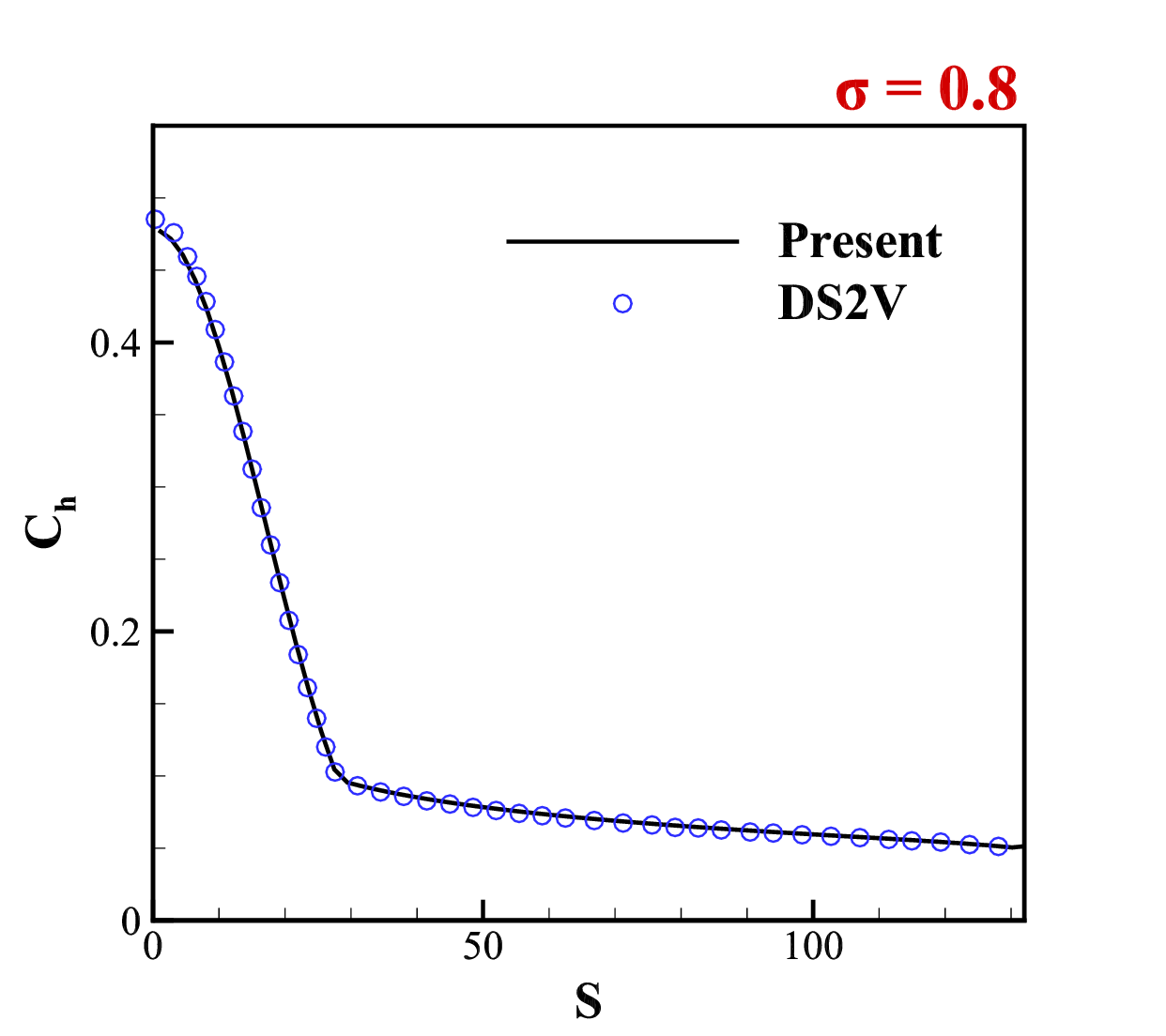}}
		\subfigure[]{\label{fig19d}\includegraphics[width=0.45\textwidth]{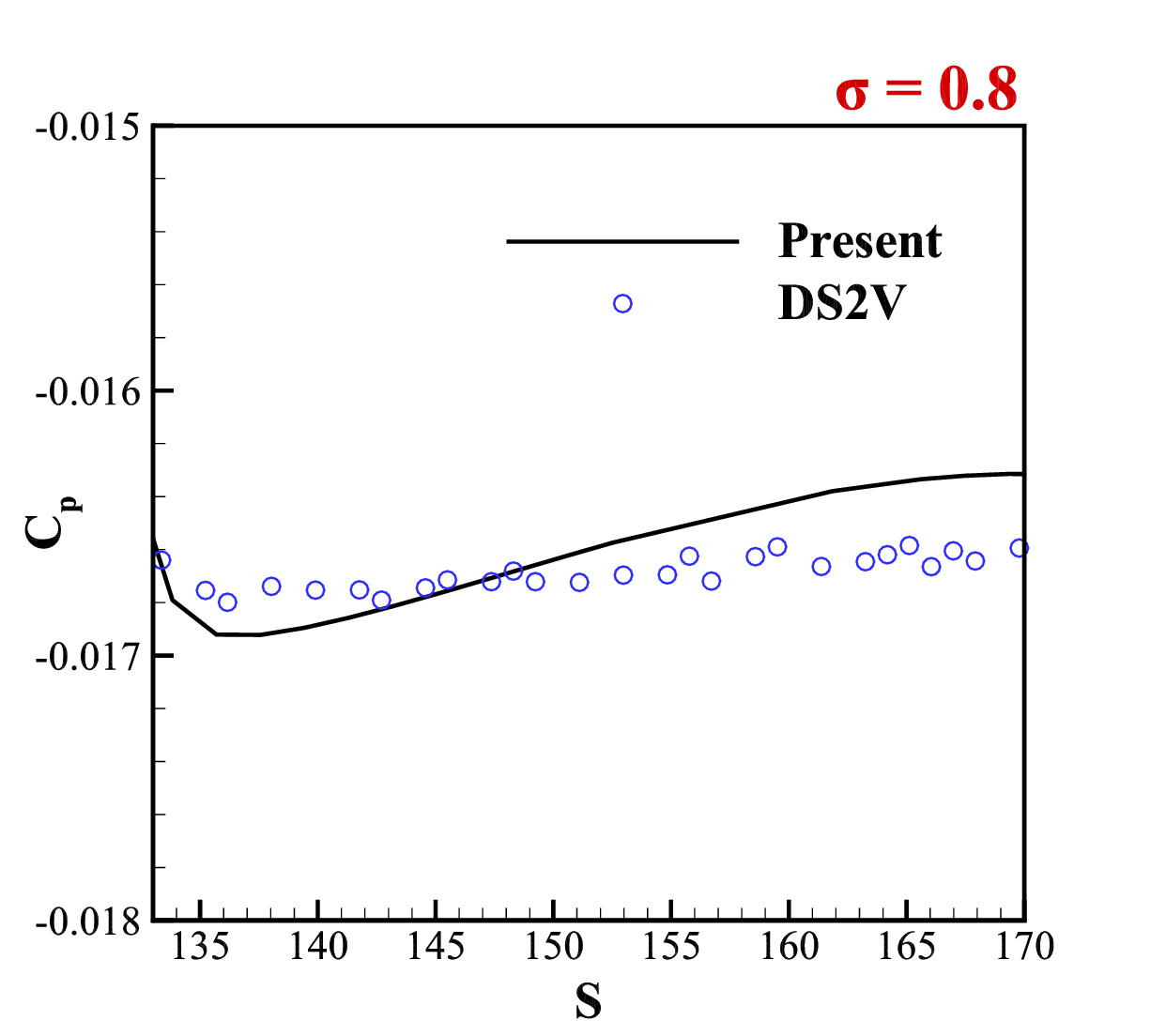}}
		\subfigure[]{\label{fig19e}\includegraphics[width=0.45\textwidth]{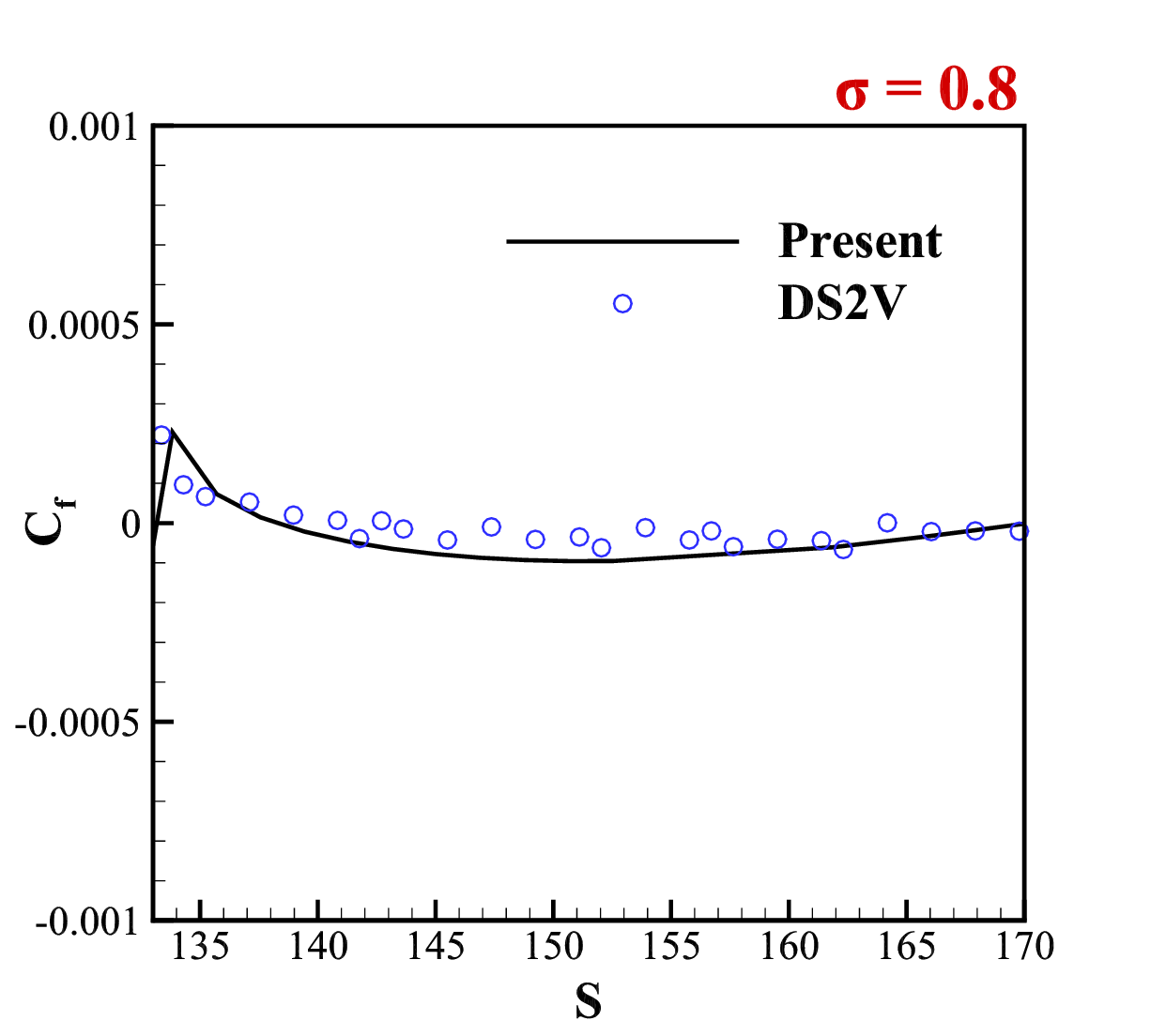}}
		\subfigure[]{\label{fig19f}\includegraphics[width=0.45\textwidth]{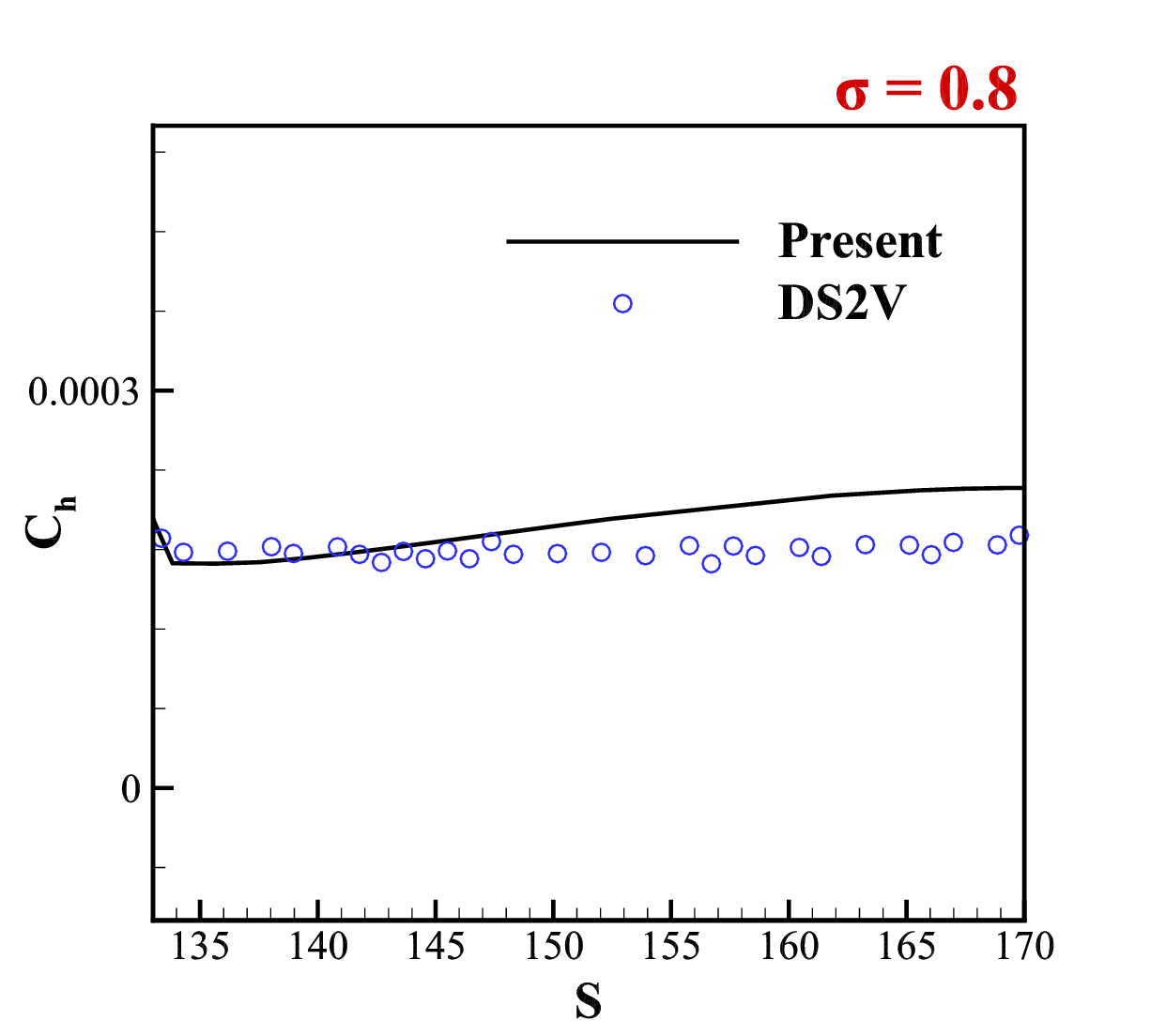}}
		\caption{\label{fig19}{Comparison of the (a) body pressure coefficient, (b) body skin friction coefficient, (c) body heat transfer, (d) bottom pressure coefficient, (e) bottom skin friction coefficient, and (f) bottom heat transfer coefficient on the surface of blunt wedge with $\sigma$ = 0.8 ($Ma$ = 8.1, $Kn$ = 0.338, $T_{\infty}$ = 189 K, $T_{w}$ = 273 K).}}
	\end{figure}

\begin{figure}
		\centering
		\subfigure[]{\label{fig20a}\includegraphics[width=0.45\textwidth]{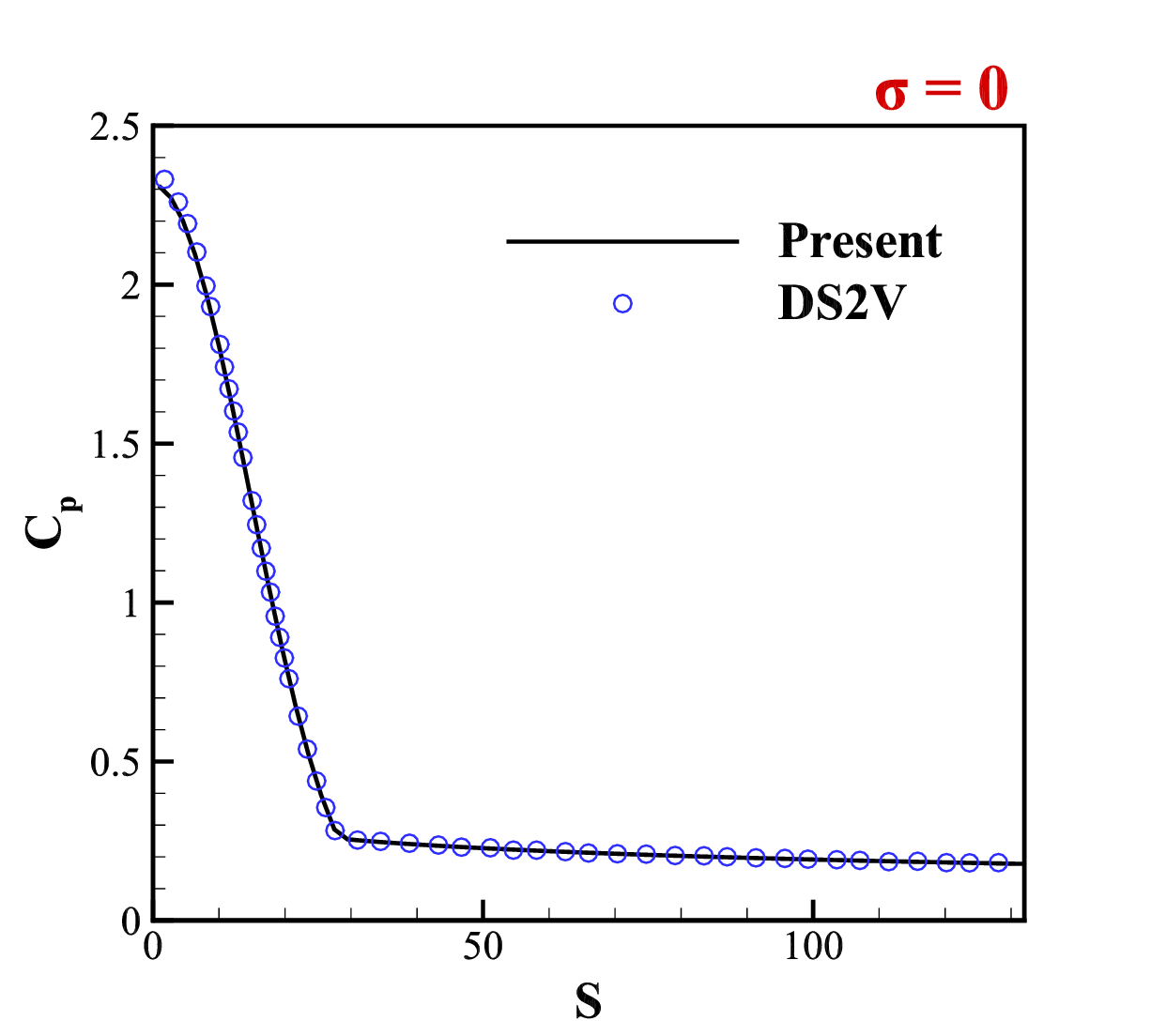}}
		\subfigure[]{\label{fig20b}\includegraphics[width=0.45\textwidth]{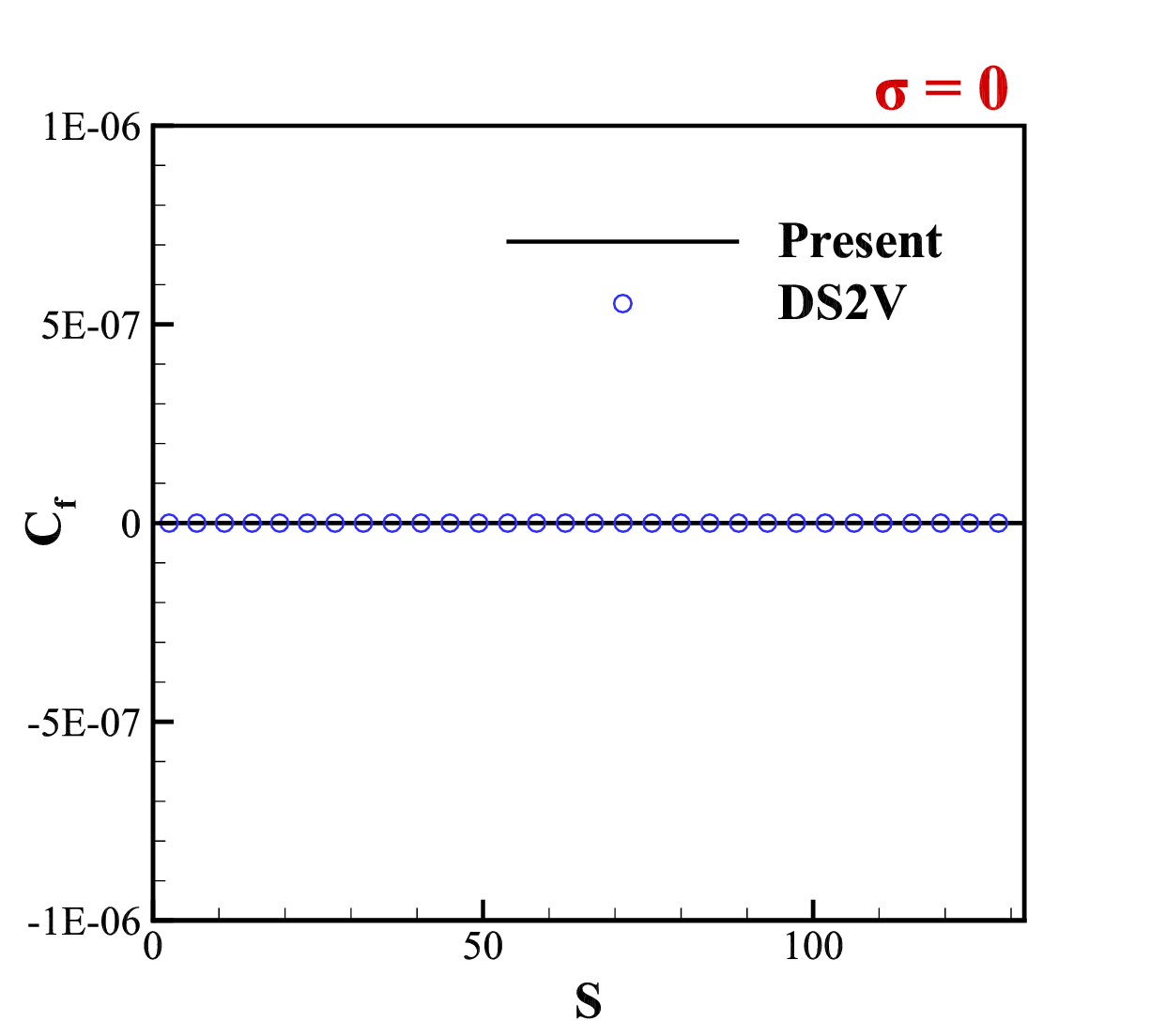}}
		\subfigure[]{\label{fig20c}\includegraphics[width=0.45\textwidth]{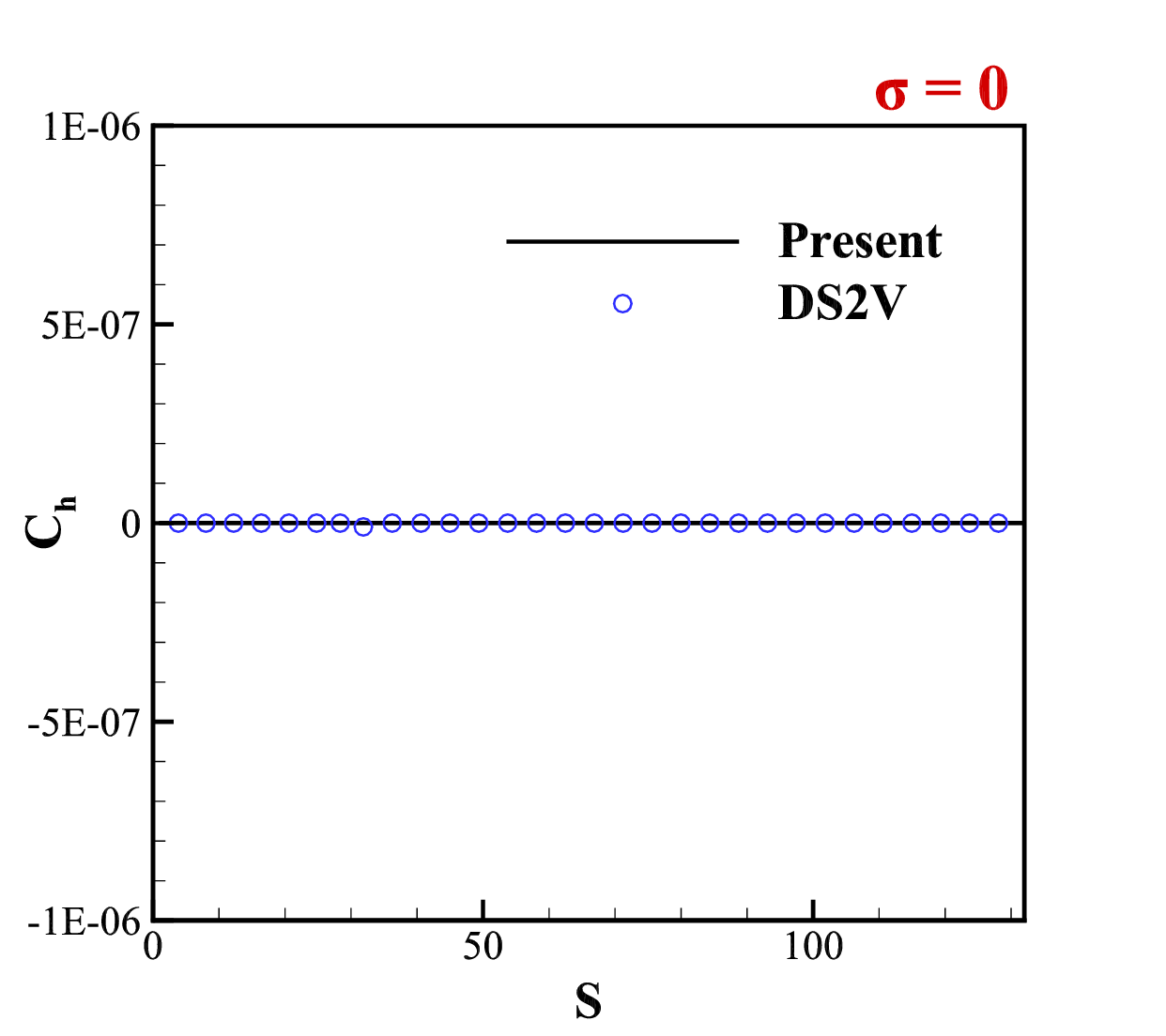}}
		\subfigure[]{\label{fig20d}\includegraphics[width=0.45\textwidth]{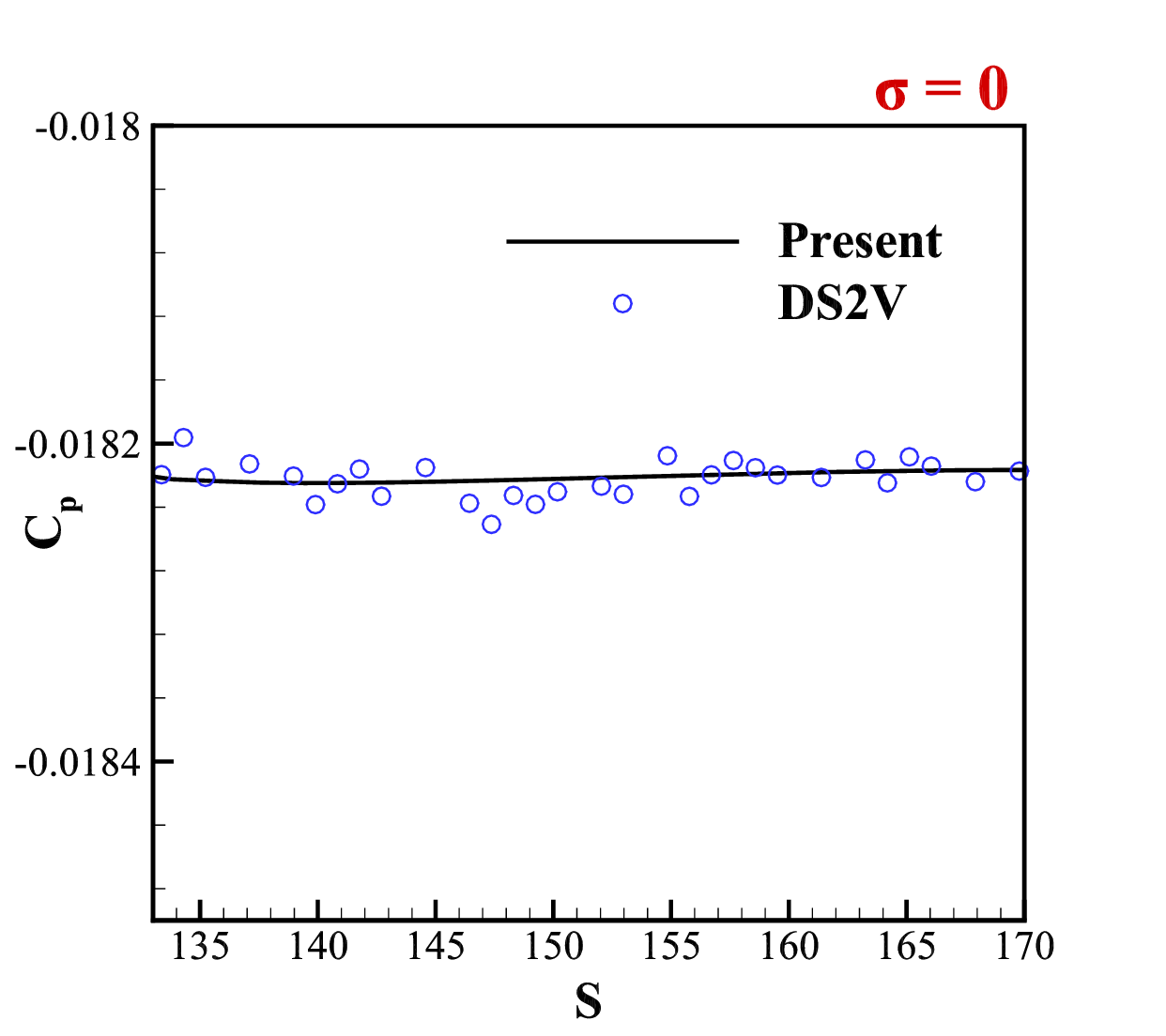}}
		\subfigure[]{\label{fig20e}\includegraphics[width=0.45\textwidth]{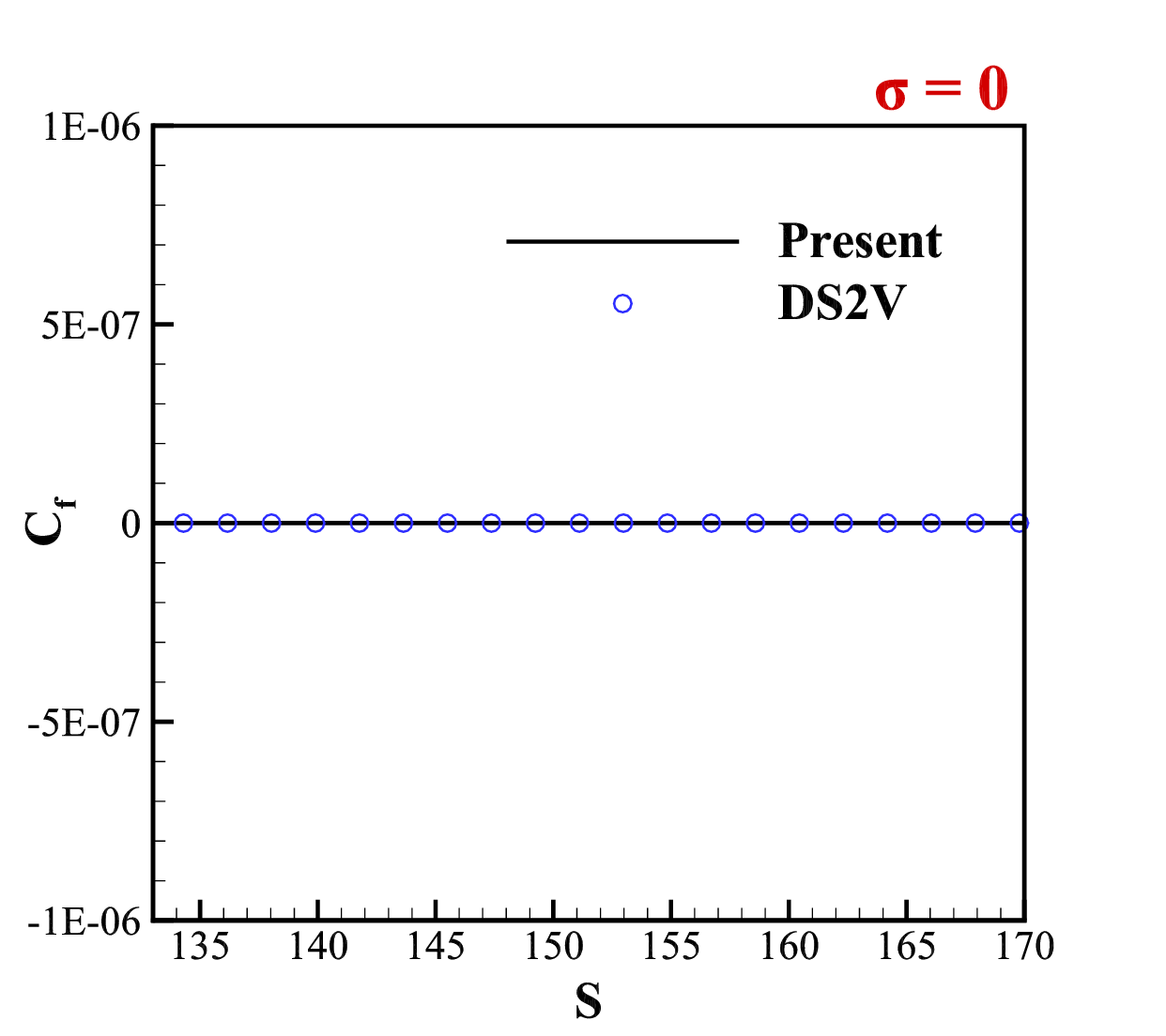}}
		\subfigure[]{\label{fig20f}\includegraphics[width=0.45\textwidth]{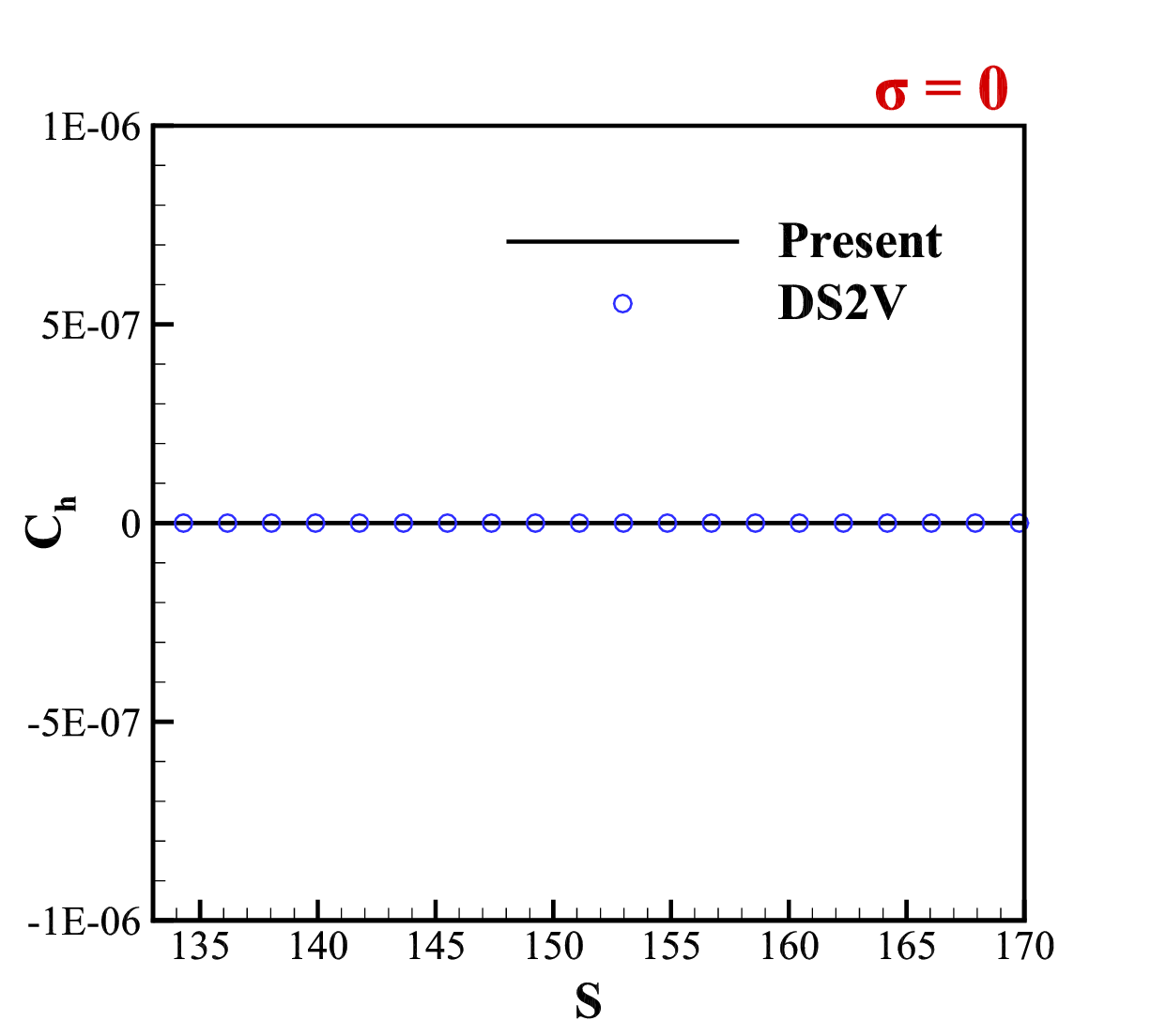}}
		\caption{\label{fig20}{Comparison of the (a) body pressure coefficient, (b) body skin friction coefficient, (c) body heat transfer, (d) bottom pressure coefficient, (e) bottom skin friction coefficient, and (f) bottom heat transfer coefficient on the surface of blunt wedge with $\sigma$ = 0 ($Ma$ = 8.1, $Kn$ = 0.338, $T_{\infty}$ = 189 K, $T_{w}$ = 273 K).}}
	\end{figure}

\begin{figure}
		\centering
		\subfigure[]{\label{fig21a}\includegraphics[width=0.45\textwidth]{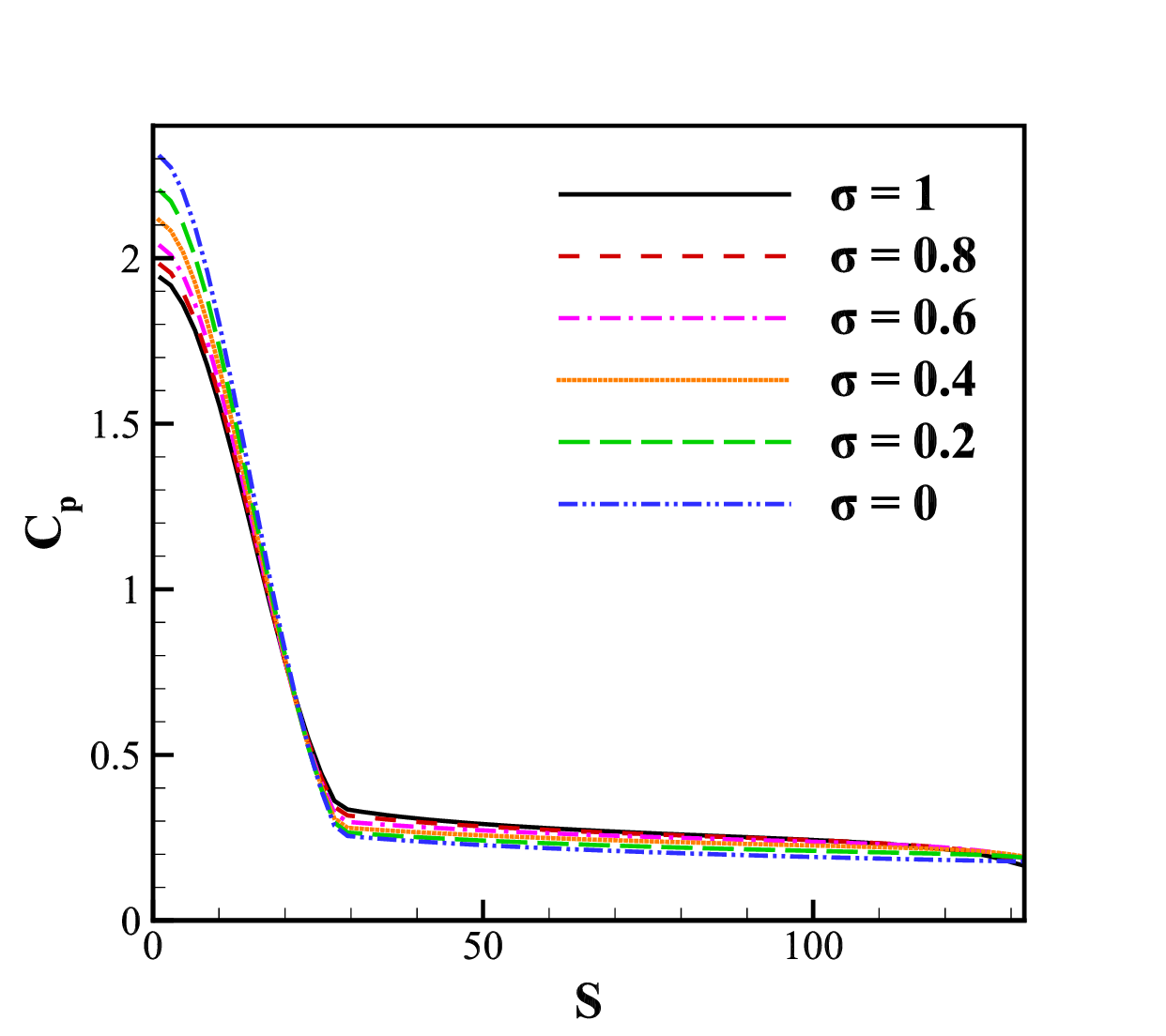}}
		\subfigure[]{\label{fig21b}\includegraphics[width=0.45\textwidth]{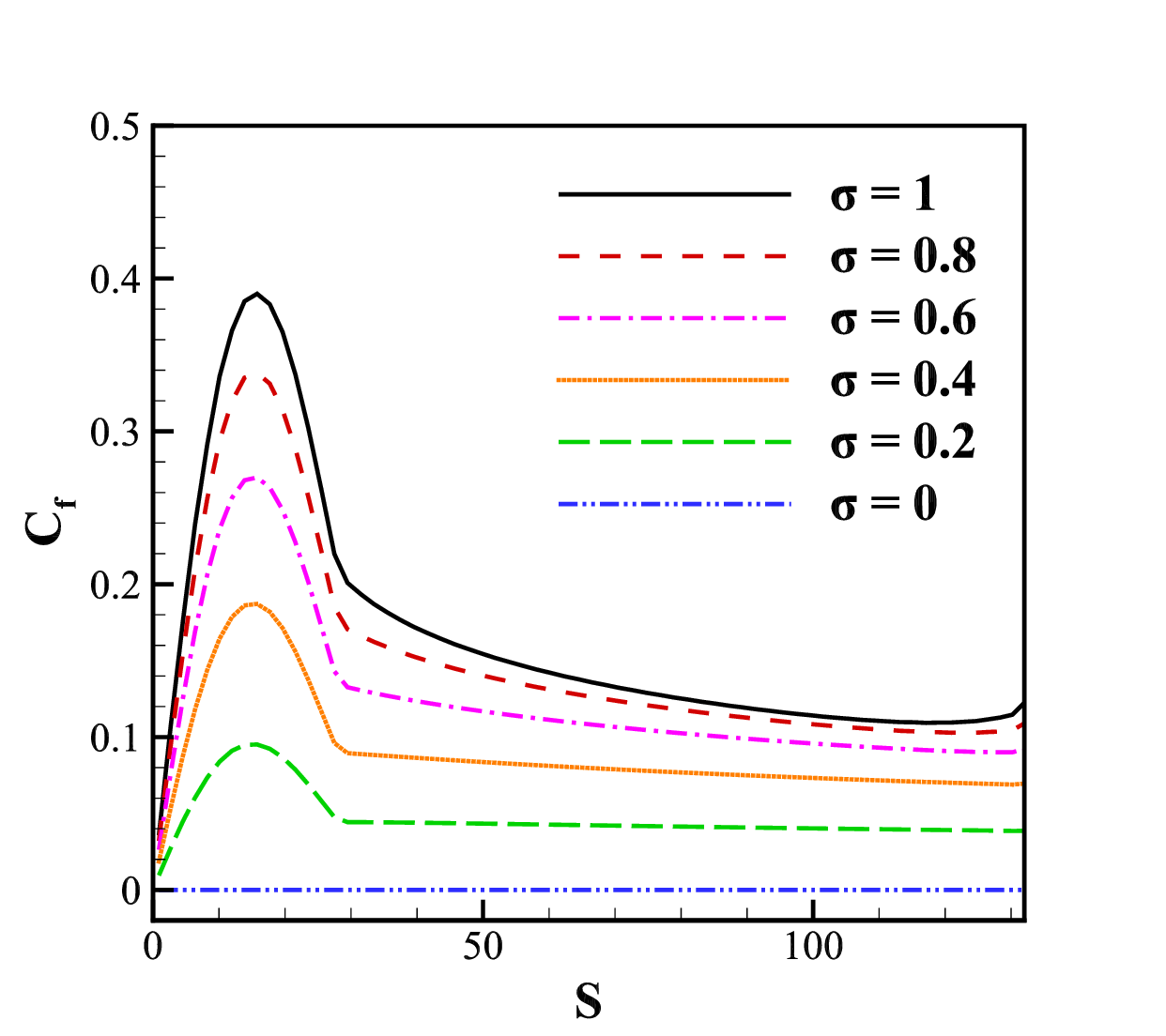}}
		\subfigure[]{\label{fig21c}\includegraphics[width=0.45\textwidth]{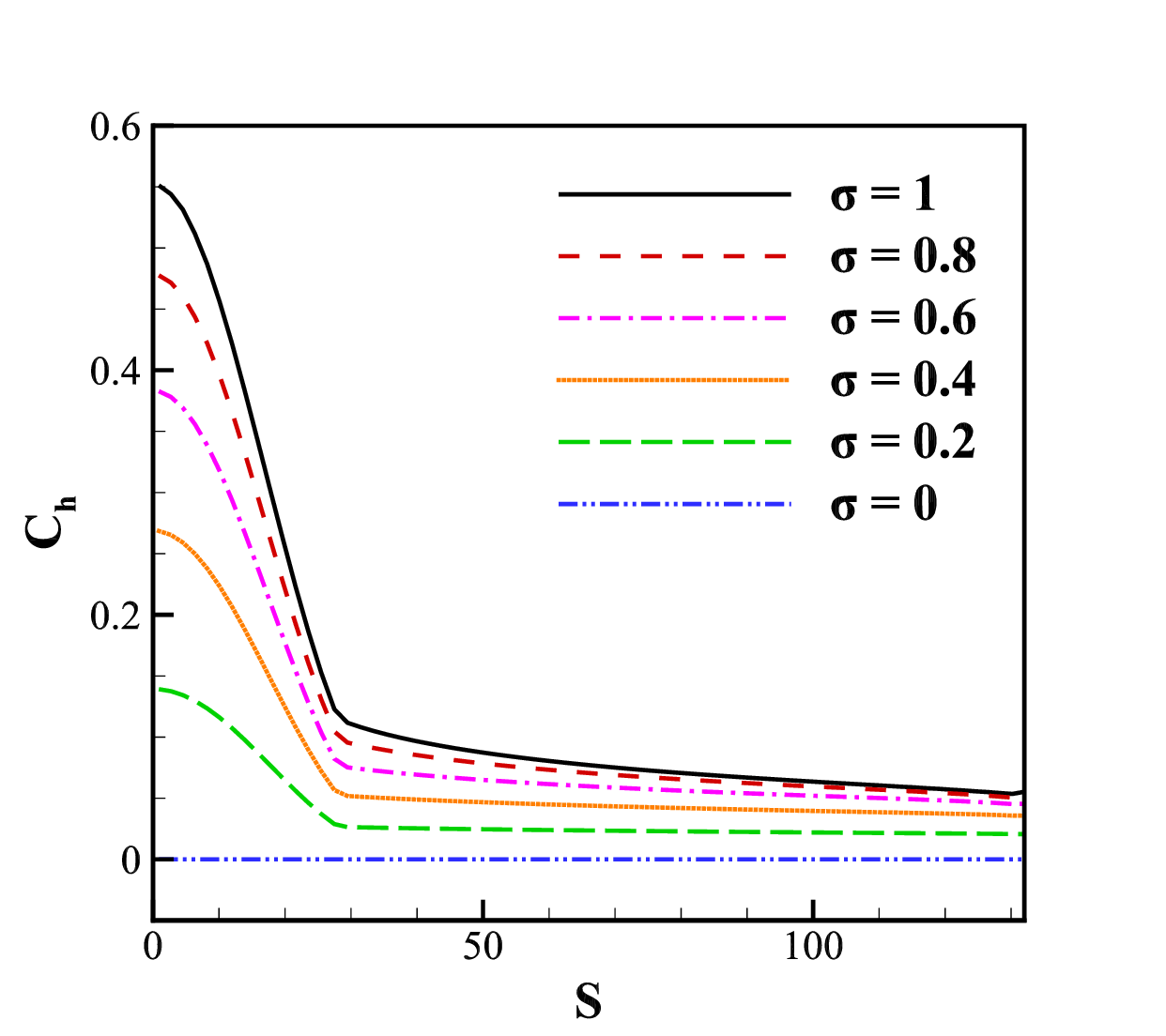}}
		\subfigure[]{\label{fig21d}\includegraphics[width=0.45\textwidth]{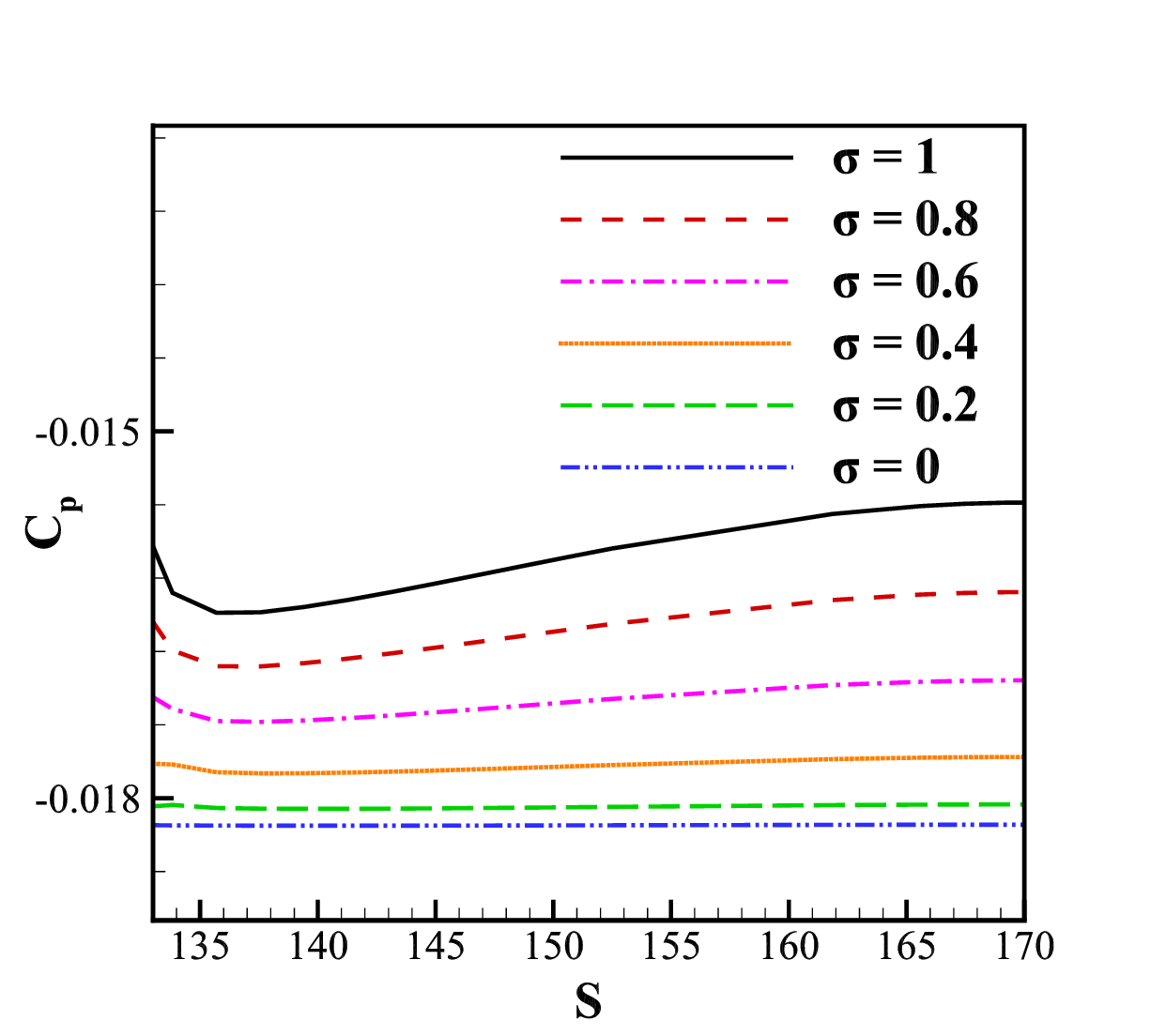}}
		\subfigure[]{\label{fig21e}\includegraphics[width=0.45\textwidth]{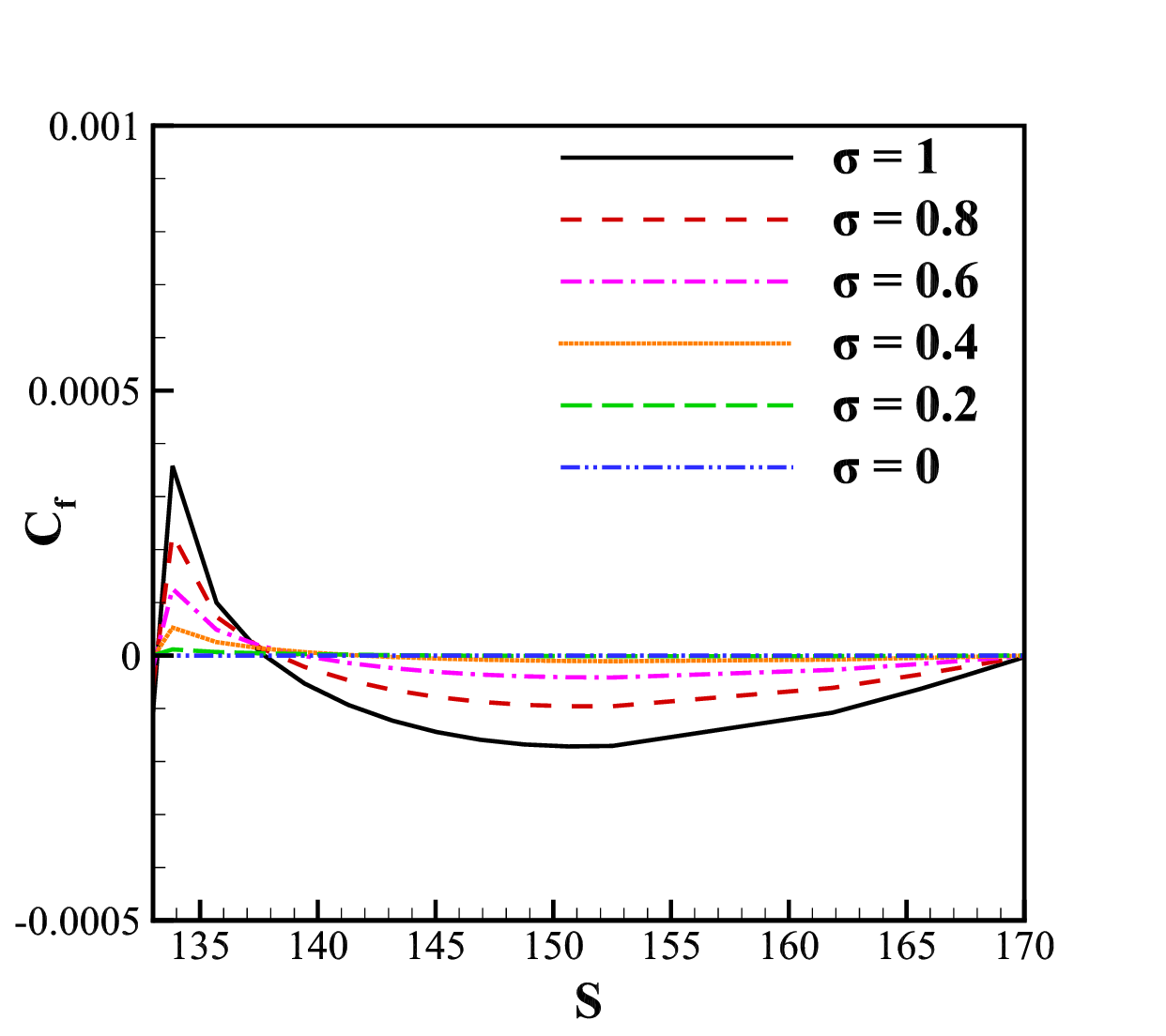}}
		\subfigure[]{\label{fig21f}\includegraphics[width=0.45\textwidth]{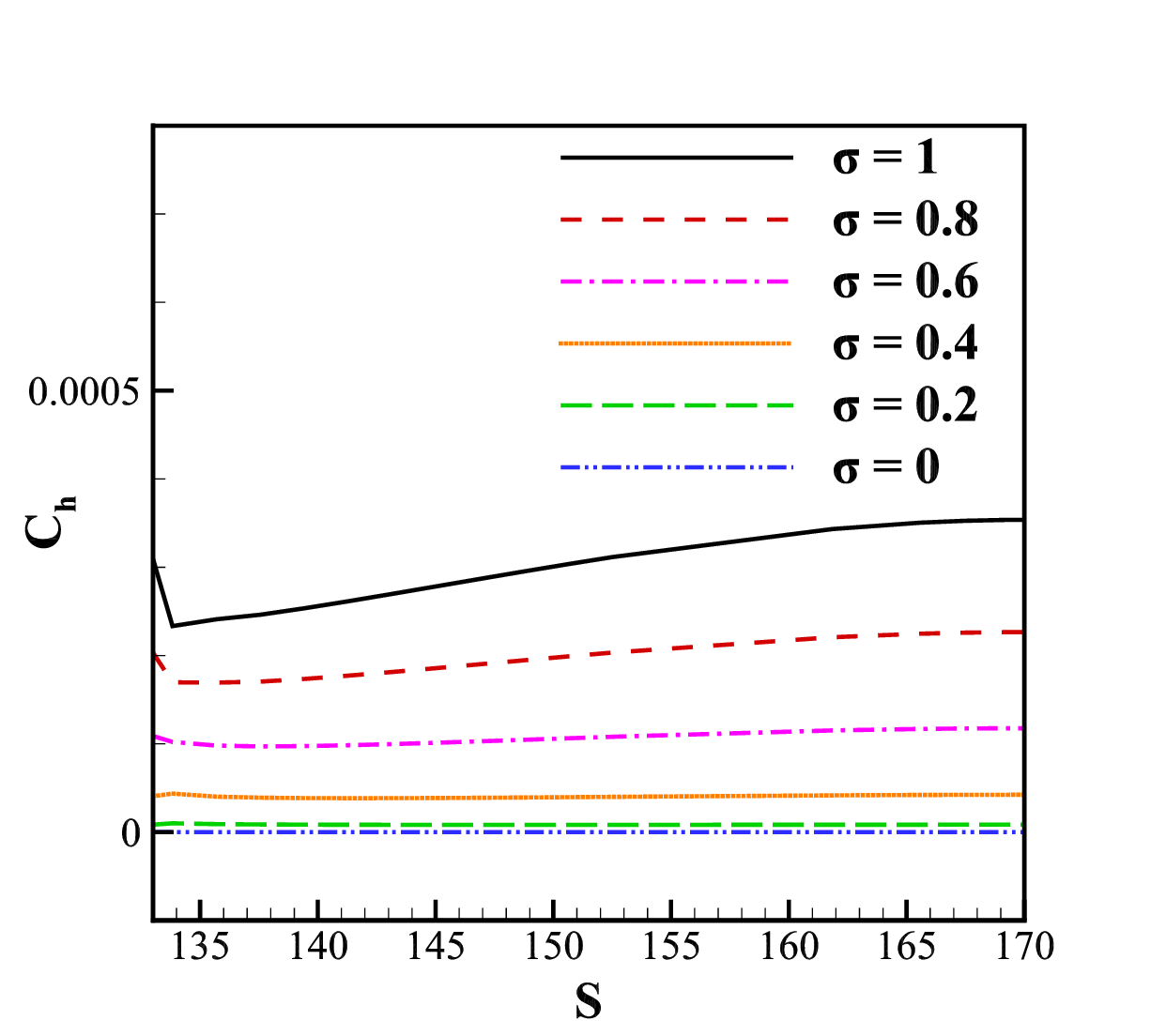}}
		\caption{\label{fig21}{Comparison of the (a) body pressure coefficient, (b) body skin friction coefficient, (c) body heat transfer, (d) bottom pressure coefficient, (e) bottom skin friction coefficient, and (f) bottom heat transfer coefficient on the surface of blunt wedge with different $\sigma$ ($Ma$ = 8.1, $Kn$ = 0.338, $T_{\infty}$ = 189 K, $T_{w}$ = 273 K).}}
	\end{figure}

	The pressure, skin friction, and heat transfer coefficients over the surface of the blunt wedge at $\sigma$ = 1 are depicted in Fig. \ref{fig18}, and the results are compared with those obtained from UGKS \cite{jiang2019implicit} and DS2V models. The parameter range $0<s<27.9$ represents the head arc of the wedge, $27.9<s<132.9$ corresponds to the body, and $132.9<s<170.3$ covers the bottom of the wedge. The comparison highlights a high degree of consistency between the findings of this study and those of UGKS and DS2V. Even variables at the bottom of the wedge, which are two or three orders of magnitude lower than at the stationary location, are predicted with high precision.	
	Results for pressure, skin friction, and heat transfer coefficients on the blunt wedge surface are also presented in Figs. \ref{fig19} and \ref{fig20} for $\sigma$ = 0.8 and $\sigma$ = 0, respectively. It is evident that the simulation results align well with those of DS2V.	
	A comprehensive analysis combining the calculation results for different $\sigma$ values is presented in Fig. \ref{fig21} to investigate the impact of the GSI model on the physical properties of the blunt wedge surface. The figure illustrates that the GSI model has a more pronounced impact on skin friction and heat flux in the head region of the blunt wedge.

	\subsection{Verification of diatomic flows}
	\subsubsection{Supersonic flow around a blunt circular cylinder in the slip regime}
	In this simulation, the diatomic gas flow is modeled, considering the degrees of freedom for translation and rotation of molecules. The accuracy of the GSI model in diatomic UGKS is validated through this simulation. The scenario involves simulating the supersonic flow around a blunt circular cylinder with a freestream Mach number of 5.0, a temperature of 273 K, and a Knudsen number of 0.1 (utilizing the cylinder's radius as the characteristic length). Nitrogen is used as the working gas, and the VHS molecular model with $\omega$ = 0.74 is employed. The freestream dimensionless density, temperature, and velocity are set to 1.0, 1.0, and 4.1833, respectively, while the wall dimensionless temperature is 1.0. The physical mesh and the unstructured discrete velocity mesh used in the simulation consist of 9480 cells and 1620 cells, respectively, as depicted in Fig. \ref{fig22}. The front end of the obtuse body is represented by a semicircle with a radius of $R$ = 0.01 m and the center coordinate at (0,0). The mesh on the surface has a HFL set to 0.05 mm, and the semicircle is discretized into 100 and 79 cells in the circumferential and radial directions, respectively.
	
	The distributions of pressure, skin friction, and heat transfer coefficients over the surface at $\sigma$ = 1, 0.8, and 0 are depicted in Figs. \ref{fig23}, \ref{fig24}, and \ref{fig25}, respectively. Notably, the present results closely align with those obtained from DS2V. The distribution of physical quantities on the surface for various $\sigma$ values is presented in Fig. \ref{fig26}, highlighting the significant impact of the GSI model on the distribution of skin friction and heat transfer on the wall.
	
	\subsubsection{Rarefied hypersonic ﬂow over a truncated flat plate in the slip regime}
	As the previous simulations have been compared with numerical results, assessing the current GSI algorithm's performance through a comparison with experimental results further strengthens its validation. Allegre et al. \cite{allegre1994rarefied} conducted an experiment with a truncated flat plate positioned at a distance from a nozzle, producing a nitrogen flow with a Mach number of 20.2 and a temperature of 13.32 K, as detailed in Table \ref{tab:plate conditions}. The plate dimensions are 100 mm in length, 100 mm in width, and 5 mm in thickness, with the wall temperature maintained at 290 K. Two angles of attack, namely 0 degrees and 10 degrees, were investigated in this experiment. Experimental measurements of pressure, heat flux, and density were reported with errors of 15$\%$, 10$\%$, and 10$\%$, respectively \cite{palharini2015benchmark}. Utilizing the VHS model with $\omega=0.74$, the gas mean free path is represented as:
	
	\begin{equation}
		{{\lambda }_{\infty }}=\frac{1}{\sqrt{2}\pi d_{ref}^{2}{{n}_{\infty }}}{{\left( \frac{{{T}_{\infty }}}{{{T}_{ref}}} \right)}^{\omega -1/2}}.
	\end{equation}
	
	The molecular diameter and relative molecular mass of nitrogen are $4.17\times {{10}^{-10}}$ m and $4.65\times {{10}^{-26}}$ kg, respectively. Consequently, the molecular gas mean free path and number density are calculated as 0.00169 m and $3.713\times {{10}^{20}}\text{ }\cdot /{{m}^{3}}$, respectively. Simultaneously, utilizing the length of the plate (100 mm) as the characteristic length, the corresponding Knudsen number is determined to be 0.0169. Tables \ref{tab:ref quantities} and \ref{tab:dimless quantities} present the reference physical quantities and dimensionless physical quantities, respectively.	
	
	The unstructured physical mesh and velocity mesh utilized for the numerical experiments comprise 8055 and 3186 cells, respectively, as illustrated in Figs. \ref{fig27a} and \ref{fig27b}. The HFL of the mesh on the surface is set to 0.1 mm. To examine the effect of the GSI model on the simulation results in this example, the AC is assigned only two values: 1 and 0.8. Notably, the latter value corresponds to literature prescriptions for a nitrogen flow over a steel plate at a temperature of $T_w$ = 300 K \cite{rader2005measurements, trott2011experimental, schouler2020survey}.	
	
	Figs. \ref{fig28} and \ref{fig29} illustrate the simulated pressure and heat transfer coefficients on the lower surface of the flat plate at AOA 0 and 10 degrees, respectively. Notably, the results obtained by both the DSMC method and the proposed method with $\sigma$ = 0.8 at AOA 0 degrees exhibit a closer agreement with the experimental results compared to those obtained at $\sigma$ = 1. Moreover, the simulation results of the proposed method align more favorably with the experimental data than the DSMC results. 
	However, for the AOA 10 degrees, a different phenomenon can be observed. Although the results of the proposed method at $\sigma$ = 0.8 are closer to the experimental values than those at $\sigma$ = 1, the DSMC results for the pressure coefficient are in better agreement with the experimental values. Meanwhile, the results of the proposed method more closely match the experimental values for the heat transfer coefficient under $\sigma$ = 1 at the leading edge of the plate than those under $\sigma$ = 0.8. Overall, these numerical experiment results show that the GSI model with an adjustable adaption factor should be used because it is more reasonable and obtains results closer to the experimental values than the results at $\sigma$ = 1.
	However, for an AOA of 10 degrees, a different phenomenon becomes evident. While the results of the proposed method at $\sigma$ = 0.8 are closer to the experimental values than those at $\sigma$ = 1, the DSMC results for the pressure coefficient show better agreement with the experimental values. Meanwhile, the results of the proposed method more closely align with the experimental values for the heat transfer coefficient under $\sigma$ = 1 at the leading edge of the plate than those under $\sigma$ = 0.8. Overall, these numerical experiment results indicate that the GSI model with an adjustable adaption factor should be employed, as it proves to be more reasonable and yields results closer to the experimental values than those obtained at $\sigma$ = 1.
		
	\begin{table}
		\centering 
		\caption{\label{tab:plate conditions}Freestream conditions for flat-plate simulations.}
		\begin{tabular}{p{120pt}<{\centering} p{100pt}<{\centering} p{80pt}<{\centering}}
			\hline
			\hline			
			Parameter & Value & Unit\\
			\hline
			Velocity($V_{\infty}$)	     & 1503	                        & m/s	\\
			\hline 
			Temperature($T_{\infty}$)	 & 13.32                     	& K	    \\
			\hline
			Density($\rho_{\infty}$)	 & $1.727\times {{10}^{-5}}$	& kg/m	\\
			\hline
			Pressure($p_{\infty}$)	     & 0.0683                   	& Pa	\\
			\hline
			\hline
		\end{tabular}
	\end{table}
	
	\begin{table}
		\centering 
		\caption{\label{tab:ref quantities}Reference physical quantities.}
		\begin{tabular}{p{120pt}<{\centering} p{100pt}<{\centering} p{80pt}<{\centering}}
			\hline
			\hline			
			Parameter & Value & Unit\\
			\hline
			Velocity($V_{ref}$)	     & $\sqrt{2{R}_{{{N}_{2}}}T_{\infty}}$	            & m/s	\\
			\hline 
			Temperature($T_{ref}$)	 & $T_{\infty}$                     	& K	    \\
			\hline
			Density($\rho_{ref}$)	 & $\rho_{\infty}$	        & kg/m	\\
			\hline
			Length($L_{ref}$)	     & 1                  	& mm	\\
			\hline
			\hline
		\end{tabular}
	\end{table}
	
	\begin{table}
		\centering 
		\caption{\label{tab:dimless quantities}Dimensionless physical quantities.}
		\begin{tabular}{p{100pt}<{\centering} p{100pt}<{\centering}}
			\hline
			\hline			
			Parameter & Value\\
			\hline
			Velocity($V$)	     & 16.9   \\
			\hline 
			Temperature($T$)	 & 1      \\
			\hline
			Density($\rho$)	     & 1      \\
			\hline
			Pressure($p$)	     & 0.5	  \\
			\hline
			Length($p$)	         & 100    \\
			\hline
			\hline
		\end{tabular}
	\end{table}	
	
	\section{Conclusion}\label{sec:conclusion}
		The GSI serves as the source of aerodynamic forces and heat transfer on flying vehicles, playing a pivotal role in influencing the accuracy of flow predictions. However, the earlier DVM practice employs only a complete energy accommodation diffuse boundary condition, which is not accurate for multi-scale and rarefied flows. The primary challenge in applying the more physical Maxwell GSI model within the DVM framework revolves around the specular boundary where the mismatch in DVS occurs.
		In the proposed GSI algorithm, the macroscopic conservation is ensured through the reflected macroscopic flux, calculated from the incident macroscopic flux. A collaboration of interpolation methods and a new numerical error correction method is employed to ensure micro-macro consistency. Notably, the interpolation process in DVS is straightforward and presently does not require a limiter. The deviation of the equilibrium distribution function is utilized to correct numerical errors, which is not only efficient but also concise.
		The proposed GSI algorithm undergoes validation through numerous multi-scale flow simulations, demonstrating consistent results with DSMC. Simulations of a truncated flat plate in the slip regime reveal that both present results and DSMC results with physical accommodation are closer to the experimental data than those with full accommodation, indicating that an incomplete accommodation GSI model is necessary. As a result, the proposed GSI algorithm enables both DVM and unified methods to employ a more precise boundary condition for predicting the behavior of multi-scale flows and other physical fields.
		Moreover, the proposed GSI algorithm is easily extendable to three-dimensional cases and other physical fields using DVM and unified methods.
	
	\section*{Acknowledgments}
		The authors thank Prof. Kun Xu in the Hong Kong University of Science and Technology and Prof. Zhaoli Guo in Huazhong University of Science and Technology for discussions of the UGKS, the DUGKS and multi-scale flow simulations. Jianfeng Chen thanks Dr. Ruifeng Yuan in Northwestern Polytechnical University for useful discussions on the unstructured velocity space. Jianfeng Chen thanks researcher/Dr. Dingwu Jiang in the China Aerodynamics Research and Development Center and Dr. Hao Jin in Northwestern Polytechnical University for their help in the use of DS2V. Sha Liu thanks Prof. Ming Fang, Prof. Lin Bi and Prof. Guohua Tu in the China Aerodynamics Research and Development Center for discussions of effective boundary conditions in rarefied flows.
		This work was supported by the National Natural Science Foundation of China (Grant Nos. 12172301 and 12072283) and the 111 Project of China (Grant No. B17037). This work is supported by the high performance computing power and technical support provided by Xi'an Future Artificial Intelligence Computing Center.
	
	\clearpage
	
	\begin{figure}
		\centering
		\subfigure[]{\label{fig21a}\includegraphics[width=0.45\textwidth]{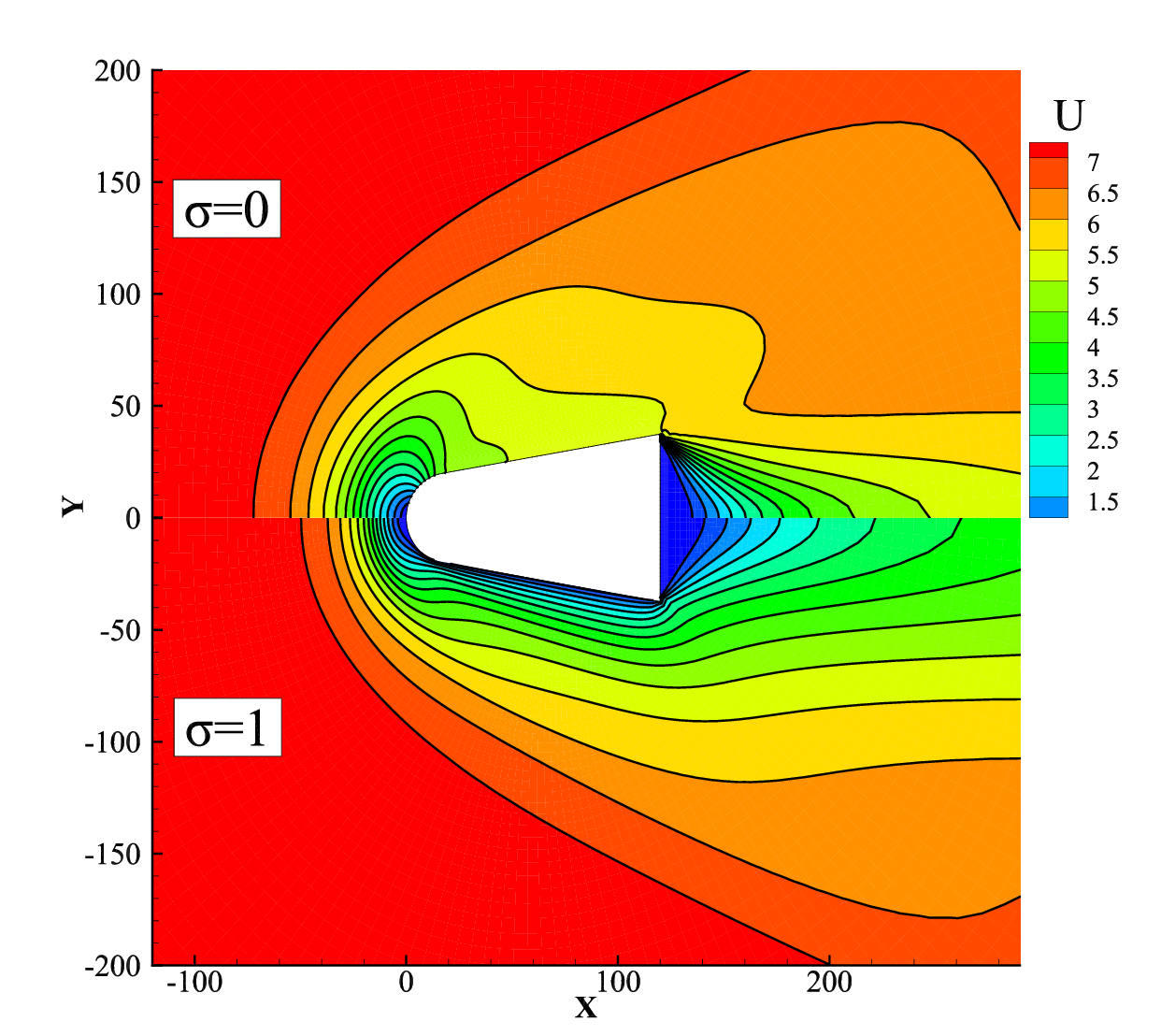}}
		\subfigure[]{\label{fig21b}\includegraphics[width=0.45\textwidth]{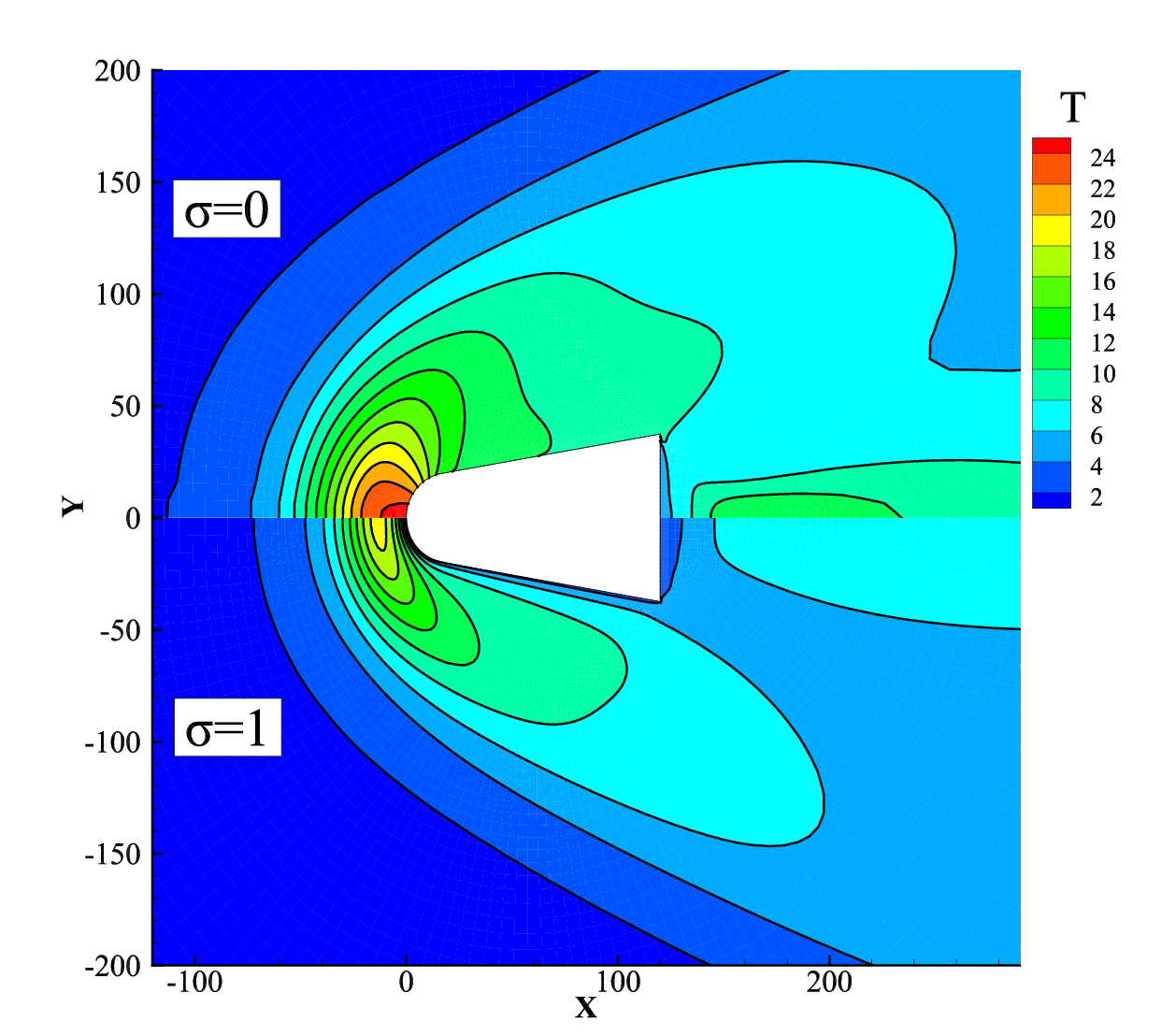}}
		\caption{\label{fig21add}{Comparison of the (a) horizontal velocity contour and (b) temperature contour from specular reflection model and the diffuse reflection model for hypersonic flows passing a blunt wedge ($Ma$ = 8.1, $Kn$ = 0.338, $T_{\infty}$ = 189 K, $T_{w}$ = 273 K).}}
	\end{figure}
	
	\begin{figure}
		\centering
		\subfigure[]{\label{fig22a}\includegraphics[width=0.45\textwidth]{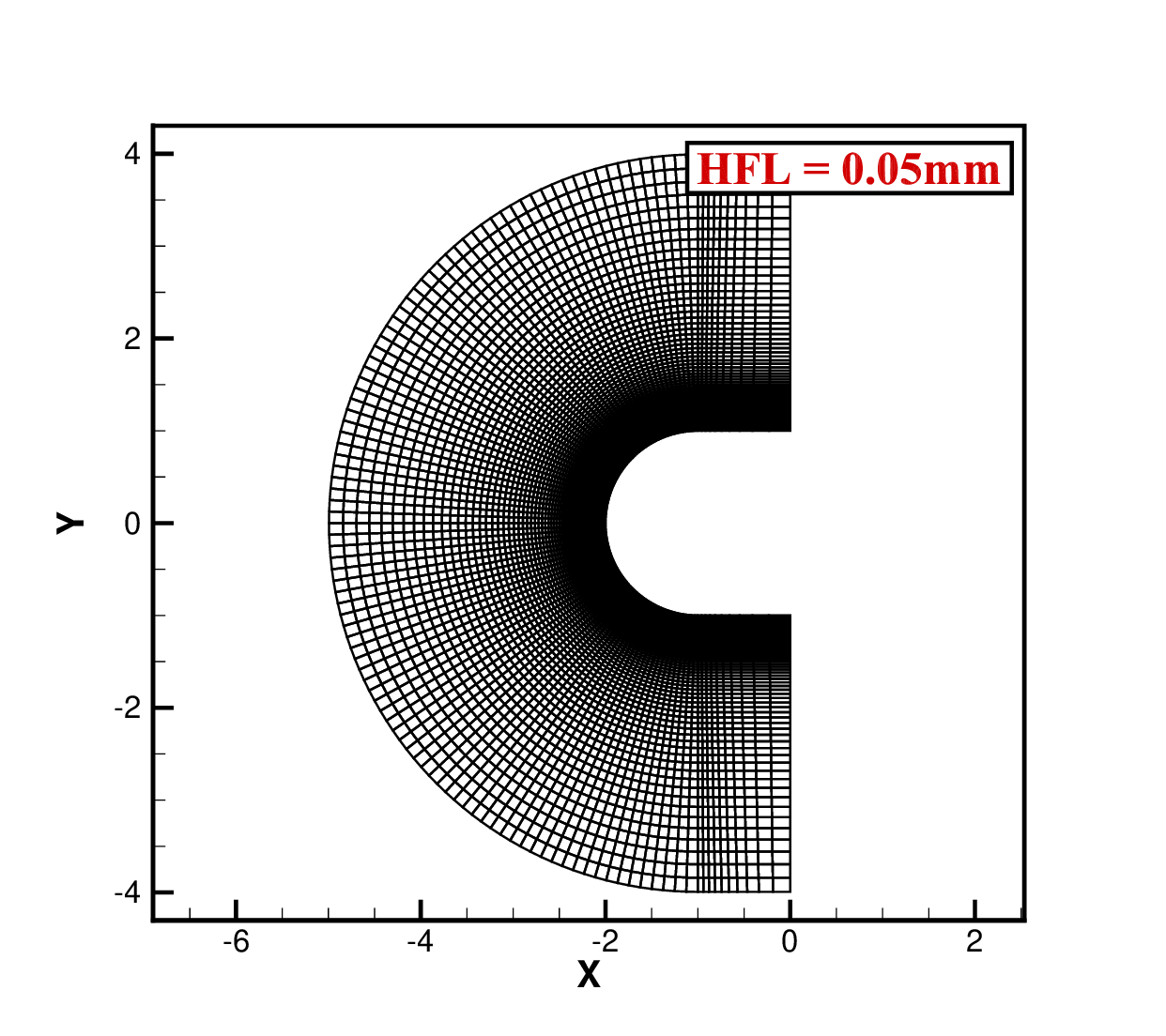}}
		\subfigure[]{\label{fig22b}\includegraphics[width=0.37\textwidth]{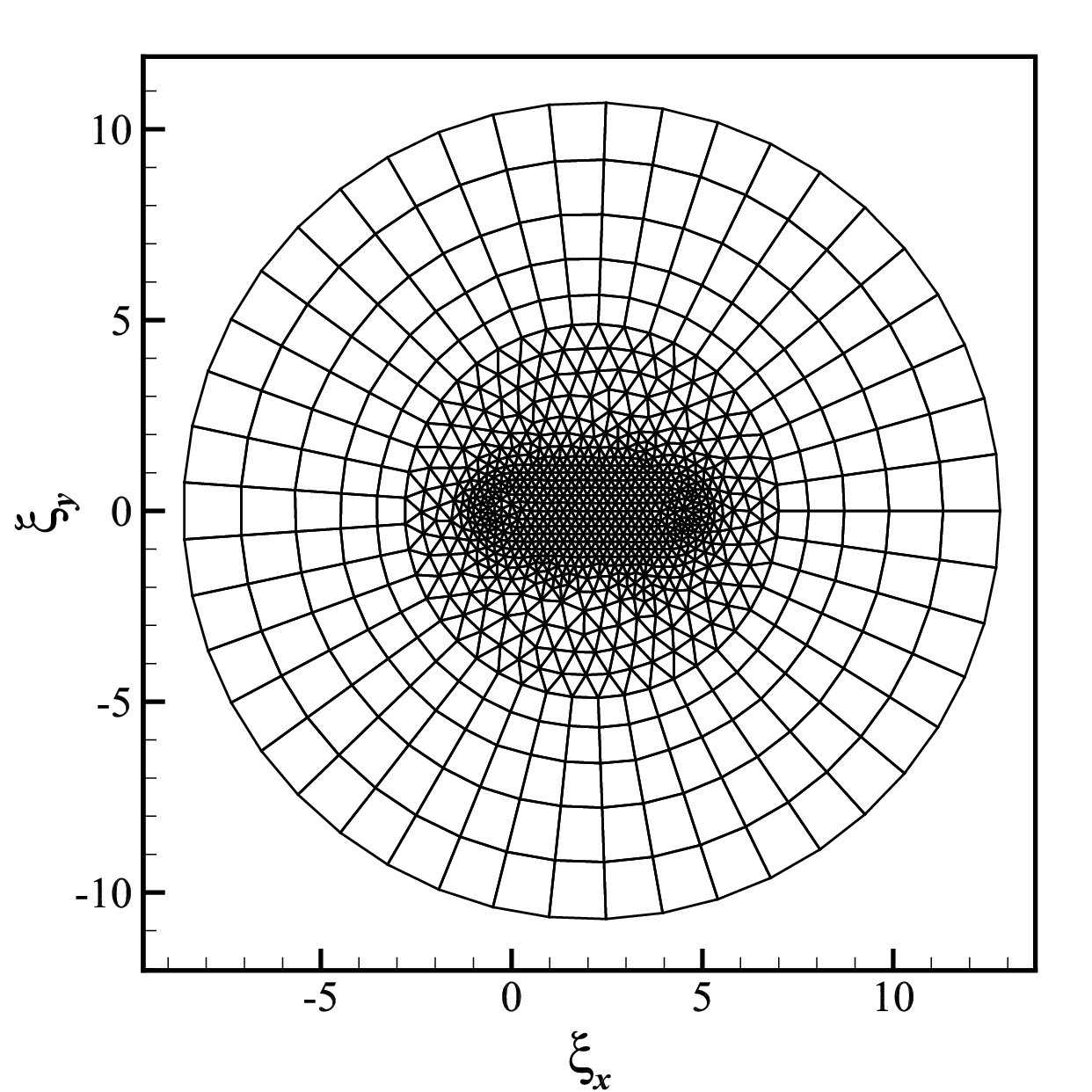}}
		\caption{\label{fig22}{The (a) unstructured physical mesh and (b) unstructured velocity mesh for the hypersonic flow passing a blunt circular cylinder ($Ma$ = 5.0, $Kn$ = 0.1, $T_{\infty}$ = 273 K, $T_{w}$ = 273 K).}}
	\end{figure}
	
	\begin{figure}
		\centering
		\subfigure[]{\label{fig23a}\includegraphics[width=0.45\textwidth]{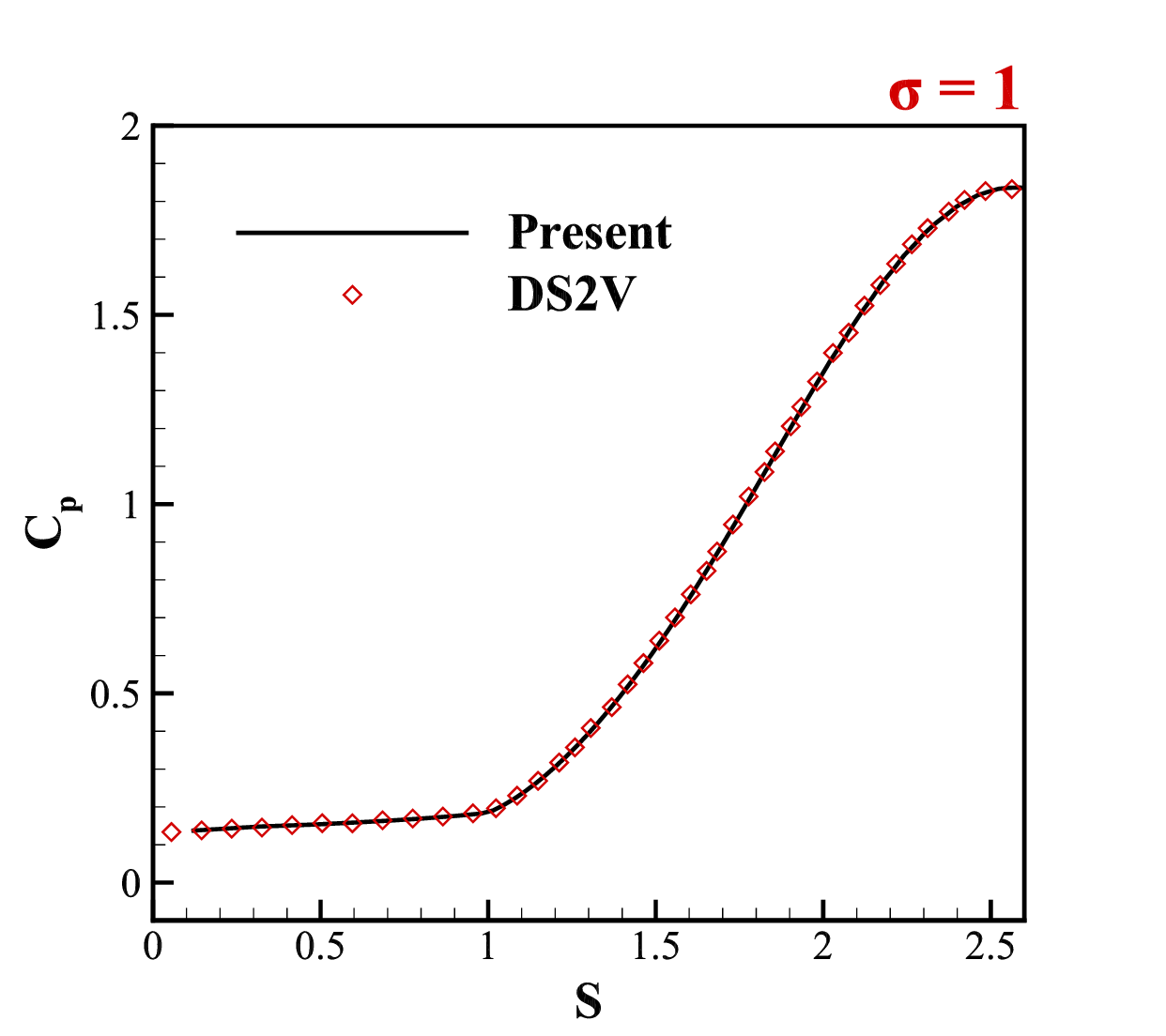}}
		\subfigure[]{\label{fig23b}\includegraphics[width=0.45\textwidth]{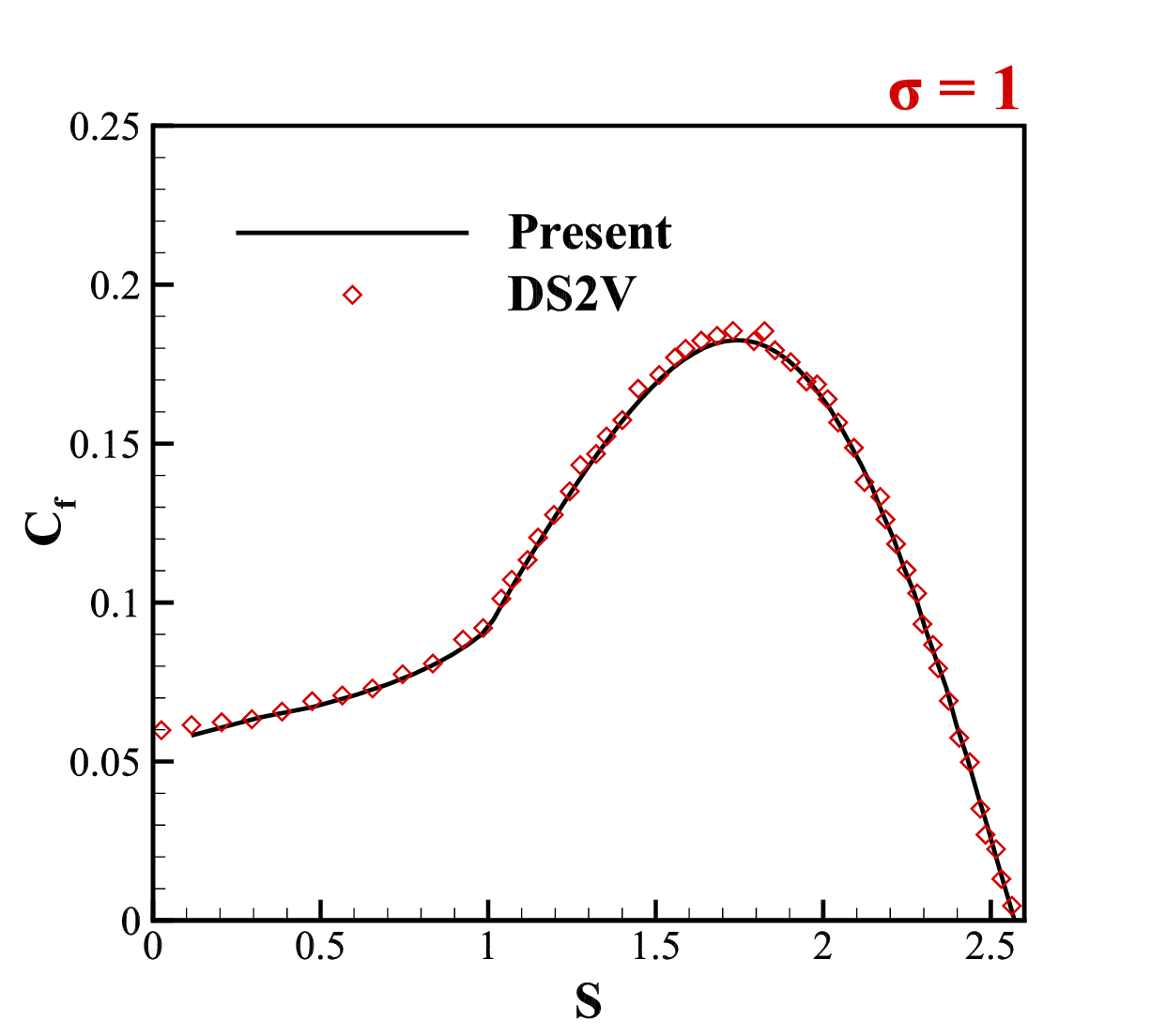}}
		\subfigure[]{\label{fig23c}\includegraphics[width=0.45\textwidth]{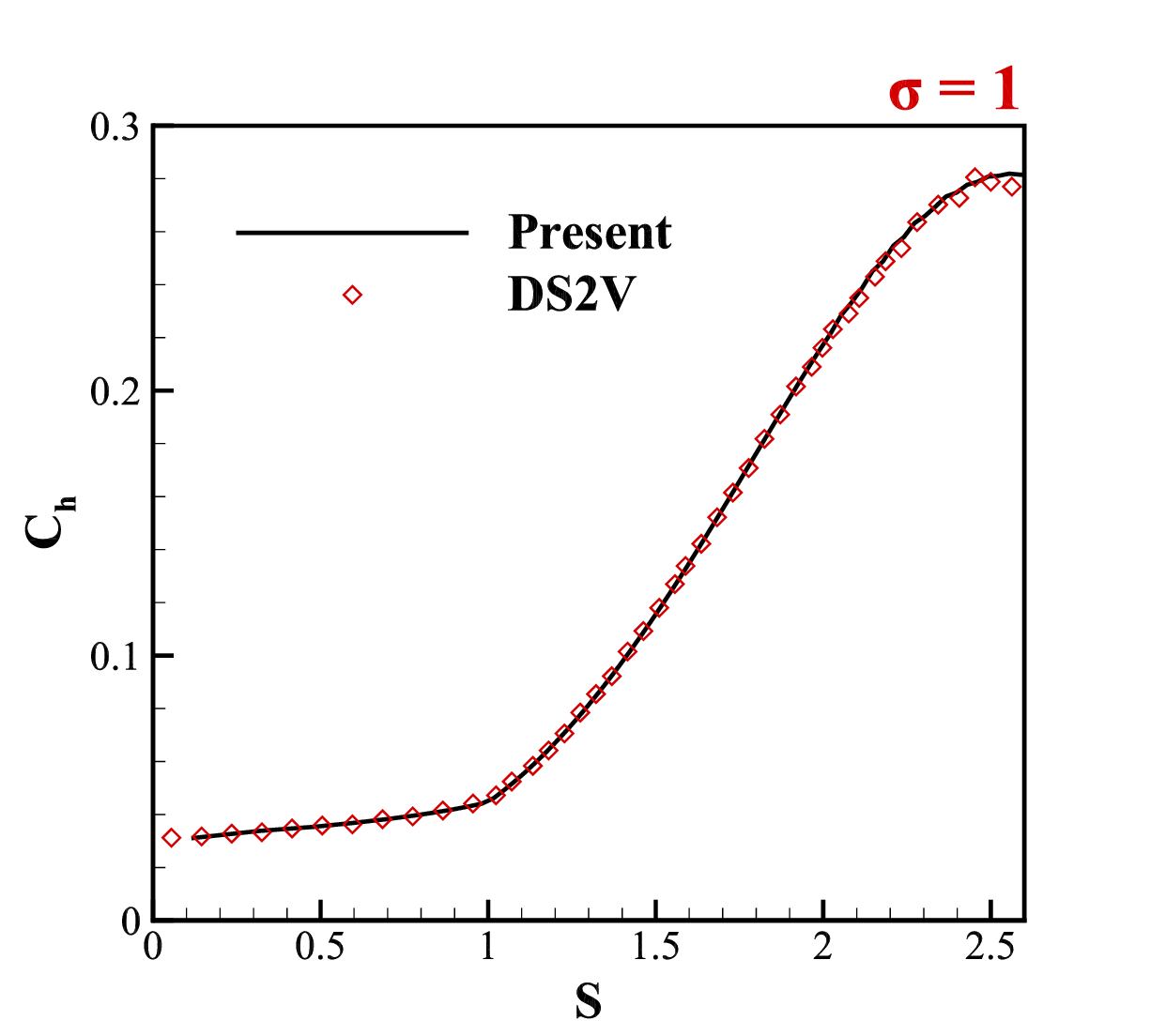}}
		\caption{\label{fig23}{Comparison of the (a) pressure coefficient, (b) skin friction coefficient, and (c) heat transfer coefficient on the surface of blunt circular cylinder with $\sigma$ = 1 ($Ma$ = 5.0, $Kn$ = 0.1, $T_{\infty}$ = 273 K, $T_{w}$ = 273 K).}}
	\end{figure}
	
	\begin{figure}
		\centering
		\subfigure[]{\label{fig24a}\includegraphics[width=0.45\textwidth]{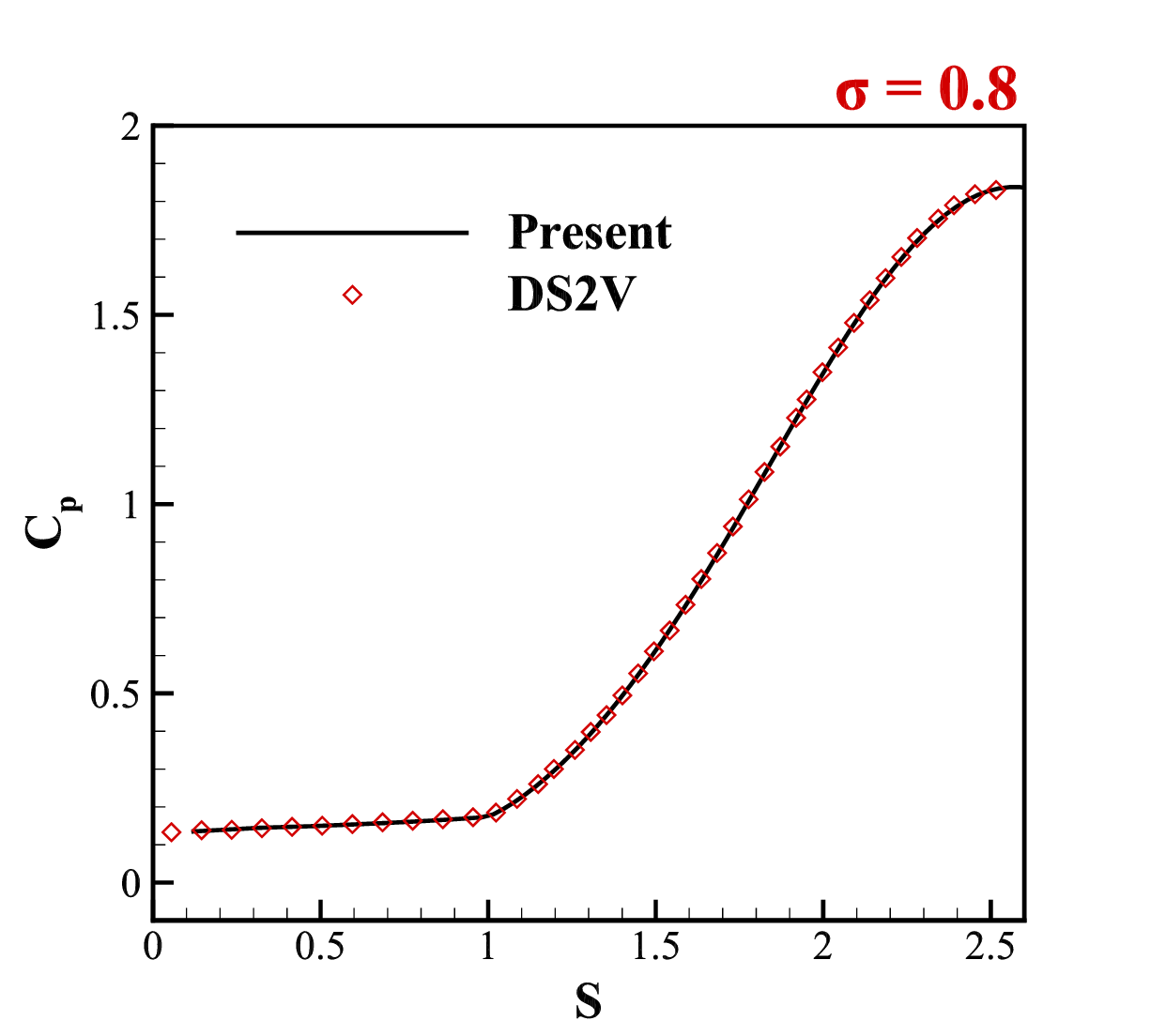}}
		\subfigure[]{\label{fig24b}\includegraphics[width=0.45\textwidth]{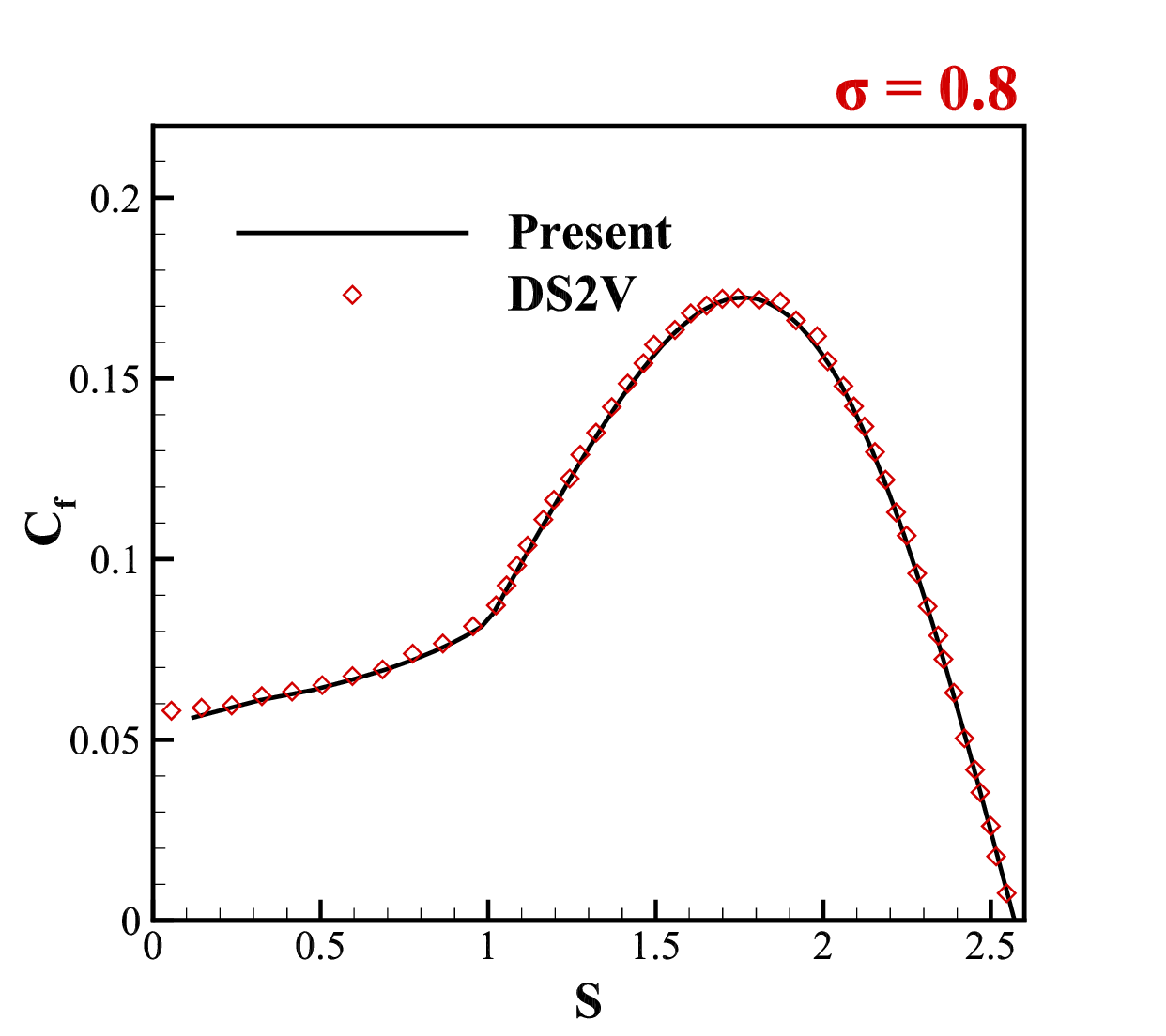}}
		\subfigure[]{\label{fig24c}\includegraphics[width=0.45\textwidth]{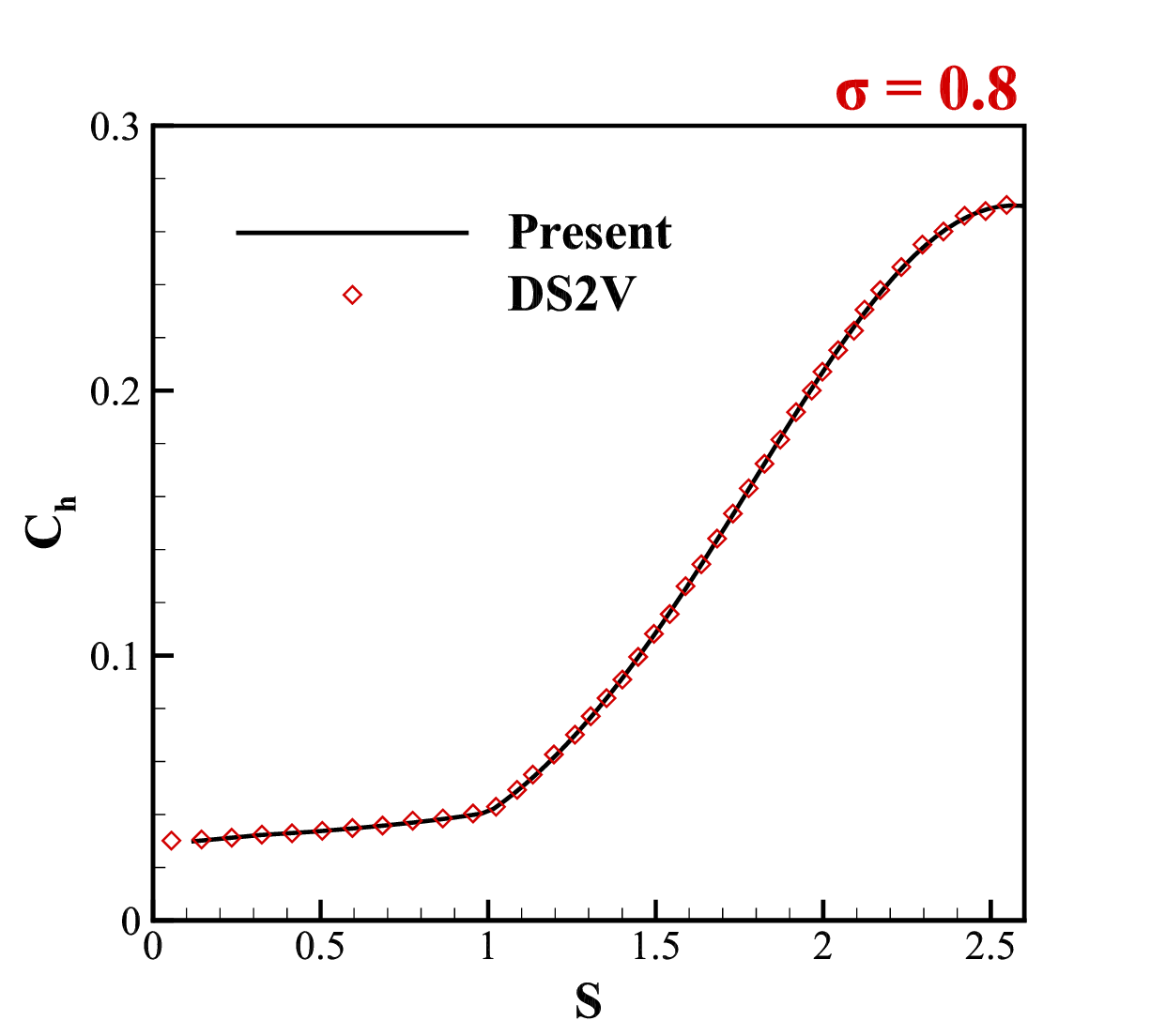}}
		\caption{\label{fig24}{Comparison of the (a) pressure coefficient, (b) skin friction coefficient, and (c) heat transfer coefficient on the surface of blunt circular cylinder with $\sigma$ = 0.8 ($Ma$ = 5.0, $Kn$ = 0.1, $T_{\infty}$ = 273 K, $T_{w}$ = 273 K).}}
	\end{figure}
	
	\begin{figure}
		\centering
		\subfigure[]{\label{fig25a}\includegraphics[width=0.45\textwidth]{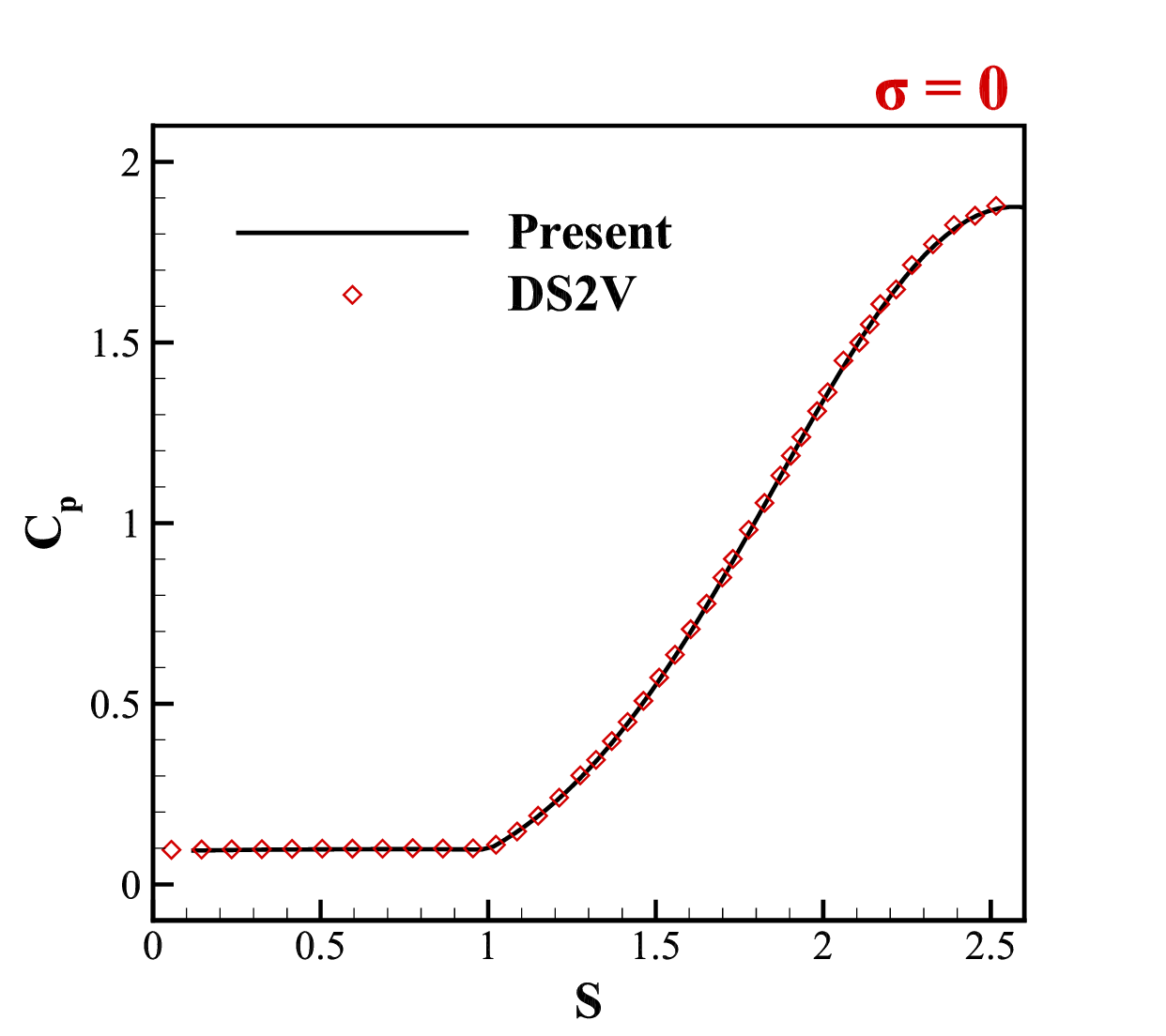}}
		\subfigure[]{\label{fig25b}\includegraphics[width=0.45\textwidth]{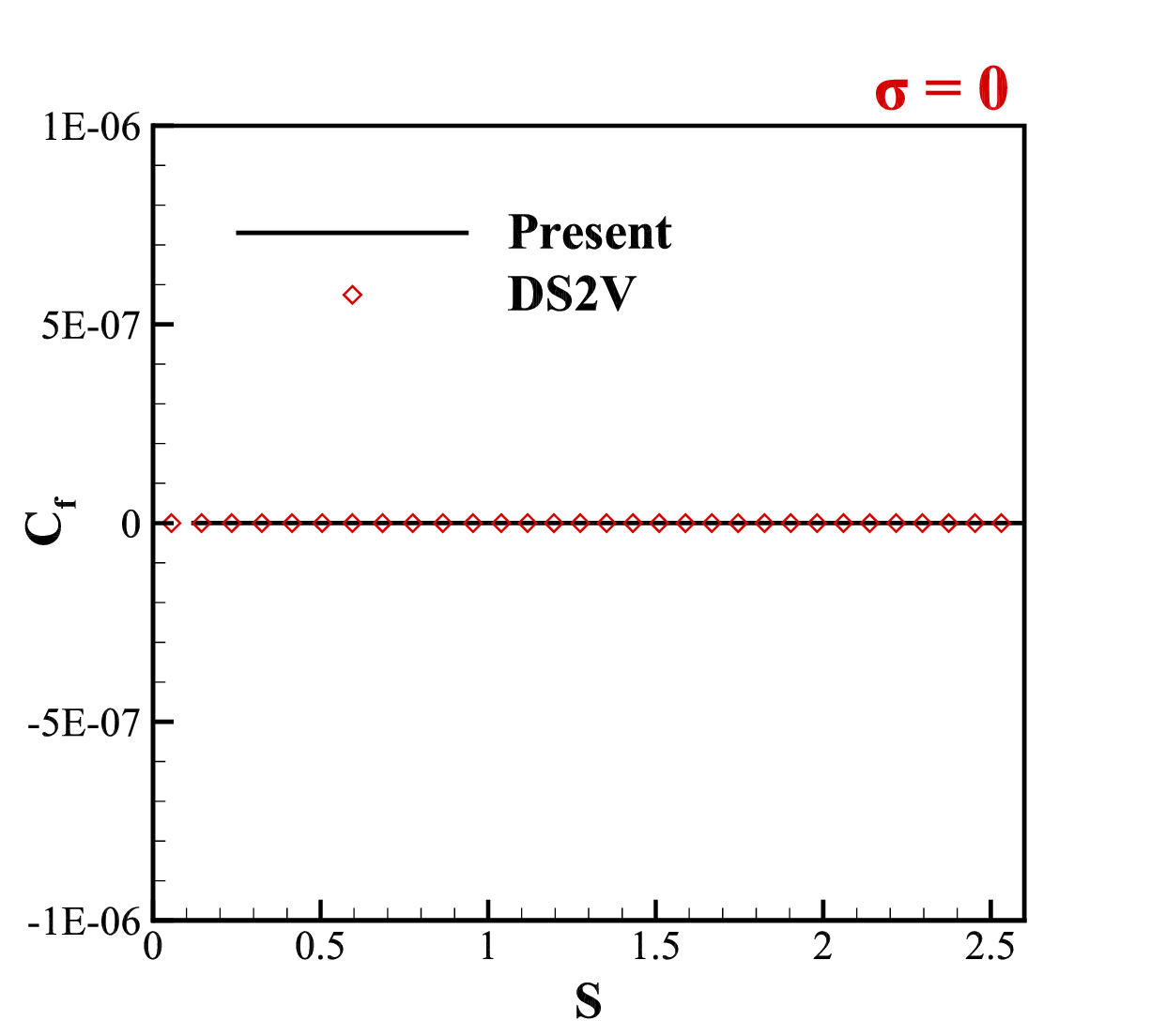}}
		\subfigure[]{\label{fig25c}\includegraphics[width=0.45\textwidth]{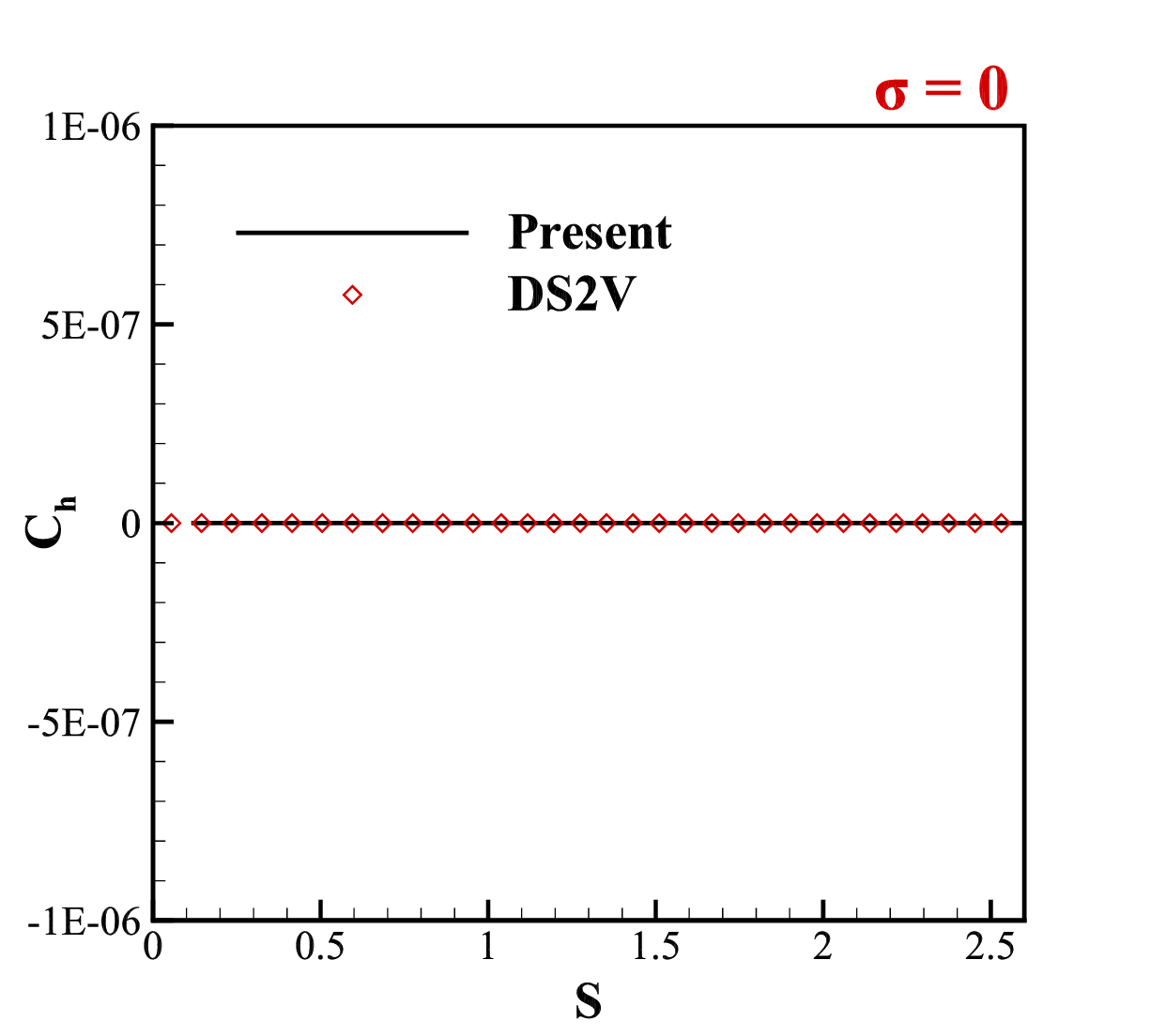}}
		\caption{\label{fig25}{Comparison of the (a) pressure coefficient, (b) skin friction coefficient, and (c) heat transfer coefficient on the surface of blunt circular cylinder with $\sigma$ = 0 ($Ma$ = 5.0, $Kn$ = 0.1, $T_{\infty}$ = 273 K, $T_{w}$ = 273 K).}}
	\end{figure}
	
	\begin{figure}
		\centering
		\subfigure[]{\label{fig26a}\includegraphics[width=0.45\textwidth]{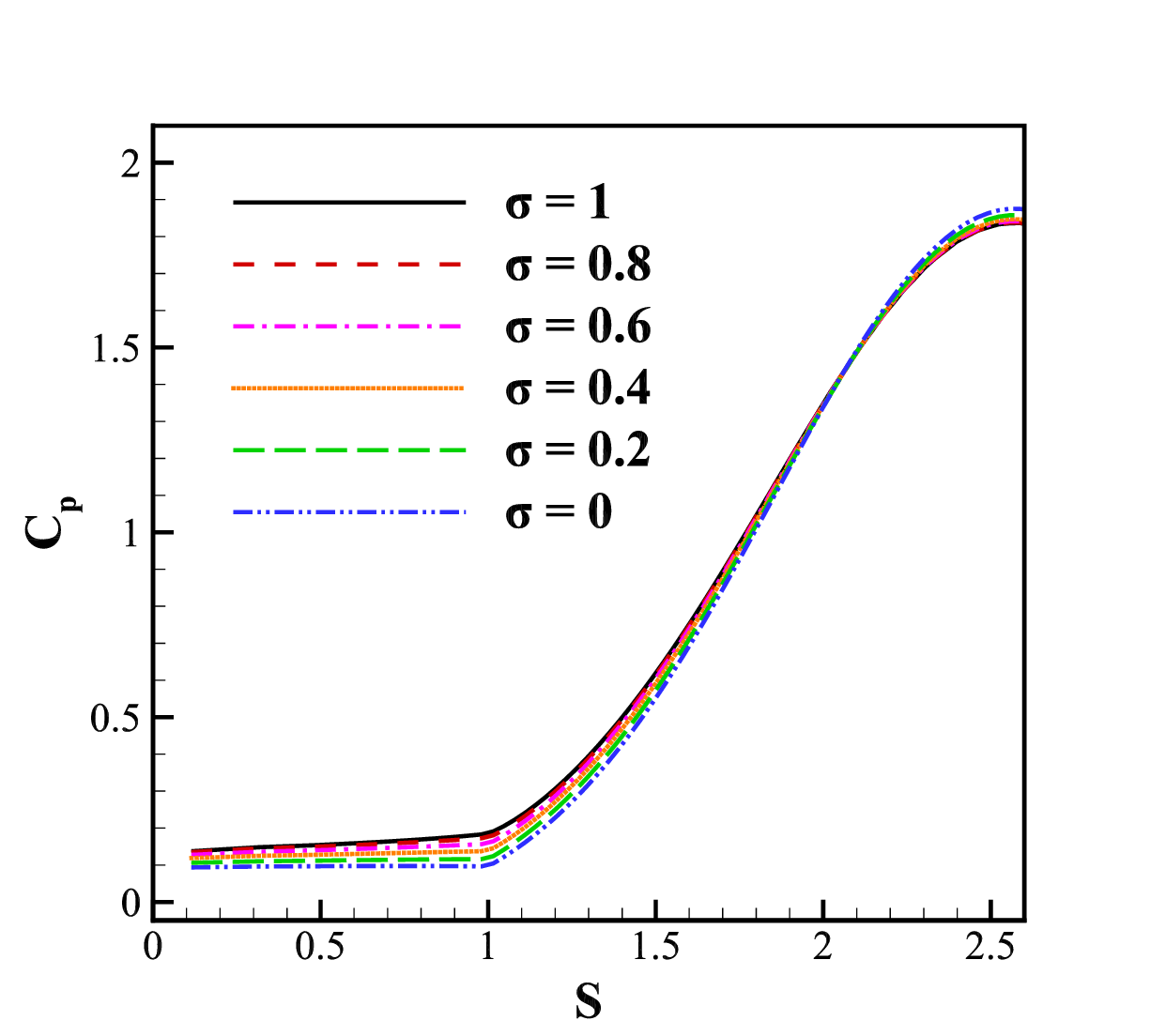}}
		\subfigure[]{\label{fig26b}\includegraphics[width=0.45\textwidth]{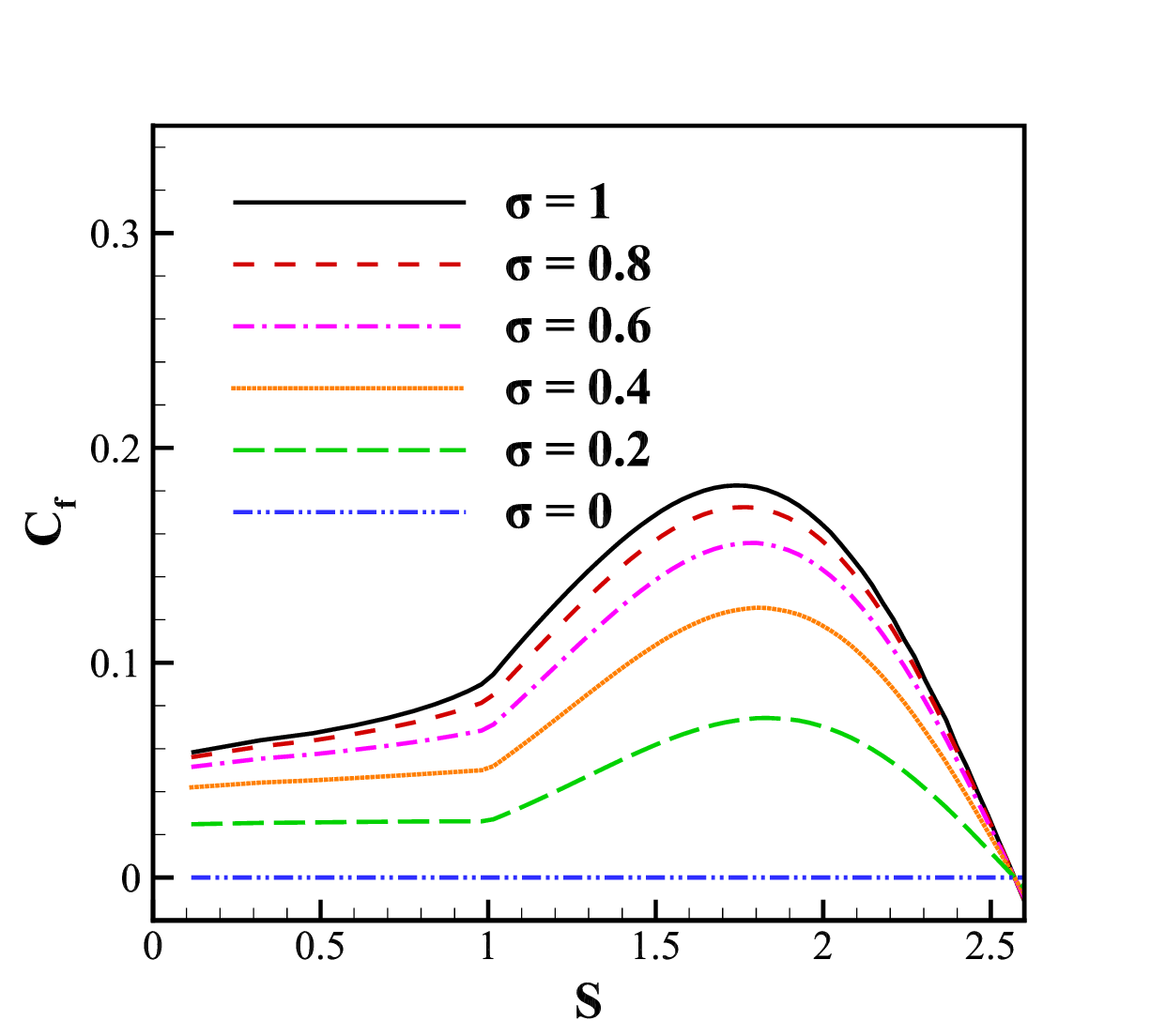}}
		\subfigure[]{\label{fig26c}\includegraphics[width=0.45\textwidth]{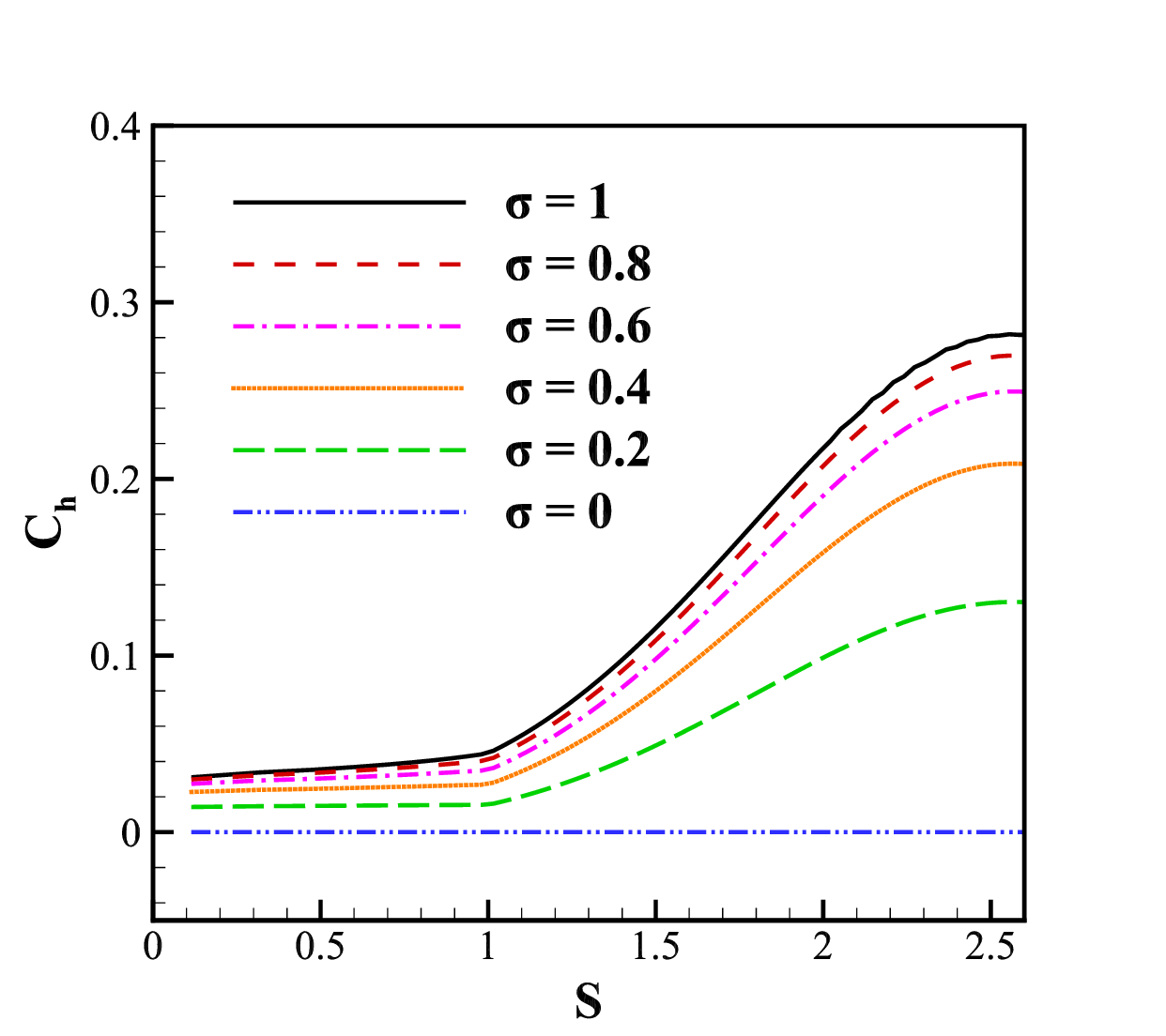}}
		\caption{\label{fig26}{Comparison of the (a) pressure coefficient, (b) skin friction coefficient, and (c) heat transfer coefficient on the surface of blunt circular cylinder with different $\sigma$ ($Ma$ = 5.0, $Kn$ = 0.1, $T_{\infty}$ = 273 K, $T_{w}$ = 273 K).}}
	\end{figure} 
	
	\begin{figure}
		\centering
		\subfigure[]{\label{fig27a}\includegraphics[width=0.45\textwidth]{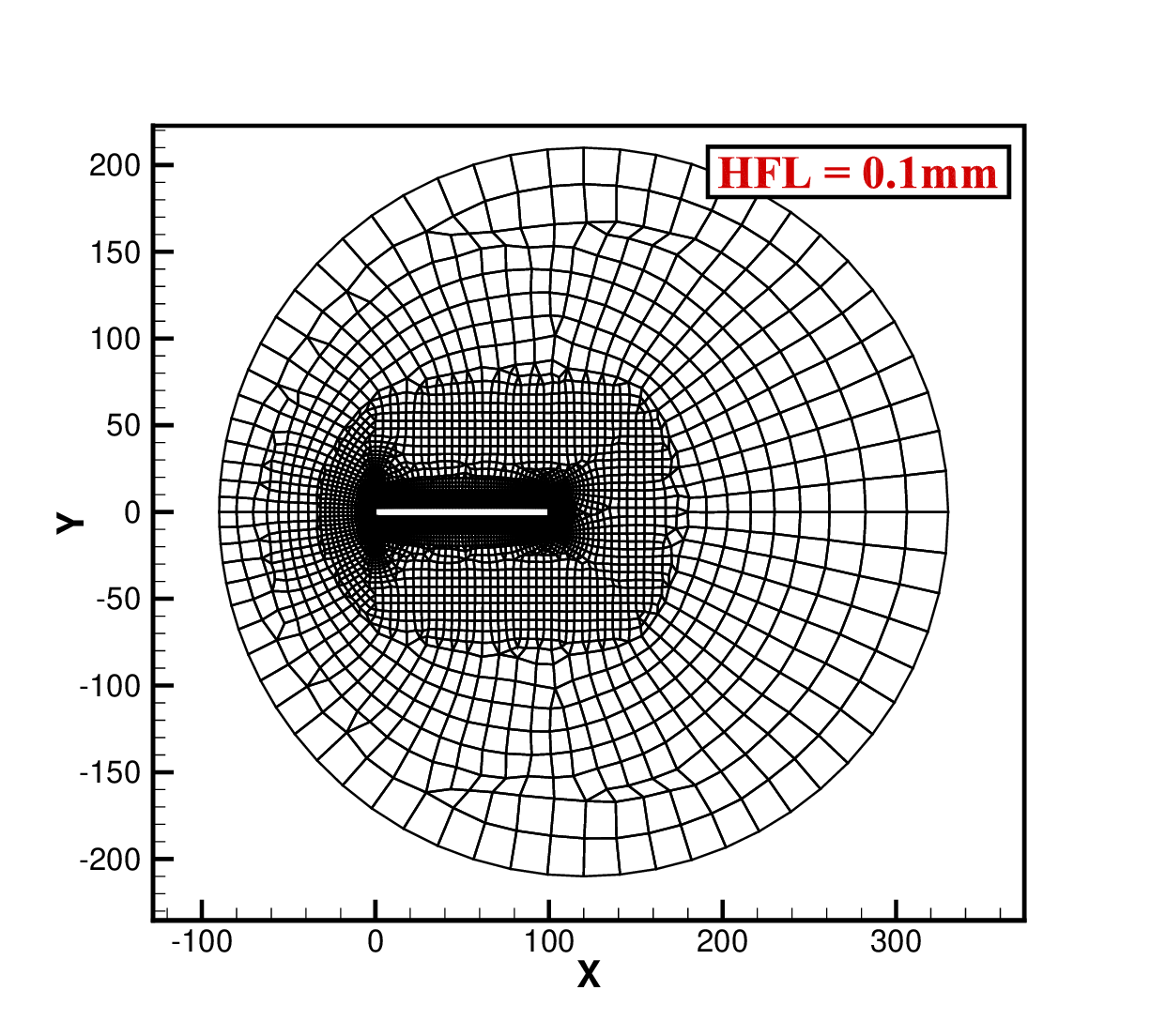}}
		\subfigure[]{\label{fig27b}\includegraphics[width=0.42\textwidth]{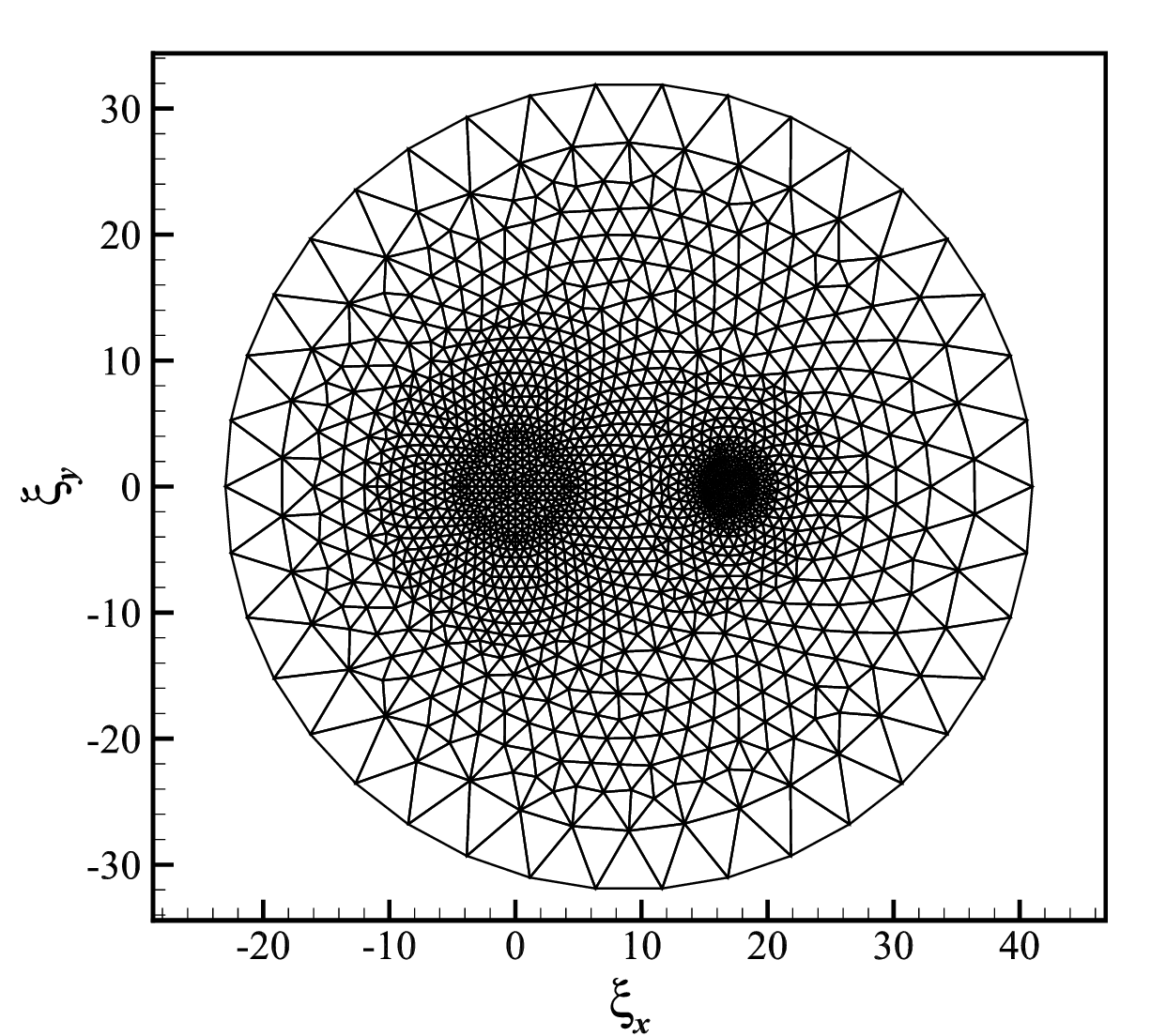}}
		\caption{\label{fig27}{The (a) unstructured physical mesh and (b) unstructured velocity mesh for the hypersonic flow over a flat plate ($Ma$ = 20.2, $Kn$ = 0.0169, $T_{\infty}$ = 13.32 K, $T_{w}$ = 290 K).}}
	\end{figure}
	
	\begin{figure}
		\centering
		\subfigure[]{\label{fig28a}\includegraphics[width=0.45\textwidth]{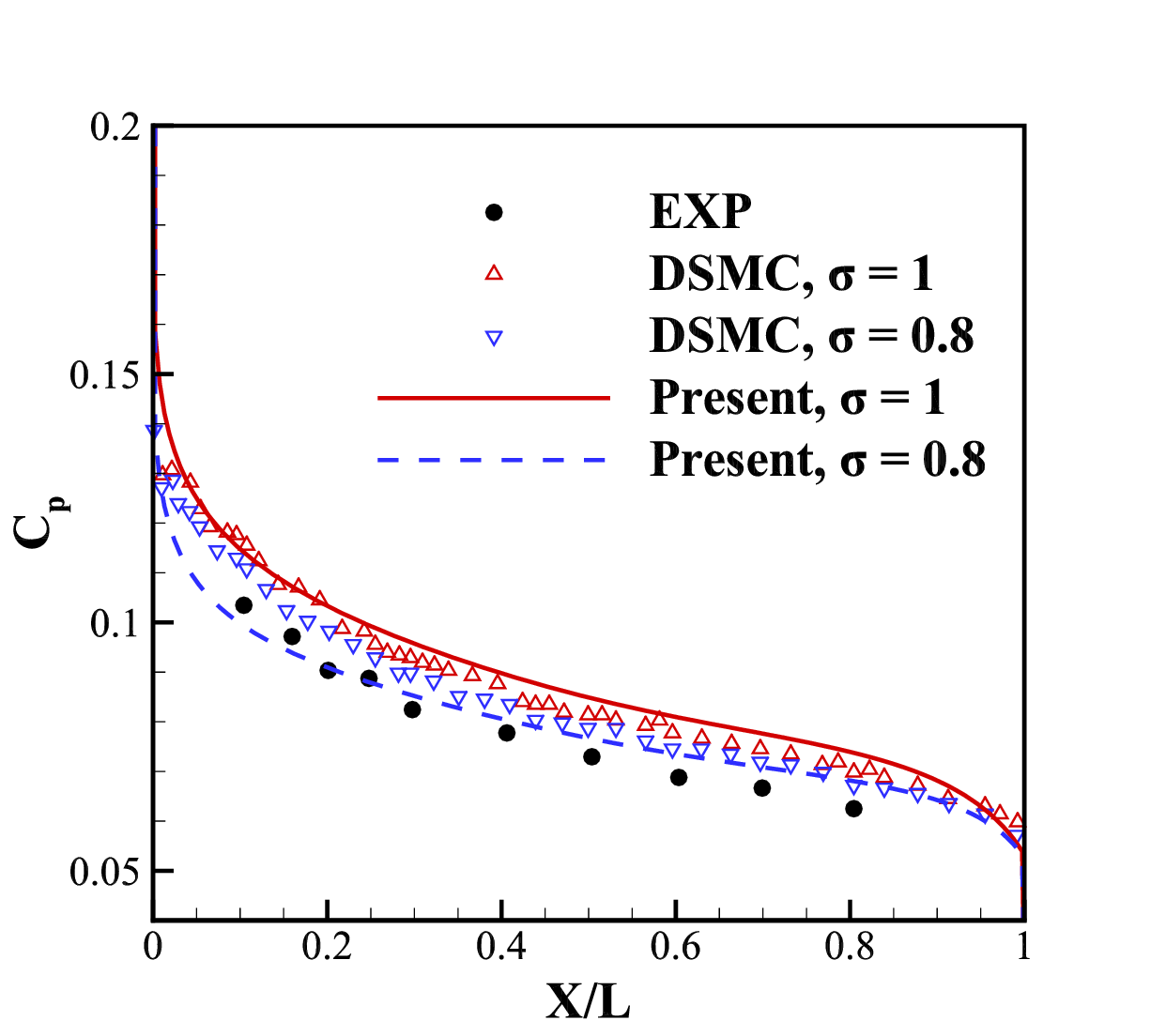}}
		\subfigure[]{\label{fig28b}\includegraphics[width=0.45\textwidth]{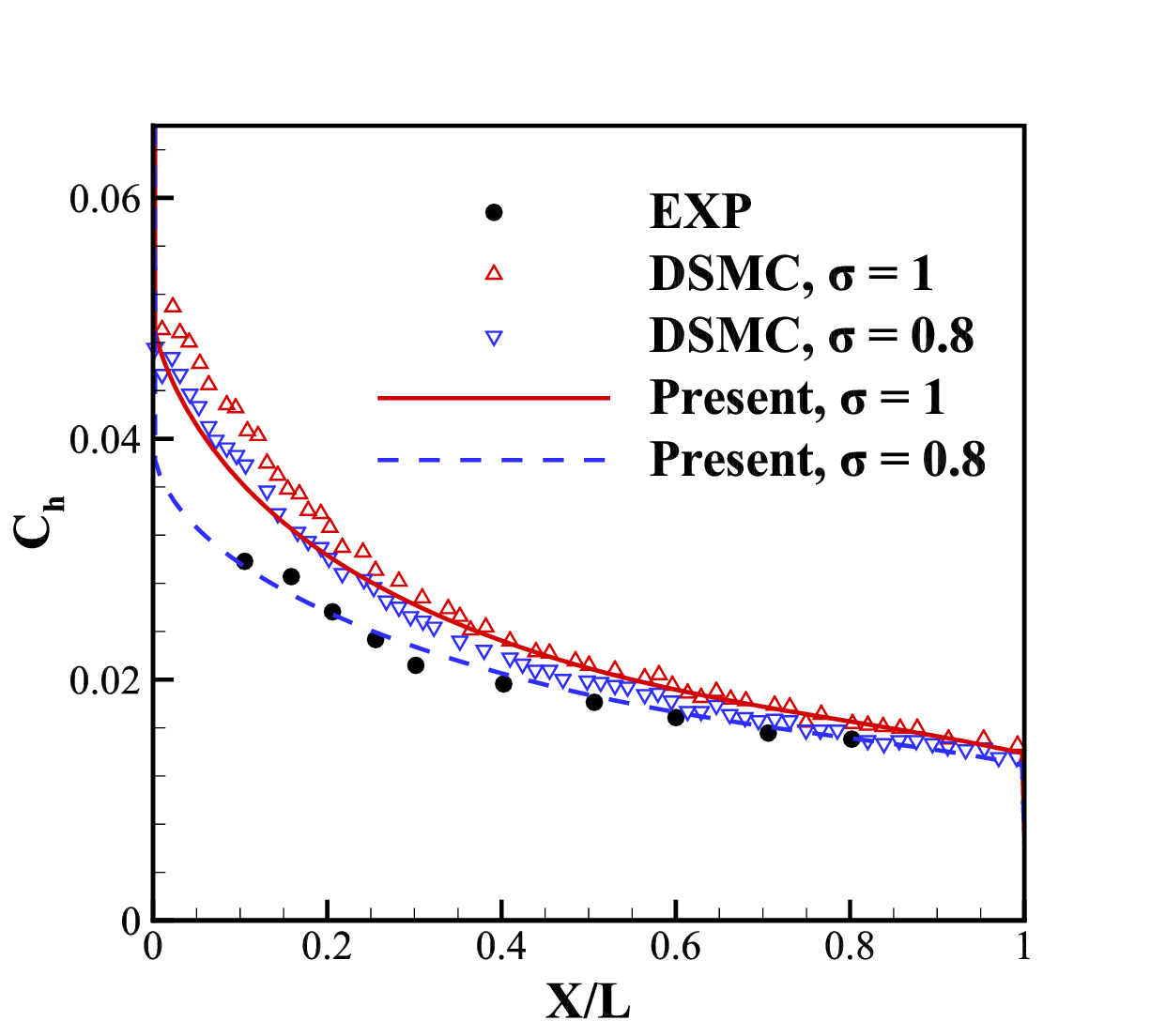}}
		\caption{\label{fig28}{Comparison of the (a) pressure coefficient and (b) heat transfer coefficient on the lower surface of truncated plate at AOA 0 degree ($Ma$ = 20.2, $Kn$ = 0.0169, $T_{\infty}$ = 13.32 K, $T_{w}$ = 290 K).}}
	\end{figure}
	
	\begin{figure}
		\centering
		\subfigure[]{\label{fig29a}\includegraphics[width=0.45\textwidth]{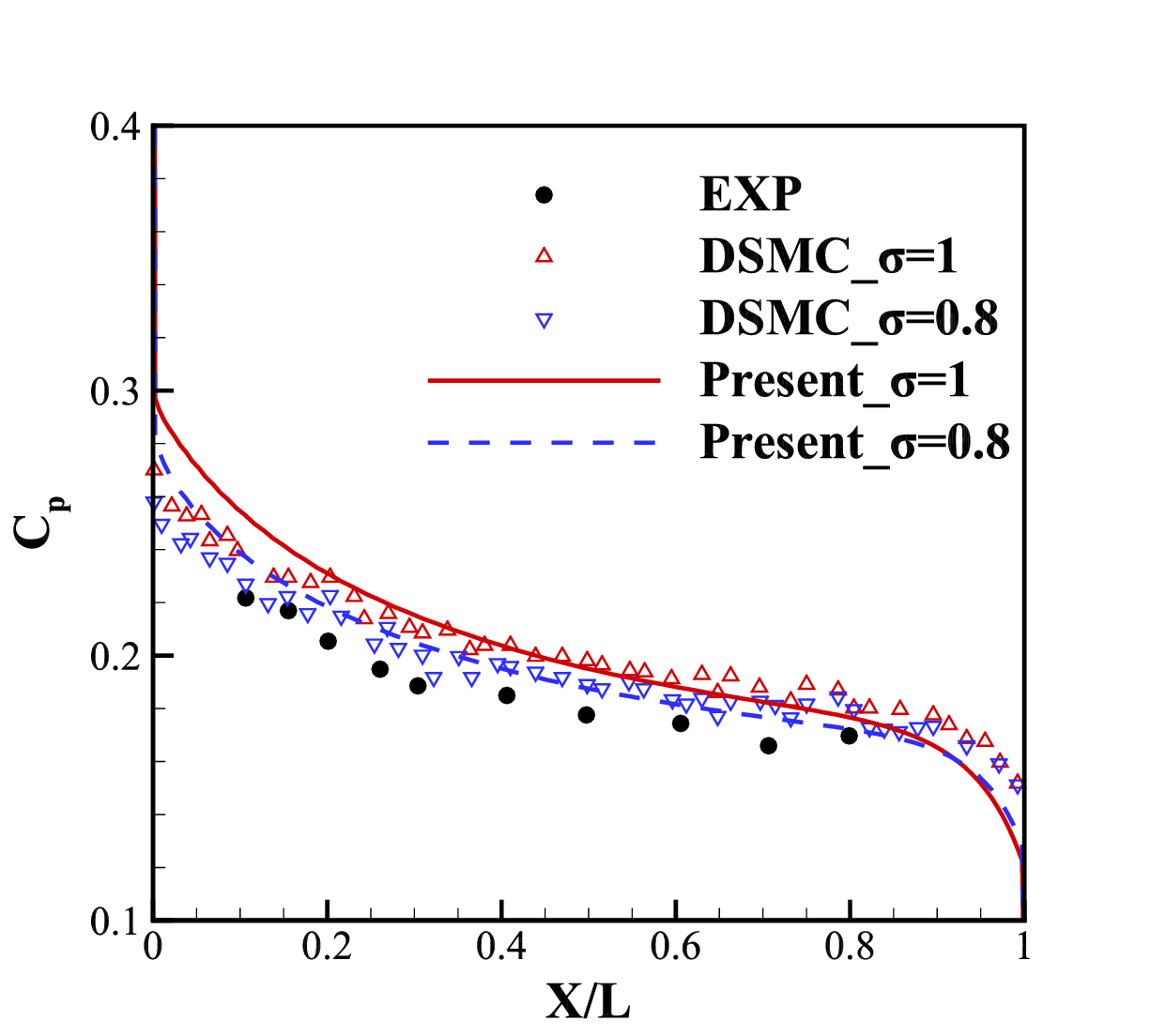}}
		\subfigure[]{\label{fig29b}\includegraphics[width=0.45\textwidth]{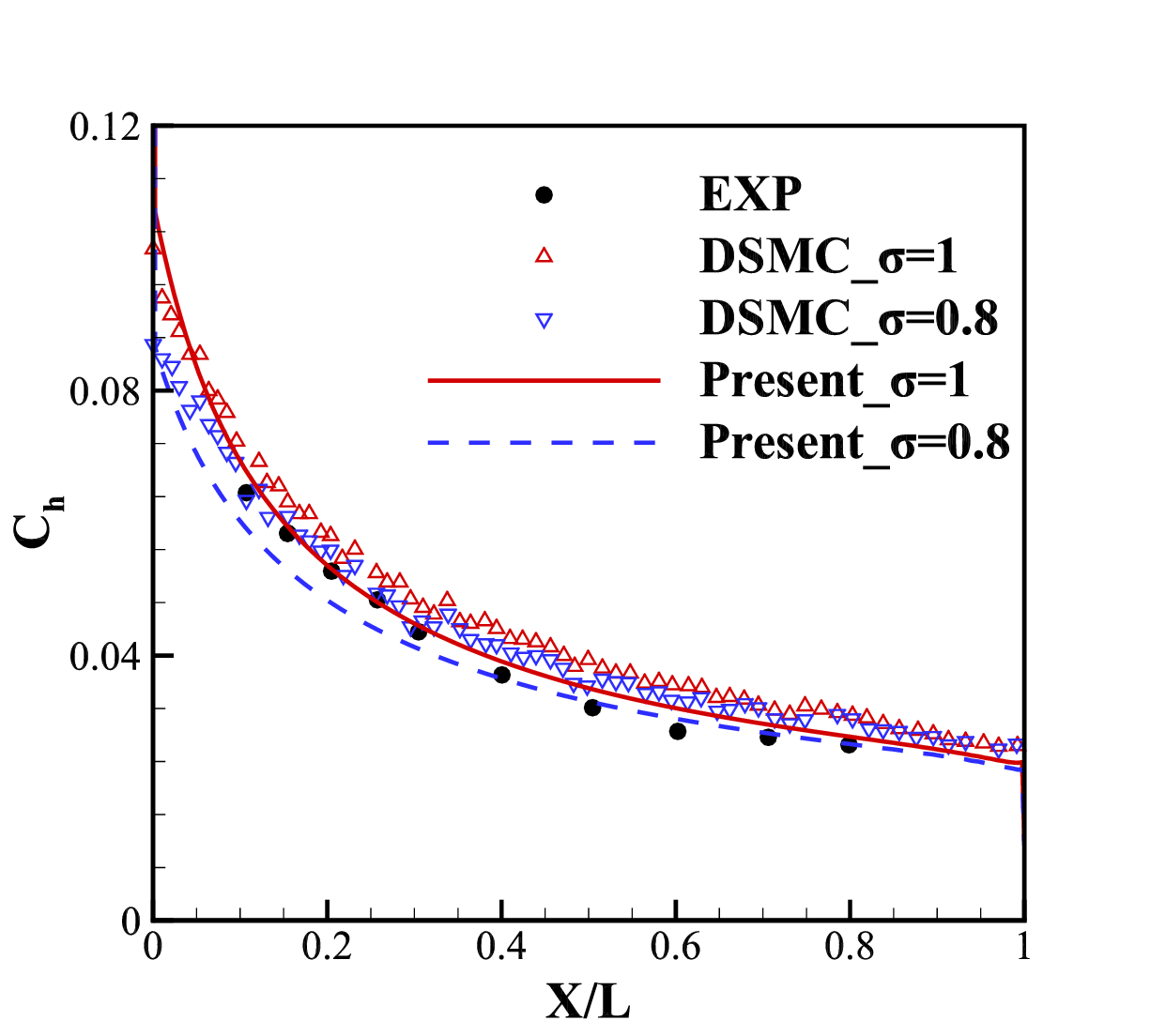}}
		\caption{\label{fig29}{Comparison of the (a) pressure coefficient and (b) heat transfer coefficient on the lower surface of truncated plate at AOA 10 degrees ($Ma$ = 20.2, $Kn$ = 0.0169, $T_{\infty}$ = 13.32 K, $T_{w}$ = 290 K).}}
	\end{figure}

	\clearpage
	\bibliography{Maxwell_BC}

\end{document}